\documentclass{CSML}

\def\dOi{13(1:9)2017}
\lmcsheading%
{\dOi}
{1--37}
{}
{}
{Jan.~20, 2016}
{Mar.~17, 2017}
{}

\usepackage[latin1]{inputenc}

\usepackage[english]{babel}

% ...

\usepackage{amsthm}

% ...

\usepackage{hyperref}

\usepackage{changebar}
\usepackage[pagewise, modulo]{lineno}%\linenumbers

\usepackage{etoolbox}

\usepackage{fixltx2e}

\usepackage{xargs}

\usepackage{xspace}

\usepackage{xstring}

\usepackage{boolexpr}

\usepackage[cmex10]{mathtools}
\usepackage{amssymb}
\usepackage{amsfonts}
\usepackage{mathrsfs}
\usepackage{latexsym}
\usepackage{textcomp}
\usepackage{pifont}

\newcommand{\argemp}[2]
	{\if&#1&\else#2\fi}

\newcommand{\argdef}[2]
	{\if&#1&#2\else#1\fi}

\newcommand{\argint}[3]
	{\if&#2&\else#1#2#3\fi}

\newcommand{\argext}[3]
	{\if&#1&#3\else#1\if&#3&\else#2#3\fi\fi}

\newcommandx{\argsubsup}[3][2=, 3=]
	{\def\argsubscript{{#2}}\def\argsuperscript{{#3}}#1}

\newcommandx{\argind}[9][2=, 3=, 4=, 5=, 6=, 7=, 8=, 9=]
	{%
	\switch[#1=]%
		\case{0}#2%
		\case{1}#3%
		\case{2}#4%
		\case{3}#5%
		\case{4}#6%
		\case{5}#7%
		\case{6}#8%
		\case{7}#9%
		\otherwise\ensuremath{\clubsuit}%
	\endswitch%
	}

\newcommand{\arga}[1]
	{#1}
\newcommand{\argb}[2]
	{\argext{\arga{#1}}{, \allowbreak}{#2}}
\newcommand{\argc}[3]
	{\argext{\argb{#1}{#2}}{, \allowbreak}{#3}}
\newcommand{\argd}[4]
	{\argext{\argc{#1}{#2}{#3}}{, \allowbreak}{#4}}
\newcommand{\arge}[5]
	{\argext{\argd{#1}{#2}{#3}{#4}}{, \allowbreak}{#5}}
\newcommand{\argf}[6]
	{\argext{\arge{#1}{#2}{#3}{#4}{#5}}{, \allowbreak}{#6}}
\newcommand{\argg}[7]
	{\argext{\argf{#1}{#2}{#3}{#4}{#5}{#6}}{, \allowbreak}{#7}}
\newcommand{\argh}[8]
	{\argext{\argg{#1}{#2}{#3}{#4}{#5}{#6}{#7}}{, \allowbreak}{#8}}
\newcommand{\argi}[9]
	{\argext{\argh{#1}{#2}{#3}{#4}{#5}{#6}{#7}{#8}}{, \allowbreak}{#9}}

\newcommand{\argj}[9]
	{%
	\def\valarga{#1}%
	\def\valargb{#2}%
	\def\valargc{#3}%
	\def\valargd{#4}%
	\def\valarge{#5}%
	\def\valargf{#6}%
	\def\valargg{#7}%
	\def\valargh{#8}%
	\def\valargi{#9}%
	\argauxj%
	}
\newcommand{\argk}[9]
	{%
	\def\valarga{#1}%
	\def\valargb{#2}%
	\def\valargc{#3}%
	\def\valargd{#4}%
	\def\valarge{#5}%
	\def\valargf{#6}%
	\def\valargg{#7}%
	\def\valargh{#8}%
	\def\valargi{#9}%
	\argauxk%
	}
\newcommand{\argl}[9]
	{%
	\def\valarga{#1}%
	\def\valargb{#2}%
	\def\valargc{#3}%
	\def\valargd{#4}%
	\def\valarge{#5}%
	\def\valargf{#6}%
	\def\valargg{#7}%
	\def\valargh{#8}%
	\def\valargi{#9}%
	\argauxl%
	}
\newcommand{\argm}[9]
	{%
	\def\valarga{#1}%
	\def\valargb{#2}%
	\def\valargc{#3}%
	\def\valargd{#4}%
	\def\valarge{#5}%
	\def\valargf{#6}%
	\def\valargg{#7}%
	\def\valargh{#8}%
	\def\valargi{#9}%
	\argauxm%
	}
\newcommand{\argn}[9]
	{%
	\def\valarga{#1}%
	\def\valargb{#2}%
	\def\valargc{#3}%
	\def\valargd{#4}%
	\def\valarge{#5}%
	\def\valargf{#6}%
	\def\valargg{#7}%
	\def\valargh{#8}%
	\def\valargi{#9}%
	\argauxn%
	}
\newcommand{\argo}[9]
	{%
	\def\valarga{#1}%
	\def\valargb{#2}%
	\def\valargc{#3}%
	\def\valargd{#4}%
	\def\valarge{#5}%
	\def\valargf{#6}%
	\def\valargg{#7}%
	\def\valargh{#8}%
	\def\valargi{#9}%
	\argauxo%
	}
\newcommand{\argp}[9]
	{%
	\def\valarga{#1}%
	\def\valargb{#2}%
	\def\valargc{#3}%
	\def\valargd{#4}%
	\def\valarge{#5}%
	\def\valargf{#6}%
	\def\valargg{#7}%
	\def\valargh{#8}%
	\def\valargi{#9}%
	\argauxp%
	}
\newcommand{\argq}[9]
	{%
	\def\valarga{#1}%
	\def\valargb{#2}%
	\def\valargc{#3}%
	\def\valargd{#4}%
	\def\valarge{#5}%
	\def\valargf{#6}%
	\def\valargg{#7}%
	\def\valargh{#8}%
	\def\valargi{#9}%
	\argauxq%
	}
\newcommand{\argr}[9]
	{%
	\def\valarga{#1}%
	\def\valargb{#2}%
	\def\valargc{#3}%
	\def\valargd{#4}%
	\def\valarge{#5}%
	\def\valargf{#6}%
	\def\valargg{#7}%
	\def\valargh{#8}%
	\def\valargi{#9}%
	\argauxr%
	}

\newcommand{\argauxj}[1]
	{%
	\argext%
		{\argi{\valarga}{\valargb}{\valargc}{\valargd}{\valarge}{\valargf}{\valargg}
			{\valargh}{\valargi}}
		{, \allowbreak}{#1}%
	}
\newcommand{\argauxk}[2]
	{\argext{\argauxj{#1}}{, \allowbreak}{#2}}
\newcommand{\argauxl}[3]
	{\argext{\argauxk{#1}{#2}}{, \allowbreak}{#3}}
\newcommand{\argauxm}[4]
	{\argext{\argauxl{#1}{#2}{#3}}{, \allowbreak}{#4}}
\newcommand{\argauxn}[5]
	{\argext{\argauxm{#1}{#2}{#3}{#4}}{, \allowbreak}{#5}}
\newcommand{\argauxo}[6]
	{\argext{\argauxn{#1}{#2}{#3}{#4}{#5}}{, \allowbreak}{#6}}
\newcommand{\argauxp}[7]
	{\argext{\argauxo{#1}{#2}{#3}{#4}{#5}{#6}}{, \allowbreak}{#7}}
\newcommand{\argauxq}[8]
	{\argext{\argauxp{#1}{#2}{#3}{#4}{#5}{#6}{#7}}{, \allowbreak}{#8}}
\newcommand{\argauxr}[9]
	{\argext{\argauxq{#1}{#2}{#3}{#4}{#5}{#6}{#7}{#8}}{, \allowbreak}{#9}}

\newcommand{\txtfnt}[2][]
	{{%
	\IfStrEq{#1}{}
		{#2}
		{%
		\StrLeft{#1}{2}[\optbgn]%
		\StrGobbleLeft{#1}{2}[\optend]%
		\IfStrEqCase{\optbgn}
			{%
			{Rm}{\rmfamily\txtfnt[\optend]{#2}}%
			{Sf}{\sffamily\txtfnt[\optend]{#2}}%
			{Tt}{\ttfamily\txtfnt[\optend]{#2}}%
			{Up}{\upshape\txtfnt[\optend]{#2}}%
			{It}{\itshape\txtfnt[\optend]{#2}}%
			{Sl}{\slshape\txtfnt[\optend]{#2}}%
			{Sc}{\scshape\txtfnt[\optend]{#2}}%
			{Md}{\mdseries\txtfnt[\optend]{#2}}%
			{Bf}{\bfseries\txtfnt[\optend]{#2}}%
			{Em}{\emph{\txtfnt[\optend]{#2}}}%
			}
			[\ensuremath{\clubsuit}]%
		}%
	}}

\newcommand{\txtsub}[2][]
	{\argemp{#2}{\ensuremath{_{\text{\txtfnt[#1]{#2}}}}}}

\newcommand{\txtsup}[2][]
	{\argemp{#2}{\ensuremath{^{\text{\txtfnt[#1]{#2}}}}}}

\newcommandx{\txt}[4][1=, 3=, 4=]
	{%
	\ensuremath{\text{%
		\txtfnt[#1]{#2}\ensuremath{\txtsub[#1]{#3}\txtsup[#1]{#4}}%
	}}%
	}

\newcommandx{\txtarg}[5][1=, 3=, 4=]
	{{\txt[#1]{#2}[#3][#4]\argint{(}{#5}{)}}}

\newcommand{\txtstyname}{RmScMd}
\newcommand{\txtname}[1][]
	{\txt[\argdef{#1}{\txtstyname}]}
\newcommand{\txtargname}[1][]
	{\txtarg[\argdef{#1}{\txtstyname}]}

\newcommand{\txtstyabr}{Em}
\newcommand{\txtabr}[1][]
	{\txt[\argdef{#1}{\txtstyabr}]}

\newcommandx{\mthfnt}[3][1=, 2=0]
	{{%
	\IfStrEqCase{#1}
		{%
		{}%
			{#3}%
		{Name}%
			{%
			\IfStrEqCase{#2}
				{%
				{0}{\mathcal{#3}}%
				{1}{\mathscr{#3}}%
				{2}{\mathfrak{#3}}%
				{3}{\mathbb{#3}}%
				}
				[\ensuremath{\clubsuit}]%
			}%
		{Set}%
			{%
			\IfStrEqCase{#2}
				{%
				{0}{\mathrm{#3}}%
				{1}{\mathsf{#3}}%
				{2}{\mathbb{#3}}%
				{3}{\mathbf{#3}}%
				}
				[\ensuremath{\clubsuit}]%
			}%
		{Fun}%
			{%
			\IfStrEqCase{#2}
				{%
				{0}{\mathsf{#3}}%
				{1}{\mathrm{#3}}%
				}
				[\ensuremath{\clubsuit}]%
			}%
		{Rel}%
			{%
			\IfStrEqCase{#2}
				{%
				{0}{\mathit{#3}}%
				{1}{\mathtt{#3}}%
				}
				[\ensuremath{\clubsuit}]%
			}%
		{Sym}%
			{%
			\IfStrEqCase{#2}
				{%
				{0}{\mathtt{#3}}%
				{1}{\mathbf{#3}}%
				}
				[\ensuremath{\clubsuit}]%
			}%
		{Elm}%
			{\mathnormal{#3}}
		}
		[\ensuremath{\clubsuit}]%
	}}

\newcommand{\mthsub}[1]
	{\argemp{#1}{\ensuremath{_{\mathnormal{#1}}}}}

\newcommand{\mthsup}[1]
	{\argemp{#1}{\ensuremath{^{\mathnormal{#1}}}}}

\newcommandx{\mth}[5][1=, 2=0, 4=, 5=]
	{{\ensuremath{\mthfnt[#1][#2]{#3}\mthsub{#4}\mthsup{#5}}}}

\newcommandx{\mtharg}[6][1=, 2=0, 4=, 5=]
	{{\mth[#1][#2]{#3}[#4][#5]\ensuremath{\argint{(}{#6}{)}}}}

\newcommand{\mthempty}
	{\mth[][]}

\newcommand{\mthstyname}{0}
\newcommand{\mthname}[1][]
	{\mth[Name][\argdef{#1}{\mthstyname}]}

\newcommand{\mthstyset}{0}
\newcommand{\mthset}[1][]
	{\mth[Set][\argdef{#1}{\mthstyset}]}
\newcommand{\mthargset}[1][]
	{\mtharg[Set][\argdef{#1}{\mthstyset}]}

\newcommand{\mthstyfun}{0}
\newcommand{\mthfun}[1][]
	{\mth[Fun][\argdef{#1}{\mthstyfun}]}
\newcommand{\mthargfun}[1][]
	{\mtharg[Fun][\argdef{#1}{\mthstyfun}]}

\newcommand{\mthstyrel}{0}
\newcommand{\mthrel}[1][]
	{\mth[Rel][\argdef{#1}{\mthstyrel}]}

\newcommand{\mthstysym}{0}
\newcommand{\mthsym}[1][]
	{\mth[Sym][\argdef{#1}{\mthstysym}]}

\newcommand{\mthstyelm}{0}
\newcommand{\mthelm}[1][]
	{\mth[Elm][\argdef{#1}{\mthstyelm}]}

\newcommandx{\AName}[4][1=, 2=, 3=, 4=]{\mthname[#4]{A#3}[#1][#2]}
\newcommandx{\BName}[4][1=, 2=, 3=, 4=]{\mthname[#4]{B#3}[#1][#2]}
\newcommandx{\CName}[4][1=, 2=, 3=, 4=]{\mthname[#4]{C#3}[#1][#2]}
\newcommandx{\DName}[4][1=, 2=, 3=, 4=]{\mthname[#4]{D#3}[#1][#2]}
\newcommandx{\EName}[4][1=, 2=, 3=, 4=]{\mthname[#4]{E#3}[#1][#2]}
\newcommandx{\FName}[4][1=, 2=, 3=, 4=]{\mthname[#4]{F#3}[#1][#2]}
\newcommandx{\GName}[4][1=, 2=, 3=, 4=]{\mthname[#4]{G#3}[#1][#2]}
\newcommandx{\HName}[4][1=, 2=, 3=, 4=]{\mthname[#4]{H#3}[#1][#2]}
\newcommandx{\IName}[4][1=, 2=, 3=, 4=]{\mthname[#4]{I#3}[#1][#2]}
\newcommandx{\JName}[4][1=, 2=, 3=, 4=]{\mthname[#4]{J#3}[#1][#2]}
\newcommandx{\KName}[4][1=, 2=, 3=, 4=]{\mthname[#4]{K#3}[#1][#2]}
\newcommandx{\LName}[4][1=, 2=, 3=, 4=]{\mthname[#4]{L#3}[#1][#2]}
\newcommandx{\MName}[4][1=, 2=, 3=, 4=]{\mthname[#4]{M#3}[#1][#2]}
\newcommandx{\NName}[4][1=, 2=, 3=, 4=]{\mthname[#4]{N#3}[#1][#2]}
\newcommandx{\OName}[4][1=, 2=, 3=, 4=]{\mthname[#4]{O#3}[#1][#2]}
\newcommandx{\PName}[4][1=, 2=, 3=, 4=]{\mthname[#4]{P#3}[#1][#2]}
\newcommandx{\QName}[4][1=, 2=, 3=, 4=]{\mthname[#4]{Q#3}[#1][#2]}
\newcommandx{\RName}[4][1=, 2=, 3=, 4=]{\mthname[#4]{R#3}[#1][#2]}
\newcommandx{\SName}[4][1=, 2=, 3=, 4=]{\mthname[#4]{S#3}[#1][#2]}
\newcommandx{\TName}[4][1=, 2=, 3=, 4=]{\mthname[#4]{T#3}[#1][#2]}
\newcommandx{\UName}[4][1=, 2=, 3=, 4=]{\mthname[#4]{U#3}[#1][#2]}
\newcommandx{\VName}[4][1=, 2=, 3=, 4=]{\mthname[#4]{V#3}[#1][#2]}
\newcommandx{\WName}[4][1=, 2=, 3=, 4=]{\mthname[#4]{W#3}[#1][#2]}
\newcommandx{\XName}[4][1=, 2=, 3=, 4=]{\mthname[#4]{X#3}[#1][#2]}
\newcommandx{\YName}[4][1=, 2=, 3=, 4=]{\mthname[#4]{Y#3}[#1][#2]}
\newcommandx{\ZName}[4][1=, 2=, 3=, 4=]{\mthname[#4]{Z#3}[#1][#2]}

\newcommandx{\ASet}[4][1=, 2=, 3=, 4=]{\mthset[#4]{A#3}[#1][#2]}
\newcommandx{\BSet}[4][1=, 2=, 3=, 4=]{\mthset[#4]{B#3}[#1][#2]}
\newcommandx{\CSet}[4][1=, 2=, 3=, 4=]{\mthset[#4]{C#3}[#1][#2]}
\newcommandx{\DSet}[4][1=, 2=, 3=, 4=]{\mthset[#4]{D#3}[#1][#2]}
\newcommandx{\ESet}[4][1=, 2=, 3=, 4=]{\mthset[#4]{E#3}[#1][#2]}
\newcommandx{\FSet}[4][1=, 2=, 3=, 4=]{\mthset[#4]{F#3}[#1][#2]}
\newcommandx{\GSet}[4][1=, 2=, 3=, 4=]{\mthset[#4]{G#3}[#1][#2]}
\newcommandx{\HSet}[4][1=, 2=, 3=, 4=]{\mthset[#4]{H#3}[#1][#2]}
\newcommandx{\ISet}[4][1=, 2=, 3=, 4=]{\mthset[#4]{I#3}[#1][#2]}
\newcommandx{\JSet}[4][1=, 2=, 3=, 4=]{\mthset[#4]{J#3}[#1][#2]}
\newcommandx{\KSet}[4][1=, 2=, 3=, 4=]{\mthset[#4]{K#3}[#1][#2]}
\newcommandx{\LSet}[4][1=, 2=, 3=, 4=]{\mthset[#4]{L#3}[#1][#2]}
\newcommandx{\MSet}[4][1=, 2=, 3=, 4=]{\mthset[#4]{M#3}[#1][#2]}
\newcommandx{\NSet}[4][1=, 2=, 3=, 4=]{\mthset[#4]{N#3}[#1][#2]}
\newcommandx{\OSet}[4][1=, 2=, 3=, 4=]{\mthset[#4]{O#3}[#1][#2]}
\newcommandx{\PSet}[4][1=, 2=, 3=, 4=]{\mthset[#4]{P#3}[#1][#2]}
\newcommandx{\QSet}[4][1=, 2=, 3=, 4=]{\mthset[#4]{Q#3}[#1][#2]}
\newcommandx{\RSet}[4][1=, 2=, 3=, 4=]{\mthset[#4]{R#3}[#1][#2]}
\newcommandx{\SSet}[4][1=, 2=, 3=, 4=]{\mthset[#4]{S#3}[#1][#2]}
\newcommandx{\TSet}[4][1=, 2=, 3=, 4=]{\mthset[#4]{T#3}[#1][#2]}
\newcommandx{\USet}[4][1=, 2=, 3=, 4=]{\mthset[#4]{U#3}[#1][#2]}
\newcommandx{\VSet}[4][1=, 2=, 3=, 4=]{\mthset[#4]{V#3}[#1][#2]}
\newcommandx{\WSet}[4][1=, 2=, 3=, 4=]{\mthset[#4]{W#3}[#1][#2]}
\newcommandx{\XSet}[4][1=, 2=, 3=, 4=]{\mthset[#4]{X#3}[#1][#2]}
\newcommandx{\YSet}[4][1=, 2=, 3=, 4=]{\mthset[#4]{Y#3}[#1][#2]}
\newcommandx{\ZSet}[4][1=, 2=, 3=, 4=]{\mthset[#4]{Z#3}[#1][#2]}

\newcommandx{\aSet}[4][1=, 2=, 3=, 4=]{\mthset[#4]{a#3}[#1][#2]}
\newcommandx{\bSet}[4][1=, 2=, 3=, 4=]{\mthset[#4]{b#3}[#1][#2]}
\newcommandx{\cSet}[4][1=, 2=, 3=, 4=]{\mthset[#4]{c#3}[#1][#2]}
\newcommandx{\dSet}[4][1=, 2=, 3=, 4=]{\mthset[#4]{d#3}[#1][#2]}
\newcommandx{\eSet}[4][1=, 2=, 3=, 4=]{\mthset[#4]{e#3}[#1][#2]}
\newcommandx{\fSet}[4][1=, 2=, 3=, 4=]{\mthset[#4]{f#3}[#1][#2]}
\newcommandx{\gSet}[4][1=, 2=, 3=, 4=]{\mthset[#4]{g#3}[#1][#2]}
\newcommandx{\hSet}[4][1=, 2=, 3=, 4=]{\mthset[#4]{h#3}[#1][#2]}
\newcommandx{\iSet}[4][1=, 2=, 3=, 4=]{\mthset[#4]{i#3}[#1][#2]}
\newcommandx{\jSet}[4][1=, 2=, 3=, 4=]{\mthset[#4]{j#3}[#1][#2]}
\newcommandx{\kSet}[4][1=, 2=, 3=, 4=]{\mthset[#4]{k#3}[#1][#2]}
\newcommandx{\lSet}[4][1=, 2=, 3=, 4=]{\mthset[#4]{l#3}[#1][#2]}
\newcommandx{\mSet}[4][1=, 2=, 3=, 4=]{\mthset[#4]{m#3}[#1][#2]}
\newcommandx{\nSet}[4][1=, 2=, 3=, 4=]{\mthset[#4]{n#3}[#1][#2]}
\newcommandx{\oSet}[4][1=, 2=, 3=, 4=]{\mthset[#4]{o#3}[#1][#2]}
\newcommandx{\pSet}[4][1=, 2=, 3=, 4=]{\mthset[#4]{p#3}[#1][#2]}
\newcommandx{\qSet}[4][1=, 2=, 3=, 4=]{\mthset[#4]{q#3}[#1][#2]}
\newcommandx{\rSet}[4][1=, 2=, 3=, 4=]{\mthset[#4]{r#3}[#1][#2]}
\newcommandx{\sSet}[4][1=, 2=, 3=, 4=]{\mthset[#4]{s#3}[#1][#2]}
\newcommandx{\tSet}[4][1=, 2=, 3=, 4=]{\mthset[#4]{t#3}[#1][#2]}
\newcommandx{\uSet}[4][1=, 2=, 3=, 4=]{\mthset[#4]{u#3}[#1][#2]}
\newcommandx{\vSet}[4][1=, 2=, 3=, 4=]{\mthset[#4]{v#3}[#1][#2]}
\newcommandx{\wSet}[4][1=, 2=, 3=, 4=]{\mthset[#4]{w#3}[#1][#2]}
\newcommandx{\xSet}[4][1=, 2=, 3=, 4=]{\mthset[#4]{x#3}[#1][#2]}
\newcommandx{\ySet}[4][1=, 2=, 3=, 4=]{\mthset[#4]{y#3}[#1][#2]}
\newcommandx{\zSet}[4][1=, 2=, 3=, 4=]{\mthset[#4]{z#3}[#1][#2]}

\newcommandx{\AFun}[4][1=, 2=, 3=, 4=]{\mthfun[#4]{A#3}[#1][#2]}
\newcommandx{\BFun}[4][1=, 2=, 3=, 4=]{\mthfun[#4]{B#3}[#1][#2]}
\newcommandx{\CFun}[4][1=, 2=, 3=, 4=]{\mthfun[#4]{C#3}[#1][#2]}
\newcommandx{\DFun}[4][1=, 2=, 3=, 4=]{\mthfun[#4]{D#3}[#1][#2]}
\newcommandx{\EFun}[4][1=, 2=, 3=, 4=]{\mthfun[#4]{E#3}[#1][#2]}
\newcommandx{\FFun}[4][1=, 2=, 3=, 4=]{\mthfun[#4]{F#3}[#1][#2]}
\newcommandx{\GFun}[4][1=, 2=, 3=, 4=]{\mthfun[#4]{G#3}[#1][#2]}
\newcommandx{\HFun}[4][1=, 2=, 3=, 4=]{\mthfun[#4]{H#3}[#1][#2]}
\newcommandx{\IFun}[4][1=, 2=, 3=, 4=]{\mthfun[#4]{I#3}[#1][#2]}
\newcommandx{\JFun}[4][1=, 2=, 3=, 4=]{\mthfun[#4]{J#3}[#1][#2]}
\newcommandx{\KFun}[4][1=, 2=, 3=, 4=]{\mthfun[#4]{K#3}[#1][#2]}
\newcommandx{\LFun}[4][1=, 2=, 3=, 4=]{\mthfun[#4]{L#3}[#1][#2]}
\newcommandx{\MFun}[4][1=, 2=, 3=, 4=]{\mthfun[#4]{M#3}[#1][#2]}
\newcommandx{\NFun}[4][1=, 2=, 3=, 4=]{\mthfun[#4]{N#3}[#1][#2]}
\newcommandx{\OFun}[4][1=, 2=, 3=, 4=]{\mthfun[#4]{O#3}[#1][#2]}
\newcommandx{\PFun}[4][1=, 2=, 3=, 4=]{\mthfun[#4]{P#3}[#1][#2]}
\newcommandx{\QFun}[4][1=, 2=, 3=, 4=]{\mthfun[#4]{Q#3}[#1][#2]}
\newcommandx{\RFun}[4][1=, 2=, 3=, 4=]{\mthfun[#4]{R#3}[#1][#2]}
\newcommandx{\SFun}[4][1=, 2=, 3=, 4=]{\mthfun[#4]{S#3}[#1][#2]}
\newcommandx{\TFun}[4][1=, 2=, 3=, 4=]{\mthfun[#4]{T#3}[#1][#2]}
\newcommandx{\UFun}[4][1=, 2=, 3=, 4=]{\mthfun[#4]{U#3}[#1][#2]}
\newcommandx{\VFun}[4][1=, 2=, 3=, 4=]{\mthfun[#4]{V#3}[#1][#2]}
\newcommandx{\WFun}[4][1=, 2=, 3=, 4=]{\mthfun[#4]{W#3}[#1][#2]}
\newcommandx{\XFun}[4][1=, 2=, 3=, 4=]{\mthfun[#4]{X#3}[#1][#2]}
\newcommandx{\YFun}[4][1=, 2=, 3=, 4=]{\mthfun[#4]{Y#3}[#1][#2]}
\newcommandx{\ZFun}[4][1=, 2=, 3=, 4=]{\mthfun[#4]{Z#3}[#1][#2]}

\newcommandx{\aFun}[4][1=, 2=, 3=, 4=]{\mthfun[#4]{a#3}[#1][#2]}
\newcommandx{\bFun}[4][1=, 2=, 3=, 4=]{\mthfun[#4]{b#3}[#1][#2]}
\newcommandx{\cFun}[4][1=, 2=, 3=, 4=]{\mthfun[#4]{c#3}[#1][#2]}
\newcommandx{\dFun}[4][1=, 2=, 3=, 4=]{\mthfun[#4]{d#3}[#1][#2]}
\newcommandx{\eFun}[4][1=, 2=, 3=, 4=]{\mthfun[#4]{e#3}[#1][#2]}
\newcommandx{\fFun}[4][1=, 2=, 3=, 4=]{\mthfun[#4]{f#3}[#1][#2]}
\newcommandx{\gFun}[4][1=, 2=, 3=, 4=]{\mthfun[#4]{g#3}[#1][#2]}
\newcommandx{\hFun}[4][1=, 2=, 3=, 4=]{\mthfun[#4]{h#3}[#1][#2]}
\newcommandx{\iFun}[4][1=, 2=, 3=, 4=]{\mthfun[#4]{i#3}[#1][#2]}
\newcommandx{\jFun}[4][1=, 2=, 3=, 4=]{\mthfun[#4]{j#3}[#1][#2]}
\newcommandx{\kFun}[4][1=, 2=, 3=, 4=]{\mthfun[#4]{k#3}[#1][#2]}
\newcommandx{\lFun}[4][1=, 2=, 3=, 4=]{\mthfun[#4]{l#3}[#1][#2]}
\newcommandx{\mFun}[4][1=, 2=, 3=, 4=]{\mthfun[#4]{m#3}[#1][#2]}
\newcommandx{\nFun}[4][1=, 2=, 3=, 4=]{\mthfun[#4]{n#3}[#1][#2]}
\newcommandx{\oFun}[4][1=, 2=, 3=, 4=]{\mthfun[#4]{o#3}[#1][#2]}
\newcommandx{\pFun}[4][1=, 2=, 3=, 4=]{\mthfun[#4]{p#3}[#1][#2]}
\newcommandx{\qFun}[4][1=, 2=, 3=, 4=]{\mthfun[#4]{q#3}[#1][#2]}
\newcommandx{\rFun}[4][1=, 2=, 3=, 4=]{\mthfun[#4]{r#3}[#1][#2]}
\newcommandx{\sFun}[4][1=, 2=, 3=, 4=]{\mthfun[#4]{s#3}[#1][#2]}
\newcommandx{\tFun}[4][1=, 2=, 3=, 4=]{\mthfun[#4]{t#3}[#1][#2]}
\newcommandx{\uFun}[4][1=, 2=, 3=, 4=]{\mthfun[#4]{u#3}[#1][#2]}
\newcommandx{\vFun}[4][1=, 2=, 3=, 4=]{\mthfun[#4]{v#3}[#1][#2]}
\newcommandx{\wFun}[4][1=, 2=, 3=, 4=]{\mthfun[#4]{w#3}[#1][#2]}
\newcommandx{\xFun}[4][1=, 2=, 3=, 4=]{\mthfun[#4]{x#3}[#1][#2]}
\newcommandx{\yFun}[4][1=, 2=, 3=, 4=]{\mthfun[#4]{y#3}[#1][#2]}
\newcommandx{\zFun}[4][1=, 2=, 3=, 4=]{\mthfun[#4]{z#3}[#1][#2]}

\newcommandx{\ARel}[4][1=, 2=, 3=, 4=]{\mthrel[#4]{A#3}[#1][#2]}
\newcommandx{\BRel}[4][1=, 2=, 3=, 4=]{\mthrel[#4]{B#3}[#1][#2]}
\newcommandx{\CRel}[4][1=, 2=, 3=, 4=]{\mthrel[#4]{C#3}[#1][#2]}
\newcommandx{\DRel}[4][1=, 2=, 3=, 4=]{\mthrel[#4]{D#3}[#1][#2]}
\newcommandx{\ERel}[4][1=, 2=, 3=, 4=]{\mthrel[#4]{E#3}[#1][#2]}
\newcommandx{\FRel}[4][1=, 2=, 3=, 4=]{\mthrel[#4]{F#3}[#1][#2]}
\newcommandx{\GRel}[4][1=, 2=, 3=, 4=]{\mthrel[#4]{G#3}[#1][#2]}
\newcommandx{\HRel}[4][1=, 2=, 3=, 4=]{\mthrel[#4]{H#3}[#1][#2]}
\newcommandx{\IRel}[4][1=, 2=, 3=, 4=]{\mthrel[#4]{I#3}[#1][#2]}
\newcommandx{\JRel}[4][1=, 2=, 3=, 4=]{\mthrel[#4]{J#3}[#1][#2]}
\newcommandx{\KRel}[4][1=, 2=, 3=, 4=]{\mthrel[#4]{K#3}[#1][#2]}
\newcommandx{\LRel}[4][1=, 2=, 3=, 4=]{\mthrel[#4]{L#3}[#1][#2]}
\newcommandx{\MRel}[4][1=, 2=, 3=, 4=]{\mthrel[#4]{M#3}[#1][#2]}
\newcommandx{\NRel}[4][1=, 2=, 3=, 4=]{\mthrel[#4]{N#3}[#1][#2]}
\newcommandx{\ORel}[4][1=, 2=, 3=, 4=]{\mthrel[#4]{O#3}[#1][#2]}
\newcommandx{\PRel}[4][1=, 2=, 3=, 4=]{\mthrel[#4]{P#3}[#1][#2]}
\newcommandx{\QRel}[4][1=, 2=, 3=, 4=]{\mthrel[#4]{Q#3}[#1][#2]}
\newcommandx{\RRel}[4][1=, 2=, 3=, 4=]{\mthrel[#4]{R#3}[#1][#2]}
\newcommandx{\SRel}[4][1=, 2=, 3=, 4=]{\mthrel[#4]{S#3}[#1][#2]}
\newcommandx{\TRel}[4][1=, 2=, 3=, 4=]{\mthrel[#4]{T#3}[#1][#2]}
\newcommandx{\URel}[4][1=, 2=, 3=, 4=]{\mthrel[#4]{U#3}[#1][#2]}
\newcommandx{\VRel}[4][1=, 2=, 3=, 4=]{\mthrel[#4]{V#3}[#1][#2]}
\newcommandx{\WRel}[4][1=, 2=, 3=, 4=]{\mthrel[#4]{W#3}[#1][#2]}
\newcommandx{\XRel}[4][1=, 2=, 3=, 4=]{\mthrel[#4]{X#3}[#1][#2]}
\newcommandx{\YRel}[4][1=, 2=, 3=, 4=]{\mthrel[#4]{Y#3}[#1][#2]}
\newcommandx{\ZRel}[4][1=, 2=, 3=, 4=]{\mthrel[#4]{Z#3}[#1][#2]}

\newcommandx{\aRel}[4][1=, 2=, 3=, 4=]{\mthrel[#4]{a#3}[#1][#2]}
\newcommandx{\bRel}[4][1=, 2=, 3=, 4=]{\mthrel[#4]{b#3}[#1][#2]}
\newcommandx{\cRel}[4][1=, 2=, 3=, 4=]{\mthrel[#4]{c#3}[#1][#2]}
\newcommandx{\dRel}[4][1=, 2=, 3=, 4=]{\mthrel[#4]{d#3}[#1][#2]}
\newcommandx{\eRel}[4][1=, 2=, 3=, 4=]{\mthrel[#4]{e#3}[#1][#2]}
\newcommandx{\fRel}[4][1=, 2=, 3=, 4=]{\mthrel[#4]{f#3}[#1][#2]}
\newcommandx{\gRel}[4][1=, 2=, 3=, 4=]{\mthrel[#4]{g#3}[#1][#2]}
\newcommandx{\hRel}[4][1=, 2=, 3=, 4=]{\mthrel[#4]{h#3}[#1][#2]}
\newcommandx{\iRel}[4][1=, 2=, 3=, 4=]{\mthrel[#4]{i#3}[#1][#2]}
\newcommandx{\jRel}[4][1=, 2=, 3=, 4=]{\mthrel[#4]{j#3}[#1][#2]}
\newcommandx{\kRel}[4][1=, 2=, 3=, 4=]{\mthrel[#4]{k#3}[#1][#2]}
\newcommandx{\lRel}[4][1=, 2=, 3=, 4=]{\mthrel[#4]{l#3}[#1][#2]}
\newcommandx{\mRel}[4][1=, 2=, 3=, 4=]{\mthrel[#4]{m#3}[#1][#2]}
\newcommandx{\nRel}[4][1=, 2=, 3=, 4=]{\mthrel[#4]{n#3}[#1][#2]}
\newcommandx{\oRel}[4][1=, 2=, 3=, 4=]{\mthrel[#4]{o#3}[#1][#2]}
\newcommandx{\pRel}[4][1=, 2=, 3=, 4=]{\mthrel[#4]{p#3}[#1][#2]}
\newcommandx{\qRel}[4][1=, 2=, 3=, 4=]{\mthrel[#4]{q#3}[#1][#2]}
\newcommandx{\rRel}[4][1=, 2=, 3=, 4=]{\mthrel[#4]{r#3}[#1][#2]}
\newcommandx{\sRel}[4][1=, 2=, 3=, 4=]{\mthrel[#4]{s#3}[#1][#2]}
\newcommandx{\tRel}[4][1=, 2=, 3=, 4=]{\mthrel[#4]{t#3}[#1][#2]}
\newcommandx{\uRel}[4][1=, 2=, 3=, 4=]{\mthrel[#4]{u#3}[#1][#2]}
\newcommandx{\vRel}[4][1=, 2=, 3=, 4=]{\mthrel[#4]{v#3}[#1][#2]}
\newcommandx{\wRel}[4][1=, 2=, 3=, 4=]{\mthrel[#4]{w#3}[#1][#2]}
\newcommandx{\xRel}[4][1=, 2=, 3=, 4=]{\mthrel[#4]{x#3}[#1][#2]}
\newcommandx{\yRel}[4][1=, 2=, 3=, 4=]{\mthrel[#4]{y#3}[#1][#2]}
\newcommandx{\zRel}[4][1=, 2=, 3=, 4=]{\mthrel[#4]{z#3}[#1][#2]}

\newcommandx{\ASym}[4][1=, 2=, 3=, 4=]{\mthsym[#4]{A#3}[#1][#2]}
\newcommandx{\BSym}[4][1=, 2=, 3=, 4=]{\mthsym[#4]{B#3}[#1][#2]}
\newcommandx{\CSym}[4][1=, 2=, 3=, 4=]{\mthsym[#4]{C#3}[#1][#2]}
\newcommandx{\DSym}[4][1=, 2=, 3=, 4=]{\mthsym[#4]{D#3}[#1][#2]}
\newcommandx{\ESym}[4][1=, 2=, 3=, 4=]{\mthsym[#4]{E#3}[#1][#2]}
\newcommandx{\FSym}[4][1=, 2=, 3=, 4=]{\mthsym[#4]{F#3}[#1][#2]}
\newcommandx{\GSym}[4][1=, 2=, 3=, 4=]{\mthsym[#4]{G#3}[#1][#2]}
\newcommandx{\HSym}[4][1=, 2=, 3=, 4=]{\mthsym[#4]{H#3}[#1][#2]}
\newcommandx{\ISym}[4][1=, 2=, 3=, 4=]{\mthsym[#4]{I#3}[#1][#2]}
\newcommandx{\JSym}[4][1=, 2=, 3=, 4=]{\mthsym[#4]{J#3}[#1][#2]}
\newcommandx{\KSym}[4][1=, 2=, 3=, 4=]{\mthsym[#4]{K#3}[#1][#2]}
\newcommandx{\LSym}[4][1=, 2=, 3=, 4=]{\mthsym[#4]{L#3}[#1][#2]}
\newcommandx{\MSym}[4][1=, 2=, 3=, 4=]{\mthsym[#4]{M#3}[#1][#2]}
\newcommandx{\NSym}[4][1=, 2=, 3=, 4=]{\mthsym[#4]{N#3}[#1][#2]}
\newcommandx{\OSym}[4][1=, 2=, 3=, 4=]{\mthsym[#4]{O#3}[#1][#2]}
\newcommandx{\PSym}[4][1=, 2=, 3=, 4=]{\mthsym[#4]{P#3}[#1][#2]}
\newcommandx{\QSym}[4][1=, 2=, 3=, 4=]{\mthsym[#4]{Q#3}[#1][#2]}
\newcommandx{\RSym}[4][1=, 2=, 3=, 4=]{\mthsym[#4]{R#3}[#1][#2]}
\newcommandx{\SSym}[4][1=, 2=, 3=, 4=]{\mthsym[#4]{S#3}[#1][#2]}
\newcommandx{\TSym}[4][1=, 2=, 3=, 4=]{\mthsym[#4]{T#3}[#1][#2]}
\newcommandx{\USym}[4][1=, 2=, 3=, 4=]{\mthsym[#4]{U#3}[#1][#2]}
\newcommandx{\VSym}[4][1=, 2=, 3=, 4=]{\mthsym[#4]{V#3}[#1][#2]}
\newcommandx{\WSym}[4][1=, 2=, 3=, 4=]{\mthsym[#4]{W#3}[#1][#2]}
\newcommandx{\XSym}[4][1=, 2=, 3=, 4=]{\mthsym[#4]{X#3}[#1][#2]}
\newcommandx{\YSym}[4][1=, 2=, 3=, 4=]{\mthsym[#4]{Y#3}[#1][#2]}
\newcommandx{\ZSym}[4][1=, 2=, 3=, 4=]{\mthsym[#4]{Z#3}[#1][#2]}

\newcommandx{\aSym}[4][1=, 2=, 3=, 4=]{\mthsym[#4]{a#3}[#1][#2]}
\newcommandx{\bSym}[4][1=, 2=, 3=, 4=]{\mthsym[#4]{b#3}[#1][#2]}
\newcommandx{\cSym}[4][1=, 2=, 3=, 4=]{\mthsym[#4]{c#3}[#1][#2]}
\newcommandx{\dSym}[4][1=, 2=, 3=, 4=]{\mthsym[#4]{d#3}[#1][#2]}
\newcommandx{\eSym}[4][1=, 2=, 3=, 4=]{\mthsym[#4]{e#3}[#1][#2]}
\newcommandx{\fSym}[4][1=, 2=, 3=, 4=]{\mthsym[#4]{f#3}[#1][#2]}
\newcommandx{\gSym}[4][1=, 2=, 3=, 4=]{\mthsym[#4]{g#3}[#1][#2]}
\newcommandx{\hSym}[4][1=, 2=, 3=, 4=]{\mthsym[#4]{h#3}[#1][#2]}
\newcommandx{\iSym}[4][1=, 2=, 3=, 4=]{\mthsym[#4]{i#3}[#1][#2]}
\newcommandx{\jSym}[4][1=, 2=, 3=, 4=]{\mthsym[#4]{j#3}[#1][#2]}
\newcommandx{\kSym}[4][1=, 2=, 3=, 4=]{\mthsym[#4]{k#3}[#1][#2]}
\newcommandx{\lSym}[4][1=, 2=, 3=, 4=]{\mthsym[#4]{l#3}[#1][#2]}
\newcommandx{\mSym}[4][1=, 2=, 3=, 4=]{\mthsym[#4]{m#3}[#1][#2]}
\newcommandx{\nSym}[4][1=, 2=, 3=, 4=]{\mthsym[#4]{n#3}[#1][#2]}
\newcommandx{\oSym}[4][1=, 2=, 3=, 4=]{\mthsym[#4]{o#3}[#1][#2]}
\newcommandx{\pSym}[4][1=, 2=, 3=, 4=]{\mthsym[#4]{p#3}[#1][#2]}
\newcommandx{\qSym}[4][1=, 2=, 3=, 4=]{\mthsym[#4]{q#3}[#1][#2]}
\newcommandx{\rSym}[4][1=, 2=, 3=, 4=]{\mthsym[#4]{r#3}[#1][#2]}
\newcommandx{\sSym}[4][1=, 2=, 3=, 4=]{\mthsym[#4]{s#3}[#1][#2]}
\newcommandx{\tSym}[4][1=, 2=, 3=, 4=]{\mthsym[#4]{t#3}[#1][#2]}
\newcommandx{\uSym}[4][1=, 2=, 3=, 4=]{\mthsym[#4]{u#3}[#1][#2]}
\newcommandx{\vSym}[4][1=, 2=, 3=, 4=]{\mthsym[#4]{v#3}[#1][#2]}
\newcommandx{\wSym}[4][1=, 2=, 3=, 4=]{\mthsym[#4]{w#3}[#1][#2]}
\newcommandx{\xSym}[4][1=, 2=, 3=, 4=]{\mthsym[#4]{x#3}[#1][#2]}
\newcommandx{\ySym}[4][1=, 2=, 3=, 4=]{\mthsym[#4]{y#3}[#1][#2]}
\newcommandx{\zSym}[4][1=, 2=, 3=, 4=]{\mthsym[#4]{z#3}[#1][#2]}

\newcommandx{\AElm}[4][1=, 2=, 3=, 4=]{\mthelm[#4]{A#3}[#1][#2]}
\newcommandx{\BElm}[4][1=, 2=, 3=, 4=]{\mthelm[#4]{B#3}[#1][#2]}
\newcommandx{\CElm}[4][1=, 2=, 3=, 4=]{\mthelm[#4]{C#3}[#1][#2]}
\newcommandx{\DElm}[4][1=, 2=, 3=, 4=]{\mthelm[#4]{D#3}[#1][#2]}
\newcommandx{\EElm}[4][1=, 2=, 3=, 4=]{\mthelm[#4]{E#3}[#1][#2]}
\newcommandx{\FElm}[4][1=, 2=, 3=, 4=]{\mthelm[#4]{F#3}[#1][#2]}
\newcommandx{\GElm}[4][1=, 2=, 3=, 4=]{\mthelm[#4]{G#3}[#1][#2]}
\newcommandx{\HElm}[4][1=, 2=, 3=, 4=]{\mthelm[#4]{H#3}[#1][#2]}
\newcommandx{\IElm}[4][1=, 2=, 3=, 4=]{\mthelm[#4]{I#3}[#1][#2]}
\newcommandx{\JElm}[4][1=, 2=, 3=, 4=]{\mthelm[#4]{J#3}[#1][#2]}
\newcommandx{\KElm}[4][1=, 2=, 3=, 4=]{\mthelm[#4]{K#3}[#1][#2]}
\newcommandx{\LElm}[4][1=, 2=, 3=, 4=]{\mthelm[#4]{L#3}[#1][#2]}
\newcommandx{\MElm}[4][1=, 2=, 3=, 4=]{\mthelm[#4]{M#3}[#1][#2]}
\newcommandx{\NElm}[4][1=, 2=, 3=, 4=]{\mthelm[#4]{N#3}[#1][#2]}
\newcommandx{\OElm}[4][1=, 2=, 3=, 4=]{\mthelm[#4]{O#3}[#1][#2]}
\newcommandx{\PElm}[4][1=, 2=, 3=, 4=]{\mthelm[#4]{P#3}[#1][#2]}
\newcommandx{\QElm}[4][1=, 2=, 3=, 4=]{\mthelm[#4]{Q#3}[#1][#2]}
\newcommandx{\RElm}[4][1=, 2=, 3=, 4=]{\mthelm[#4]{R#3}[#1][#2]}
\newcommandx{\SElm}[4][1=, 2=, 3=, 4=]{\mthelm[#4]{S#3}[#1][#2]}
\newcommandx{\TElm}[4][1=, 2=, 3=, 4=]{\mthelm[#4]{T#3}[#1][#2]}
\newcommandx{\UElm}[4][1=, 2=, 3=, 4=]{\mthelm[#4]{U#3}[#1][#2]}
\newcommandx{\VElm}[4][1=, 2=, 3=, 4=]{\mthelm[#4]{V#3}[#1][#2]}
\newcommandx{\WElm}[4][1=, 2=, 3=, 4=]{\mthelm[#4]{W#3}[#1][#2]}
\newcommandx{\XElm}[4][1=, 2=, 3=, 4=]{\mthelm[#4]{X#3}[#1][#2]}
\newcommandx{\YElm}[4][1=, 2=, 3=, 4=]{\mthelm[#4]{Y#3}[#1][#2]}
\newcommandx{\ZElm}[4][1=, 2=, 3=, 4=]{\mthelm[#4]{Z#3}[#1][#2]}

\newcommandx{\aElm}[4][1=, 2=, 3=, 4=]{\mthelm[#4]{a#3}[#1][#2]}
\newcommandx{\bElm}[4][1=, 2=, 3=, 4=]{\mthelm[#4]{b#3}[#1][#2]}
\newcommandx{\cElm}[4][1=, 2=, 3=, 4=]{\mthelm[#4]{c#3}[#1][#2]}
\newcommandx{\dElm}[4][1=, 2=, 3=, 4=]{\mthelm[#4]{d#3}[#1][#2]}
\newcommandx{\eElm}[4][1=, 2=, 3=, 4=]{\mthelm[#4]{e#3}[#1][#2]}
\newcommandx{\fElm}[4][1=, 2=, 3=, 4=]{\mthelm[#4]{f#3}[#1][#2]}
\newcommandx{\gElm}[4][1=, 2=, 3=, 4=]{\mthelm[#4]{g#3}[#1][#2]}
\newcommandx{\hElm}[4][1=, 2=, 3=, 4=]{\mthelm[#4]{h#3}[#1][#2]}
\newcommandx{\iElm}[4][1=, 2=, 3=, 4=]{\mthelm[#4]{i#3}[#1][#2]}
\newcommandx{\jElm}[4][1=, 2=, 3=, 4=]{\mthelm[#4]{j#3}[#1][#2]}
\newcommandx{\kElm}[4][1=, 2=, 3=, 4=]{\mthelm[#4]{k#3}[#1][#2]}
\newcommandx{\lElm}[4][1=, 2=, 3=, 4=]{\mthelm[#4]{l#3}[#1][#2]}
\newcommandx{\mElm}[4][1=, 2=, 3=, 4=]{\mthelm[#4]{m#3}[#1][#2]}
\newcommandx{\nElm}[4][1=, 2=, 3=, 4=]{\mthelm[#4]{n#3}[#1][#2]}
\newcommandx{\oElm}[4][1=, 2=, 3=, 4=]{\mthelm[#4]{o#3}[#1][#2]}
\newcommandx{\pElm}[4][1=, 2=, 3=, 4=]{\mthelm[#4]{p#3}[#1][#2]}
\newcommandx{\qElm}[4][1=, 2=, 3=, 4=]{\mthelm[#4]{q#3}[#1][#2]}
\newcommandx{\rElm}[4][1=, 2=, 3=, 4=]{\mthelm[#4]{r#3}[#1][#2]}
\newcommandx{\sElm}[4][1=, 2=, 3=, 4=]{\mthelm[#4]{s#3}[#1][#2]}
\newcommandx{\tElm}[4][1=, 2=, 3=, 4=]{\mthelm[#4]{t#3}[#1][#2]}
\newcommandx{\uElm}[4][1=, 2=, 3=, 4=]{\mthelm[#4]{u#3}[#1][#2]}
\newcommandx{\vElm}[4][1=, 2=, 3=, 4=]{\mthelm[#4]{v#3}[#1][#2]}
\newcommandx{\wElm}[4][1=, 2=, 3=, 4=]{\mthelm[#4]{w#3}[#1][#2]}
\newcommandx{\xElm}[4][1=, 2=, 3=, 4=]{\mthelm[#4]{x#3}[#1][#2]}
\newcommandx{\yElm}[4][1=, 2=, 3=, 4=]{\mthelm[#4]{y#3}[#1][#2]}
\newcommandx{\zElm}[4][1=, 2=, 3=, 4=]{\mthelm[#4]{z#3}[#1][#2]}

\newcommand{\aka}
	{\txtabr{a.k.a.}\xspace}

\newcommand{\eg}
	{\txtabr{e.g.}\xspace}

\newcommand{\ie}
	{\txtabr{i.e.}\xspace}

\newcommand{\wrt}
	{\txtabr{w.r.t.}\xspace}

\newcommand{\wlogx}
	{\txtabr{w.l.o.g.}\xspace}

\newcommandx{\defeq}
	{\ensuremath{\triangleq}}
\newcommandx{\seteq}
	{\ensuremath{:=}}

\renewcommand{\implies}
	{\ensuremath{\Rightarrow}}

\newcommand{\fst}
	{\mthargfun{fst}}
\newcommand{\lst}
	{\mthargfun{lst}}

\newcommand{\prj}
	{\ensuremath{\downarrow}}

\newcommand{\dual}[1]
	{\mthempty{\overline{#1}}}
\newcommand{\adj}[1]
	{\mthempty{\widetilde{#1}}}
\newcommand{\der}[1]
	{\mthempty{\widehat{#1}}}

\newcommand{\tuple}[1]
	{\ensuremath{\!\argint{\langle}{#1}{\rangle}}}

\newcommand{\tupleb}[2]
	{\tuple{\argb{#1}{#2}}}
\newcommand{\tuplec}[3]
	{\tuple{\argc{#1}{#2}{#3}}}
\newcommand{\tupled}[4]
	{\tuple{\argd{#1}{#2}{#3}{#4}}}
\newcommand{\tuplee}[5]
	{\tuple{\arge{#1}{#2}{#3}{#4}{#5}}}
\newcommand{\tuplef}[6]
	{\tuple{\argf{#1}{#2}{#3}{#4}{#5}{#6}}}
\newcommand{\tupleg}[7]
	{\tuple{\argg{#1}{#2}{#3}{#4}{#5}{#6}{#7}}}
\newcommand{\tupleh}[8]
	{\tuple{\argh{#1}{#2}{#3}{#4}{#5}{#6}{#7}{#8}}}
\newcommand{\tuplei}[9]
	{\tuple{\argi{#1}{#2}{#3}{#4}{#5}{#6}{#7}{#8}{#9}}}

\newcommand{\tuplej}[9]
	{%
	\def\defarga{#1}%
	\def\defargb{#2}%
	\def\defargc{#3}%
	\def\defargd{#4}%
	\def\defarge{#5}%
	\def\defargf{#6}%
	\def\defargg{#7}%
	\def\defargh{#8}%
	\def\defargi{#9}%
	\tupleauxj%
	}
\newcommand{\tuplek}[9]
	{%
	\def\defarga{#1}%
	\def\defargb{#2}%
	\def\defargc{#3}%
	\def\defargd{#4}%
	\def\defarge{#5}%
	\def\defargf{#6}%
	\def\defargg{#7}%
	\def\defargh{#8}%
	\def\defargi{#9}%
	\tupleauxk%
	}
\newcommand{\tuplel}[9]
	{%
	\def\defarga{#1}%
	\def\defargb{#2}%
	\def\defargc{#3}%
	\def\defargd{#4}%
	\def\defarge{#5}%
	\def\defargf{#6}%
	\def\defargg{#7}%
	\def\defargh{#8}%
	\def\defargi{#9}%
	\tupleauxl%
	}
\newcommand{\tuplem}[9]
	{%
	\def\defarga{#1}%
	\def\defargb{#2}%
	\def\defargc{#3}%
	\def\defargd{#4}%
	\def\defarge{#5}%
	\def\defargf{#6}%
	\def\defargg{#7}%
	\def\defargh{#8}%
	\def\defargi{#9}%
	\tupleauxm%
	}
\newcommand{\tuplen}[9]
	{%
	\def\defarga{#1}%
	\def\defargb{#2}%
	\def\defargc{#3}%
	\def\defargd{#4}%
	\def\defarge{#5}%
	\def\defargf{#6}%
	\def\defargg{#7}%
	\def\defargh{#8}%
	\def\defargi{#9}%
	\tupleauxn%
	}
\newcommand{\tupleo}[9]
	{%
	\def\defarga{#1}%
	\def\defargb{#2}%
	\def\defargc{#3}%
	\def\defargd{#4}%
	\def\defarge{#5}%
	\def\defargf{#6}%
	\def\defargg{#7}%
	\def\defargh{#8}%
	\def\defargi{#9}%
	\tupleauxo%
	}
\newcommand{\tuplep}[9]
	{%
	\def\defarga{#1}%
	\def\defargb{#2}%
	\def\defargc{#3}%
	\def\defargd{#4}%
	\def\defarge{#5}%
	\def\defargf{#6}%
	\def\defargg{#7}%
	\def\defargh{#8}%
	\def\defargi{#9}%
	\tupleauxp%
	}
\newcommand{\tupleq}[9]
	{%
	\def\defarga{#1}%
	\def\defargb{#2}%
	\def\defargc{#3}%
	\def\defargd{#4}%
	\def\defarge{#5}%
	\def\defargf{#6}%
	\def\defargg{#7}%
	\def\defargh{#8}%
	\def\defargi{#9}%
	\tupleauxq%
	}
\newcommand{\tupler}[9]
	{%
	\def\defarga{#1}%
	\def\defargb{#2}%
	\def\defargc{#3}%
	\def\defargd{#4}%
	\def\defarge{#5}%
	\def\defargf{#6}%
	\def\defargg{#7}%
	\def\defargh{#8}%
	\def\defargi{#9}%
	\tupleauxr%
	}

\newcommand{\tupleauxj}[1]
	{%
	\tuple{\argj{\defarga}{\defargb}{\defargc}{\defargd}{\defarge}{\defargf}%
		{\defargg}{\defargh}{\defargi}{#1}}%
	}
\newcommand{\tupleauxk}[2]
	{%
	\tuple{\argk{\defarga}{\defargb}{\defargc}{\defargd}{\defarge}{\defargf}%
		{\defargg}{\defargh}{\defargi}{#1}{#2}}%
	}
\newcommand{\tupleauxl}[3]
	{%
	\tuple{\argl{\defarga}{\defargb}{\defargc}{\defargd}{\defarge}{\defargf}%
		{\defargg}{\defargh}{\defargi}{#1}{#2}{#3}}%
	}
\newcommand{\tupleauxm}[4]
	{%
	\tuple{\argm{\defarga}{\defargb}{\defargc}{\defargd}{\defarge}{\defargf}%
		{\defargg}{\defargh}{\defargi}{#1}{#2}{#3}{#4}}%
	}
\newcommand{\tupleauxn}[5]
	{%
	\tuple{\argn{\defarga}{\defargb}{\defargc}{\defargd}{\defarge}{\defargf}%
		{\defargg}{\defargh}{\defargi}{#1}{#2}{#3}{#4}{#5}}%
	}
\newcommand{\tupleauxo}[6]
	{%
	\tuple{\argo{\defarga}{\defargb}{\defargc}{\defargd}{\defarge}{\defargf}%
		{\defargg}{\defargh}{\defargi}{#1}{#2}{#3}{#4}{#5}{#6}}%
	}
\newcommand{\tupleauxp}[7]
	{%
	\tuple{\argp{\defarga}{\defargb}{\defargc}{\defargd}{\defarge}{\defargf}%
		{\defargg}{\defargh}{\defargi}{#1}{#2}{#3}{#4}{#5}{#6}{#7}}%
	}
\newcommand{\tupleauxq}[8]
	{%
	\tuple{\argq{\defarga}{\defargb}{\defargc}{\defargd}{\defarge}{\defargf}%
		{\defargg}{\defargh}{\defargi}{#1}{#2}{#3}{#4}{#5}{#6}{#7}{#8}}%
	}
\newcommand{\tupleauxr}[9]
	{%
	\tuple{\argr{\defarga}{\defargb}{\defargc}{\defargd}{\defarge}{\defargf}%
		{\defargg}{\defargh}{\defargi}{#1}{#2}{#3}{#4}{#5}{#6}{#7}{#8}{#9}}%
	}

\newcommand{\tuplebx}[2]
	{%
	\def\defarga{#1}%
	\def\defargb{#2}%
	\argsubsup{\tupleauxbx}%
	}
\newcommand{\tuplecx}[3]
	{%
	\def\defarga{#1}%
	\def\defargb{#2}%
	\def\defargc{#3}%
	\argsubsup{\tupleauxcx}%
	}
\newcommand{\tupledx}[4]
	{%
	\def\defarga{#1}%
	\def\defargb{#2}%
	\def\defargc{#3}%
	\def\defargd{#4}%
	\argsubsup{\tupleauxdx}%
	}
\newcommand{\tupleex}[5]
	{%
	\def\defarga{#1}%
	\def\defargb{#2}%
	\def\defargc{#3}%
	\def\defargd{#4}%
	\def\defarge{#5}%
	\argsubsup{\tupleauxex}%
	}
\newcommand{\tuplefx}[6]
	{%
	\def\defarga{#1}%
	\def\defargb{#2}%
	\def\defargc{#3}%
	\def\defargd{#4}%
	\def\defarge{#5}%
	\def\defargf{#6}%
	\argsubsup{\tupleauxfx}%
	}
\newcommand{\tuplegx}[7]
	{%
	\def\defarga{#1}%
	\def\defargb{#2}%
	\def\defargc{#3}%
	\def\defargd{#4}%
	\def\defarge{#5}%
	\def\defargf{#6}%
	\def\defargg{#7}%
	\argsubsup{\tupleauxgx}%
	}
\newcommand{\tuplehx}[8]
	{%
	\def\defarga{#1}%
	\def\defargb{#2}%
	\def\defargc{#3}%
	\def\defargd{#4}%
	\def\defarge{#5}%
	\def\defargf{#6}%
	\def\defargg{#7}%
	\def\defargh{#8}%
	\argsubsup{\tupleauxhx}%
	}
\newcommand{\tupleix}[9]
	{%
	\def\defarga{#1}%
	\def\defargb{#2}%
	\def\defargc{#3}%
	\def\defargd{#4}%
	\def\defarge{#5}%
	\def\defargf{#6}%
	\def\defargg{#7}%
	\def\defargh{#8}%
	\def\defargi{#9}%
	\argsubsup{\tupleauxix}%
	}

\newcommandx{\tupleauxbx}[2][1=, 2=]
	{%
	\tupleb
		{\argdef{#1}{\defarga[\argsubscript][\argsuperscript]}}
		{\argdef{#2}{\defargb[\argsubscript][\argsuperscript]}}%
	}
\newcommandx{\tupleauxcx}[3][1=, 2=, 3=]
	{%
	\tuplec
		{\argdef{#1}{\defarga[\argsubscript][\argsuperscript]}}
		{\argdef{#2}{\defargb[\argsubscript][\argsuperscript]}}
		{\argdef{#3}{\defargc[\argsubscript][\argsuperscript]}}%
	}
\newcommandx{\tupleauxdx}[4][1=, 2=, 3=, 4=]
	{%
	\tupled
		{\argdef{#1}{\defarga[\argsubscript][\argsuperscript]}}
		{\argdef{#2}{\defargb[\argsubscript][\argsuperscript]}}
		{\argdef{#3}{\defargc[\argsubscript][\argsuperscript]}}
		{\argdef{#4}{\defargd[\argsubscript][\argsuperscript]}}%
	}
\newcommandx{\tupleauxex}[5][1=, 2=, 3=, 4=, 5=]
	{%
	\tuplee
		{\argdef{#1}{\defarga[\argsubscript][\argsuperscript]}}
		{\argdef{#2}{\defargb[\argsubscript][\argsuperscript]}}
		{\argdef{#3}{\defargc[\argsubscript][\argsuperscript]}}
		{\argdef{#4}{\defargd[\argsubscript][\argsuperscript]}}
		{\argdef{#5}{\defarge[\argsubscript][\argsuperscript]}}%
	}
\newcommandx{\tupleauxfx}[6][1=, 2=, 3=, 4=, 5=, 6=]
	{%
	\tuplef
		{\argdef{#1}{\defarga[\argsubscript][\argsuperscript]}}
		{\argdef{#2}{\defargb[\argsubscript][\argsuperscript]}}
		{\argdef{#3}{\defargc[\argsubscript][\argsuperscript]}}
		{\argdef{#4}{\defargd[\argsubscript][\argsuperscript]}}
		{\argdef{#5}{\defarge[\argsubscript][\argsuperscript]}}
		{\argdef{#6}{\defargf[\argsubscript][\argsuperscript]}}%
	}
\newcommandx{\tupleauxgx}[7][1=, 2=, 3=, 4=, 5=, 6=, 7=]
	{%
	\tupleg
		{\argdef{#1}{\defarga[\argsubscript][\argsuperscript]}}
		{\argdef{#2}{\defargb[\argsubscript][\argsuperscript]}}
		{\argdef{#3}{\defargc[\argsubscript][\argsuperscript]}}
		{\argdef{#4}{\defargd[\argsubscript][\argsuperscript]}}
		{\argdef{#5}{\defarge[\argsubscript][\argsuperscript]}}
		{\argdef{#6}{\defargf[\argsubscript][\argsuperscript]}}
		{\argdef{#7}{\defargg[\argsubscript][\argsuperscript]}}%
	}
\newcommandx{\tupleauxhx}[8][1=, 2=, 3=, 4=, 5=, 6=, 7=, 8=]
	{%
	\tupleh
		{\argdef{#1}{\defarga[\argsubscript][\argsuperscript]}}
		{\argdef{#2}{\defargb[\argsubscript][\argsuperscript]}}
		{\argdef{#3}{\defargc[\argsubscript][\argsuperscript]}}
		{\argdef{#4}{\defargd[\argsubscript][\argsuperscript]}}
		{\argdef{#5}{\defarge[\argsubscript][\argsuperscript]}}
		{\argdef{#6}{\defargf[\argsubscript][\argsuperscript]}}
		{\argdef{#7}{\defargg[\argsubscript][\argsuperscript]}}
		{\argdef{#8}{\defargh[\argsubscript][\argsuperscript]}}%
	}
\newcommandx{\tupleauxix}[9][1=, 2=, 3=, 4=, 5=, 6=, 7=, 8=, 9=]
	{%
	\tuplei
		{\argdef{#1}{\defarga[\argsubscript][\argsuperscript]}}
		{\argdef{#2}{\defargb[\argsubscript][\argsuperscript]}}
		{\argdef{#3}{\defargc[\argsubscript][\argsuperscript]}}
		{\argdef{#4}{\defargd[\argsubscript][\argsuperscript]}}
		{\argdef{#5}{\defarge[\argsubscript][\argsuperscript]}}
		{\argdef{#6}{\defargf[\argsubscript][\argsuperscript]}}
		{\argdef{#7}{\defargg[\argsubscript][\argsuperscript]}}
		{\argdef{#8}{\defargh[\argsubscript][\argsuperscript]}}
		{\argdef{#9}{\defargi[\argsubscript][\argsuperscript]}}%
	}

\newcommand{\tuplejx}[9]
	{%
	\def\tuplearga{#1}%
	\def\tupleargb{#2}%
	\def\tupleargc{#3}%
	\def\tupleargd{#4}%
	\def\tuplearge{#5}%
	\def\tupleargf{#6}%
	\def\tupleargg{#7}%
	\def\tupleargh{#8}%
	\def\tupleargi{#9}%
	\argsubsup{\tupleauxjx}%
	}

\newcommand{\tupleauxjx}[1]
	{%
	\def\tupleargj{#1}%
	\argsubsup{\tupleauxxjx}%
	}

\newcommandx{\tupleauxxjx}[9][1=, 2=, 3=, 4=, 5=, 6=, 7=, 8=, 9=]
	{%
	\def\optarga{#1}%
	\def\optargb{#2}%
	\def\optargc{#3}%
	\def\optargd{#4}%
	\def\optarge{#5}%
	\def\optargf{#6}%
	\def\optargg{#7}%
	\def\optargh{#8}%
	\def\optargi{#9}%
	\tupleauxxxjx%
	}
\newcommandx{\tupleauxxkx}[9][1=, 2=, 3=, 4=, 5=, 6=, 7=, 8=, 9=]
	{%
	\def\optarga{#1}%
	\def\optargb{#2}%
	\def\optargc{#3}%
	\def\optargd{#4}%
	\def\optarge{#5}%
	\def\optargf{#6}%
	\def\optargg{#7}%
	\def\optargh{#8}%
	\def\optargi{#9}%
	\tupleauxxxkx%
	}
\newcommandx{\tupleauxxlx}[9][1=, 2=, 3=, 4=, 5=, 6=, 7=, 8=, 9=]
	{%
	\def\optarga{#1}%
	\def\optargb{#2}%
	\def\optargc{#3}%
	\def\optargd{#4}%
	\def\optarge{#5}%
	\def\optargf{#6}%
	\def\optargg{#7}%
	\def\optargh{#8}%
	\def\optargi{#9}%
	\tupleauxxxlx%
	}
\newcommandx{\tupleauxxmx}[9][1=, 2=, 3=, 4=, 5=, 6=, 7=, 8=, 9=]
	{%
	\def\optarga{#1}%
	\def\optargb{#2}%
	\def\optargc{#3}%
	\def\optargd{#4}%
	\def\optarge{#5}%
	\def\optargf{#6}%
	\def\optargg{#7}%
	\def\optargh{#8}%
	\def\optargi{#9}%
	\tupleauxxxmx%
	}
\newcommandx{\tupleauxxnx}[9][1=, 2=, 3=, 4=, 5=, 6=, 7=, 8=, 9=]
	{%
	\def\optarga{#1}%
	\def\optargb{#2}%
	\def\optargc{#3}%
	\def\optargd{#4}%
	\def\optarge{#5}%
	\def\optargf{#6}%
	\def\optargg{#7}%
	\def\optargh{#8}%
	\def\optargi{#9}%
	\tupleauxxxnx%
	}
\newcommandx{\tupleauxxox}[9][1=, 2=, 3=, 4=, 5=, 6=, 7=, 8=, 9=]
	{%
	\def\optarga{#1}%
	\def\optargb{#2}%
	\def\optargc{#3}%
	\def\optargd{#4}%
	\def\optarge{#5}%
	\def\optargf{#6}%
	\def\optargg{#7}%
	\def\optargh{#8}%
	\def\optargi{#9}%
	\tupleauxxxox%
	}
\newcommandx{\tupleauxxpx}[9][1=, 2=, 3=, 4=, 5=, 6=, 7=, 8=, 9=]
	{%
	\def\optarga{#1}%
	\def\optargb{#2}%
	\def\optargc{#3}%
	\def\optargd{#4}%
	\def\optarge{#5}%
	\def\optargf{#6}%
	\def\optargg{#7}%
	\def\optargh{#8}%
	\def\optargi{#9}%
	\tupleauxxxpx%
	}
\newcommandx{\tupleauxxqx}[9][1=, 2=, 3=, 4=, 5=, 6=, 7=, 8=, 9=]
	{%
	\def\optarga{#1}%
	\def\optargb{#2}%
	\def\optargc{#3}%
	\def\optargd{#4}%
	\def\optarge{#5}%
	\def\optargf{#6}%
	\def\optargg{#7}%
	\def\optargh{#8}%
	\def\optargi{#9}%
	\tupleauxxxqx%
	}
\newcommandx{\tupleauxxrx}[9][1=, 2=, 3=, 4=, 5=, 6=, 7=, 8=, 9=]
	{%
	\def\optarga{#1}%
	\def\optargb{#2}%
	\def\optargc{#3}%
	\def\optargd{#4}%
	\def\optarge{#5}%
	\def\optargf{#6}%
	\def\optargg{#7}%
	\def\optargh{#8}%
	\def\optargi{#9}%
	\tupleauxxxrx%
	}

\newcommandx{\tupleauxxxjx}[1][1=]
	{%
	\tuplej
		{\argdef{\optarga}{\tuplearga[\argsubscript][\argsuperscript]}}
		{\argdef{\optargb}{\tupleargb[\argsubscript][\argsuperscript]}}
		{\argdef{\optargc}{\tupleargc[\argsubscript][\argsuperscript]}}
		{\argdef{\optargd}{\tupleargd[\argsubscript][\argsuperscript]}}
		{\argdef{\optarge}{\tuplearge[\argsubscript][\argsuperscript]}}
		{\argdef{\optargf}{\tupleargf[\argsubscript][\argsuperscript]}}
		{\argdef{\optargg}{\tupleargg[\argsubscript][\argsuperscript]}}
		{\argdef{\optargh}{\tupleargh[\argsubscript][\argsuperscript]}}
		{\argdef{\optargi}{\tupleargi[\argsubscript][\argsuperscript]}}
		{\argdef{#1}{\tupleargj[\argsubscript][\argsuperscript]}}%
	}
\newcommandx{\tupleauxxxkx}[2][1=, 2=]
	{%
	\tuplek
		{\argdef{\optarga}{\tuplearga[\argsubscript][\argsuperscript]}}
		{\argdef{\optargb}{\tupleargb[\argsubscript][\argsuperscript]}}
		{\argdef{\optargc}{\tupleargc[\argsubscript][\argsuperscript]}}
		{\argdef{\optargd}{\tupleargd[\argsubscript][\argsuperscript]}}
		{\argdef{\optarge}{\tuplearge[\argsubscript][\argsuperscript]}}
		{\argdef{\optargf}{\tupleargf[\argsubscript][\argsuperscript]}}
		{\argdef{\optargg}{\tupleargg[\argsubscript][\argsuperscript]}}
		{\argdef{\optargh}{\tupleargh[\argsubscript][\argsuperscript]}}
		{\argdef{\optargi}{\tupleargi[\argsubscript][\argsuperscript]}}
		{\argdef{#1}{\tupleargj[\argsubscript][\argsuperscript]}}
		{\argdef{#2}{\tupleargk[\argsubscript][\argsuperscript]}}
	}
\newcommandx{\tupleauxxxlx}[3][1=, 2=, 3=]
	{%
	\tuplel
		{\argdef{\optarga}{\tuplearga[\argsubscript][\argsuperscript]}}
		{\argdef{\optargb}{\tupleargb[\argsubscript][\argsuperscript]}}
		{\argdef{\optargc}{\tupleargc[\argsubscript][\argsuperscript]}}
		{\argdef{\optargd}{\tupleargd[\argsubscript][\argsuperscript]}}
		{\argdef{\optarge}{\tuplearge[\argsubscript][\argsuperscript]}}
		{\argdef{\optargf}{\tupleargf[\argsubscript][\argsuperscript]}}
		{\argdef{\optargg}{\tupleargg[\argsubscript][\argsuperscript]}}
		{\argdef{\optargh}{\tupleargh[\argsubscript][\argsuperscript]}}
		{\argdef{\optargi}{\tupleargi[\argsubscript][\argsuperscript]}}
		{\argdef{#1}{\tupleargj[\argsubscript][\argsuperscript]}}
		{\argdef{#2}{\tupleargk[\argsubscript][\argsuperscript]}}
		{\argdef{#3}{\tupleargl[\argsubscript][\argsuperscript]}}
	}
\newcommandx{\tupleauxxxmx}[4][1=, 2=, 3=, 4=]
	{%
	\tuplem
		{\argdef{\optarga}{\tuplearga[\argsubscript][\argsuperscript]}}
		{\argdef{\optargb}{\tupleargb[\argsubscript][\argsuperscript]}}
		{\argdef{\optargc}{\tupleargc[\argsubscript][\argsuperscript]}}
		{\argdef{\optargd}{\tupleargd[\argsubscript][\argsuperscript]}}
		{\argdef{\optarge}{\tuplearge[\argsubscript][\argsuperscript]}}
		{\argdef{\optargf}{\tupleargf[\argsubscript][\argsuperscript]}}
		{\argdef{\optargg}{\tupleargg[\argsubscript][\argsuperscript]}}
		{\argdef{\optargh}{\tupleargh[\argsubscript][\argsuperscript]}}
		{\argdef{\optargi}{\tupleargi[\argsubscript][\argsuperscript]}}
		{\argdef{#1}{\tupleargj[\argsubscript][\argsuperscript]}}
		{\argdef{#2}{\tupleargk[\argsubscript][\argsuperscript]}}
		{\argdef{#3}{\tupleargl[\argsubscript][\argsuperscript]}}
		{\argdef{#4}{\tupleargm[\argsubscript][\argsuperscript]}}
	}
\newcommandx{\tupleauxxxnx}[5][1=, 2=, 3=, 4=, 5=]
	{%
	\tuplen
		{\argdef{\optarga}{\tuplearga[\argsubscript][\argsuperscript]}}
		{\argdef{\optargb}{\tupleargb[\argsubscript][\argsuperscript]}}
		{\argdef{\optargc}{\tupleargc[\argsubscript][\argsuperscript]}}
		{\argdef{\optargd}{\tupleargd[\argsubscript][\argsuperscript]}}
		{\argdef{\optarge}{\tuplearge[\argsubscript][\argsuperscript]}}
		{\argdef{\optargf}{\tupleargf[\argsubscript][\argsuperscript]}}
		{\argdef{\optargg}{\tupleargg[\argsubscript][\argsuperscript]}}
		{\argdef{\optargh}{\tupleargh[\argsubscript][\argsuperscript]}}
		{\argdef{\optargi}{\tupleargi[\argsubscript][\argsuperscript]}}
		{\argdef{#1}{\tupleargj[\argsubscript][\argsuperscript]}}
		{\argdef{#2}{\tupleargk[\argsubscript][\argsuperscript]}}
		{\argdef{#3}{\tupleargl[\argsubscript][\argsuperscript]}}
		{\argdef{#4}{\tupleargm[\argsubscript][\argsuperscript]}}
		{\argdef{#5}{\tupleargn[\argsubscript][\argsuperscript]}}
	}
\newcommandx{\tupleauxxxox}[6][1=, 2=, 3=, 4=, 5=, 6=]
	{%
	\tupleo
		{\argdef{\optarga}{\tuplearga[\argsubscript][\argsuperscript]}}
		{\argdef{\optargb}{\tupleargb[\argsubscript][\argsuperscript]}}
		{\argdef{\optargc}{\tupleargc[\argsubscript][\argsuperscript]}}
		{\argdef{\optargd}{\tupleargd[\argsubscript][\argsuperscript]}}
		{\argdef{\optarge}{\tuplearge[\argsubscript][\argsuperscript]}}
		{\argdef{\optargf}{\tupleargf[\argsubscript][\argsuperscript]}}
		{\argdef{\optargg}{\tupleargg[\argsubscript][\argsuperscript]}}
		{\argdef{\optargh}{\tupleargh[\argsubscript][\argsuperscript]}}
		{\argdef{\optargi}{\tupleargi[\argsubscript][\argsuperscript]}}
		{\argdef{#1}{\tupleargj[\argsubscript][\argsuperscript]}}
		{\argdef{#2}{\tupleargk[\argsubscript][\argsuperscript]}}
		{\argdef{#3}{\tupleargl[\argsubscript][\argsuperscript]}}
		{\argdef{#4}{\tupleargm[\argsubscript][\argsuperscript]}}
		{\argdef{#5}{\tupleargn[\argsubscript][\argsuperscript]}}
		{\argdef{#6}{\tupleargo[\argsubscript][\argsuperscript]}}
	}
\newcommandx{\tupleauxxxpx}[7][1=, 2=, 3=, 4=, 5=, 6=, 7=]
	{%
	\tuplep
		{\argdef{\optarga}{\tuplearga[\argsubscript][\argsuperscript]}}
		{\argdef{\optargb}{\tupleargb[\argsubscript][\argsuperscript]}}
		{\argdef{\optargc}{\tupleargc[\argsubscript][\argsuperscript]}}
		{\argdef{\optargd}{\tupleargd[\argsubscript][\argsuperscript]}}
		{\argdef{\optarge}{\tuplearge[\argsubscript][\argsuperscript]}}
		{\argdef{\optargf}{\tupleargf[\argsubscript][\argsuperscript]}}
		{\argdef{\optargg}{\tupleargg[\argsubscript][\argsuperscript]}}
		{\argdef{\optargh}{\tupleargh[\argsubscript][\argsuperscript]}}
		{\argdef{\optargi}{\tupleargi[\argsubscript][\argsuperscript]}}
		{\argdef{#1}{\tupleargj[\argsubscript][\argsuperscript]}}
		{\argdef{#2}{\tupleargk[\argsubscript][\argsuperscript]}}
		{\argdef{#3}{\tupleargl[\argsubscript][\argsuperscript]}}
		{\argdef{#4}{\tupleargm[\argsubscript][\argsuperscript]}}
		{\argdef{#5}{\tupleargn[\argsubscript][\argsuperscript]}}
		{\argdef{#6}{\tupleargo[\argsubscript][\argsuperscript]}}
		{\argdef{#7}{\tupleargp[\argsubscript][\argsuperscript]}}
	}
\newcommandx{\tupleauxxxqx}[8][1=, 2=, 3=, 4=, 5=, 6=, 7=, 8=]
	{%
	\tupleq
		{\argdef{\optarga}{\tuplearga[\argsubscript][\argsuperscript]}}
		{\argdef{\optargb}{\tupleargb[\argsubscript][\argsuperscript]}}
		{\argdef{\optargc}{\tupleargc[\argsubscript][\argsuperscript]}}
		{\argdef{\optargd}{\tupleargd[\argsubscript][\argsuperscript]}}
		{\argdef{\optarge}{\tuplearge[\argsubscript][\argsuperscript]}}
		{\argdef{\optargf}{\tupleargf[\argsubscript][\argsuperscript]}}
		{\argdef{\optargg}{\tupleargg[\argsubscript][\argsuperscript]}}
		{\argdef{\optargh}{\tupleargh[\argsubscript][\argsuperscript]}}
		{\argdef{\optargi}{\tupleargi[\argsubscript][\argsuperscript]}}
		{\argdef{#1}{\tupleargj[\argsubscript][\argsuperscript]}}
		{\argdef{#2}{\tupleargk[\argsubscript][\argsuperscript]}}
		{\argdef{#3}{\tupleargl[\argsubscript][\argsuperscript]}}
		{\argdef{#4}{\tupleargm[\argsubscript][\argsuperscript]}}
		{\argdef{#5}{\tupleargn[\argsubscript][\argsuperscript]}}
		{\argdef{#6}{\tupleargo[\argsubscript][\argsuperscript]}}
		{\argdef{#7}{\tupleargp[\argsubscript][\argsuperscript]}}
		{\argdef{#8}{\tupleargq[\argsubscript][\argsuperscript]}}
	}
\newcommandx{\tupleauxxxrx}[9][1=, 2=, 3=, 4=, 5=, 6=, 7=, 8=, 9=]
	{%
	\tupler
		{\argdef{\optarga}{\tuplearga[\argsubscript][\argsuperscript]}}
		{\argdef{\optargb}{\tupleargb[\argsubscript][\argsuperscript]}}
		{\argdef{\optargc}{\tupleargc[\argsubscript][\argsuperscript]}}
		{\argdef{\optargd}{\tupleargd[\argsubscript][\argsuperscript]}}
		{\argdef{\optarge}{\tuplearge[\argsubscript][\argsuperscript]}}
		{\argdef{\optargf}{\tupleargf[\argsubscript][\argsuperscript]}}
		{\argdef{\optargg}{\tupleargg[\argsubscript][\argsuperscript]}}
		{\argdef{\optargh}{\tupleargh[\argsubscript][\argsuperscript]}}
		{\argdef{\optargi}{\tupleargi[\argsubscript][\argsuperscript]}}
		{\argdef{#1}{\tupleargj[\argsubscript][\argsuperscript]}}
		{\argdef{#2}{\tupleargk[\argsubscript][\argsuperscript]}}
		{\argdef{#3}{\tupleargl[\argsubscript][\argsuperscript]}}
		{\argdef{#4}{\tupleargm[\argsubscript][\argsuperscript]}}
		{\argdef{#5}{\tupleargn[\argsubscript][\argsuperscript]}}
		{\argdef{#6}{\tupleargo[\argsubscript][\argsuperscript]}}
		{\argdef{#7}{\tupleargp[\argsubscript][\argsuperscript]}}
		{\argdef{#8}{\tupleargq[\argsubscript][\argsuperscript]}}
		{\argdef{#9}{\tupleargr[\argsubscript][\argsuperscript]}}%
	}

\newcommand{\set}[2]
	{\ensuremath{\argint{\{}{\argext{#1}{\allowbreak:\allowbreak}{#2}}{\}}}}

\newcommand{\pow}[1]
	{\ensuremath{2^{#1}}}

\newcommand{\card}[1]
	{\mthempty{\argint{\vert}{#1}{\vert}}}

\newcommand{\dom}
	{\mthargfun{dom}}

\newcommand{\rng}
	{\mthargfun{rng}}

\newcommand{\cmp}
	{\ensuremath{\circ}}

\newcommand{\rst}
	{\mthempty{\upharpoonright}}

\newcommand{\class}[2]
	{\ensuremath{(\argext{#1}{\allowbreak\!/\!\allowbreak}{#2})}}

\newcommandx{\pto}[2][1=, 2=]
	{\ensuremath{\rightharpoonup}}

\newcommandx{\cto}[2][1=, 2=]
	{\:\mthempty{\to}[#1][#2]\:}
\newcommandx{\cpto}[2][1=, 2=]
	{\:\mthempty{\pto}[#1][#2]\:}

\newcommand{\emptyfun}
	{\mthempty{\varnothing}}

\newcommand{\flip}[1]
	{\mthempty{\widehat{#1}}}

\newcommand{\AOmicron}
	{\mthargset{O}}

\newcommand{\SetN}
	{\mthset[2]{N}}

\newcommand{\numcc}[2]
	{\mthempty{[\argb{#1}{#2}]}}
\newcommand{\numco}[2]
	{\mthempty{[\argb{#1}{#2}[\:\!}}

\newcommand{\argset}{Ar}
\newcommandx{\ArgSet}[3][1=, 2=, 3=]
	{\mthset{\argset#3}[#1][#2]}

\newcommand{\argsym}{a}
\newcommandx{\argSym}[3][1=, 2=, 3=]
	{\mthsym{\argsym#3}[#1][#2]}

\newcommand{\argelm}{a}
\newcommandx{\argElm}[3][1=, 2=, 3=]
	{\mthelm{\argelm#3}[#1][#2]}

\newcommand{\relset}{Rl}
\newcommandx{\RelSet}[3][1=, 2=, 3=]
	{\mthset{\relset#3}[#1][#2]}

\newcommand{\relsym}{r}
\newcommandx{\relSym}[3][1=, 2=, 3=]
	{\mthsym{\relsym#3}[#1][#2]}

\newcommand{\relelm}{r}
\newcommandx{\relElm}[3][1=, 2=, 3=]
	{\mthelm{\relelm#3}[#1][#2]}

\newcommand{\argfun}{ar}
\newcommandx{\argFun}[4][1=, 2=, 3=, 4=]
	{\mthargfun{\argfun#4}[#1][#2]{#3}}

\newcommand{\lansig}{LS}
\newcommandx{\LanSig}[5][1=, 2=, 3=, 4=, 5=]
	{\txtargname{\lansig#5{\small\argint{$[$}{#1}{$]$}}}[#2][#3]{#4}\xspace}

\newcommand{\lansigname}{L}
\newcommand{\LanSigName}
	{\mthname{\lansigname}}

\newcommand{\lansigcls}{LS}
\newcommandx{\LanSigCls}[5][1=, 2=, 3=, 4=, 5=]
	{\mthset[#5]{\lansigcls#4\text{\txtname{\small\argint{$[$}{#1}{$]$}}}}[#2]%
	[#3]}

\newcommand{\LanSigStr}
	{\tuplecx{\ArgSet}{\RelSet}{\argFun}}

\newcommand{\domset}{Dm}
\newcommandx{\DomSet}[3][1=, 2=, 3=]
	{\mthset{\domset#3}[#1][#2]}

\newcommand{\domsym}{d}
\newcommandx{\domSym}[3][1=, 2=, 3=]
	{\mthsym{\domsym#3}[#1][#2]}

\newcommand{\domelm}{d}
\newcommandx{\domElm}[3][1=, 2=, 3=]
	{\mthelm{\domelm#3}[#1][#2]}

\newcommand{\relfun}{rl}
\newcommandx{\relFun}[4][1=, 2=, 3=, 4=]
	{\mthargfun{\relfun#4}[#1][#2]{#3}}

\newcommand{\relstr}{RS}
\newcommandx{\RelStr}[5][1=, 2=, 3=, 4=, 5=]
	{\txtargname{\relstr#5{\small\argint{$[$}{#1}{$]$}}}[#2][#3]{#4}\xspace}

\newcommand{\relstrname}{R}
\newcommand{\RelStrName}
	{\mthname{\relstrname}}

\newcommand{\relstrcls}{RS}
\newcommandx{\RelStrCls}[5][1=, 2=, 3=, 4=, 5=]
	{\mthset[#5]{\relstrcls#4\text{\txtname{\small\argint{$[$}{#1}{$]$}}}}[#2]%
	[#3]}

\newcommand{\RelStrStr}
	{\tuplebx{\DomSet}{\relFun}}

\newcommandx{\ordFun}[3][1=, 2=, 3=]
	{\mthempty{\argint{\left\vert}{#3}{\right\vert}}[#1][#2]}

\newcommandx{\sizFun}[3][1=, 2=, 3=]
	{\mthempty{\argint{\left\Vert}{#3}{\right\Vert}}[#1][#2]}

\newcommand{\verset}{Vr}
\newcommandx{\VerSet}[3][1=, 2=, 3=]
	{\mthset{\verset#3}[#1][#2]}

\newcommand{\versym}{v}
\newcommandx{\verSym}[3][1=, 2=, 3=]
	{\mthsym{\versym#3}[#1][#2]}

\newcommand{\verelm}{v}
\newcommandx{\verElm}[3][1=, 2=, 3=]
	{\mthelm{\verelm#3}[#1][#2]}

\newcommand{\edgrel}{Ed}
\newcommandx{\EdgRel}[3][1=, 2=, 3=]
	{\mthrel{\edgrel#3}[#1][#2]}

\newcommand{\edgsym}{e}
\newcommandx{\edgSym}[3][1=, 2=, 3=]
	{\mthsym{\edgsym#3}[#1][#2]}

\newcommand{\edgelm}{e}
\newcommandx{\edgElm}[3][1=, 2=, 3=]
	{\mthelm{\edgelm#3}[#1][#2]}

\newcommand{\orgfun}{or}
\newcommandx{\orgFun}[4][1=, 2=, 3=, 4=]
	{\mthargfun{\orgfun#4}[#1][#2]{#3}}

\newcommand{\desfun}{ds}
\newcommandx{\desFun}[4][1=, 2=, 3=, 4=]
	{\mthargfun{\desfun#4}[#1][#2]{#3}}

\newcommand{\grp}{Gr}
\newcommandx{\Grp}[5][1=, 2=, 3=, 4=, 5=]
	{\txtargname{\grp#5{\small\argint{$[$}{#1}{$]$}}}[#2][#3]{#4}\xspace}

\newcommand{\grpcls}{Gr}
\newcommandx{\GrpCls}[5][1=, 2=, 3=, 4=, 5=]
	{\mthset[#5]{\grpcls#4\text{\small\txtname{\argint{$[$}{#1}{$]$}}}}[#2][#3]}

\newcommand{\pthset}{Pth}
\newcommandx{\PthSet}[3][1=, 2=, 3=]
	{\mthset{\pthset#3}[#1][#2]}

\newcommand{\pthsym}{\pi}
\newcommandx{\pthSym}[3][1=, 2=, 3=]
	{\mthsym{\pthsym#3}[#1][#2]}

\newcommand{\pthelm}{\pi}
\newcommandx{\pthElm}[3][1=, 2=, 3=]
	{\mthelm{\pthelm#3}[#1][#2]}

\newcommand{\apset}{AP}
\newcommandx{\APSet}[3][1=, 2=, 3=]
	{\mthset{\apset#3}[#1][#2]}

\newcommand{\apsym}{p}
\newcommandx{\apSym}[3][1=, 2=, 3=]
	{\mthsym{\apsym#3}[#1][#2]}

\newcommand{\apelm}{p}
\newcommandx{\apElm}[3][1=, 2=, 3=]
	{\mthelm{\apelm#3}[#1][#2]}

\newcommand{\apfun}{ap}
\newcommandx{\apFun}[4][1=, 2=, 3=, 4=]
	{\mthargfun{\apfun#4}[#1][#2]{#3}}

\newcommand{\labgrp}{L\grp}
\newcommandx{\LabGrp}[5][1=, 2=, 3=, 4=, 5=]
	{\txtargname{\labgrp#5{\small\argint{$[$}{#1}{$]$}}}[#2][#3]{#4}\xspace}

\newcommand{\labgrpcls}{L\grpcls}
\newcommandx{\LabGrpCls}[5][1=, 2=, 3=, 4=, 5=]
	{\mthset[#5]{\labgrpcls#4\text{\small\txtname{\argint{$[$}{#1}{$]$}}}}[#2]%
	[#3]}

\newcommand{\trcset}{Trc}
\newcommandx{\TrcSet}[3][1=, 2=, 3=]
	{\mthset{\trcset#3}[#1][#2]}

\newcommand{\trcsym}{\varrho}
\newcommandx{\trcSym}[3][1=, 2=, 3=]
	{\mthsym{\trcsym#3}[#1][#2]}

\newcommand{\trcelm}{\varrho}
\newcommandx{\trcElm}[3][1=, 2=, 3=]
	{\mthelm{\trcelm#3}[#1][#2]}

\newcommand{\colset}{Cl}
\newcommandx{\ColSet}[3][1=, 2=, 3=]
	{\mthset{\colset#3}[#1][#2]}

\newcommand{\colsym}{c}
\newcommandx{\colSym}[3][1=, 2=, 3=]
	{\mthsym{\colsym#3}[#1][#2]}

\newcommand{\colelm}{c}
\newcommandx{\colElm}[3][1=, 2=, 3=]
	{\mthelm{\colelm#3}[#1][#2]}

\newcommand{\colfun}{cl}
\newcommandx{\colFun}[4][1=, 2=, 3=, 4=]
	{\mthargfun{\colfun#4}[#1][#2]{#3}}

\newcommand{\colgrp}{C\grp}
\newcommandx{\ColGrp}[5][1=, 2=, 3=, 4=, 5=]
	{\txtargname{\colgrp#5{\small\argint{$[$}{#1}{$]$}}}[#2][#3]{#4}\xspace}

\newcommand{\colgrpcls}{C\grpcls}
\newcommandx{\ColGrpCls}[5][1=, 2=, 3=, 4=, 5=]
	{\mthset[#5]{\colgrpcls#4\text{\small\txtname{\argint{$[$}{#1}{$]$}}}}[#2]%
	[#3]}

\newcommand{\wghset}{Wg}
\newcommandx{\WghSet}[3][1=, 2=, 3=]
	{\mthset{\wghset#3}[#1][#2]}

\newcommand{\wghsym}{w}
\newcommandx{\wghSym}[3][1=, 2=, 3=]
	{\mthsym{\wghsym#3}[#1][#2]}

\newcommand{\wghelm}{w}
\newcommandx{\wghElm}[3][1=, 2=, 3=]
	{\mthelm{\wghelm#3}[#1][#2]}

\newcommand{\wghfun}{wg}
\newcommandx{\wghFun}[4][1=, 2=, 3=, 4=]
	{\mthargfun{\wghfun#4}[#1][#2]{#3}}

\newcommand{\wghgrp}{W\grp}
\newcommandx{\WghGrp}[5][1=, 2=, 3=, 4=, 5=]
	{\txtargname{\wghgrp#5{\small\argint{$[$}{#1}{$]$}}}[#2][#3]{#4}\xspace}

\newcommand{\wghgrpcls}{W\grpcls}
\newcommandx{\WghGrpCls}[5][1=, 2=, 3=, 4=, 5=]
	{\mthset[#5]{\wghgrpcls#4\text{\small\txtname{\argint{$[$}{#1}{$]$}}}}[#2]%
	[#3]}

\newcommand{\gamkin}{2PT}

\newcommand{\plrset}{Pl}
\newcommandx{\PlrSet}[3][1=, 2=, 3=]
	{\mthset{\plrset#3}[#1][#2]}

\newcommand{\plrsym}{p}
\newcommandx{\plrSym}[3][1=, 2=, 3=]
	{\mthsym{\plrsym#3}[#1][#2]}

\newcommand{\plrelm}{p}
\newcommandx{\plrElm}[3][1=, 2=, 3=]
	{\mthelm{\plrelm#3}[#1][#2]}

\newcommand{\agnset}{Ag}
\newcommandx{\AgnSet}[3][1=, 2=, 3=]
	{\mthset{\agnset#3}[#1][#2]}

\newcommand{\agnsym}{a}
\newcommandx{\agnSym}[3][1=, 2=, 3=]
	{\mthsym{\agnsym#3}[#1][#2]}

\newcommand{\agnelm}{a}
\newcommandx{\agnElm}[3][1=, 2=, 3=]
	{\mthelm{\agnelm#3}[#1][#2]}

\newcommand{\movset}{Mv}
\newcommandx{\MovSet}[3][1=, 2=, 3=]
	{\mthset{\movset#3}[#1][#2]}

\newcommand{\movrel}{Mv}
\newcommandx{\MovRel}[3][1=, 2=, 3=]
	{\mthrel{\movrel#3}[#1][#2]}

\newcommand{\movsym}{m}
\newcommandx{\movSym}[3][1=, 2=, 3=]
	{\mthsym{\movsym#3}[#1][#2]}

\newcommand{\movelm}{m}
\newcommandx{\movElm}[3][1=, 2=, 3=]
	{\mthelm{\movelm#3}[#1][#2]}

\newcommand{\actset}{Ac}
\newcommandx{\ActSet}[3][1=, 2=, 3=]
	{\mthset{\actset#3}[#1][#2]}

\newcommand{\actrel}{Ac}
\newcommandx{\ActRel}[3][1=, 2=, 3=]
	{\mthrel{\actrel#3}[#1][#2]}

\newcommand{\actsym}{c}
\newcommandx{\actSym}[3][1=, 2=, 3=]
	{\mthsym{\actsym#3}[#1][#2]}

\newcommand{\actelm}{c}
\newcommandx{\actElm}[3][1=, 2=, 3=]
	{\mthelm{\actelm#3}[#1][#2]}

\newcommand{\decset}{Dc}
\newcommandx{\DecSet}[3][1=, 2=, 3=]
	{\mthset{\decset#3}[#1][#2]}

\newcommand{\decsym}{\delta}
\newcommandx{\decSym}[4][1=, 2=, 3=, 4=]
	{\mthargfun{\decsym#4}[#1][#2]{#3}}

\newcommand{\decelm}{\delta}
\newcommandx{\decElm}[4][1=, 2=, 3=, 4=]
	{\mthargfun{\decelm#4}[#1][#2]{#3}}

\newcommand{\posset}{Ps}
\newcommandx{\PosSet}[3][1=, 2=, 3=]
	{\mthset{\posset#3}[#1][#2]}
\newcommandx{\FPosSet}[3][1=, 2=, 3=]
	{\mthset{\posset#3}[0#1][#2]}
\newcommandx{\SPosSet}[3][1=, 2=, 3=]
	{\mthset{\posset#3}[1#1][#2]}

\newcommand{\possym}{v}
\newcommandx{\posSym}[3][1=, 2=, 3=]
	{\mthsym{\possym#3}[#1][#2]}

\newcommand{\poselm}{v}
\newcommandx{\posElm}[3][1=, 2=, 3=]
	{\mthelm{\poselm#3}[#1][#2]}

\newcommand{\sttset}{St}
\newcommandx{\SttSet}[3][1=, 2=, 3=]
	{\mthset{\sttset#3}[#1][#2]}
\newcommandx{\FSttSet}[3][1=, 2=, 3=]
	{\mthset{\sttset#3}[0#1][#2]}
\newcommandx{\SSttSet}[3][1=, 2=, 3=]
	{\mthset{\sttset#3}[1#1][#2]}

\newcommand{\sttsym}{s}
\newcommandx{\sttSym}[3][1=, 2=, 3=]
	{\mthsym{\sttsym#3}[#1][#2]}

\newcommand{\sttelm}{s}
\newcommandx{\sttElm}[3][1=, 2=, 3=]
	{\mthelm{\sttelm#3}[#1][#2]}

\newcommand{\plrfun}{pl}
\newcommandx{\plrFun}[4][1=, 2=, 3=, 4=]
	{\mthargfun{\plrfun#4}[#1][#2]{#3}}

\newcommand{\agnfun}{ag}
\newcommandx{\agnFun}[4][1=, 2=, 3=, 4=]
	{\mthargfun{\agnfun#4}[#1][#2]{#3}}

\newcommand{\movfun}{mv}
\newcommandx{\movFun}[4][1=, 2=, 3=, 4=]
	{\mthargfun{\movfun#4}[#1][#2]{#3}}

\newcommand{\actfun}{ac}
\newcommandx{\actFun}[4][1=, 2=, 3=, 4=]
	{\mthargfun{\actfun#4}[#1][#2]{#3}}

\newcommand{\decfun}{dc}
\newcommandx{\decFun}[4][1=, 2=, 3=, 4=]
	{\mthargfun{\decfun#4}[#1][#2]{#3}}

\newcommand{\trnfun}{tr}
\newcommandx{\trnFun}[4][1=, 2=, 3=, 4=]
	{\mthargfun{\trnfun#4}[#1][#2]{#3}}

\newcommand{\arn}{Ar}
\newcommandx{\Arn}[5][1=, 2=, 3=, 4=, 5=]
	{\txtargname{\arn#5{\small\argint{$[$}{#1}{$]$}}}[#2][#3]{#4}\xspace}

\newcommand{\arnname}{A}
\newcommand{\ArnName}
	{\mthname{\arnname}}

\newcommand{\arncls}{Ar}
\newcommandx{\ArnCls}[5][1=, 2=, 3=, 4=, 5=]
	{\mthset[#5]{\arncls#4\text{\small\txtname{\argint{$[$}{#1}{$]$}}}}[#2][#3]}

\newcommand{\ArnStr}[1][]
	{%
	\IfStrEqCase{\argdef{#1}{\gamkin}}
		{%
		{2PT}
			{\tuplecx{\FPosSet}{\SPosSet}{\MovRel}}%
		{MPC0}
			{\tupledx{\PlrSet}{\MovSet}{\PosSet}{\trnFun}}%
		{MPC1}
			{\tupleex{\PlrSet}{\MovSet}{\PosSet}{\decFun}{\trnFun}}%
		{MPC2}
			{\tuplefx{\PlrSet}{\MovSet}{\PosSet}{\plrFun}{\movFun}{\trnFun}}%
		{MPC3}
			{\tuplegx{\PlrSet}{\MovSet}{\PosSet}{\plrFun}{\movFun}{\decFun}{\trnFun}}%
		{2AT}
			{\tuplecx{\FSttSet}{\SSttSet}{\ActRel}}%
		{MAC0}
			{\tupledx{\AgnSet}{\ActSet}{\SttSet}{\trnFun}}%
		{MAC1}
			{\tupleex{\AgnSet}{\ActSet}{\SttSet}{\decFun}{\trnFun}}%
		{MAC2}
			{\tuplefx{\AgnSet}{\ActSet}{\SttSet}{\agnFun}{\actFun}{\trnFun}}%
		{MAC3}
			{\tuplegx{\AgnSet}{\ActSet}{\SttSet}{\agnFun}{\actFun}{\decFun}{\trnFun}}%
		}
		[\ensuremath{\clubsuit}]%
	}

\newcommand{\hstset}{Hst}
\newcommandx{\HstSet}[3][1=, 2=, 3=]
	{\mthset{\hstset#3}[#1][#2]}

\newcommand{\hstsym}{\rho}
\newcommandx{\hstSym}[3][1=, 2=, 3=]
	{\mthsym{\hstsym#3}[#1][#2]}

\newcommand{\hstelm}{\rho}
\newcommandx{\hstElm}[3][1=, 2=, 3=]
	{\mthelm{\hstelm#3}[#1][#2]}

\newcommand{\strset}{Str}
\newcommandx{\StrSet}[3][1=, 2=, 3=]
	{\mthset{\strset#3}[#1][#2]}

\newcommand{\strsym}{\sigma}
\newcommandx{\strSym}[4][1=, 2=, 3=, 4=]
	{\mthargfun{\strsym#4}[#1][#2]{#3}}

\newcommand{\strelm}{\sigma}
\newcommandx{\strElm}[4][1=, 2=, 3=, 4=]
	{\mthargfun{\strelm#4}[#1][#2]{#3}}

\newcommand{\prfset}{Prf}
\newcommandx{\PrfSet}[3][1=, 2=, 3=]
	{\mthset{\prfset#3}[#1][#2]}

\newcommand{\prfsym}{\xi}
\newcommandx{\prfSym}[4][1=, 2=, 3=, 4=]
	{\mthargfun{\prfsym#4}[#1][#2]{#3}}

\newcommandx{\prfElm}[4][1=, 2=, 3=, 4=]
	{\mthargfun{\prfsym#4}[#1][#2]{#3}}

\newcommand{\playfun}{play}
\newcommandx{\playFun}[4][1=, 2=, 3=, 4=]
	{\mthargfun{\playfun#4}[#1][#2]{#3}}

\newcommand{\labarn}{L\arn}
\newcommandx{\LabArn}[5][1=, 2=, 3=, 4=, 5=]
	{\txtargname{\labarn#5{\small\argint{$[$}{#1}{$]$}}}[#2][#3]{#4}\xspace}

\newcommand{\labarncls}{L\arncls}
\newcommandx{\LabArnCls}[5][1=, 2=, 3=, 4=, 5=]
	{\mthset[#5]{\labarncls#4\text{\small\txtname{\argint{$[$}{#1}{$]$}}}}[#2]%
	[#3]}

\newcommand{\colarn}{C\arn}
\newcommandx{\ColArn}[5][1=, 2=, 3=, 4=, 5=]
	{\txtargname{\colarn#5{\small\argint{$[$}{#1}{$]$}}}[#2][#3]{#4}\xspace}

\newcommand{\colarncls}{C\arncls}
\newcommandx{\ColArnCls}[5][1=, 2=, 3=, 4=, 5=]
	{\mthset[#5]{\colarncls#4\text{\small\txtname{\argint{$[$}{#1}{$]$}}}}[#2]%
	[#3]}

\newcommand{\wgharn}{W\arn}
\newcommandx{\WghArn}[5][1=, 2=, 3=, 4=, 5=]
	{\txtargname{\wgharn#5{\small\argint{$[$}{#1}{$]$}}}[#2][#3]{#4}\xspace}

\newcommand{\wgharncls}{W\arncls}
\newcommandx{\WghArnCls}[5][1=, 2=, 3=, 4=, 5=]
	{\mthset[#5]{\wgharncls#4\text{\small\txtname{\argint{$[$}{#1}{$]$}}}}[#2]%
	[#3]}

\newcommand{\winset}{Wn}
\newcommandx{\WinSet}[3][1=, 2=, 3=]
	{\mthset{\winset#3}[#1][#2]}

\newcommand{\prdset}{Pr}
\newcommandx{\PrdSet}[3][1=, 2=, 3=]
	{\mthset{\prdset#3}[#1][#2]}

\newcommand{\prdsym}{p}
\newcommandx{\prdSym}[3][1=, 2=, 3=]
	{\mthsym{\prdsym#3}[#1][#2]}

\newcommand{\prdelm}{p}
\newcommandx{\prdElm}[3][1=, 2=, 3=]
	{\mthelm{\prdelm#3}[#1][#2]}

\newcommand{\prdfun}{pr}
\newcommandx{\prdFun}[4][1=, 2=, 3=, 4=]
	{\mthargfun{\prdfun#4}[#1][#2]{#3}}

\newcommand{\extname}{E}
\newcommand{\ExtName}
	{\mthname{\extname}}

\newcommand{\extcls}{Ex}
\newcommandx{\ExtCls}[5][1=, 2=, 3=, 4=, 5=]
	{\mthset[#5]{\extcls#4\text{\small\txtname{\argint{$[$}{#1}{$]$}}}}[#2][#3]}

\newcommand{\tarset}{Tr}
\newcommandx{\TarSet}[3][1=, 2=, 3=]
	{\mthset{\tarset#3}[#1][#2]}

\newcommand{\tarsym}{t}
\newcommandx{\tarSym}[3][1=, 2=, 3=]
	{\mthsym{\tarsym#3}[#1][#2]}

\newcommand{\tarelm}{t}
\newcommandx{\tarElm}[3][1=, 2=, 3=]
	{\mthelm{\tarelm#3}[#1][#2]}

\newcommand{\schrel}{\models}
\newcommandx{\schRel}[4][1=, 2=, 3=, 4=]
	{\mthrel{\schrel#3}[#1][#2]}

\newcommand{\schcls}{Sc}
\newcommandx{\SchCls}[5][1=, 2=, 3=, 4=, 5=]
	{\mthset[#5]{\schcls#4\text{\small\txtname{\argint{$[$}{#1}{$]$}}}}[#2][#3]}

\newcommand{\gamcls}{Gm}
\newcommandx{\GamCls}[5][1=, 2=, 3=, 4=, 5=]
	{\mthset[#5]{\gamcls#4\text{\small\txtname{\argint{$[$}{#1}{$]$}}}}[#2][#3]}

\newcommandx{\GamStr}[1][1=]
	{%
	\StrLeft{\argdef{#1}{\gamkin}}{2}[\optgamkin]%
	\IfStrEqCase{\optgamkin}
		{%
		{2P}
			{\gamstrauxtp}%
		{MP}
			{\gamstrauxmp}%
		{2A}
			{\gamstrauxta}%
		{MA}
			{\gamstrauxma}%
		}
		[\ensuremath{\clubsuit}]%
	}
\newcommandx{\gamstrauxtp}[5][1=, 2=, 3=, 4=, 5=]
	{\tuplecx{\ArnName}{\posElm}{\WinSet}[#3][#4][#5][#1][#2]}
\newcommandx{\gamstrauxmp}[5][1=, 2=, 3=, 4=, 5=]
	{\tuplecx{\ExtName}{\posElm}{\tarElm}[#3][#4][#5][#1][#2]}
\newcommandx{\gamstrauxta}[5][1=, 2=, 3=, 4=, 5=]
	{\tuplecx{\ArnName}{\sttElm}{\WinSet}[#3][#4][#5][#1][#2]}
\newcommandx{\gamstrauxma}[5][1=, 2=, 3=, 4=, 5=]
	{\tuplecx{\ExtName}{\sttElm}{\tarElm}[#3][#4][#5][#1][#2]}

\newcommand{\worset}{W}
\newcommandx{\WorSet}[3][1=, 2=, 3=]
	{\mthset{\worset#3}[#1][#2]}

\newcommand{\worsym}{w}
\newcommandx{\worSym}[3][1=, 2=, 3=]
	{\mthsym{\worsym#3}[#1][#2]}

\newcommand{\worelm}{w}
\newcommandx{\worElm}[3][1=, 2=, 3=]
	{\mthelm{\worelm#3}[#1][#2]}

\newcommand{\trnrel}{R}
\newcommandx{\TrnRel}[3][1=, 2=, 3=]
	{\mthrel{\trnrel#3}[#1][#2]}

\newcommand{\trnsym}{r}
\newcommandx{\trnSym}[3][1=, 2=, 3=]
	{\mthsym{\trnsym#3}[#1][#2]}

\newcommand{\trnelm}{r}
\newcommandx{\trnElm}[3][1=, 2=, 3=]
	{\mthelm{\trnelm#3}[#1][#2]}

\newcommand{\labfun}{L}
\newcommandx{\labFun}[4][1=, 2=, 3=, 4=]
	{\mthargfun{\labfun#4}[#1][#2]{#3}}

\newcommand{\krpstr}{KS}
\newcommandx{\KrpStr}[5][1=, 2=, 3=, 4=, 5=]
	{\txtargname{\krpstr#5{\small\argint{$[$}{#1}{$]$}}}[#2][#3]{#4}\xspace}

\newcommand{\krpstrcls}{KS}
\newcommandx{\KrpStrCls}[5][1=, 2=, 3=, 4=, 5=]
	{\mthset[#5]{\krpstrcls#4\text{\small\txtname{\argint{$[$}{#1}{$]$}}}}[#2]%
	[#3]}

\newcommand{\trkset}{Trk}
\newcommandx{\TrkSet}[3][1=, 2=, 3=]
	{\mthset{\trkset#3}[#1][#2]}

\newcommand{\trksym}{\rho}
\newcommandx{\trkSym}[3][1=, 2=, 3=]
	{\mthsym{\trksym#3}[#1][#2]}

\newcommand{\trkelm}{\rho}
\newcommandx{\trkElm}[3][1=, 2=, 3=]
	{\mthelm{\trkelm#3}[#1][#2]}

\newcommand{\krptree}{KT}
\newcommandx{\KrpTree}[5][1=, 2=, 3=, 4=, 5=]
	{\txtargname{\krptree#5{\small\argint{$[$}{#1}{$]$}}}[#2][#3]{#4}\xspace}

\newcommand{\krptreecls}{KT}
\newcommandx{\KrpTreeCls}[5][1=, 2=, 3=, 4=, 5=]
	{\mthset[#5]{\krptreecls#4\text{\small\txtname{\argint{$[$}{#1}{$]$}}}}[#2]%
	[#3]}

\newcommand{\dirset}{Dir}
\newcommandx{\DirSet}[3][1=, 2=, 3=]
	{\mthset{\dirset#3}[#1][#2]}

\newcommand{\dirsym}{d}
\newcommandx{\dirSym}[3][1=, 2=, 3=]
	{\mthsym{\dirsym#3}[#1][#2]}

\newcommand{\direlm}{d}
\newcommandx{\dirElm}[3][1=, 2=, 3=]
	{\mthelm{\direlm#3}[#1][#2]}

\newcommand{\unwfun}{unw}
\newcommandx{\unwFun}[4][1=, 2=, 3=, 4=]
	{\mthargfun{\unwfun#4}[#1][#2]{#3}}

\newcommand{\congamstrkin}{MAC0}

\newcommand{\congamstr}{CGS}
\newcommandx{\ConGamStr}[5][1=, 2=, 3=, 4=, 5=]
	{\txtargname{\congamstr#5{\small\argint{$[$}{#1}{$]$}}}[#2][#3]{#4}\xspace}
\newcommand{\CGS}{\ConGamStr}

\newcommand{\congamstrname}{G}
\newcommand{\ConGamStrName}
	{\mthname{\congamstrname}}

\newcommandx{\ConGamStrCls}[5][1=, 2=, 3=, 4=, 5=]
	{\mthset[#5]{\arncls#4\text{\small\txtname{\argint{$[$}{#1}{$]$}}}}[#2][#3]}

\newcommandx{\ConGamStrStr}[1][1=]
	{%
	\IfStrEqCase{\argdef{#1}{\congamstrkin}}
		{%
		{IP}
			{\congamstrstrauxip}%
		{2PT}
			{\congamstrstrauxpt}%
		{MPC0}
			{\congamstrstrauxpca}%
		{MPC1}
			{\congamstrstrauxpcb}%
		{MPC2}
			{\congamstrstrauxpcc}%
		{MPC3}
			{\congamstrstrauxpcd}%
		{IA}
			{\congamstrstrauxia}%
		{2AT}
			{\congamstrstrauxat}%
		{MAC0}
			{\congamstrstrauxaca}%
		{MAC1}
			{\congamstrstrauxacb}%
		{MAC2}
			{\congamstrstrauxacc}%
		{MAC3}
			{\congamstrstrauxacd}%
		}
		[\ensuremath{\clubsuit}]%
	}
\newcommandx{\congamstrstrauxip}[3][1=, 2=, 3=]
	{%
	\def\defini{#1}%
	\def\defsubscr{#2}%
	\def\defsupscr{#3}%
	\congamstrstrauxxip%
	}
\newcommandx{\congamstrstrauxxip}[3][1=, 2=, 3=]
	{%
	\tupledx{\ArnName}{\APSet}{\apFun}{\posElm}%
		[\defsubscr][\defsupscr][#1][#2][#3][\defini]%
	}
\newcommandx{\congamstrstrauxpt}[3][1=, 2=, 3=]
	{%
	\def\defini{#1}%
	\def\defsubscr{#2}%
	\def\defsupscr{#3}%
	\congamstrstrauxxpt%
	}
\newcommandx{\congamstrstrauxxpt}[5][1=, 2=, 3=, 4=, 5=]
	{%
	\tuplefx{\APSet}{\FPosSet}{\SPosSet}{\MovRel}{\apFun}{\posElm}%
		[\defsubscr][\defsupscr][#1][#2][#3][#4][#5][\defini]%
	}
\newcommandx{\congamstrstrauxpca}[3][1=, 2=, 3=]
	{%
	\def\defini{#1}%
	\def\defsubscr{#2}%
	\def\defsupscr{#3}%
	\congamstrstrauxxpca%
	}
\newcommandx{\congamstrstrauxxpca}[6][1=, 2=, 3=, 4=, 5=, 6=]
	{%
	\tuplegx{\APSet}{\PlrSet}{\MovSet}{\PosSet}{\trnFun}{\apFun}{\posElm}%
		[\defsubscr][\defsupscr][#1][#2][#3][#4][#5][#6][\defini]%
	}
\newcommandx{\congamstrstrauxpcb}[3][1=, 2=, 3=]
	{%
	\def\defini{#1}%
	\def\defsubscr{#2}%
	\def\defsupscr{#3}%
	\congamstrstrauxxpcb%
	}
\newcommandx{\congamstrstrauxxpcb}[7][1=, 2=, 3=, 4=, 5=, 6=, 7=]
	{%
	\tuplehx{\APSet}{\PlrSet}{\MovSet}{\PosSet}{\decFun}{\trnFun}{\apFun}%
		{\posElm}%
		[\defsubscr][\defsupscr][#1][#2][#3][#4][#5][#6][#7][\defini]%
	}
\newcommandx{\congamstrstrauxpcc}[3][1=, 2=, 3=]
	{%
	\def\defini{#1}%
	\def\defsubscr{#2}%
	\def\defsupscr{#3}%
	\congamstrstrauxxpcc%
	}
\newcommandx{\congamstrstrauxxpcc}[8][1=, 2=, 3=, 4=, 5=, 6=, 7=, 8=]
	{%
	\tupleix{\APSet}{\PlrSet}{\MovSet}{\PosSet}{\plrFun}{\movFun}{\trnFun}%
		{\apFun}{\posElm}%
		[\defsubscr][\defsupscr][#1][#2][#3][#4][#5][#6][#7][#8][\defini]%
	}
\newcommandx{\congamstrstrauxpcd}[3][1=, 2=, 3=]
	{%
	\def\defini{#1}%
	\def\defsubscr{#2}%
	\def\defsupscr{#3}%
	\congamstrstrauxxpcd%
	}
\newcommandx{\congamstrstrauxxpcd}[9][1=, 2=, 3=, 4=, 5=, 6=, 7=, 8=, 9=]
	{%
	\tuplejx{\APSet}{\PlrSet}{\MovSet}{\PosSet}{\plrFun}{\movFun}{\decFun}%
	{\trnFun}{\apFun}{\posElm}%
		[\defsubscr][\defsupscr][#1][#2][#3][#4][#5][#6][#7][#8][#9][\defini]%
	}
\newcommandx{\congamstrstrauxia}[3][1=, 2=, 3=]
	{%
	\def\defini{#1}%
	\def\defsubscr{#2}%
	\def\defsupscr{#3}%
	\congamstrstrauxxia%
	}
\newcommandx{\congamstrstrauxxia}[3][1=, 2=, 3=]
	{%
	\tupledx{\ArnName}{\APSet}{\apFun}{\sttElm}%
		[\defsubscr][\defsupscr][#1][#2][#3][\defini]%
	}
\newcommandx{\congamstrstrauxat}[3][1=, 2=, 3=]
	{%
	\def\defini{#1}%
	\def\defsubscr{#2}%
	\def\defsupscr{#3}%
	\congamstrstrauxxat%
	}
\newcommandx{\congamstrstrauxxat}[5][1=, 2=, 3=, 4=, 5=]
	{%
	\tuplefx{\APSet}{\FSttSet}{\SSttSet}{\ActRel}{\apFun}{\sttElm}%
		[\defsubscr][\defsupscr][#1][#2][#3][#4][#5][\defini]%
	}
\newcommandx{\congamstrstrauxaca}[3][1=, 2=, 3=]
	{%
	\def\defini{#1}%
	\def\defsubscr{#2}%
	\def\defsupscr{#3}%
	\congamstrstrauxxaca%
	}
\newcommandx{\congamstrstrauxxaca}[6][1=, 2=, 3=, 4=, 5=, 6=]
	{%
	\tuplegx{\APSet}{\AgnSet}{\ActSet}{\SttSet}{\trnFun}{\apFun}{\sttElm[0]}%
		[\defsubscr][\defsupscr][#1][#2][#3][#4][#5][#6][\defini]%
	}
\newcommandx{\congamstrstrauxacb}[3][1=, 2=, 3=]
	{%
	\def\defini{#1}%
	\def\defsubscr{#2}%
	\def\defsupscr{#3}%
	\congamstrstrauxxacb%
	}
\newcommandx{\congamstrstrauxxacb}[7][1=, 2=, 3=, 4=, 5=, 6=, 7=]
	{%
	\tuplehx{\APSet}{\AgnSet}{\ActSet}{\SttSet}{\decFun}{\trnFun}{\apFun}%
		{\sttElm}%
		[\defsubscr][\defsupscr][#1][#2][#3][#4][#5][#6][#7][\defini]%
	}
\newcommandx{\congamstrstrauxacc}[3][1=, 2=, 3=]
	{%
	\def\defini{#1}%
	\def\defsubscr{#2}%
	\def\defsupscr{#3}%
	\congamstrstrauxxacc%
	}
\newcommandx{\congamstrstrauxxacc}[8][1=, 2=, 3=, 4=, 5=, 6=, 7=, 8=]
	{%
	\tupleix{\APSet}{\AgnSet}{\ActSet}{\SttSet}{\agnFun}{\actFun}{\trnFun}%
		{\apFun}{\sttElm}%
		[\defsubscr][\defsupscr][#1][#2][#3][#4][#5][#6][#7][#8][\defini]%
	}
\newcommandx{\congamstrstrauxacd}[3][1=, 2=, 3=]
	{%
	\def\defini{#1}%
	\def\defsubscr{#2}%
	\def\defsupscr{#3}%
	\congamstrstrauxxacd%
	}
\newcommandx{\congamstrstrauxxacd}[9][1=, 2=, 3=, 4=, 5=, 6=, 7=, 8=, 9=]
	{%
	\tuplejx{\APSet}{\AgnSet}{\ActSet}{\SttSet}{\agnFun}{\actFun}{\decFun}%
		{\trnFun}{\apFun}{\sttElm}%
		[\defsubscr][\defsupscr][#1][#2][#3][#4][#5][#6][#7][#8][#9][\defini]%
	}
\newcommand{\CGSStr}{\ConGamStrStr}

\newcommand{\trntabkin}{D}

\newcommand{\symset}{Sm}
\newcommandx{\SymSet}[3][1=, 2=, 3=]
	{\mthset{\symset#3}[#1][#2]}

\newcommand{\symsym}{\ell}
\newcommandx{\symSym}[3][1=, 2=, 3=]
	{\mthsym{\symsym#3}[#1][#2]}

\newcommand{\symelm}{\ell}
\newcommandx{\symElm}[3][1=, 2=, 3=]
	{\mthelm{\symelm#3}[#1][#2]}

\newcommand{\DSttSet}[1][]
	{\SttSet[\Delta#1]}

\newcommand{\ESttSet}[1][]
	{\SttSet[\exists#1]}

\newcommand{\ASttSet}[1][]
	{\SttSet[\forall#1]}

\newcommand{\trntab}{tt}
\newcommandx{\TrnTab}[5][1=, 2=, 3=, 4=, 5=]
	{\txtargname{\trntab#5{\small\argint{$[$}{#1}{$]$}}}[#2][#3]{#4}\xspace}

\newcommand{\trntabcls}{TT}
\newcommandx{\TrnTabCls}[5][1=, 2=, 3=, 4=, 5=]
	{\mthset[#5]{\trntabcls#4\text{\txtname{\small\argint{$[$}{#1}{$]$}}}}[#2]%
	[#3]}

\newcommand{\TrnTabStr}[1][]
	{%
	\IfStrEqCase{\argdef{#1}{\trntabkin}}
		{%
		{D}{\tuplecx{\SymSet}{\SttSet}{\trnFun}}%
		{N}{\tupledx{\SymSet}{\DSttSet}{\ESttSet}{\trnFun}}%
		{U}{\tupledx{\SymSet}{\DSttSet}{\ASttSet}{\trnFun}}%
		{A}{\tupleex{\SymSet}{\DSttSet}{\ESttSet}{\ASttSet}{\trnFun}}%
		}
		[\ensuremath{\clubsuit}]%
	}

\newcommandx{\FOL}[5][1=, 2=, 3=, 4=, 5=]
	{\txtargname{FOL#5{\small\argint{$[$}{#1}{$]$}}}[#2][#3]{#4}\xspace}

\newcommand{\OBFOL}[1][]
	{\FOL[\argb{1b}{#1}]}

\newcommand{\pnf}
	{\txtabr{pnf}\xspace}

\newcommand{\enf}
	{\txtabr{enf}\xspace}

\newcommand{\Qnt}
	{\mthsym{Qn}}

\newcommand{\Opr}
	{\mthsym{Op}}

\newcommand{\Tt}
	{\mthsym{t}}

\newcommand{\Ff}
	{\mthsym{f}}

\newcommand{\varset}{Vr}
\newcommandx{\VarSet}[3][1=, 2=, 3=]
	{\mthset{\varset#3}[#1][#2]}

\newcommand{\varsym}{x}
\newcommandx{\varSym}[3][1=, 2=, 3=]
	{\mthsym{\varsym#3}[#1][#2]}

\newcommand{\varelm}{x}
\newcommandx{\varElm}[3][1=, 2=, 3=]
	{\mthelm{\varelm#3}[#1][#2]}

\newcommand{\varfun}{vr}
\newcommandx{\varFun}[4][1=, 2=, 3=, 4=]
	{\mthargfun{\varfun#4}[#1][#2]{#3}}

\newcommand{\qntset}{Qn}
\newcommandx{\QntSet}[3][1=, 2=, 3=]
	{\mthset{\qntset#3}[#1][#2]}

\newcommand{\qntsym}{\wp}
\newcommandx{\qntSym}[3][1=, 2=, 3=]
	{\mthsym{\qntsym#3}[#1][#2]}

\newcommand{\qntelm}{\wp}
\newcommandx{\qntElm}[3][1=, 2=, 3=]
	{\mthelm{\qntelm#3}[#1][#2]}

\newcommand{\bndset}{Bn}
\newcommandx{\BndSet}[3][1=, 2=, 3=]
	{\mthset{\bndset#3}[#1][#2]}

\newcommand{\bndsym}{\flat}
\newcommandx{\bndSym}[3][1=, 2=, 3=]
	{\mthsym{\bndsym#3}[#1][#2]}

\newcommand{\bndelm}{\flat}
\newcommandx{\bndElm}[3][1=, 2=, 3=]
	{\mthelm{\bndelm#3}[#1][#2]}

\newcommand{\depset}{\Delta}
\newcommandx{\DepSet}[3][1=, 2=, 3=]
	{\mthset{\depset#3}[#1][#2]}

\newcommand{\asgset}{Asg}
\newcommandx{\AsgSet}[3][1=, 2=, 3=]
	{\mthset{\asgset#3}[#1][#2]}

\newcommand{\asgfun}{\chi}
\newcommandx{\asgFun}[4][1=, 2=, 3=, 4=]
	{\mthargfun{\asgfun#4}[#1][#2]{#3}}

\newcommand{\smset}{SF}
\newcommandx{\SMSet}[3][1=, 2=, 3=]
	{\mthset{\smset#3}[#1][#2]}

\newcommand{\bsmset}{BSF}
\newcommandx{\BSMSet}[3][1=, 2=, 3=]
	{\mthset{\bsmset#3}[#1][#2]}

\newcommand{\smfun}{\delta}
\newcommandx{\smFun}[4][1=, 2=, 3=, 4=]
	{\mthargfun{\smfun#4}[#1][#2]{#3}}

\newcommand{\cmset}{CM}
\newcommandx{\CMSet}[3][1=, 2=, 3=]
	{\mthset{\cmset#3}[#1][#2]}

\newcommand{\cmfun}{\gamma}
\newcommandx{\cmFun}[4][1=, 2=, 3=, 4=]
	{\mthargfun{\cmfun#4}[#1][#2]{#3}}

\newcommand{\schset}{Sch}
\newcommandx{\SchSet}[3][1=, 2=, 3=]
	{\mthset{\schset#3}[#1][#2]}

\newcommand{\schsym}{\sigma}
\newcommandx{\schSym}[3][1=, 2=, 3=]
	{\mthsym{\schsym#3}[#1][#2]}

\newcommand{\schelm}{\sigma}
\newcommandx{\schElm}[3][1=, 2=, 3=]
	{\mthelm{\schelm#3}[#1][#2]}

\newcommand{\entset}{Ent}
\newcommandx{\EntSet}[4][1=, 2=, 3=, 4=]
	{\mthset{\entset#4}[#1][#2]{#3}}

\newcommand{\entfun}{ent}
\newcommandx{\entFun}[4][1=, 2=, 3=, 4=]
	{\mthargfun{\entfun#4}[#1][#2]{#3}}

\newcommandx{\SOL}[5][1=, 2=, 3=, 4=, 5=]
	{\txtargname{SOL#5{\small\argint{$[$}{#1}{$]$}}}[#2][#3]{#4}\xspace}

\newcommandx{\TL}[5][1=, 2=, 3=, 4=, 5=]
	{\txtargname{TL#5{\small\argint{$[$}{#1}{$]$}}}[#2][#3]{#4}\xspace}

\newcommandx{\PL}[5][1=, 2=, 3=, 4=, 5=]
	{\txtargname{PL#5{\small\argint{$[$}{#1}{$]$}}}[#2][#3]{#4}\xspace}

\newcommand{\fvarset}{FVr}
\newcommandx{\FVarSet}[3][1=, 2=, 3=]
	{\mthset{\fvarset#3}[#1][#2]}

\newcommand{\fvarsym}{x}
\newcommandx{\fvarSym}[3][1=, 2=, 3=]
	{\mthsym{\fvarsym#3}[#1][#2]}

\newcommand{\fvarelm}{x}
\newcommandx{\fvarElm}[3][1=, 2=, 3=]
	{\mthelm{\fvarelm#3}[#1][#2]}

\newcommand{\fvarfun}{fvr}
\newcommandx{\fvarFun}[4][1=, 2=, 3=, 4=]
	{\mthargfun{\fvarfun#4}[#1][#2]{#3}}

\newcommand{\svarset}{SVr}
\newcommandx{\SVarSet}[3][1=, 2=, 3=]
	{\mthset{\svarset#3}[#1][#2]}

\newcommand{\svarsym}{X}
\newcommandx{\svarSym}[3][1=, 2=, 3=]
	{\mthsym{\svarsym#3}[#1][#2]}

\newcommand{\svarelm}{X}
\newcommandx{\svarElm}[3][1=, 2=, 3=]
	{\mthelm{\svarelm#3}[#1][#2]}

\newcommand{\svarfun}{svr}
\newcommandx{\svarFun}[4][1=, 2=, 3=, 4=]
	{\mthargfun{\svarfun#4}[#1][#2]{#3}}

\newcommandx{\MuCalculus}[5][1=, 2=, 3=, 4=, 5=]
	{\txtargname{$\mu$Calculus#5{\small\argint{$[$}{#1}{$]$}}}[#2][#3]{#4}\xspace}

\newcommandx{\LTL}[5][1=, 2=, 3=, 4=, 5=]
	{\txtargname{LTL#5{\small\argint{$[$}{#1}{$]$}}}[#2][#3]{#4}\xspace}

\newcommandx{\PTL}[5][1=, 2=, 3=, 4=, 5=]
	{\txtargname{PTL#5{\small\argint{$[$}{#1}{$]$}}}[#2][#3]{#4}\xspace}

\newcommand{\X}
	{\mthsym{X}}

\newcommand{\F}
	{\mthsym{F}}

\newcommand{\G}
	{\mthsym{G}}

\newcommand{\U}
	{\mthsym{U}}

\newcommand{\R}
	{\mthsym{R}}

\newcommandx{\CTL}[5][1=, 2=, 3=, 4=, 5=]
	{\txtargname{CTL#5{\small\argint{$[$}{#1}{$]$}}}[#2][#3]{#4}\xspace}

\newcommandx{\CTLP}[5][1=, 2=, 3=, 4=, 5=]
	{\txtargname{CTL$^{+}$#5{\small\argint{$[$}{#1}{$]$}}}[#2][#3]{#4}\xspace}

\newcommandx{\CTLS}[5][1=, 2=, 3=, 4=, 5=]
	{\txtargname{CTL$^{\star}$#5{\small\argint{$[$}{#1}{$]$}}}[#2][#3]{#4}\xspace}

\newcommand{\E}
	{\mthsym{E}}

\newcommand{\A}
	{\mthsym{A}}

\newcommandx{\STL}[5][1=, 2=, 3=, 4=, 5=]
	{\txtargname{STL#5{\small\argint{$[$}{#1}{$]$}}}[#2][#3]{#4}\xspace}

\newcommandx{\STLP}[5][1=, 2=, 3=, 4=, 5=]
	{\txtargname{STL$^{+}$#5{\small\argint{$[$}{#1}{$]$}}}[#2][#3]{#4}\xspace}

\newcommandx{\STLS}[5][1=, 2=, 3=, 4=, 5=]
	{\txtargname{STL$^{\star}$#5{\small\argint{$[$}{#1}{$]$}}}[#2][#3]{#4}\xspace}

\newcommandx{\ATL}[5][1=, 2=, 3=, 4=, 5=]
	{\txtargname{ATL#5{\small\argint{$[$}{#1}{$]$}}}[#2][#3]{#4}\xspace}

\newcommandx{\ATLP}[5][1=, 2=, 3=, 4=, 5=]
	{\txtargname{ATL$^{+}$#5{\small\argint{$[$}{#1}{$]$}}}[#2][#3]{#4}\xspace}

\newcommandx{\ATLS}[5][1=, 2=, 3=, 4=, 5=]
	{\txtargname{ATL$^{\star}$#5{\small\argint{$[$}{#1}{$]$}}}[#2][#3]{#4}\xspace}

\newcommandx{\SL}[5][1=, 2=, 3=, 4=, 5=]
	{\txtargname{SL#5{\small\argint{$[$}{#1}{$]$}}}[#2][#3]{#4}\xspace}

\newcommand{\OGSL}[1][]
	{\SL[\argb{1g}{#1}]}

\newcommand{\BGSL}[1][]
	{\SL[\argb{bg}{#1}]}

\newcommand{\NGSL}[1][]
	{\SL[\argb{ng}{#1}]}

\newcommand{\EExs}[1]
	{\ensuremath{%
	\argint{\mbox{$\langle\!\langle$}}{#1}{\mbox{$\rangle\!\rangle$}}%
	}}

\newcommand{\AAll}[1]
	{\ensuremath{\argint{\mbox{$[\:\!\![$}}{#1}{\mbox{$]\:\!\!]$}}}}

\newcommandx{\EF}[5][1=, 2=, 3=, 4=, 5=]
	{\txtargname{EF#5{\small\argint{$[$}{#1}{$]$}}}[#2][#3]{#4}\xspace}

\newcommandx{\SG}[5][1=, 2=, 3=, 4=, 5=]
	{\txtargname{SG#5{\small\argint{$[$}{#1}{$]$}}}[#2][#3]{#4}\xspace}

\newcommandx{\LogTime}[4][1=, 2=, 3=, 4=]
	{\txtargname{LogTime#4}[#2][#3]{#1}\xspace}
\newcommandx{\LogTimeH}[4][1=, 2=, 3=, 4=]
	{\LogTime[#1][#2][#3][#4]-\HComp}
\newcommandx{\LogTimeE}[4][1=, 2=, 3=, 4=]
	{\LogTime[#1][#2][#3][#4]-\EComp}
\newcommandx{\LogTimeC}[4][1=, 2=, 3=, 4=]
	{\LogTime[#1][#2][#3][#4]-\CComp}

\newcommand{\NLogTime}
	{\txtname{N}\LogTime}
\newcommandx{\NLogTimeH}[4][1=, 2=, 3=, 4=]
	{\NLogTime[#1][#2][#3][#4]-\HComp}
\newcommandx{\NLogTimeE}[4][1=, 2=, 3=, 4=]
	{\NLogTime[#1][#2][#3][#4]-\EComp}
\newcommandx{\NLogTimeC}[4][1=, 2=, 3=, 4=]
	{\NLogTime[#1][#2][#3][#4]-\CComp}

\newcommand{\CoNLogTime}
	{\txtname{Co}\NLogTime}
\newcommandx{\CoNLogTimeH}[4][1=, 2=, 3=, 4=]
	{\CoNLogTime[#1][#2][#3][#4]-\HComp}
\newcommandx{\CoNLogTimeE}[4][1=, 2=, 3=, 4=]
	{\CoNLogTime[#1][#2][#3][#4]-\EComp}
\newcommandx{\CoNLogTimeC}[4][1=, 2=, 3=, 4=]
	{\CoNLogTime[#1][#2][#3][#4]-\CComp}

\newcommandx{\ALogTime}
	{\txtname{A}\LogTime}
\newcommandx{\ALogTimeH}[4][1=, 2=, 3=, 4=]
	{\ALogTime[#1][#2][#3][#4]-\HComp}
\newcommandx{\ALogTimeE}[4][1=, 2=, 3=, 4=]
	{\ALogTime[#1][#2][#3][#4]-\EComp}
\newcommandx{\ALogTimeC}[4][1=, 2=, 3=, 4=]
	{\ALogTime[#1][#2][#3][#4]-\CComp}

\newcommandx{\LogSpace}[4][1=, 2=, 3=, 4=]
	{\txtargname{LogSpace#4}[#2][#3]{#1}\xspace}
\newcommandx{\LogSpaceH}[4][1=, 2=, 3=, 4=]
	{\LogSpace[#1][#2][#3][#4]-\HComp}
\newcommandx{\LogSpaceE}[4][1=, 2=, 3=, 4=]
	{\LogSpace[#1][#2][#3][#4]-\EComp}
\newcommandx{\LogSpaceC}[4][1=, 2=, 3=, 4=]
	{\LogSpace[#1][#2][#3][#4]-\CComp}

\newcommandx{\NLogSpace}
	{\txtname{N}\LogSpace}
\newcommandx{\NLogSpaceH}[4][1=, 2=, 3=, 4=]
	{\NLogSpace[#1][#2][#3][#4]-\HComp}
\newcommandx{\NLogSpaceE}[4][1=, 2=, 3=, 4=]
	{\NLogSpace[#1][#2][#3][#4]-\EComp}
\newcommandx{\NLogSpaceC}[4][1=, 2=, 3=, 4=]
	{\NLogSpace[#1][#2][#3][#4]-\CComp}

\newcommandx{\CoNLogSpace}
	{\txtname{Co}\NLogSpace}
\newcommandx{\CoNLogSpaceH}[4][1=, 2=, 3=, 4=]
	{\CoNLogSpace[#1][#2][#3][#4]-\HComp}
\newcommandx{\CoNLogSpaceE}[4][1=, 2=, 3=, 4=]
	{\CoNLogSpace[#1][#2][#3][#4]-\EComp}
\newcommandx{\CoNLogSpaceC}[4][1=, 2=, 3=, 4=]
	{\CoNLogSpace[#1][#2][#3][#4]-\CComp}

\newcommandx{\ALogSpace}
	{\txtname{A}\LogSpace}
\newcommandx{\ALogSpaceH}[4][1=, 2=, 3=, 4=]
	{\ALogSpace[#1][#2][#3][#4]-\HComp}
\newcommandx{\ALogSpaceE}[4][1=, 2=, 3=, 4=]
	{\ALogSpace[#1][#2][#3][#4]-\EComp}
\newcommandx{\ALogSpaceC}[4][1=, 2=, 3=, 4=]
	{\ALogSpace[#1][#2][#3][#4]-\CComp}

\newcommandx{\PTime}[4][1=, 2=, 3=, 4=]
	{\txtargname{PTime#4}[#2][#3]{#1}\xspace}
\newcommandx{\PTimeH}[4][1=, 2=, 3=, 4=]
	{\PTime[#1][#2][#3][#4]-\HComp}
\newcommandx{\PTimeE}[4][1=, 2=, 3=, 4=]
	{\PTime[#1][#2][#3][#4]-\EComp}
\newcommandx{\PTimeC}[4][1=, 2=, 3=, 4=]
	{\PTime[#1][#2][#3][#4]-\CComp}

\newcommandx{\UPTime}
	{\txtname{U}\PTime}
\newcommandx{\UPTimeH}[4][1=, 2=, 3=, 4=]
	{\UPTime[#1][#2][#3][#4]-\HComp}
\newcommandx{\UPTimeE}[4][1=, 2=, 3=, 4=]
	{\UPTime[#1][#2][#3][#4]-\EComp}
\newcommandx{\UPTimeC}[4][1=, 2=, 3=, 4=]
	{\UPTime[#1][#2][#3][#4]-\CComp}

\newcommandx{\CoUPTime}
	{\txtname{Co}\UPTime}
\newcommandx{\CoUPTimeH}[4][1=, 2=, 3=, 4=]
	{\CoUPTime[#1][#2][#3][#4]-\HComp}
\newcommandx{\CoUPTimeE}[4][1=, 2=, 3=, 4=]
	{\CoUPTime[#1][#2][#3][#4]-\EComp}
\newcommandx{\CoUPTimeC}[4][1=, 2=, 3=, 4=]
	{\CoUPTime[#1][#2][#3][#4]-\CComp}

\newcommandx{\NPTime}
	{\txtname{N}\PTime}
\newcommandx{\NPTimeH}[4][1=, 2=, 3=, 4=]
	{\NPTime[#1][#2][#3][#4]-\HComp}
\newcommandx{\NPTimeE}[4][1=, 2=, 3=, 4=]
	{\NPTime[#1][#2][#3][#4]-\EComp}
\newcommandx{\NPTimeC}[4][1=, 2=, 3=, 4=]
	{\NPTime[#1][#2][#3][#4]-\CComp}

\newcommandx{\CoNPTime}
	{\txtname{Co}\NPTime}
\newcommandx{\CoNPTimeH}[4][1=, 2=, 3=, 4=]
	{\CoNPTime[#1][#2][#3][#4]-\HComp}
\newcommandx{\CoNPTimeE}[4][1=, 2=, 3=, 4=]
	{\CoNPTime[#1][#2][#3][#4]-\EComp}
\newcommandx{\CoNPTimeC}[4][1=, 2=, 3=, 4=]
	{\CoNPTime[#1][#2][#3][#4]-\CComp}

\newcommandx{\APTime}
	{\txtname{A}\PTime}
\newcommandx{\APTimeH}[4][1=, 2=, 3=, 4=]
	{\APTime[#1][#2][#3][#4]-\HComp}
\newcommandx{\APTimeE}[4][1=, 2=, 3=, 4=]
	{\APTime[#1][#2][#3][#4]-\EComp}
\newcommandx{\APTimeC}[4][1=, 2=, 3=, 4=]
	{\APTime[#1][#2][#3][#4]-\CComp}

\newcommandx{\PSpace}[4][1=, 2=, 3=, 4=]
	{\txtargname{PSpace#4}[#2][#3]{#1}\xspace}
\newcommandx{\PSpaceH}[4][1=, 2=, 3=, 4=]
	{\PSpace[#1][#2][#3][#4]-\HComp}
\newcommandx{\PSpaceE}[4][1=, 2=, 3=, 4=]
	{\PSpace[#1][#2][#3][#4]-\EComp}
\newcommandx{\PSpaceC}[4][1=, 2=, 3=, 4=]
	{\PSpace[#1][#2][#3][#4]-\CComp}

\newcommandx{\NPSpace}
	{\txtname{N}\PSpace}
\newcommandx{\NPSpaceH}[4][1=, 2=, 3=, 4=]
	{\NPSpace[#1][#2][#3][#4]-\HComp}
\newcommandx{\NPSpaceE}[4][1=, 2=, 3=, 4=]
	{\NPSpace[#1][#2][#3][#4]-\EComp}
\newcommandx{\NPSpaceC}[4][1=, 2=, 3=, 4=]
	{\NPSpace[#1][#2][#3][#4]-\CComp}

\newcommandx{\CoNPSpace}
	{\txtname{Co}\NPSpace}
\newcommandx{\CoNPSpaceH}[4][1=, 2=, 3=, 4=]
	{\CoNPSpace[#1][#2][#3][#4]-\HComp}
\newcommandx{\CoNPSpaceE}[4][1=, 2=, 3=, 4=]
	{\CoNPSpace[#1][#2][#3][#4]-\EComp}
\newcommandx{\CoNPSpaceC}[4][1=, 2=, 3=, 4=]
	{\CoNPSpace[#1][#2][#3][#4]-\CComp}

\newcommandx{\APSpace}
	{\txtname{A}\PSpace}
\newcommandx{\APSpaceH}[4][1=, 2=, 3=, 4=]
	{\APSpace[#1][#2][#3][#4]-\HComp}
\newcommandx{\APSpaceE}[4][1=, 2=, 3=, 4=]
	{\APSpace[#1][#2][#3][#4]-\EComp}
\newcommandx{\APSpaceC}[4][1=, 2=, 3=, 4=]
	{\APSpace[#1][#2][#3][#4]-\CComp}

\newcommandx{\ExpTime}[4][1=, 2=, 3=, 4=]
	{\txtargname{ExpTime#4}[#2][#3]{#1}\xspace}
\newcommandx{\ExpTimeH}[4][1=, 2=, 3=, 4=]
	{\ExpTime[#1][#2][#3][#4]-\HComp}
\newcommandx{\ExpTimeE}[4][1=, 2=, 3=, 4=]
	{\ExpTime[#1][#2][#3][#4]-\EComp}
\newcommandx{\ExpTimeC}[4][1=, 2=, 3=, 4=]
	{\ExpTime[#1][#2][#3][#4]-\CComp}

\newcommandx{\NExpTime}
	{\txtname{N}\ExpTime}
\newcommandx{\NExpTimeH}[4][1=, 2=, 3=, 4=]
	{\NExpTime[#1][#2][#3][#4]-\HComp}
\newcommandx{\NExpTimeE}[4][1=, 2=, 3=, 4=]
	{\NExpTime[#1][#2][#3][#4]-\EComp}
\newcommandx{\NExpTimeC}[4][1=, 2=, 3=, 4=]
	{\NExpTime[#1][#2][#3][#4]-\CComp}

\newcommandx{\CoNExpTime}
	{\txtname{Co}\NExpTime}
\newcommandx{\CoNExpTimeH}[4][1=, 2=, 3=, 4=]
	{\CoNExpTime[#1][#2][#3][#4]-\HComp}
\newcommandx{\CoNExpTimeE}[4][1=, 2=, 3=, 4=]
	{\CoNExpTime[#1][#2][#3][#4]-\EComp}
\newcommandx{\CoNExpTimeC}[4][1=, 2=, 3=, 4=]
	{\CoNExpTime[#1][#2][#3][#4]-\CComp}

\newcommandx{\AExpTime}
	{\txtname{A}\ExpTime}
\newcommandx{\AExpTimeH}[4][1=, 2=, 3=, 4=]
	{\AExpTime[#1][#2][#3][#4]-\HComp}
\newcommandx{\AExpTimeE}[4][1=, 2=, 3=, 4=]
	{\AExpTime[#1][#2][#3][#4]-\EComp}
\newcommandx{\AExpTimeC}[4][1=, 2=, 3=, 4=]
	{\AExpTime[#1][#2][#3][#4]-\CComp}

\newcommandx{\ExpSpace}[4][1=, 2=, 3=, 4=]
	{\txtargname{ExpSpace#4}[#2][#3]{#1}\xspace}
\newcommandx{\ExpSpaceH}[4][1=, 2=, 3=, 4=]
	{\ExpSpace[#1][#2][#3][#4]-\HComp}
\newcommandx{\ExpSpaceE}[4][1=, 2=, 3=, 4=]
	{\ExpSpace[#1][#2][#3][#4]-\EComp}
\newcommandx{\ExpSpaceC}[4][1=, 2=, 3=, 4=]
	{\ExpSpace[#1][#2][#3][#4]-\CComp}

\newcommandx{\NExpSpace}
	{\txtname{N}\ExpSpace}
\newcommandx{\NExpSpaceH}[4][1=, 2=, 3=, 4=]
	{\NExpSpace[#1][#2][#3][#4]-\HComp}
\newcommandx{\NExpSpaceE}[4][1=, 2=, 3=, 4=]
	{\NExpSpace[#1][#2][#3][#4]-\EComp}
\newcommandx{\NExpSpaceC}[4][1=, 2=, 3=, 4=]
	{\NExpSpace[#1][#2][#3][#4]-\CComp}

\newcommandx{\CoNExpSpace}
	{\txtname{Co}\NExpSpace}
\newcommandx{\CoNExpSpaceH}[4][1=, 2=, 3=, 4=]
	{\CoNExpSpace[#1][#2][#3][#4]-\HComp}
\newcommandx{\CoNExpSpaceE}[4][1=, 2=, 3=, 4=]
	{\CoNExpSpace[#1][#2][#3][#4]-\EComp}
\newcommandx{\CoNExpSpaceC}[4][1=, 2=, 3=, 4=]
	{\CoNExpSpace[#1][#2][#3][#4]-\CComp}

\newcommandx{\AExpSpace}
	{\txtname{A}\ExpSpace}
\newcommandx{\AExpSpaceH}[4][1=, 2=, 3=, 4=]
	{\AExpSpace[#1][#2][#3][#4]-\HComp}
\newcommandx{\AExpSpaceE}[4][1=, 2=, 3=, 4=]
	{\AExpSpace[#1][#2][#3][#4]-\EComp}
\newcommandx{\AExpSpaceC}[4][1=, 2=, 3=, 4=]
	{\AExpSpace[#1][#2][#3][#4]-\CComp}

\newcommandx{\NonElmTime}[4][1=, 2=, 3=, 4=]
	{\txtargname{NonElementaryTime#4}[#2][#3]{#1}\xspace}
\newcommandx{\NonElmTimeH}[4][1=, 2=, 3=, 4=]
	{\NonElmTime[#1][#2][#3][#4]-\HComp}
\newcommandx{\NonElmTimeE}[4][1=, 2=, 3=, 4=]
	{\NonElmTime[#1][#2][#3][#4]-\EComp}
\newcommandx{\NonElmTimeC}[4][1=, 2=, 3=, 4=]
	{\NonElmTime[#1][#2][#3][#4]-\CComp}

\newcommandx{\NonElmSpace}[4][1=, 2=, 3=, 4=]
	{\txtargname{NonElementarySpace#4}[#2][#3]{#1}\xspace}
\newcommandx{\NonElmSpaceH}[4][1=, 2=, 3=, 4=]
	{\NonElmSpace[#1][#2][#3][#4]-\HComp}
\newcommandx{\NonElmSpaceE}[4][1=, 2=, 3=, 4=]
	{\NonElmSpace[#1][#2][#3][#4]-\EComp}
\newcommandx{\NonElmSpaceC}[4][1=, 2=, 3=, 4=]
	{\NonElmSpace[#1][#2][#3][#4]-\CComp}

\newcommandx{\DLHier}[4][2=, 3=, 4=]
	{\mthargset[0]{\Delta#4}[#1][#3]{#2}\xspace}
\newcommandx{\DLHierH}[4][2=, 3=, 4=]
	{\DLHier{#1}[#2][#3][#4]-\HComp}
\newcommandx{\DLHierE}[4][2=, 3=, 4=]
	{\DLHier{#1}[#2][#3][#4]-\EComp}
\newcommandx{\DLHierC}[4][2=, 3=, 4=]
	{\DLHier{#1}[#2][#3][#4]-\CComp}

\newcommandx{\ELHier}[4][2=, 3=, 4=]
	{\mthargset[0]{\Sigma#4}[#1][#3]{#2}\xspace}
\newcommandx{\ELHierH}[4][2=, 3=, 4=]
	{\ELHier{#1}[#2][#3][#4]-\HComp}
\newcommandx{\ELHierE}[4][2=, 3=, 4=]
	{\ELHier{#1}[#2][#3][#4]-\EComp}
\newcommandx{\ELHierC}[4][2=, 3=, 4=]
	{\ELHier{#1}[#2][#3][#4]-\CComp}

\newcommandx{\ULHier}[4][2=, 3=, 4=]
	{\mthargset[0]{\Pi#4}[#1][#3]{#2}\xspace}
\newcommandx{\ULHierH}[4][2=, 3=, 4=]
	{\ULHier{#1}[#2][#3][#4]-\HComp}
\newcommandx{\ULHierE}[4][2=, 3=, 4=]
	{\ULHier{#1}[#2][#3][#4]-\EComp}
\newcommandx{\ULHierC}[4][2=, 3=, 4=]
	{\ULHier{#1}[#2][#3][#4]-\CComp}

\newcommandx{\DBHier}[4][2=, 3=, 4=]
	{\mthargset[3]{\Delta#4}[#1][#3]{#2}\xspace}
\newcommandx{\DBHierH}[4][2=, 3=, 4=]
	{\DBHier{#1}[#2][#3][#4]-\HComp}
\newcommandx{\DBHierE}[4][2=, 3=, 4=]
	{\DBHier{#1}[#2][#3][#4]-\EComp}
\newcommandx{\DBHierC}[4][2=, 3=, 4=]
	{\DBHier{#1}[#2][#3][#4]-\CComp}

\newcommandx{\EBHier}[4][2=, 3=, 4=]
	{\mthargset[3]{\Sigma#4}[#1][#3]{#2}\xspace}
\newcommandx{\EBHierH}[4][2=, 3=, 4=]
	{\EBHier{#1}[#2][#3][#4]-\HComp}
\newcommandx{\EBHierE}[4][2=, 3=, 4=]
	{\EBHier{#1}[#2][#3][#4]-\EComp}
\newcommandx{\EBHierC}[4][2=, 3=, 4=]
	{\EBHier{#1}[#2][#3][#4]-\CComp}

\newcommandx{\UBHier}[4][2=, 3=, 4=]
	{\mthargset[3]{\Pi#4}[#1][#3]{#2}\xspace}
\newcommandx{\UBHierH}[4][2=, 3=, 4=]
	{\UBHier{#1}[#2][#3][#4]-\HComp}
\newcommandx{\UBHierE}[4][2=, 3=, 4=]
	{\UBHier{#1}[#2][#3][#4]-\EComp}
\newcommandx{\UBHierC}[4][2=, 3=, 4=]
	{\UBHier{#1}[#2][#3][#4]-\CComp}

\newcommandx{\DPolHier}[4][2=, 3=, 4=]
	{\DLHier{#1}[#2][\argb{\mathrm{P}}{#3}][#4]}
\newcommandx{\DPolHierH}[4][2=, 3=, 4=]
	{\DPolHier{#1}[#2][#3][#4]-\HComp}
\newcommandx{\DPolHierE}[4][2=, 3=, 4=]
	{\DPolHier{#1}[#2][#3][#4]-\EComp}
\newcommandx{\DPolHierC}[4][2=, 3=, 4=]
	{\DPolHier{#1}[#2][#3][#4]-\CComp}

\newcommandx{\EPolHier}[4][2=, 3=, 4=]
	{\ELHier{#1}[#2][\argb{\mathrm{P}}{#3}][#4]}
\newcommandx{\EPolHierH}[4][2=, 3=, 4=]
	{\EPolHier{#1}[#2][#3][#4]-\HComp}
\newcommandx{\EPolHierE}[4][2=, 3=, 4=]
	{\EPolHier{#1}[#2][#3][#4]-\EComp}
\newcommandx{\EPolHierC}[4][2=, 3=, 4=]
	{\EPolHier{#1}[#2][#3][#4]-\CComp}

\newcommandx{\UPolHier}[4][2=, 3=, 4=]
	{\ULHier{#1}[#2][\argb{\mathrm{P}}{#3}][#4]}
\newcommandx{\UPolHierH}[4][2=, 3=, 4=]
	{\UPolHier{#1}[#2][#3][#4]-\HComp}
\newcommandx{\UPolHierE}[4][2=, 3=, 4=]
	{\UPolHier{#1}[#2][#3][#4]-\EComp}
\newcommandx{\UPolHierC}[4][2=, 3=, 4=]
	{\UPolHier{#1}[#2][#3][#4]-\CComp}

\newcommandx{\DAriHier}[4][2=, 3=, 4=]
	{\DLHier{#1}[#2][\argb{0}{#3}][#4]}
\newcommandx{\DAriHierH}[4][2=, 3=, 4=]
	{\DAriHier{#1}[#2][#3][#4]-\HComp}
\newcommandx{\DAriHierE}[4][2=, 3=, 4=]
	{\DAriHier{#1}[#2][#3][#4]-\EComp}
\newcommandx{\DAriHierC}[4][2=, 3=, 4=]
	{\DAriHier{#1}[#2][#3][#4]-\CComp}

\newcommandx{\EAriHier}[4][2=, 3=, 4=]
	{\ELHier{#1}[#2][\argb{0}{#3}][#4]}
\newcommandx{\EAriHierH}[4][2=, 3=, 4=]
	{\EAriHier{#1}[#2][#3][#4]-\HComp}
\newcommandx{\EAriHierE}[4][2=, 3=, 4=]
	{\EAriHier{#1}[#2][#3][#4]-\EComp}
\newcommandx{\EAriHierC}[4][2=, 3=, 4=]
	{\EAriHier{#1}[#2][#3][#4]-\CComp}

\newcommandx{\UAriHier}[4][2=, 3=, 4=]
	{\ULHier{#1}[#2][\argb{0}{#3}][#4]}
\newcommandx{\UAriHierH}[4][2=, 3=, 4=]
	{\UAriHier{#1}[#2][#3][#4]-\HComp}
\newcommandx{\UAriHierE}[4][2=, 3=, 4=]
	{\UAriHier{#1}[#2][#3][#4]-\EComp}
\newcommandx{\UAriHierC}[4][2=, 3=, 4=]
	{\UAriHier{#1}[#2][#3][#4]-\CComp}

\newcommandx{\DAnaHier}[4][2=, 3=, 4=]
	{\DLHier{#1}[#2][\argb{1}{#3}][#4]}
\newcommandx{\DAnaHierH}[4][2=, 3=, 4=]
	{\DAnaHier{#1}[#2][#3][#4]-\HComp}
\newcommandx{\DAnaHierE}[4][2=, 3=, 4=]
	{\DAnaHier{#1}[#2][#3][#4]-\EComp}
\newcommandx{\DAnaHierC}[4][2=, 3=, 4=]
	{\DAnaHier{#1}[#2][#3][#4]-\CComp}

\newcommandx{\EAnaHier}[4][2=, 3=, 4=]
	{\ELHier{#1}[#2][\argb{1}{#3}][#4]}
\newcommandx{\EAnaHierH}[4][2=, 3=, 4=]
	{\EAnaHier{#1}[#2][#3][#4]-\HComp}
\newcommandx{\EAnaHierE}[4][2=, 3=, 4=]
	{\EAnaHier{#1}[#2][#3][#4]-\EComp}
\newcommandx{\EAnaHierC}[4][2=, 3=, 4=]
	{\EAnaHier{#1}[#2][#3][#4]-\CComp}

\newcommandx{\UAnaHier}[4][2=, 3=, 4=]
	{\ULHier{#1}[#2][\argb{1}{#3}][#4]}
\newcommandx{\UAnaHierH}[4][2=, 3=, 4=]
	{\UAnaHier{#1}[#2][#3][#4]-\HComp}
\newcommandx{\UAnaHierE}[4][2=, 3=, 4=]
	{\UAnaHier{#1}[#2][#3][#4]-\EComp}
\newcommandx{\UAnaHierC}[4][2=, 3=, 4=]
	{\UAnaHier{#1}[#2][#3][#4]-\CComp}

\newcommandx{\DBorHier}[4][2=, 3=, 4=]
	{\DBHier{#1}[#2][\argb{\mathrm{B}}{#3}][#4]}
\newcommandx{\DBorHierH}[4][2=, 3=, 4=]
	{\DBorHier{#1}[#2][#3][#4]-\HComp}
\newcommandx{\DBorHierE}[4][2=, 3=, 4=]
	{\DBorHier{#1}[#2][#3][#4]-\EComp}
\newcommandx{\DBorHierC}[4][2=, 3=, 4=]
	{\DBorHier{#1}[#2][#3][#4]-\CComp}

\newcommandx{\EBorHier}[4][2=, 3=, 4=]
	{\EBHier{#1}[#2][\argb{\mathrm{B}}{#3}][#4]}
\newcommandx{\EBorHierH}[4][2=, 3=, 4=]
	{\EBorHier{#1}[#2][#3][#4]-\HComp}
\newcommandx{\EBorHierE}[4][2=, 3=, 4=]
	{\EBorHier{#1}[#2][#3][#4]-\EComp}
\newcommandx{\EBorHierC}[4][2=, 3=, 4=]
	{\EBorHier{#1}[#2][#3][#4]-\CComp}

\newcommandx{\UBorHier}[4][2=, 3=, 4=]
	{\UBHier{#1}[#2][\argb{\mathrm{B}}{#3}][#4]}
\newcommandx{\UBorHierH}[4][2=, 3=, 4=]
	{\UBorHier{#1}[#2][#3][#4]-\HComp}
\newcommandx{\UBorHierE}[4][2=, 3=, 4=]
	{\UBorHier{#1}[#2][#3][#4]-\EComp}
\newcommandx{\UBorHierC}[4][2=, 3=, 4=]
	{\UBorHier{#1}[#2][#3][#4]-\CComp}

\newcommand{\HComp}
	{\txtname{hard}\xspace}

\newcommand{\EComp}
	{\txtname{easy}\xspace}

\newcommand{\CComp}
	{\txtname{complete}\xspace}

\usepackage{varioref}
\newcommand{\CHPSL}{\txtname{CHP}-\SL}

\providecommand{\strFun}[1][]{\mthfun{f}}

\providecommand{\DSet}[1][]{\mthset[#1]{D}}

\providecommand{\HRel}[1][]{\mthrel[#1]{H}}

\providecommand{\VRel}[1][]{\mthrel[#1]{V}}

\newcommand{\DSTuple}[4]
	{
	\ifx&#4&
		\tuplec{#1}{#2}{#3}
	\else
		\tupled{#1}{#2}{#3}{#4}
	\fi
	}

\newcommand{\DSDef}[2][]
	{
	\DSTuple {\DSet[#2]} {\HRel[#2]} {\VRel[#2]} {#1}
	}

\newcommand{\DSStruct}[1][]
	{
	\DSDef [#1] {}
	}

\providecommand{\dsmFun}[1][]{\mthfun[#1]{\partial}}

\newcommand{\QPSet}[1][]{\mthset[#1]{Qnt}}

\newcommand{\qpSym}[1][]{\mthsym[#1]{\wp}}

\newcommand{\qpElm}[1][]{\mthelm[#1]{\wp}}

\newcommand{\QPVSet}[1][]{\mthset[#1]{V}}

\newcommand{\QPEVSet}[2][]{\EExs{#2}}

\newcommand{\QPAVSet}[2][]{\AAll{#2}}

\newcommand{\qpordRel}[1][]{\mthrel[#1]{<}}

\newcommand{\qpdepRel}[1][]{\mthrel[#1]{\rightsquigarrow}}

\newcommand{\QPDepSet}[1][]{\mthset[#1]{Dep}}

\newcommand{\BPSet}[1][]{\mthset[#1]{Bnd}}

\newcommand{\bpSym}[1][]{\mthsym[#1]{\flat}}

\newcommand{\bpElm}[1][]{\mthelm[#1]{\flat}}

\providecommand{\ValSet}[1][]{\mthset[#1]{Val}}

\providecommand{\valFun}[1][]{\mthfun[#1]{v}}

\newcommand{\skodepfun}{\mthsym{Sdf}\xspace}

\providecommand{\LabSet}[1][]{\mthset[#1]{\Sigma}}

\providecommand{\DirSet}[1][]{\mthset[#1]{\Delta}}

\providecommand{\TSet}[1][]{\mthset[#1]{T}}

\providecommand{\vFun}[1][]{\mthfun[#1]{v}}

\newcommand{\LTTuple}[4]
	{
	\ifx&#1&
		\ifx&#2&
			\tupleb{#3}{#4}
		\else
			\tuplec{#2}{#3}{#4}
		\fi
	\else
		\ifx&#2&
			\tuplec{#1}{#3}{#4}
		\else
			\tupled{#1}{#2}{#3}{#4}
		\fi
	\fi
	}

\newcommand{\LTDef}[3][]
	{
	\LTTuple {#2} {#3} {\TSet[#1]} {\vFun[#1]}
	}

\newcommand{\LTStruct}[1][]
	{
	\LTDef [#1] {} {}
	}

\newcommand{\NTA}{\txtname{Nta}}
\newcommand{\UTA}{\txtname{Uta}}
\newcommand{\ATA}{\txtname{Ata}}

\newcommand{\UCT}{\txtname{Uct}}
\newcommand{\ACT}{\txtname{Act}}

\newcommand{\NPT}{\txtname{Npt}}
\newcommand{\UPT}{\txtname{Upt}}
\newcommand{\APT}{\txtname{Apt}}

\providecommand{\SymSet}[1][]{\mthset[#1]{\Sigma}}

\providecommand{\QSet}[1][]{\mthset[#1]{Q}}

\providecommand{\PSet}[1][]{\mthset[#1]{P}}

\providecommand{\atFun}[1][]{\mthfun[#1]{\delta}}

\newcommand{\TATuple}[6]
	{
	\ifx&#2&
		\ifx&#6&
			\tupled{#1}{#3}{#4}{#5}
		\else
			\tuplee{#1}{#3}{#4}{#5}{#6}
		\fi
	\else
		\ifx&#6&
			\tuplee{#1}{#2}{#3}{#4}{#5}
		\else
			\tuplef{#1}{#2}{#3}{#4}{#5}{#6}
		\fi
	\fi
	}

\newcommand{\TMTuple}[8]
	{
	\ifx&#3&
		\ifx&#8&
			\tuplef{#1}{#2}{#4}{#5}{#6}{#7}
		\else
			\tupleg{#1}{#2}{#4}{#5}{#6}{#7}{#8}
		\fi
	\else
		\ifx&#8&
			\tupleg{#1}{#2}{#3}{#4}{#5}{#6}{#7}
		\else
			\tupleh{#1}{#2}{#3}{#4}{#5}{#6}{#7}{#8}
		\fi
	\fi
	}

\newcommand{\TADef}[4][q_{0}]
	{
	\TATuple {\LabSet} {#2} {\QSet[#4]} {\atFun[#4]} {#1} {#3}
	}

\newcommand{\TAStruct}[2][q_{0}]
	{
	\TADef [#1] {#2} {\aleph} {}
	}

\newcommand{\ATAStruct}[1][q_{0}]
	{
	\TAStruct [#1] {\DirSet}
	}

\newcommand{\NBW}{\txtname{Nbw}}

\newcommand{\UCW}{\txtname{Ucw}}

\providecommand{\SymSet}[1][]{\mthset[#1]{\Sigma}}

\providecommand{\QSet}[1][]{\mthset[#1]{Q}}

\providecommand{\PSet}[1][]{\mthset[#1]{P}}

\providecommand{\atFun}[1][]{\mthfun[#1]{\delta}}

\newcommand{\WATuple}[5]
	{
	\ifx&#5&
		\tupled{#1}{#2}{#3}{#4}
	\else
		\tuplee{#1}{#2}{#3}{#4}{#5}
	\fi
	}

\newcommand{\WMTuple}[7]
	{
	\ifx&#7&
		\tuplef{#1}{#2}{#3}{#4}{#5}{#6}
	\else
		\tupleg{#1}{#2}{#3}{#4}{#5}{#6}{#7}
	\fi
	}

\newcounter{explevautnot}

\providecommand{\autknd}{APT}

\providecommandx{\NFA}[5][1=, 2=, 3=, 4=, 5=]
	{\txtargname{Nfa#5{\small\argint{$[$}{#1}{$]$}}}[#2][#3]{#4}\xspace}

\providecommandx{\APT}[5][1=, 2=, 3=, 4=, 5=]
	{\txtargname{Apt#5{\small\argint{$[$}{#1}{$]$}}}[#2][#3]{#4}\xspace}

\providecommandx{\AlphSet}[3][1=, 2=, 3=]
	{\mthset{\Sigma#3}[#1][#2]}

\providecommandx{\alphSym}[3][1=, 2=, 3=]
	{\mthsym{\sigma#3}[#1][#2]}

\providecommandx{\alphElm}[3][1=, 2=, 3=]
	{\mthelm{\sigma#3}[#1][#2]}

\providecommandx{\AutStSet}[3][1=, 2=, 3=]
	{\mthset{Q#3}[#1][#2]}

\providecommandx{\autstSym}[3][1=, 2=, 3=]
	{\mthsym{q#3}[#1][#2]}

\providecommandx{\autstElm}[3][1=, 2=, 3=]
	{\mthelm{q#3}[#1][#2]}

\providecommandx{\trFun}[3][1=, 2=, 3=]
	{\mthfun{\delta#3}[#1][#2]}

\providecommandx{\BoolCom}{\BName[][+]}

\providecommand{\accCon}[1][]
	{%
	\IfStrEqCase{\argdef{#1}{\autknd}}
		{%
		{NFA}{\mthset{F}}%
		{APT}{\mthset{\beta}}%
		{NBA}{\mthset{\alpha}}%
		}
		[\ensuremath{\clubsuit}]%
	}

\providecommandx{\runElm}[4][1=, 2=, 3=, 4=]
	{\mthsym{\chi#4}[#1][#2]{#3}}

\providecommandx{\runSym}[4][1=, 2=, 3=, 4=]
	{\mthsym{\chi#4}[#1][#2]{#3}}

\providecommandx{\runStr}
	{\tupleb{\TSet}{\rFun}}

\providecommandx{\trkElm}[4][1=, 2=, 3=, 4=]
	{\mthelm{\eta#4}[#1][#2]{#3}}

\providecommandx{\trkSym}[4][1=, 2=, 3=, 4=]
	{\mthsym{\eta#4}[#1][#2]{#3}}

\providecommandx{\PthSet}[3][1=, 2=, 3=]
	{\mthset{Pth#3}[#1][#2]}

\providecommandx{\pthSym}[3][1=, 2=, 3=]
	{\mthsym{\varpi#3}[#1][#2]}

\providecommandx{\pthElm}[3][1=, 2=, 3=]
	{\mthelm{\varpi#3}[#1][#2]}

\providecommandx{\AutName}
	{\mthname{A}}

\providecommandx{\AutCls}[5][1=, 2=, 3=, 4=, 5=]
	{\mthset[#5]{Aut#4\text{\small\txtname{\argint{$[$}{#1}{$]$}}}}[#2][#3]}

\providecommand{\AutStr}[1][]
	{%
	\IfStrEqCase{\argdef{#1}{\autknd}}
		{%
		{NFA}{\tupled{\AlphSet}{\AutStSet}{\trFun}{\accCon[NFA]}}%
		{APT}{\tupled{\AlphSet}{\AutStSet}{\trFun}{\accCon[APT]}}%
		{NBA}{\tupled{\AlphSet}{\AutStSet}{\trFun}{\accCon[NBA]}}%
		}
		[\ensuremath{\clubsuit}]%
	}

\newcommand{\SComp}[2]{$\Sigma_{\arga{#1}}^{\arga{#2}}$}
\newcommand{\SCompH}[2]{\SComp{#1}{#2}-\HComp}
\newcommand{\SCompC}[2]{\SComp{#1}{#2}-\CComp}

\newcommand{\sub}[1]{\mthfun{sub}(#1)}

\newcommand{\snt}[1]{\mthfun{snt}(#1)}

\newcommand{\psnt}[1]{\mthfun{psnt}(#1)}

\newcommand{\free}[1]{\mthfun{free}(#1)}

\newcommand{\playElm}{\pi}

\providecommand{\bmodels}{\models_{\txtname{B}}}

\newcommandx{\DT}[5][1=, 2=, 3=, 4=, 5=]
	{\txtargname{DT#5{\small\argint{$[$}{#1}{$]$}}}[#2][#3]{#4}\xspace}

\providecommand{\LangSet}[1][]{\mthset[#1]{L}}

\newcommand{\headFun}[1][]{\mthfun[#1]{head}}

\newcommand{\bodyFun}[1][]{\mthfun[#1]{body}}

\newcommand{\mask}{\mthfun{mask}}

\renewcommand{\root}{\varepsilon}

\renewcommand{\prj}{\mthfun{prj}}

\renewcommand{\smfun}{\theta}

\newcommand{\congamtree}{CGT}
\newcommandx{\ConGamTree}[5][1=, 2=, 3=, 4=, 5=]
	{\txtargname{\congamtree#5{\small\argint{$[$}{#1}{$]$}}}[#2][#3]{#4}\xspace}
\newcommand{\CGT}{\ConGamTree}

\newcommandx{\NonElm}[4][1=, 2=, 3=, 4=]
	{\txtargname{NonElementary#4}[#2][#3]{#1}\xspace}
\newcommandx{\NonElmH}[4][1=, 2=, 3=, 4=]
	{\NonElm[#1][#2][#3][#4]-\HComp}
\newcommandx{\NonElmE}[4][1=, 2=, 3=, 4=]
	{\NonElm[#1][#2][#3][#4]-\EComp}
\newcommandx{\NonElmC}[4][1=, 2=, 3=, 4=]
	{\NonElm[#1][#2][#3][#4]-\CComp}

\providecommand{\BoolSet}[1][]{\mathcal{B}}

\providecommand{\PBoolSet}[1][]{\mathcal{B}_{#1}^{+}}

\providecommand{\LangSet}[1][]{\mthset[#1]{L}}

\providecommand{\infFun}[1][]{\mthfun[#1]{inf}}

\usepackage{tikz}
\usetikzlibrary{arrows,shapes}

\usepackage{wrapfig}
\usepackage[caption=false,font=footnotesize]{subfig}

\tikzstyle{every node} =
	[draw = none, fill = white, thin]
\tikzstyle{every edge} +=
	[black, thick]

\tikzstyle{noall} =
	[draw = none, fill = none]
\tikzstyle{nodraw} =
	[draw = none, fill = white]
\tikzstyle{nofill} =
	[draw = black, fill = none]

\tikzstyle{cnode} =
	[circle, draw = black]
\tikzstyle{snode} =
	[regular polygon, regular polygon sides = 4, draw = gray!50]
\tikzstyle{lnode} =
	[diamond, draw = gray!75]
\tikzstyle{pnode} =
	[regular polygon, regular polygon sides = 5, draw = gray]

\AfterEndPreamble
	{

\newcommand{\figexmppd}
	{
	\begin{wrapfigure}[25]{r}{0.450\textwidth}
 		\vspace{-1.75em}
		\begin{center}
			\footnotesize
			\mbox{\scalebox{1.00}[1.00]{
			\begin{tikzpicture}
				[node distance = 3.25cm, bend angle = 25, shorten >= 2pt, shorten <=
				2pt]
				\node [cnode]
							(SI)
							{$\stackrel{\sSym[0]}{\emptyset}$};
				\node [cnode]
							(SA)
							[below left of = SI]
							{$\stackrel{\sSym[ {\ASym[1]} ]}{\fSym[ {\ASym[1]} ]}$};
				\node [cnode]
							(SAJ)
							[above left of = SI]
							{$\stackrel{\sSym[ {\ASym[1]} j]}{\fSym[ {\ASym[1]} ]}$};
				\node [cnode]
							(SJ)
							[below of = SI]
							{$\stackrel{\sSym[j]}{\emptyset}$};
				\node [cnode]
							(SAB)
							[above of = SI, node distance = 4cm]
							{$\stackrel{\sSym[ {\ASym[1] \ASym[2]} ]}{\fSym[ {\ASym[1]} ],
							\fSym[ {\ASym[2]} ]}$};
				\node [cnode]
							(SB)
							[below right of = SI]
							{$\stackrel{\sSym[ {\ASym[2]} ]}{\fSym[ {\ASym[2]} ]}$};
				\node [cnode]
							(SBJ)
							[above right of = SI]
							{$\stackrel{\sSym[ {\ASym[2]} j]}{\fSym[ {\ASym[2]} ]}$};
				\path[-stealth']
					(SI)	edge	[]
											node [nodraw] {$101$}
											(SA)
								edge	[]
											node [nodraw] {$100$}
											(SAJ)
								edge	[]
											node [nodraw] {$11*$}
											(SJ)
								edge	[]
											node [nodraw] {$011$}
											(SB)
								edge	[]
											node [nodraw] {$010$}
											(SBJ)
								edge	[loop left]
											node [nodraw] {$00*$}
											()
					(SA)	edge	[loop below]
											node [nodraw] {$***$}
											()
					(SAJ)	edge	[]
											node [pos = 0.45, nodraw] {$**0$}
											(SAB)
								edge	[loop above]
											node [nodraw] {$**1$}
											()
					(SJ)	edge	[loop below]
											node [nodraw] {$***$}
											()
					(SAB)	edge	[loop above]
											node [nodraw] {$***$}
											()
					(SB)	edge	[loop below]
											node [nodraw] {$***$}
											()
					(SBJ)	edge	[]
											node [pos = 0.45, nodraw] {$**0$}
											(SAB)
								edge	[loop above]
											node [nodraw] {$**1$}
											()
					;
			\end{tikzpicture}
			}}
			\caption{\label{fig:exm:ppd} The \CGS\ $\GName[P\!P\!D]$.}
		\end{center}
		\vspace{-15pt}
	\end{wrapfigure}
	}

\newcommand{\figexmpsvar}
	{
	\begin{wrapfigure}[17]{r}{0.325\textwidth}
		\vspace{-0.75em}
		\begin{center}
			\footnotesize
			\mbox{\scalebox{0.80}[0.80]{
			\begin{tikzpicture}
				[node distance = 3.25cm, bend angle = 25, shorten >= 2pt, shorten <=
				2pt]
				\node [cnode]
							(SI)
							{$\stackrel{\sSym[i]}{\emptyset}$};
				\node [cnode]
							(R1)
							[below left of = SI]
							{$\stackrel{\sSym[1]}{\rSym[1]}$};
				\node [cnode]
							(R2)
							[below right of = SI]
							{$\stackrel{\sSym[2]}{\rSym[2]}$};
				\node [cnode]
							(R12)
							[below of = SI]
							{$\stackrel{\sSym[1, 2]}{\rSym[1], \rSym[2]}$};
				\node [cnode]
							(SA)
							[below of = R1]
							{$\stackrel{\sSym'_{1}}{\gSym[1]}$};
				\node [cnode]
							(SB)
							[below of = R2]
							{$\stackrel{\sSym'_{2}}{\gSym[2]}$};
				\path[-stealth']
					(SI)	edge	[bend right]
											node [nodraw] {$10*$}
											(R1)
								edge	[bend left]
											node [nodraw] {$01*$}
											(R2)
								edge	[]
											node [nodraw] {$11*$}
											(R12)
								edge	[loop above]
											node [nodraw] {$00*$}
											()
					(R1)	edge	[]
											node [nodraw] {$***$}
											(SA)
					(R2)	edge	[]
											node [nodraw] {$***$}
											(SB)
					(R12)	edge	[]
											node [nodraw] {$**0$}
											(SA)
								edge	[]
											node [nodraw] {$**1$}
											(SB)
					(SA)	edge	[bend left]
											node [nodraw] {$01*$}
											(SB)
								edge	[]
											node [nodraw] {$00*, 111$}
											(SI)
								edge [loop below]
											node [nodraw] {$10*, 110$}
											()
					(SB)	edge	[bend left]
											node [nodraw] {$10*$}
											(SA)
								edge	[]
											node [nodraw] {$00*, 111$}
											(SI)
								edge [loop below]
											node [nodraw] {$01*, 110$}
											()
					;
			\end{tikzpicture}
			}}
			\vspace{-2em}
			\caption{\label{fig:exm:ps} The \CGS\ $\GName[P\!S]$.}
		\end{center}
		\vspace{-2em}
	\end{wrapfigure}
	}

\newcommand{\figrasvii}
	{
	\begin{wrapfigure}[6]{r}{0.350\textwidth} % [6]{r}{0.400\textwidth}
		\vspace{-5em}
		\begin{center}
			\footnotesize
			\mbox{\scalebox{1.00}[1.00]{
			\begin{tikzpicture}
				[node distance = 2.75cm, bend angle = 15, shorten >= 2pt, shorten <=
					2pt]
				\node [cnode]
							(S0)
							{$\stackrel{\sSym[0]}{\emptyset}$};
				\node [cnode]
							(S1)
							[below left of = S0]
							{$\stackrel{\sSym[1]}{\pElm}$};
				\node [cnode]
							(S2)
							[below right of = S0]
							{$\stackrel{\sSym[2]}{\emptyset}$};
				\path[-stealth']
					(S0)	edge	[]
											node [nodraw] {$\PSet$}
											(S1)
								edge	[]
											node [nodraw] {$** \setminus \PSet$}
											(S2)
					(S1)	edge	[loop right]
											node [nodraw] {$**$}
											()
					(S2)	edge	[loop left]
											node [nodraw] {$**$}
											()
					;
			\end{tikzpicture}
			}}
			\caption{\label{fig:ordsat:sl} The \CGS\ $\GName^{\star}$ model
				of $\varphi^{ord}$.}
		\end{center}
		\vspace{-15pt}
	\end{wrapfigure}
	}

\newcommand{\figdefdecunw}
	{
	\begin{figure}[htbp]
		\begin{center}
			\footnotesize
			\mbox{
				{
				\label{fig:def:decunw:orig}
				\scalebox{0.80}[0.65]{
				\begin{tikzpicture}
					[node distance = 2.5cm, bend angle = 20, shorten >= 2pt, shorten <=
					2pt]
					\node [noall]
								(S0)
								{$\stackrel{\sSym[0]}{\emptyset}$};
					\node [noall]
								(S1)
								[below of = S0]
								{$\stackrel{\sSym[1]}{\emptyset}$};
					\node [noall]
								(S2)
								[below left of = S1]
								{$\stackrel{\sSym[2]}{\pSym}$};
					\node [noall]
								(S3)
								[below right of = S1]
								{$\stackrel{\sSym[3]}{\emptyset}$};
					\path[-stealth']
						(S0)	edge	[]
												node [nodraw] {$**$}
												(S1)
						(S1)	edge	[]
												node [nodraw] {$0*$}
												(S2)
									edge	[]
												node [nodraw] {$1*$}
												(S3)
						(S2)	edge	[loop right]
												node [nodraw] {$**$}
												()
						(S3)	edge	[loop left]
												node [nodraw] {$**$}
												()
						;
				\end{tikzpicture}
				}}
				{
				\label{fig:def:decunw:unw}
				\scalebox{0.65}[0.65]{
				\begin{tikzpicture}
					[node distance = 2.25cm, bend angle = 20, shorten >= 2pt, shorten <=
					2pt]
					\node [noall]
								(S0)
								{$\stackrel{\epsilon}{\emptyset}$};
					\node [noall]
								(S1-01)
								[below left of = S0]
								{$\stackrel{01}{\emptyset}$};
					\node [noall]
								(S1-00)
								[left of = S1-01]
								{$\stackrel{00}{\emptyset}$};
					\node [noall]
								(S1-10)
								[below right of = S0]
								{$\stackrel{10}{\emptyset}$};
					\node [noall]
								(S1-11)
								[right of = S1-10]
								{$\stackrel{11}{\emptyset}$};
					\node [noall]
								(S3-00-11)
								[below of = S1-01]
								{$\stackrel{00 \cdot 11}{\emptyset}$};
					\node [noall]
								(S3-00-10)
								[left of = S3-00-11]
								{$\stackrel{00 \cdot 10}{\emptyset}$};
					\node [noall]
								(S2-00-01)
								[left of = S3-00-10]
								{$\stackrel{00 \cdot 01}{\pSym}$};
					\node [noall]
								(S2-00-00)
								[left of = S2-00-01]
								{$\stackrel{00 \cdot 00}{\pSym}$};
					\node [noall]
								(D)
								[below right of = S1-01]
								{$\cdots$};
					\node [noall]
								(S2-11-00)
								[below of = S1-10]
								{$\stackrel{11 \cdot 00}{\pSym}$};
					\node [noall]
								(S2-11-01)
								[right of = S2-11-00]
								{$\stackrel{11 \cdot 01}{\pSym}$};
					\node [noall]
								(S3-11-10)
								[right of = S2-11-01]
								{$\stackrel{11 \cdot 10}{\emptyset}$};
					\node [noall]
								(S3-11-11)
								[right of = S3-11-10]
								{$\stackrel{11 \cdot 11}{\emptyset}$};
					\path[-stealth']
						(S0)		edge	[]
													node [nodraw] {$00$}
													(S1-00)
										edge	[]
													node [nodraw] {$01$}
													(S1-01)
										edge	[]
													node [nodraw] {$10$}
													(S1-10)
										edge	[]
													node [nodraw] {$11$}
													(S1-11)
						(S1-00)	edge	[]
													node [nodraw] {$00$}
													(S2-00-00)
										edge	[]
													node [nodraw] {$01$}
													(S2-00-01)
										edge	[]
													node [nodraw] {$10$}
													(S3-00-10)
										edge	[]
													node [nodraw] {$11$}
													(S3-00-11)
						(S1-11)	edge	[]
													node [nodraw] {$00$}
													(S2-11-00)
										edge	[]
													node [nodraw] {$01$}
													(S2-11-01)
										edge	[]
													node [nodraw] {$10$}
													(S3-11-10)
										edge	[]
													node [nodraw] {$11$}
													(S3-11-11)
						;
				\end{tikzpicture}
				}}
			}
			\caption{\label{fig:def:decunw} A \CGS\ and part of its
								decision-unwinding.}
		\end{center}
		\vspace{-1em}
	\end{figure}
	}

\newcommand{\figrasviii}
	{
	\begin{wrapfigure}[16]{r}{0.425\textwidth}
		\begin{center}
			\footnotesize
			\mbox{\scalebox{1.00}[0.80]{
			\begin{tikzpicture}
				[node distance = 2.50cm, bend angle = 45, shorten >= 2pt, shorten <=
				2pt]
				\node [nofill, circle]
							(S0)
							{$\stackrel{\sSym[0]}{\emptyset}$};
				\node [nofill, circle]
							(S1)
							[below of = S0]
							{$\stackrel{(\pSym, \tSym[1])}{\pSym, \tSym[1]}$};
				\node [nofill, circle]
							(S2)
							[right of = S1]
							{$\stackrel{(\pSym, \tSym[2])}{\pSym, \tSym[2]}$};
				\node [nofill, circle]
							(S3)
							[below of = S1]
							{$\stackrel{(\neg \pSym, \tSym[3])}{\tSym[3]}$};
				\node [nofill, circle]
							(S4)
							[right of = S3]
							{$\stackrel{(\pSym, \tSym[4])}{\pSym, \tSym[4]}$};
				\path[-stealth']
					(S0)	edge	[bend right]
											node [nodraw, pos = 0.70] {$00$}
											(S1)
								edge	[]
											node [nodraw] {$01$}
											(S2)
								edge	[bend right]
											node [nodraw] {$10$}
											(S3)
								edge	[]
											node [nodraw] {$11$}
											(S4)
					(S1)	edge	[loop above]
											node [nodraw] {$**$}
											()
					(S2)	edge	[loop above]
											node [nodraw] {$**$}
											()
					(S3)	edge	[loop above]
											node [nodraw] {$**$}
											()
					(S4)	edge	[loop above]
											node [nodraw] {$**$}
											()
					;
			\end{tikzpicture}
			}}
			\caption{\label{fig:satisfiability[sl]} Part of the \CGS\
				$\GName[\dsmFun]^{\star}$ model of $\varphi^{dom}$, where $\dsmFun(0,
				0) = \tSym[1]$, $\dsmFun(0, 1) = \tSym[2]$, $\dsmFun(1, 0) = \tSym[3]$,
				and $\dsmFun(1, 1) = \tSym[4]$.}
		\end{center}
		\vspace{-15pt}
	\end{wrapfigure}
	}

	}

\begin{document}

	\title
		[Reasoning About Strategies: On the Satisfiability Problem]
		{Reasoning About Strategies: \\ On the Satisfiability
		Problem\rsuper*}

	\author[F. Mogavero]{Fabio Mogavero\rsuper a}
	\address{{\lsuper{a,c}}University of Oxford}
        \email{fabio.mogavero@cs.ox.ac.uk}

	\author[A. Murano]{Aniello Murano\rsuper b}
	\address{{\lsuper b}Universit\`a degli Studi di Napoli Federico II}
        \email{aniello.murano@unina.it}

	\author[G. Perelli]{Giuseppe Perelli\rsuper c}
        \address{\vspace{-18 pt}}
        \thanks{{\lsuper c}The author thanks the support of the ERC
          Advanced Grant RACE (291528) at Oxford.}
        \email{giuseppe.perelli@cs.ox.ac.uk}

	\author[M.Y. Vardi]{Moshe Y. Vardi\rsuper d} 
	\address{{\lsuper d}Rice University}
        \thanks{{\lsuper d}Work supported in part by NSF grants CCF-1319459 and IIS-1527668, by NSF Expeditions in Computing project "ExCAPE: Expeditions in Computer Augmented Program Engineering", and by BSF grant 9800096.}
	\email{vardi@cs.rice.edu}


	\titlecomment{{\lsuper*}This work is partially based on the
	articles~\cite{MMV10b} and~\cite{MMPV12} appearing in FST\&TCS'10 and
	CONCUR'12, respectively.}

%%****************************************************************************%%
%%                                                                            %%
%% Reasoning About Strategies: On the Satisfiability Problem                  %%
%%                                                                            %%
%% Abstract.tex                                                               %%
%%                                                                            %%
%% Revision 0                                                                 %%
%%                                                                            %%
%% Copyright (C) 2014, Fabio Mogavero, Aniello Murano, Giuseppe Perelli, and  %%
%%                     Moshe Y. Vardi.                                        %%
%% All rights reserved.                                                       %%
%%                                                                            %%
%%****************************************************************************%%

% Begin of file Abstract.tex

\begin{abstract}

% 	\cbstart
	\emph{Strategy Logic} (\SL, for short) has been introduced by Mogavero,
	Murano, and Vardi as a useful formalism for reasoning explicitly about
	strategies, as first-order objects, in multi-agent concurrent games.
	This logic turns out to be very powerful, subsuming all major previously
	studied modal logics for strategic reasoning, including \ATL, \ATLS, and the
	like.
	Unfortunately, due to its high expressiveness, \SL has a non-elementarily
	decidable model-checking problem and the satisfiability question
	is undecidable, specifically \SCompH{1}{1}.
	
	In order to obtain a decidable sublogic, we introduce and study here
	\emph{One-Goal Strategy Logic} (\OGSL, for short).
	This is a syntactic fragment of \SL, strictly subsuming \ATLS, which
	encompasses formulas in prenex normal form having a single temporal goal at a
	time, for every strategy quantification of agents.
% 	\OGSL\ is known to have an elementarily decidable model-checking problem.
%	Here
	We prove that, unlike \SL, \OGSL has the bounded tree-model property and its
	satisfiability problem is decidable in 2\ExpTime, thus not harder than the one
	for \ATLS.
% 	\cbend

\end{abstract}

% End of file Abstract.tex

	\maketitle

%%****************************************************************************%%
%%                                                                            %%
%% Reasoning About Strategies: On the Satisfiability Problem                  %%
%%                                                                            %%
%% Introduction.tex                                                           %%
%%                                                                            %%
%% Revision 0                                                                 %%
%%                                                                            %%
%% Copyright (C) 2014, Fabio Mogavero, Aniello Murano, Giuseppe Perelli, and  %%
%%                     MoshMMe Y. Vardi.                                        %%
%% All rights reserved.                                                       %%
%%                                                                            %%
%%****************************************************************************%%

% Begin of file Introduction.tex

\begin{section}{Introduction}

	In open-system verification~\cite{CGP02,KVW01}, an important area of
  research is the study of modal logics for strategic reasoning in the setting
  of multi-agent games~\cite{AHK02,JH04,Pau02,AGJ07,HWW07,BJ14,JM14}.
	An important contribution in this field has been the development of
	\emph{Alternating-Time Temporal Logic} (\ATLS, for short), introduced by
	Alur, Henzinger, and Kupferman~\cite{AHK02}.
	\ATLS\ allows reasoning about strategic behavior of agents with temporal
	goals.
	Formally, it is obtained as a generalization of the branching-time temporal
	logic \CTLS~\cite{EH86}, where the path quantifiers \emph{there exists}
	``$\E$'' and \emph{for all} ``$\A$'' are replaced with strategic modalities
	of the form ``$\EExs{\ASet}$'' and ``$\AAll{\ASet}$'', for a set $\ASet$ of
	\emph{agents}.
	Such strategic modalities are used to express cooperation and competition
	among agents in order to achieve certain temporal goals.
	In particular, these modalities express selective quantifications over those
	paths that are the results of infinite games between a coalition and its
	complement.
	\ATLS\ formulas are interpreted over \emph{concurrent game structures} (\CGS,
	for short)~\cite{AHK02}, which model interacting processes.
	Given a \CGS\ $\GName$ and a set $\ASet$ of agents, the \ATLS\ formula
	$\EExs{\ASet} \psi$ holds at a state $\sElm$ of $\GName$ if there is a set of
	strategies for the agents in $\ASet$ such that, no matter which strategies
	are 	executed by the agents not in $\ASet$, the resulting outcome of the
	interaction in $\GName$ satisfies $\psi$ at $\sElm$.
	Several decision problems have been investigated about \ATLS; both its
	model-checking and satisfiability problems are decidable in
	2\ExpTime~\cite{Sch08}.
	The complexity of the latter is just like the one for \CTLS~\cite{EJ88,EJ99}.
	\\ \indent
	Despite its powerful expressiveness, \ATLS\ suffers from the strong limitation
	that strategies are treated only implicitly through modalities that refer to
	games between competing coalitions.
	To overcome this problem, Chatterjee, Henzinger, and Piterman introduced
	\emph{Strategy Logic} (\CHPSL, for short)~\cite{CHP07,CHP10}, a logic that
	treats strategies in \emph{two-player turn-based games} as \emph{first-order
	objects}.
	The explicit treatment of strategies in this logic allows the expression of
	many properties not expressible in \ATLS.
	Although the model-checking problem of \CHPSL\ is known to be decidable with
	a non-elementary upper bound, it is not known whether the satisfiability
	problem is decidable as well~\cite{CHP10}.
	While the basic idea exploited in~\cite{CHP10} of explicitly quantifying over
	strategies is powerful and useful~\cite{FKL10}, \CHPSL\ still suffers from
	various limitations.
	In particular, it is limited to two-player turn-based games.
	Furthermore, \CHPSL\ does not allow different players to share the same
	strategy, suggesting that strategies have yet to become truly first-class
	objects in this logic.
	For example, it is impossible to describe the classic strategy-stealing
	argument of combinatorial games such as Hex and the like.
	\\ \indent
	These considerations led us to introduce and investigate a new \emph{Strategy
	Logic}, denoted \SL, as a more general framework than \CHPSL, for explicit
	reasoning about strategies in multi-agent concurrent
	games~\cite{MMV10b,MMPV14}.
	Syntactically, \SL\ extends the linear-time temporal-logic \LTL~\cite{Pnu77}
	by means of \emph{strategy quantifiers}, the existential $\EExs{\xElm}$ and
	the universal $\AAll{\xElm}$, as well as \emph{agent binding} $(\aElm,
	\xElm)$, where $\aElm$ is an agent and $\xElm$ a variable.
	Intuitively, these elements can be read as \emph{``there exists a strategy
	$\xElm$''}, \emph{``for all strategies $\xElm$''}, and \emph{``bind agent
	$\aElm$ to the strategy associated with $\xElm$''}, respectively.
	For example, in a \CGS\ $\GName$ with agents $\alpha$, $\beta$, and $\gamma$,
	consider the property ``$\alpha$ and $\beta$ have a common strategy to avoid a
	failure''.
	This property can be expressed by the \SL\ formula $\EExs{\xSym} \AAll{\ySym}
	(\alpha, \xSym) (\beta, \xSym) (\gamma, \ySym) (\G \neg \mathit{fail})$.
	The variable $\xSym$ is used to select a strategy for the agents $\alpha$ and
	$\beta$, while $\ySym$ is used to select another one for agent $\gamma$ such
	that their composition, after the binding, results in a play where
	$\mathit{fail}$ is never met.
%	Further examples, motivations, and results can be found in~\cite{MMPV14}.
%	%
%	\\ \indent
%	%
	In~\cite{MMPV14} it has been showed that \SL\ is very expressive and can
	represent several solution concepts. However, this high expressiveness comes
	at a price.
	Indeed, it has been shown in~\cite{MMPV14} that the model-checking problem is
	non-elementarily decidable. In particular, this problem is $k$-\ExpSpaceH in
	the alternation number $k$ of quantifications in the specification.
	\\ \indent
	In this paper we investigate the satisfiability problem and some
	basic model-theoretic properties for \SL. Regarding the former, as main
	result we show that \SL is \emph{highly undecidable}, precisely,
	\SCompH{1}{1}. Regarding the latter, we show that \SL\ does not have the
	bounded-tree model property.
	\\ \indent
	The contrast between the undecidability of the satisfiability problem for \SL\
	and the elementary decidability of the same problem for \ATLS, provides
	motivation for an investigation of decidable fragments of \SL\ that subsume
	\ATLS.
% 	In particular, we would like to understand why \SL\ is computationally more
% 	difficult than \ATLS.
	%
	\\ \indent
	We introduce here the syntactic fragment \emph{One-Goal Strategy Logic}
	(\OGSL, for short), which encompasses formulas in a special prenex normal form
	having a single temporal goal at a time.
	For goal we mean an \SL\ formula of the type $\bpElm \psi$, where $\bpElm$ is
	a binding prefix of the form $(\alpha_{1}, \xElm[1]), \ldots, (\alpha_{n},
	\xElm[n])$ containing all the involved agents and $\psi$ is
	a formula in which every agent is not bounded to any variable, as for example an \LTL specification.
	%	an agent-full
%	formula, as for example an \LTL specification.
% 	This means that every temporal formula $\psi$ is prefixed with a
% 	quantification-binding prefix that quantifies over a tuple of strategies and
% 	bind strategies to all agents.
	With \OGSL\ one can express, for instance, visibility constraints on
	strategies among agents, i.e., only some agents from a coalition have
	knowledge of the strategies taken by those in the opponent coalition.
	Also, one can describe the fact that, in the Hex game, the
	strategy-stealing argument does not let the player who adopts it to win.
	Observe that the above properties cannot be expressed neither in \ATLS\
	nor in \CHPSL.
	\\ \indent
	In~\cite{MMPV14}, we showed that \OGSL\ is strictly more
	expressive that \ATLS, yet its model-checking problem is 2\ExpTimeC, just like
	the one for \ATLS, while the same problem for \SL\ is non-elementarily
	decidable.
	Our main result here is that the satisfiability problem for \OGSL\ is also
	2\ExpTimeC.
	Thus, in spite of its expressiveness, \OGSL\ has the same computational
	properties of \ATLS, which suggests that the one-goal restriction is the key
	to the elementary complexity of the latter logic too.
	\\ \indent
	To achieve our main result, we use a fundamental property of the semantics of
	\OGSL\, called \emph{behavioral}\footnote{We use this term as it has a
	direct correspondence with the ``behavioral''
	concept used in game theory \cite{Mey97,MMS13,MMS14b}.}, which allows us to
	simplify reasoning about strategies by reducing it to a set of reasonings
	about actions.
	This intrinsic characteristic of \OGSL\ means that, to choose an existentially quantified strategy, we do not need to know the entire structure of universally-quantified strategies, as it is the case for \SL, but only their values on the histories of interest.
	Technically, to formally describe this property, we make use of the machinery
	of \emph{dependence maps}, which is introduced to define a Skolemization
	procedure for \SL, inspired by the one in first-order logic.
	By exploiting the behavioral property, one can show that \OGSL\ satisfies
	the \emph{bounded tree-model property}\footnote{In~\cite{MMPV12}, we indeed
	make use of a non-trivial proof to show this. In this paper, instead, we avoid
	the burden by making use of a recent result proved in~\cite{MP15b}.}.
	This allows us to efficiently make use of a \emph{tree automata-theoretic
	approach}~\cite{Var96,VW86a} to solve the satisfiability problem.
	Given a formula $\varphi$, we build an \emph{alternating co-B\"uchi tree
	automaton}~\cite{KVW00,MS95}, whose size is only exponential in the size of
	$\varphi$, accepting all bounded-branching tree models of the formula
	with a suitable width. Then, together with the complexity of
	automata-nonemptiness checking, we get
	that the satisfiability procedure for \OGSL\ is 2\ExpTime.
	We believe that our proof techniques are of independent interest and
	applicable to other logics as well.
	\\ \indent
	\paragraph{\textbf{Related works.}}
		Several works have focused on extensions of \ATL\ and \ATLS\ to incorporate
		more powerful strategic constructs.
		Among them, we recall \emph{Alternating-Time
                  $\mu$\textsc{Calcu\-lus}} (AMuCalculus, for
%		Among them, we recall \emph{Alternating-Time \MuCalculus} (AMuCalculus, for
		short) and \emph{Game Logic} (GL, for short)~\cite{AHK02},
		\emph{Quantified Decision Modality \MuCalculus} (QDMuCalculus, for
		short)~\cite{Pin07}, \emph{Coordination Logic} (CL, for short)~\cite{FS10},
		(\emph{\ATL\ with plausibility} (ATLP, for short)~\cite{BJD08},
		(\emph{\ATL\ with Irrevocable strategies } (I\ATL, for short)~\cite{AGJ07},
		(\emph{Memoryful \ATLS\ } (m\ATLS, for short)~\cite{MMV10a,MMV16},
		\emph{Basic Strategy-Interaction Logic} (BSIL, for short)~\cite{WHY11}
		\emph{Temporal Cooperation Logic} (TCL, for short)~\cite{HSW13},
		\emph{Alternating-time Temporal Logic with Explicit Actions} (ATLEA, for
		short)~\cite{HLW13} and some extensions of \ATLS\ considered
		in~\cite{BLLM09}.
		AMuCalculus\ and QDMuCalculus\ are intrinsically different from \SL\ (as
		well as from \CHPSL\ and \ATLS) as they are obtained by extending the
		propositional $\mu$-calculus~\cite{Koz83} with strategic modalities.
		CL is similar to QDMuCalculus\ but with \LTL\ temporal operators instead
		of explicit fixpoint constructors.
		GL is strictly included in \CHPSL, in the case of two-player turn-based
		games, but it does not use any explicit treatment of strategies, as well as the extensions of \ATLS\ introduced in~\cite{BLLM09}, which consider restrictions on the memory for strategy quantifiers.
		ATLP enables to express rationality assumptions of intelligent agents in
		\ATL.
		In IATL, the semantics of the logic \ATL\ is changed in a way that, in the
		evaluation of the goal, agents can be forced to keep the strategy they have
		chosen in the past in order to reach the state where a goal is evaluated.
		m\ATLS\ enriches \ATLS\ by giving the ability to agents to ``relent'' and
		change their goals and strategies depending on the history of the play.
		BSIL allows to specify behaviors of a system that can cooperate with
		several strategies of the environment for different requirements.
		TCL extends \ATL\ by allowing successive definitions of agent strategies,
		with the aim of using the collaborative power of groups of agents to enforce
		different temporal objectives.
		ATLEA introduces explicit actions in the logic \ATL\ to reason about
		abilities of agents under commitments to play precise actions.
		Thus, all above logics are different from \SL.
%		, which we recall it aims
%		to be a minimal but powerful logic to reason about strategic behavior in
%		multi-agent systems.

		At roughly the time we have conceived Strategy Logic, another
		generalization of \ATLS, named \emph{\ATLS\ with Strategy Contexts}, which
		turns out to be very expressive but a proper sublogic of \SL, has been
		considered in~\cite{DLM10} (see also \cite{DLM12,TW12,LM13,LM15} for more recent works).
		In this logic, a quantification over strategies does not reset the
		strategies previously quantified but allows to maintain them in a particular
		context in order to be reused.
		This makes the logic much more expressive than \ATLS.
%		On the other hand, as it does not allow agents to share the same strategy,
%		it seems not comparable with \OGSL.

		Recently, several extensions of \SL\ have been also investigated.
		\emph{Updatable Strategy Logic} (USL, for short) has been considered
		in~\cite{BCC13,CBC15} where, in addition to \SL, an agent can refine its own
		strategies by means of an "unbinder" operator, which explicitly deletes
		the binding of a strategy to an agent.
		In~\cite{Bel14,CLMM14}, an epistemic extension of \SL\ with modal operators
		for individual knowledge has been considered, showing that the complexity of
		model checking for this logic is not worse than the one for (non-epistemic)
		\SL.
		%In \cite{MMMS15}, an extension of \SL\ with graded modalities has been
		%introduced and investigated over a fragment of \OGSL.
		Last but not least, in \cite{CLM15} a BDD-based model
		checker for the verification of systems against specifications expressed in
		\OGSL has been introduced (see also \cite{CLMM14} for an introduction to
		the conceived tool).

		Finally, worth of mention are the works handling the synthesis question
		of specifications expressed in the logic \CHPSL, as well as logics related
		to \SL.
		Among the others, we report the works~\cite{CDFR14,FKL10,KPV14,GHW14,KPV16}.

		\paragraph{Outline}
      In Section~\ref{sec:sl}, we first introduce the syntax of \SL, as well as
      the notion of \emph{Concurrent Game Structure}, on which the logic is
      interpreted.
      We also provide examples to show useful applications of the logic in the
      context of formal verification.
      In Section~\ref{sec:hard} we show that the satisfiability problem for \SL
      is highly undecidable.
      We do this by first proving that the logic does not have the bounded-model
      property and then providing a reduction of the satisfiability problem from
      the \emph{recurrent domino problem}, which has been proved to be
      undecidable by Harel in~\cite{Har84}.
      Given this negative result, in Section~\ref{sec:what}, we investigate on
      the theoretical properties that make \ATLS decidable.
      We first recall the definition of two syntactic fragments of \SL, which we
      call \emph{Boolean-Goal Strategy Logic} (\BGSL) and \emph{One-Goal
      Strategy Logic} (\OGSL), showing that the former retains all the negative
      properties of \SL, while the latter satisfies a fundamental property,
      namely the \emph{behavioral} semantics, that turns out to be fundamental,
      in Section~\ref{sec:dec}, to prove that \OGSL enjoys the bounded-model
      property and a decidable satisfiability problem.
      In order to prove the last result, we employ an automata-theoretic
      approach, from which we derive a 2\ExpTime procedure.

\end{section}

% End of file Introduction.tex

% 	\input{Preliminaries}

%%****************************************************************************%%
%%                                                                            %%
%% Reasoning About Strategies: On the Satisfiability Problem                  %%
%%                                                                            %%
%% SectionI.tex                                                               %%
%%                                                                            %%
%% Revision 0                                                                 %%
%%                                                                            %%
%% Copyright (C) 2014, Fabio Mogavero, Aniello Murano, Giuseppe Perelli, and  %%
%%                     Moshe Y. Vardi.                                        %%
%% All rights reserved.                                                       %%
%%                                                                            %%
%%****************************************************************************%%

% Begin of file SectionI.tex

\begin{section}{Strategy Logic}
	\label{sec:sl}

	\emph{Strategy Logic}~\cite{MMV10b} (\SL, for short) is an extension	of the
	classic linear-time temporal logic \LTL~\cite{Pnu77} along with the	concepts
	of strategy quantifications and agent binding, which formalism allows to
	express strategic plans over temporal goals.
	The main distinctive feature of this formalism \wrt other logics with the same
	aim resides in the decoupling strategy instantiations, done through the
	quantifications, from their applications, by means
	of bindings.
	Consequently, the logic is not simply propositional but predicative, since we
	treat strategies as a first order concept via the use of
	agents and variables as explicit syntactic elements.
	This fact allows us to write Boolean combinations and nesting of complex
	predicates, each one representing a different temporal goal, linked together
	by some common strategic choices.

	The section is organized as follows.
	In Subsection~\ref{sec:sl;sub:undfrm}, we recall the definition of concurrent
	game structure used to interpret \SL, whose syntax is reported in
	Subsection~\ref{sec:sl;sub:syn}.
	Then, in Subsection~\ref{sec:sl;sub:basnon}, we give, among the others, the
	notions of strategy, assignment, and play, which are finally used to define
	the semantics of the logic in Subsection~\ref{sec:sl;sub:sem}.

	\begin{subsection}{Underlying framework}
		\label{sec:sl;sub:undfrm}

		As semantic framework for \SL, we use the \emph{graph-based model} for
		\emph{multi-player games} named \emph{concurrent game
		structure}~\cite{AHK02}, which is a generalization of
		\emph{Kripke structures}~\cite{Kri63} and \emph{labeled transition
		systems}~\cite{Kel76}.
		It allows to model \emph{multi-agent systems} viewed as extensive form
		games, in which players perform \emph{concurrent actions} to trigger
		different transitions over the graph.

		\begin{defi}[Concurrent Game Structures]
			\label{def:cgs}
			A \emph{concurrent game structure} (\ConGamStr, for short) is a
			tuple $\ConGamStrName \defeq \ConGamStrStr$, where $\APSet$ and $\AgnSet$
			are finite non-empty sets of \emph{atomic propositions} and \emph{agents},
			$\ActSet$ and $\SttSet$ are enumerable non-empty sets of \emph{actions}
			and \emph{states}, $\sttElm[0] \in \SttSet$ is a designated \emph{initial
			state}, and $\apFun : \SttSet \to \pow{\APSet}$ is a \emph{labeling
			function} that maps each state to the set of atomic propositions true in
			that state.
			Let $\DecSet \defeq \ActSet^{\AgnSet}$ be the set of \emph{decisions}, \aka \emph{action profiles} in the literature,
			\ie, functions from $\AgnSet$ to $\ActSet$ representing the choices of an
			action for each agent.~\footnote{In the following, we use both $\XSet \to
			\YSet$ and $\YSet^{\XSet}$ to denote the set of functions from the domain
			$\XSet$ to the codomain $\YSet$.}
			Then, $\trnFun: \SttSet \times \DecSet \to \SttSet$ is a \emph{transition
			function} mapping a pair of a state and a decision to a state.
		\end{defi}

		\begin{rem}
			The reader might note that the definition of \ConGamStr given here differs to the one provided in the literature, for example in~\cite{AHK02}, from the fact that set of actions is not specified for each agent.
			On one hand, this has the advantage of simplifying the notation and the technical development of the results.
			On the other hand, this looks less general from the mechanism design point of view, in which the roles of the agents are not always symmetric, and it turns to be useful to assign a specific set of action per each agent.
			However, by means of a suitable mapping of actions, it is not hard to show that the general case of \ConGamStr described in~\cite{AHK02} can be accounted in Definition~\ref{def:cgs}.
		\end{rem}

		To get familiar with the concept of \ConGamStr, we present here some running
		example of simple concurrent games.

		\par\figexmppd\
		First, we analyze an extended version of the well-known prisoner's
		dilemma~\cite{OR94} in which also the actions of the police are taken into
		account.
		\begin{exa}[Prisoners and Police's Dilemma]
			\label{exm:ppd}
			In the \emph{prisoner's dilemma} (PD, for short), two accomplices are
			interrogated in separated rooms by the police, which offers them the same
			agreement.
			If one defects, i.e., testifies for the prosecution against the other,
			while the other cooperates, i.e., remains silent, the defector goes free
			and the silent accomplice goes to jail.
			If both cooperate, they remain free, but will be surely interrogated in
			the next future waiting for a defection.
			On the other hand, if they both defect, both go to jail.
			In the \emph{prisoner and police's dilemma} (PPD, for short), apart from
			the classic agreement of the PD, the two accomplices also know that they
			can try to gain a better sentence by the judge if one spontaneously
			defects without being interrogated by police, since he is considered a
			``good willing man''.
			In this case, indeed, if the other cooperates, the defector becomes
			definitely free, while the other goes to jail with the possibility to
			eventually be released.
			It is important, however, that neither of them defects, otherwise the
			police can subtly act as they were interrogated.
			Moreover, differently from the PD, they are not free during the time in
			which they can be interrogated.
			This complex situation, can be modeled by the \ConGamStr
			$\ConGamStrName[P\!P\!D] \defeq \ConGamStrStr$ depicted in
			Figure~\ref{fig:exm:ppd}, where there are three agents in $\AgnSet \defeq
			\{ \ASym[1], \ASym[2], \PSym \}$, with $\PSym$ being the police, and all
			of them can execute the two abstract actions in $\ActSet \defeq \{ 0, 1
			\}$.
			For the accomplices, $0$ and $1$ have the meaning of ``cooperate''
			and ``defect''.
			For the police, on the contrary, they mean ``wait'' and ``interrogate'' in
			all the states but those in which one of the accomplices can eventually be
			released, where the meaning is ``release'' and ``maintain'', instead.
			The set of states for the game is given by $\{\sSym[i], \sSym[ {\ASym[1]
			}], \sSym[ {\ASym[2] }], \sSym[j], \sSym[ {\ASym[1]} j], \sSym[ {\ASym[2]}
			j], \sSym[ {\ASym[1] \ASym[2]} ] \}$.
			The idle state $\sSym[i]$ denotes the situation in which the two prisoners are waiting to be interrogated by police.
			They can even decide to defect before interrogation.
			The states $\sSym[ {\ASym[1] }]$ and $\sSym[ {\ASym[2] }]$ denotes the
			situation in which only one prisoner becomes definitely free.
			Moreover, the states $\sSym[ {\ASym[1]} j]$ and $\sSym[ {\ASym[2]} j]$
			indicate when one of the prisoners is free while the other in the jail is
			waiting for his release.
			Finally, $\sSym[ {\ASym[1] \ASym[2]} ]$ denotes the state in which both prisoners have gained definitely the freedom.
			To represent the different meaning of these states, we use the atomic
			propositions $\fSym[ {\ASym[i]} ]$ to denote that the prisoner $\ASym[i]$
			is free.
			Both the labeling function $\apFun$ and the transition function $\trnFun$
			can be extracted from the figure, where the agents $\ASym[1]$, $\ASym[2]$, and $\PSym$ control the first, second and third components of the triple actions over the edges, respectively.
		\end{exa}

		In addition to PPD, we model a very simple \emph{preemptive scheduling}
		protocol for the access of processes to a shared resource.

		\par\figexmpsvar\
		\begin{exa}[Preemptive Scheduling]
			\label{exm:ps}
			Consider the following \emph{preemptive scheduling} protocol (PS, for
			short) describing the access rules of two processes to a shared
			resource in a preemptive way.
			When the resource is free and only one process asks for it, this process
			directly receives the grant.
			Instead, if there is a competition of requests, it is the scheduler
			that, in a nondeterministic way, determines who can access to the
			resource.
			Finally, in case one process owns the resource while the other asks for it, the scheduler can choose whether to apply a preemption.
			These rules are formalized in the \ConGamStr  $\ConGamStrName[P\!S] \defeq
			\ConGamStrStr$ of Figure~\ref{fig:exm:ps}, where the agents
			``Process-1'', ``Process-2'' and ``Scheduler'' in $\AgnSet \defeq \{
			\PSym[1], \PSym[2], \SSym \}$ can choose between the two abstract actions
			in $\ActSet \defeq \{ 0, 1 \}$.
			The processes use the actions $0$ to not send any request and $1$ to send a request to the scheduler, while the scheduler uses them in order to decide who can have the access to the resource in a situation of competition.
			There are five states $\SttSet \defeq \{ \sSym[i], \sSym[1], \sSym[2],
			\sSym[1, 2], \sSym'_{1}, \sSym'_{2} \}$ in which the protocol can reside:
			the idle state $\sSym[i]$ in which the resource is free; the three
			states $\sSym[1]$, $\sSym[2]$, and $\sSym[1, 2]$ in which $\PSym[1]$,
			$\PSym[2]$ or both are requesting the resource; the two states
			$\sSym'_{1}$ and $\sSym'_{2}$ in which the resource has been finally
			granted to $\PSym[1]$ and $\PSym[2]$, respectively.
			To represent all information associated, we use the atomic propositions in
			$\APSet \defeq \{ \rSym[1], \rSym[2], \gSym[1], \gSym[2] \}$, where
			$\rSym[i]$ represents the request of $\PSym[i]$, while $\gSym[i]$ the fact
			that the resource has been granted to $\PSym[i]$.
		\end{exa}

	\end{subsection}

	\begin{subsection}{Syntax}
		\label{sec:sl;sub:syn}

		\emph{Strategy Logic} (\SL, for short) syntactically extends \LTL by means
		of two \emph{strategy quantifiers}, the existential $\EExs{\xElm}$ and the
		universal $\AAll{\xElm}$, and the \emph{agent binding} $(\aElm, \xElm)$,
		where $\aElm$ is an agent and $\xElm$ a variable.
		Intuitively, these new elements can be read as \emph{``there exists a
		strategy $\xElm$''}, \emph{``for all strategies $\xElm$''}, and \emph{``bind
		agent $\aElm$ to the strategy associated with the variable $\xElm$''},
		respectively.
		The formal syntax of \SL follows.
		\begin{defi}[\SL Syntax]
			\label{def:sl(syntax)}
			\SL \emph{formulas} are built inductively from the sets of atomic
			propositions $\APSet$, variables $\VarSet$, and agents $\AgnSet$, by using
			the following grammar, where $\pElm \in \APSet$, $\xElm \in \VarSet$, and
			$\aElm \in \AgnSet$:
			\begin{center}
				$\varphi ::= \pElm \mid \neg \varphi \mid \varphi \wedge \varphi \mid
				\varphi \vee \varphi \mid \X \varphi \mid \varphi \:\U \varphi \mid
				\varphi \:\R \varphi \mid \EExs{\xElm} \varphi \mid \AAll{\xElm} \varphi
				\mid (\aElm, \xElm) \varphi$.
			\end{center}
			\SL denotes the infinite set of formulas generated by the above rules.
		\end{defi}
		Observe that, by construction, \LTL is a proper syntactic fragment of \SL,
		\ie, $\LTL \subset \SL$.
		In order to abbreviate the writing of formulas, we use the boolean values
		true $\Tt$ and false $\Ff$ and the well-known temporal operators future $\F
		\varphi \defeq \Tt\: \U \varphi$ and globally $\G \varphi \defeq \Ff\: \R
		\varphi$.
		Moreover, we use the italic letters $\xElm, \yElm, \zElm, \ldots$, possibly
		with indexes, as meta-variables on the variables $\xSym, \ySym, \zSym,
		\ldots$ in $\VarSet$.

		A first classic notion related to the syntax of \SL is that of
		\emph{subformula}, \ie, a syntactic expression that is
		part of an a priori given formula.
		%
% 		By $\mthfun{sub} : \SL \to \pow{\SL}$ we formally denote the function
% 		returning the set of subformulas of an \SL formula.
%
		By $\sub{\varphi}$ we formally denote the set of subformulas of an
		\SL formula $\varphi$.
		For instance, consider $\varphi = \EExs{\xSym} (\alpha, \xSym) (\F \pSym)$.
		Then, it is immediate to see that $\sub{\varphi} = \{ \varphi, (\alpha,
		\xSym) (\F \pSym), (\F \pSym), \pSym, \Tt \}$.

		Usually, in predicative logics, we need the concepts of \emph{free} and
		\emph{bound} placeholders, to correctly define the meaning of a formula.
		In \SL, we have two different kind of placeholders: variables and agents.
		The former is used in the strategy quantifications, the latter to commit an
		agent, by means of bindings, to adhere to a strategy.
		Consequently, we need to differentiate the sets of free variables and
		free agents of an \SL formula $\varphi$.
		The first contains the variables that are not in a scope of a
		quantification.
		The second, instead, contains the agents for which there is no related
		binding in the scope of a temporal operator.
		A formula without any free variable (resp., agent) is named
		\emph{variable-closed} (resp., \emph{agent-closed}).
		A formula that is both variable- and agent-closed, is named
		\emph{sentence}.
		For a given \SL formula $\varphi$, by $\free{\varphi}$ we denote the set of
		both free variables and agents occurring in $\varphi$.
		The formal definition of $\free{\cdot}$, which we report in the following, has been given in~\cite{MMPV14}.

		\begin{defi}[\SL\ Free Agents/Variables]
			\label{def:sl(freeagvar)}
			The set of \emph{free agents/variables} of an \SL\ formula is given by the
			function $\mthfun{free} : \SL \to \pow{\AgnSet \cup \VarSet}$ defined as
			follows:
			\begin{enumerate}[label=(\roman*)]
				\setlength{\itemsep}{2pt}
				\item\label{def:sl(freeagvar:ap)}
				$\free{\pElm} \defeq \emptyset$, where $\pElm \in \APSet$;
				\item\label{def:sl(freeagvar:neg)}
				$\free{\neg \varphi} \defeq \free{\varphi}$;
				\item\label{def:sl(freeagvar:conjdisj)}
				$\free{\varphi_{1} \Opr \varphi_{2}} \defeq \free{\varphi_{1}} \cup
				\free{\varphi_{2}}$, where $\Opr\! \in \{ \wedge, \vee \}$;
				\item\label{def:sl(freeagvar:next)}
				$\free{\X \varphi} \defeq \AgnSet \cup \free{\varphi}$;
				\item\label{def:sl(freeagvar:untilrelease)}
				$\free{\varphi_{1} \Opr \varphi_{2}} \defeq \AgnSet \cup
				\free{\varphi_{1}} \cup \free{\varphi_{2}}$, where $\Opr\! \in \{
				\U\!\!, \R\! \}$;
				\item\label{def:sl(freeagvar:qnt)}
				$\free{\Qnt \varphi} \defeq \free{\varphi} \setminus \{ \xElm \}$,
				where $\Qnt\! \in \set{ \EExs{\xElm}, \AAll{\xElm} }{ \xElm \in
					\VarSet }$;
				\item\label{def:sl(freeagvar:bndprs)}
				$\free{(\aElm, \xElm) \varphi} \defeq \free{\varphi}$, if $\aElm
				\not\in \free{\varphi}$, where $\aElm \in \AgnSet$ and $\xElm \in
				\VarSet$;
				\item\label{def:sl(freeagvar:bndrem)}
				$\free{(\aElm, \xElm) \varphi} \defeq (\free{\varphi} \setminus \{
				\aElm \}) \cup \{ \xElm \}$, if $\aElm \in \free{\varphi}$, where
				$\aElm \in \AgnSet$ and $\xElm \in \VarSet$.
			\end{enumerate}
			%
%			A formula $\varphi$ without free agents (resp., variables), i.e., with
%			$\free{\varphi} \cap \AgnSet = \emptyset$ (resp., $\free{\varphi} \cap
%			\VarSet = \emptyset$), is named \emph{agent-closed} (resp.,
%			\emph{variable-closed}).
%			If $\varphi$ is both agent- and variable-closed, it is referred to as a
%			\emph{sentence}.
%			The function $\mthfun{snt} : \SL \to \pow{\SL}$ returns the set of
%			\emph{subsentences} $\snt{\varphi} \defeq \set{ \phi \in \sub{\varphi} }{
%			\free{\phi} = \emptyset }$, for each \SL\ formula $\varphi$.
		\end{defi}
		Observe that, on one hand, free agents are introduced in
		Items~(\ref{def:sl(freeagvar:next)})
		and~(\ref{def:sl(freeagvar:untilrelease)})
		and removed in Item~(\ref{def:sl(freeagvar:bndrem)}).
		On the other hand, free variables are introduced in
		Item~(\ref{def:sl(freeagvar:bndrem)}) and removed in
		Item~(\ref{def:sl(freeagvar:qnt)}).

		As an example, let $\varphi = \EExs{\xSym} (\alpha, \xSym) (\beta,
		\ySym) (\F \pSym)$ be a formula on the agents $\AgnSet = \{ \alpha, \beta,
		\gamma \}$.
		Then, we have $\free{\varphi} = \{ \gamma, \ySym \}$, since $\gamma$ is an
		agent without any binding after $\F \pSym$ and $\ySym$ has no quantification
		at all.
		Consider also the formulas $(\alpha, \zSym) \varphi$ and $(\gamma, \zSym)
		\varphi$, where the subformula $\varphi$ is the same as above.
		Then, we have $\free{(\alpha, \zSym) \varphi} = \{ \gamma, \ySym, \zSym \}$ and $\free{(\gamma, \zSym) \varphi} = \{ \ySym, \zSym \}$, since $\alpha$ is not free in $\varphi$ but $\gamma$ is, \ie, $\alpha \notin \free{\varphi}$ and $\gamma \in \free{\varphi}$.
		So, $(\gamma, \zSym) \varphi$ is agent-closed while $(\alpha, \zSym) \varphi$ is not.

		In order to practice with the syntax of \SL, we now describe few examples of
		some game-theoretic properties, which cannot be expressed neither in \ATLS
		nor in \CHPSL. We clarify this point later in the paper.
		The interpretation of these formulas is quite intuitive.
		Leastwise, the reader can rely on the formal semantics, which is given later in the paper.

		The first we introduce is the well-known concept of \emph{Nash Equilibrium}
		in concurrent infinite games with Boolean payoffs.

		\begin{exa}[Nash Equilibrium]
			\label{exm:ne}
			Consider the $n$ agents $\alpha_{1}, \ldots, \alpha_{n}$ of a game, each
			of them having, respectively, a possibly different temporal goal described
			by one of the \LTL formulas $\psi_{1}, \!\ldots\!, \psi_{n}$.
			Then, we can express the existence of a strategy profile $(\xSym[1],
			\ldots, \xSym[n])$ that is a \emph{Nash equilibrium} (NE, for short) for
			$\alpha_{1}, \ldots, \alpha_{n}$ \wrt $\psi_{1}, \ldots, \psi_{n}$ by
			using the \SL sentence $\varphi_{\!N\!\!E} \!\defeq\! \EExs{\xSym[1]}
			\cdots \EExs{\xSym[n]} (\alpha_{1}, \xSym[1]) \cdots (\alpha_{n},
			\xSym[n]) \: \psi_{\!N\!\!E}$, where $\psi_{\!N\!\!E} \!\defeq\!
			\bigwedge_{i = 1}^{n} (\EExs{\ySym} (\alpha_{i}, \ySym) \psi_{i})
			\rightarrow \psi_{i}$ is a variable-closed formula.
			Informally, this asserts that every agent $\alpha_{i}$ has $\xSym[i]$ as
			one of the best strategy \wrt the goal $\psi_{i}$, once all the other
			strategies of the remaining agents $\alpha_{j}$, with $j \neq i$, have
			been fixed to $\xSym[j]$.
			Note that here we are only considering equilibria under deterministic
			strategies.
		\end{exa}

		In a game in which not all agents are peers, we can have one or more of them
		that may vary the payoff of the others, without having a
		personal aim, \ie, without looking for the maximization of their own
		payoffs.
		Such situations can usually arise when we have games with arbiters or
		similar characters, like supervisors or government authorities, that have to
		be fair, \ie, they have to lay down an \emph{equity governance}.
		\begin{exa}[Equity Governance]
			\label{exm:eg}
			Consider a game similar to the one described in the previous
			example, in which there is, in addition, a
                        supervisor agent $\beta$ that does not
			have a specific goal.
			However, the peers want him to be fair \wrt their own goals, \ie, the
			supervisor has to use a strategy that must not prefer one agent
			over another.
			This concept is called \emph{equity governance} (EG, for short).
			In order to formalize it, we can use the \SL sentence $\varphi_{\!E\!G}
			\defeq \AAll{\xSym[1]} \cdots  \AAll{\xSym[n]} (\alpha_{1}, \xSym[1])
			\cdots (\alpha_{n}, \xSym[n]) \EExs{\ySym} (\beta, \ySym) \:
			\psi_{\!E\!G}$, where $\psi_{\!E\!G} \defeq \bigwedge_{i,j = 1, i < j}^{n}
			(\EExs{\zSym[1]} (\beta, \zSym[1]) \psi_{i}) \wedge (\EExs{\zSym[2]}
			(\beta, \zSym[2]) \psi_{j}) \rightarrow (\psi_{i} \leftrightarrow
			\psi_{j})$.
			Informally, the $\psi_{\!E\!G}$ subformula asserts that, if there are two
			strategies $\zSym[1]$ and $\zSym[2]$ for $\beta$ that allow $\alpha_{i}$
			and $\alpha_{j}$ to achieve their own goals $\psi_{i}$ and $\psi_{j}$,
			separately, then the unique strategy $\ySym$ previously chosen by the
			supervisor has to ensures the achievement of either both the goals or none
			of them.
			Note that the sentence $\varphi_{\!E\!G}$ requires the existence of an EG
			strategy $\ySym$ for $\beta$, in dependence of the strategies $\xSym[1],
			\ldots, \xSym[n]$ chosen by the peers.
			To verify the existence of a uniform EG, we may use the \SL sentence
			$\varphi_{\!U\!\!E\!G} \defeq \EExs{\ySym} (\beta, \ySym) \:
			\psi'_{\!E\!G}$, with $\psi'_{\!E\!G} \defeq \AAll{\xSym[1]} \cdot
			\AAll{\xSym[n]} (\alpha_{1}, \xSym[1]) \cdots (\alpha_{n}, \xSym[n]) \:
			\psi_{\!E\!G}$, whose difference \wrt $\varphi_{\!E\!G}$ resides only in
			the alternation of quantifiers.
			Finally, to verify the existence of a uniform EG that allows also the
			existence of an NE for the peers, we can use the \SL sentence
			$\varphi_{\!U\!\!E\!G\!+\!N\!\!E} \defeq \EExs{\ySym} (\beta, \ySym)
			(\psi'_{\!E\!G} \wedge \varphi_{\!N\!\!E})$.
		\end{exa}

		Usually, the fairness of a supervisor does not ensure that the whole game can advance, \ie, that the peers can achieve their respective goals.
		Indeed, there are games like the zero-sum ones in which the agents have
		opposite goals that cannot be achieved at the same time.
		However, there are different kind of games, as the PPD or the PS of
		Examples~\ref{exm:ppd} and~\ref{exm:ps}, in which a supervisor can try to
		help all peers in their intent, by applying an \emph{advancement
		governance}.
		\begin{exa}[Advancement Governance]
			\label{exm:ag}
			Consider the game described in the previous example of EG.
			Here, we want to consider an \emph{advancement governance} (AG, for short)
			for the supervisor, \ie, a strategy for $\beta$ that allows the peers to
			achieve their own goals, if they have the will and possibility to do so.
			Formally, this concept can be expressed by using the \SL sentence
			$\varphi_{\!A\!G} \defeq \AAll{\xSym[1]} \cdot \AAll{\xSym[n]}
			(\alpha_{1}, \xSym[1]) \cdots (\alpha_{n}, \xSym[n]) \EExs{\ySym} (\beta,
			\ySym) \: \psi_{\!A\!G}$, where $\psi_{\!A\!G} \defeq (\bigwedge_{i =
			1}^{n} (\EExs{\zSym} (\beta, \zSym) \psi_{i}) \rightarrow \psi_{i})$.
			Intuitively, the $\psi_{\!A\!G}$ subformula expresses the fact that, if
			$\beta$ has a strategy $\zSym$ able to force a goal $\psi_{i}$, once the
			strategies of the peers $\alpha_{1}, \ldots, \alpha_{n}$ have been fixed,
			then his a priori choice $\ySym$ \wrt the goals has to force $\psi_{i}$ as
			well.
			As in the case of EG, we can have an uniform version of AG, by using the
			\SL sentence $\varphi_{\!U\!\!A\!G} \defeq \EExs{\ySym} (\beta, \ySym) \:
			\psi'_{\!A\!G}$, where $\psi'_{\!A\!G} \defeq \AAll{\xSym[1]} \cdots
			\AAll{\xSym[n]} (\alpha_{1}, \xSym[1]) \cdots (\alpha_{n}, \xSym[n]) \:
			\psi_{\!A\!G}$.
		\end{exa}

		Differently from the previous examples, one can consider the case in which
		the authority agent has his own goal to be satisfied, provided the other
		agents to be in a certain equilibrium.
		In the context of system design~\cite{PR89}, \emph{rational
		synthesis}~\cite{FKL10,KPV14,KPV16} is a recent improvement of the classical reactive one.
		In this setting, the adversarial environment is not a monolithic block, but
		a set of agent components, each of them having their own goal.
		In the next example, we show that the most typical instances of a rational
		synthesis problem can be represented in \SL.

		\begin{exa}[Rational Synthesis]

			Consider a solution concept that is representable in \SL by means of a
			suitable formula $\psi_{SC}$, \eg, NE, and a temporal goal
			$\psi_{\beta}$ for the system agent.
			Here, we look for a \emph{rational synthesis} solution for the
			players, \ie, a strategy profile $(\yElm, \xElm[1], \ldots, \xElm[n])$
			such that, if $\beta$ acts according to $\yElm$, then $\psi_{\beta}$ is
			satisfied and $(\xElm[1], \ldots, \xElm[n])$ is in an equilibrium
			according to the solution concept considered, \ie, $\psi_{SC}$ is
			satisfied.
			Formally, this concept can be expressed by using the \SL sentence
			$\varphi_{RS} = \EExs{\yElm} \EExs{\xElm[1]} \ldots \EExs{\xElm[n]}
			(\beta, \yElm) (\aElm[1], \xElm[1]) \ldots (\aElm[n], \xElm[n])
			(\psi_{0} \wedge \psi_{SC})$.
			As an example, $\psi_{SC}$ can be the formula $\psi_{NE}$ of
			Example~\ref{exm:ne}.
			In this case, we obtain the rational synthesis problem for NE.
		\end{exa}

	\end{subsection}

	\begin{subsection}{Basic notions}
		\label{sec:sl;sub:basnon}

		Before continuing with the formal description of \SL, we need to introduce
		some basic notions related to \ConGamStr{s}, such as those of \emph{track},
		\emph{path}, \emph{strategy}, and the like.
%		They are the natural analogous of the ones for Kripke structures and, more generally, for game arenas.
		All these notions have been already introduced in~\cite{MMV10a}.
		However, for the sake of completeness, as well as for their importance in
		the definition of \SL semantics, we fully report them in this section.

% 		We recall that a description of the used mathematical notation is reported
% 		in Appendix~\ref{app:mthnot}.

		We start with the notions of \emph{track} and \emph{path}.
		Intuitively, tracks and paths of a \ConGamStr are legal sequences of
		reachable states that can be respectively seen as partial and complete
		descriptions of possible outcomes of the game modeled by the structure
		itself.
		%
% 		\begin{defi}[Tracks and Paths]
% 			\label{def:trkpth}
			Formally, a \emph{track} (resp., \emph{path}) in a \ConGamStr
			$\ConGamStrName$ is a finite (resp., an infinite) sequence of states
			$\trkElm \in \SttSet^{*}$ (resp., $\pthElm \in \SttSet^{\omega}$) such
			that, for all $i \in \numco{0}{\card{\trkElm} - 1}$ (resp., $i \in
			\SetN$), there exists a decision $\decElm \in \DecSet$ such that
			$(\trkElm)_{i + 1} = \trnFun((\trkElm)_{i}, \decElm)$ (resp.,
			$(\pthElm)_{i + 1} = \trnFun((\pthElm)_{i}, \decElm)$)~\footnote{The
			notation $(\wElm)_{i} \in \Sigma$ indicates the \emph{element} of index $i
			\in \numco{0}{\card{\wElm}}$ of a non-empty sequence $\wElm \in
			\Sigma^{\infty}$, where $\Sigma^{\infty} = \Sigma^{*} \cup \Sigma^{\omega}$.}.
			A track $\trkElm$ is \emph{non-trivial} if it has non-zero length, \ie,
			$\card{\trkElm} > 0$ that is $\trkElm \neq \epsilon$~\footnote{The
			Greek letter $\epsilon$ stands for the \emph{empty sequence}.}.
			The set $\TrkSet \subseteq \SttSet^{+}$ (resp., $\PthSet \subseteq
			\SttSet^{\omega}$) contains all non-trivial tracks (resp., paths).
			Moreover, $\TrkSet(\sttElm) \defeq \set{ \trkElm \in \TrkSet }{
			\fst{\trkElm} = \sttElm }$ (resp., $\PthSet(\sttElm) \defeq \set{ \pthElm
			\in \PthSet }{ \fst{\pthElm} = \sttElm }$) indicates the subsets of tracks
			(resp., paths) starting at a state $\sttElm \in \SttSet$~\footnote{By
			$\fst{\wElm} \defeq (\wElm)_{0}$ we denote the \emph{first element} of
			an infinite sequence $\wElm \in \Sigma^{\infty}$.}.
% 		\end{defi}
%
		In some cases, to avoid any ambiguity, we use subscripts like $\TrkSet[\GName]$, $\PthSet[\GName]$, and so on, to denote the fact that we are referring to the set of tracks, paths, and the like, in $\GName$.

		As an example, consider the \ConGamStr $\ConGamStrName[\!P\!S]$ in
		Figure~\ref{fig:exm:ps}.
		Then $\trkElm = \sttElm[i] \cdot \sttElm[1] \cdot \sttElm[1]' \cdot
		\sttElm[1]' \cdot \sttElm[i]$ and $\pthElm = (\sttElm[i] \cdot \sttElm[1, 2]
		\cdot \sttElm[1]' \cdot \sttElm[i] \cdot \sttElm[1, 2] \cdot
		\sttElm[2]')^{\omega}$ are a track and a path, respectively.
		Moreover, we have that $\TrkSet[ {\ConGamStrName[\!P\!S]} ] = \SttSet[
		{\ConGamStrName[\!P\!S]} ]^{*}$ and $\PthSet[ {\ConGamStrName[\!P\!S]} ]
		= \SttSet[ {\ConGamStrName[\!P\!S]} ]^{\omega}$.

		At this point, we can define the concept of \emph{strategy}.
		Intuitively, a strategy is a scheme for an agent that contains all choices
		of actions as in dependence of the history of the current outcome.
		However, observe that here there is not an a priori connection between a
		strategy and an agent, since the same strategy can be used by more than one
		agent at the same time.
		%
% 		\begin{defi}[Strategies]
% 			\label{def:str}
			Formally, a \emph{strategy} in a \ConGamStr $\ConGamStrName$ is a
%			partial
			function $\strFun : \TrkSet \to \ActSet$ that maps each non-trivial track
			to an action.
			%
%			For a state $\sttElm \in \SttSet$, a strategy $\strFun$ is said to be
%			\emph{$\sttElm$-total} if it is defined on all tracks starting in
%			$\sttElm$, \ie, $\dom{\strFun} = \TrkSet(\sttElm)$.
			%
			The set $\StrSet \defeq \TrkSet \to \ActSet$
%			(resp., $\StrSet(\sttElm) \defeq \TrkSet(\sttElm) \to \ActSet$)
			contains all
%			(resp., $\sttElm$-total)
			strategies.
% 		\end{defi}

		An example of strategy in the \ConGamStr $\ConGamStrName[\!P\!S]$ is given
		by the function $\strFun[][1] \in \StrSet$ assigning the action $0$ to all
		the tracks in which the state $\sttElm[i]$ occurs an odd number of times
		and the action $1$, otherwise.
		Another example of strategy is the function $\strFun[][2] \in \StrSet$
		assigning the action $1$ on all possible tracks of the \ConGamStr.

		We now introduce the notion of \emph{assignment}.
		Intuitively, an assignment gives a valuation of variables with strategies,
		where the latter are used to determine the behavior of agents in the game.
		With more detail, as in the case of first order logic, we use this concept
		as a technical tool to quantify over strategies associated with variables,
		independently of agents to which they are related to.
		So, assignments are used precisely as a way to define a correspondence
		between variables and agents via strategies.
		\begin{defi}[Assignments]
			\label{def:asg}
			An \emph{assignment} in a \CGS\ $\GName$ is a partial function $\asgFun :
			\VarSet \cup \AgnSet \pto \StrSet$ mapping variables and agents in its
			domain to a strategy.
			An assignment $\asgFun$ is \emph{complete} if it is defined on all agents,
			\ie, $\AgnSet \subseteq \dom{\asgFun}$.
%			For a state $\sttElm \in \SttSet$, it is said that $\asgFun$ is
%			\emph{$\sttElm$-total} if all strategies $\asgFun(\lElm)$ are
%			$\sttElm$-total, for $\lElm \in \dom{\asgFun}$.
			%
			The set $\AsgSet \defeq \VarSet \cup \AgnSet \pto \StrSet$
%			(resp., $\AsgSet(\sttElm) \defeq \VarSet \cup \AgnSet \pto \StrSet(\sttElm)$)
			contains all
%			(resp., $\sttElm$-total)
			assignments.
			Moreover, $\AsgSet(\XSet) \defeq \XSet \to \StrSet$
%			(resp., $\AsgSet(\XSet, \sttElm) \defeq \XSet \to \StrSet(\sttElm)$)
			indicates the subset of \emph{$\XSet$-defined}
%			(resp., $\sttElm$-total)
			assignments, \ie,
%			(resp., $\sttElm$-total)
			assignments defined on the set $\XSet \subseteq \VarSet \cup \AgnSet$.
		\end{defi}

		As an example of assignment, consider the \ConGamStr $\ConGamStrName[\!P\!S]$ of Example~\ref{exm:ps} in which $\AgnSet = \{\PSym[1], \PSym[2]\}$ and the function $\asgFun[1] \in \AsgSet$ in , with $\dom{\asgFun[1]} = \{\PSym[1], \xSym \}$, such that $\asgFun[1](\PSym[1]) = \strFun[][1]$ and $\asgFun[1](\xSym) = \strFun[][2]$.
		As another example, consider the assignment $\asgFun[2] \in \AsgSet$ in the same \ConGamStr, with $\dom{\asgFun[2]} = \AgnSet$, such that $\asgFun[2](\SSym) = \strFun[][1]$ and $\asgFun[2](\PSym[1]) = \asgFun[2](\PSym[2]) = \strFun[][2]$.
		Note that $\asgFun[2]$ is complete, while $\asgFun[1]$ is not.

		Given an assignment $\asgFun \in \AsgSet$, an agent or variable $\lElm \in
		\VarSet \cup \AgnSet$, and a strategy $\strFun \in \StrSet$, we need to
		describe the \emph{redefinition} of $\asgFun$, \ie, a new assignment equal
		to the first one on all elements of its domain but $\lElm$, on which it
		assumes the value $\strFun$.
		%
% 		\begin{defi}[Assignment Redefinition]
% 			\label{def:asgrdf}
% 			Let $\asgFun \in \AsgSet$ be an assignment, $\strFun \in \StrSet$ a
% 			strategy and $\lElm \in \VarSet \cup \AgnSet$ either an agent or a
% 			variable.
			Formally, with $\asgFun{[\lElm \mapsto \strFun]} \in \AsgSet$ we denote
			the new assignment defined on $\dom{\asgFun{[\lElm \mapsto \strFun]}}
			\defeq \dom{\asgFun} \cup \{ \lElm \}$ that returns $\strFun$ on $\lElm$
			and is equal to $\asgFun$ on the remaining part of its domain, \ie,
			$\asgFun{[\lElm \mapsto \strFun]}(\lElm) \defeq \strFun$ and
			$\asgFun{[\lElm \mapsto \strFun]}(\lElm') \defeq \asgFun(\lElm')$, for all
			$\lElm' \in \dom{\asgFun} \setminus \{ \lElm \}$.
% 		\end{defi}
		%
		Intuitively, if we have to add or update a strategy that needs to be bound
		by an agent or variable, we can simply take the old assignment and redefine
		it by using the above notation.
		%
%		It is worth observing that, if $\asgFun$ and $\strFun$ are $\sttElm$-total then $\asgFun{[\lElm \mapsto \strFun]}$ is $\sttElm$-total, as well.

		Now, we can formalize the concept of \emph{play} in a game.
		Intuitively, a play is the unique outcome of the game determined by all
		agent strategies participating to it.
		\begin{defi}[Plays]
			\label{def:play}
			A path $\playElm \in \PthSet(\sttElm)$ starting at a state $\sttElm \in
			\SttSet$ is a \emph{play} \wrt a complete
%			$\sttElm$-total
			assignment $\asgFun \in \AsgSet(\sttElm)$ (\emph{$(\asgFun, \sttElm)$-play}, for short) if, for all $i \in \SetN$, it holds that $(\playElm)_{i + 1} = \trnFun((\playElm)_{i}, \decElm)$, where $\decElm(\aElm) \defeq \asgFun(\aElm)((\playElm)_{\leq i})$, for each $\aElm \in \AgnSet$~\footnote{The notation $(\wElm)_{\leq i} \in \Sigma^{*}$ indicates the \emph{prefix} up to index $i \in \numcc{0}{\card{\wElm}}$ of a non-empty sequence $\wElm \in \Sigma^{\infty}$.}.
			The partial function $\playFun : \AsgSet \times \SttSet \pto \PthSet$,
			with $\dom{\playFun} \defeq \set{ (\asgFun, \sttElm) }{ \AgnSet \subseteq
			\dom{\asgFun} \land \asgFun \in \AsgSet(\sttElm) \land \sttElm \in \SttSet
			}$, returns the $(\asgFun, \sttElm)$-play $\playFun(\asgFun, \sttElm) \in
			\PthSet(\sttElm)$, for all pairs $(\asgFun, \sttElm)$ in its domain.
		\end{defi}

		As last example, consider again the \ConGamStr $\ConGamStrName[\!P\!S]$ and
		the complete
%		$\sttElm[0]$-total
		assignment $\asgFun[2]$ defined above.
		Then, we have that $\playFun(\asgFun[2], \sttElm[0]) = (\sttElm[i] \cdot
		\sttElm[1, 2] \cdot \sttElm[1]' \cdot \sttElm[i] \cdot \sttElm[1, 2] \cdot
		\sttElm[2]')^{\omega}$.

		Finally, we give the definition of global translation of a complete
		assignment associated with a state, which is used to capture, at a certain
		step of the play, what is the current state and its updated assignment.
		%
% 		\begin{defi}[Global Translation]
% 			\label{def:gbltrn}
% 			For a given state $\sttElm \in \SttSet$ and a complete $\sttElm$-total
% 			assignment $\asgFun \in \AsgSet(\sttElm)$, the \emph{$i$-th global
% 			translation} of $(\asgFun, \sttElm)$, with $i \in \SetN$, is the pair of
% 		a
% 			complete assignment and a state $(\asgFun, \sttElm)^{i} \defeq
% 			((\asgFun)_{(\playElm)_{\leq i}}, (\playElm)_{i})$, where $\playElm =
% 			\playFun(\asgFun, \sttElm)$.
% 		\end{defi}

		\begin{defi}[Global Translation]
			\label{def:gbltrn}
			For a given state $\sttElm \in \SttSet$ and a complete
%			$\sttElm$-total
			assignment $\asgFun \in \AsgSet$, an \emph{$i$-th global translation} of $(\asgFun, \sttElm)$, with $i \in \SetN$, is a pair of a complete assignment and a state $(\asgFun, \sttElm)^{i} \defeq ((\asgFun)_{(\playElm)_{\leq i}}, (\playElm)_{i})$, where $\playElm = \playFun(\asgFun, \sttElm)$ and $(\asgFun)_{(\playElm)_{\leq i}}$ denotes
%			the $(\playElm)_{i}$-total
			an assignment
%			, with $\dom{(\asgFun)_{(\playElm)_{\leq i}}} = \dom{\asgFun}$,
			such that, for all $\lElm \in \dom{\asgFun}$,
%			$\dom{(\asgFun)_{(\playElm)_{\leq i}}(\lElm)} = \set{\trkElm \in \TrkSet((\playElm)_{i})}{(\playElm)_{\leq i} \cdot \trkElm \in \dom{\asgFun(\lElm)}}$ and
		$(\asgFun)_{(\playElm)_{\leq
			i}}(\lElm)(\trkElm) = \asgFun(\lElm)((\playElm)_{\leq i} \cdot \trkElm)$,
			for all $\trkElm \in \dom{(\asgFun)_{(\playElm)_{\leq i}}(\lElm)}$.
		\end{defi}

		Intuitively, an $i$-th global translation of $(\sttElm, \asgFun)$ is meant to return a pair of a state and a complete assignment $(\asgFun, \sttElm)^{i}$ for which the play $\playFun((\asgFun, \sttElm)^{i})$ generated corresponds to $\playFun((\asgFun, \sttElm))_{\geq i}$, \ie, the suffix from the $i$-th element of the play $\playFun(\sttElm, \asgFun)$.
		This property will be used below to correctly define the semantics of the temporal operators in \SL.

	\end{subsection}

	\begin{subsection}{Semantics}
		\label{sec:sl;sub:sem}

		As already reported at the beginning of this section, just like \ATLS\ and
		differently from \CHPSL, the semantics of \SL is defined \wrt concurrent
		game structures.
		For an \SL formula $\varphi$, a \CGS\ $\GName$, a state $\sElm$ in it,
		and an
%		$\sElm$-total
		assignment $\asgFun$ with $\free{\varphi} \subseteq
		\dom{\asgFun}$, we write $\GName, \asgFun, \sElm \models \varphi$ to
		indicate that the formula $\varphi$ holds at $\sElm$ in $\GName$ under
		$\asgFun$.
		The semantics of \SL formulas involving the atomic propositions, the
		Boolean connectives $\neg$, $\wedge$, and $\vee$, as well as the temporal
		operators $\X$, $\U$, and $\R$ is defined as usual in \LTL.
		The novel part resides in the formalization of the meaning of strategy
		quantifications $\EExs{\xElm}$ and $\AAll{\xElm}$ and agent binding $(\aElm,
		\xElm)$.
		\begin{defi}[\SL Semantics]
			\label{def:semantics}
			Given a \CGS\ $\GName$, for all \SL formulas $\varphi$, states $\sElm \in
			\SttSet$, and
%			$\sElm$-total
			assignments $\asgFun \in \AsgSet$ with $\free{\varphi} \subseteq \dom{\asgFun}$, the modeling relation $\GName, \asgFun, \sElm \models \varphi$ is inductively defined as follows.
			\begin{enumerate}
				\item\label{def:semantics:ap}
					$\GName, \asgFun, \sElm \models \pElm$ if $\pElm \in \apFun(\sElm)$,
					with $\pElm \in \APSet$.
				\item\label{def:semantics:bool}
					For all formulas $\varphi$, $\varphi_{1}$, and $\varphi_{2}$, it holds
					that:
					\begin{enumerate}
						\item\label{def:semantics:neg}
							$\GName, \asgFun, \sElm \models \neg \varphi$ if not $\GName,
							\asgFun, \sElm \models \varphi$, that is $\GName, \asgFun, \sElm
							\not\models \varphi$;
						\item\label{def:semantics:conj}
							$\GName, \asgFun, \sElm \models \varphi_{1} \wedge \varphi_{2}$ if
							$\GName, \asgFun, \sElm \models \varphi_{1}$ and $\GName, \asgFun,
							\sElm \models \varphi_{2}$;
						\item\label{def:semantics:disj}
							$\GName, \asgFun, \sElm \models \varphi_{1} \vee \varphi_{2}$ if
							$\GName, \asgFun, \sElm \models \varphi_{1}$ or $\GName, \asgFun,
							\sElm \models \varphi_{2}$.
					\end{enumerate}
				\item\label{def:semantics:qnt}
					For a variable $\xElm \in \VarSet$ and a formula $\varphi$, it holds
					that:
					\begin{enumerate}
						\item\label{def:semantics:eqnt}
							$\GName, \asgFun, \sElm \models\! \EExs{\xElm} \varphi$ if there
							is a strategy $\strFun \in \StrSet$ such that $\GName, \asgFun{[\xElm \mapsto \strFun]}, \sElm \models\!
							\varphi$;
						\item\label{def:semantics:aqnt}
							$\GName, \asgFun, \sElm \models \AAll{\xElm} \varphi$ if, for all strategies $\strFun \in \StrSet$, it holds that $\GName, \asgFun{[\xElm \mapsto \strFun]}, \sElm \models \varphi$.
					\end{enumerate}
				\item\label{def:semantics:bnd}
					For an agent $\aElm \in \AgnSet$, a variable $\xElm \in \VarSet$, and
					a formula $\varphi$, it holds that $\GName, \asgFun, \sElm \models
					(\aElm, \xElm) \varphi$ if $\GName, \asgFun{[\aElm \mapsto
					\asgFun(\xElm)]}, \sElm \models \varphi$.
				\item\label{def:semantics:path}
					Finally, if the assignment $\asgFun$ is complete, for all formulas
					$\varphi$, $\varphi_{1}$, and $\varphi_{2}$, it holds that:
					\begin{enumerate}
						\item\label{def:semantics:next}
							$\GName, \asgFun, \sElm \models \X \varphi$ if $\GName, (\asgFun,
							\sElm)^{1} \models \varphi$;
						\item\label{def:semantics:until}
							$\GName, \asgFun, \sElm \models \varphi_{1} \U \varphi_{2}$ if
							there is an index $i \in \SetN$ with $k \leq i$ such that $\GName,
							(\asgFun, \sElm)^{i} \models \varphi_{2}$ and, for all indexes $j
							\in \SetN$ with $k \leq j < i$, it holds that $\GName, (\asgFun,
							\sElm)^{j} \models \varphi_{1}$;
						\item\label{def:semantics:release}
							$\GName, \asgFun, \sElm \models \varphi_{1} \R \varphi_{2}$ if,
							for all indexes $i \in \SetN$ with $k \leq i$, it holds that
							$\GName, (\asgFun, \sElm)^{i} \models \varphi_{2}$ or there is an
							index $j \in \SetN$ with $k \leq j < i$ such that $\GName,
							(\asgFun, \sElm)^{j} \models \varphi_{1}$.
					\end{enumerate}
			\end{enumerate}
		\end{defi}
		\noindent Intuitively, at Items~\ref{def:semantics:eqnt}
		and~\ref{def:semantics:aqnt}, respectively, we evaluate the existential
		$\EExs{\xElm}$ and universal $\AAll{\xElm}$ quantifiers over strategies, by
		associating them to the variable $\xElm$.
		Moreover, at Item~\ref{def:semantics:bnd}, by means of an agent binding
		$(\aElm, \xElm)$, we commit the agent $\aElm$ to a strategy associated with
		the variable $\xElm$.
		It is evident that, due to Items~\ref{def:semantics:next},
		\ref{def:semantics:until}, and~\ref{def:semantics:release}, the \LTL
		semantics is simply embedded into the \SL one.

		In order to complete the description of the semantics, we now give the
		classic notions of \emph{model} and \emph{satisfiability} of an \SL
		sentence.
		%
% 		\begin{defi}[\SL Satisfiability]
% 			\label{def:sl(sat)}
			We say that a \CGS\ $\GName$ is a \emph{model} of an \SL sentence
			$\varphi$, in symbols $\GName \models \varphi$, if $\GName, \emptyfun,
			\sElm[0] \models \varphi$.~\footnote{The symbol $\emptyfun$ stands for the
			empty function.}
			In general, we also say that $\GName$ is a \emph{model} for $\varphi$ on
			$\sElm \in \SttSet$, in symbols $\GName, \sElm \models \varphi$, if
			$\GName, \emptyfun, \sElm \models \varphi$.
			An \SL sentence $\varphi$ is \emph{satisfiable} if there is a model for
			it.
% 		\end{defi}

		It remains to formalize the concepts of \emph{implication} and
		\emph{equivalence} between \SL formulas, which are useful to describe
		transformations preserving the meaning of a specification.
		%
% 		\begin{defi}[\SL Implication and Equivalence]
% 			\label{def:sl(impeqv)}
			Given two \SL formulas $\varphi_{1}$ and $\varphi_{2}$, with
			$\free{\varphi_{1}} = \free{\varphi_{2}}$, we say that $\varphi_{1}$
			\emph{implies} $\varphi_{2}$, in symbols $\varphi_{1} \implies
			\varphi_{2}$, if, for all \CGS{s} $\GName$, states $\sElm \in \SttSet$,
			and $\free{\varphi_{1}}$-defined assignments $\asgFun \in
			\AsgSet(\free{\varphi_{1}}, \sElm)$, it holds that if $\GName, \asgFun,
			\sElm \models \varphi_{1}$ then $\GName, \asgFun, \sElm \models
			\varphi_{2}$.
			Accordingly, we say that $\varphi_{1}$ is \emph{equivalent} to
			$\varphi_{2}$, in symbols $\varphi_{1} \equiv \varphi_{2}$, if both
			$\varphi_{1} \implies \varphi_{2}$ and $\varphi_{2} \implies \varphi_{1}$
			hold.
% 		\end{defi}

		In the rest of the paper, especially when we describe a decision procedure,
		we may consider formulas in \emph{existential normal form} (\enf, for short)
		and \emph{positive normal form} (\pnf, for short), \ie, formulas in which
		only existential quantifiers appear or, respectively, the negation is
		applied solely to atomic propositions.
		In fact, it is to this aim that we have considered in the syntax of \SL
		both the Boolean connectives $\wedge$ and $\vee$, the temporal operators
		$\U$, and $\R$, and the strategy quantifiers $\EExs{\cdot}$ and
		$\AAll{\cdot}$.
		Indeed, all formulas can be linearly translated in \enf and \pnf by using De
		Morgan's laws together with the following equivalences, which directly
		follow from the semantics of the logic: $\neg \X \varphi \equiv \X \neg
		\varphi$, $\neg (\varphi_{1} \U \varphi_{2}) \equiv (\neg \varphi_{1}) \R
		(\neg \varphi_{2})$, $\neg \EExs{x} \varphi \equiv \AAll{x} \neg \varphi$,
		and $\neg (\aElm, \xElm) \varphi \equiv (\aElm, \xElm) \neg \varphi$.

		At this point, in order to better understand the meaning of the \SL
		semantics, we discuss some examples of formulas interpreted over the
		\ConGamStr{s} previously described.

		\begin{exa}[Law and Order]
			Consider the \ConGamStr $\GName[P\!P\!D]$ given in Example~\ref{exm:ppd}.
			It is easy to see that Police can ensure at least one prisoner to never be
			free.
			Indeed the formula $\varphi_{1} = \EExs{\yElm} \AAll{\xElm[1]}
			\AAll{\xElm[2]} (\PSym, \yElm) (\ASym[1], \xElm[1]) (\ASym[2], \xElm[2])
			((\F \G \neg \fSym[ {\ASym[1]} ]) \vee (\F \G \neg \fSym[ {\ASym[2]} ]))$
			is satisfied over $\GName[P\!P\!D]$.
			A way to see this is to consider the strategy $\strFun$ for $\PSym$ given
			by $\strFun(\trkElm) = 0$,  for all $\trkElm \in \TrkSet$, which allows
			agent $\PSym$ to always avoid the state $\sttElm[ {\ASym[1], \ASym[2]}
			]$.
			On the other hand, the formula $\varphi_{2} = \AAll{\xElm[1]}
			\AAll{\xElm[2]} \EExs{\yElm} (\PSym, \yElm) (\ASym[1], \xElm[1])
			(\ASym[2], \xElm[2]) (\F \G (\neg \fSym[ {\ASym[1]} ] \wedge \neg \fSym[
			{\ASym[2]} ]))$ is not satisfied over $\GName[P\!P\!D]$.
			Indeed, if agents $\ASym[1]$ and $\ASym[2]$ use strategies $\strFun[][1]$
			and $\strFun[][2]$, respectively, such that $\strFun[][2](\sttElm[i]) = 1
			- \strFun[][1](\sttElm[i])$, we have that, whatever agent $\PSym$ does, at
			least one of them gains freedom.
		\end{exa}

		\begin{exa}[Fair Scheduler]
			Consider the \ConGamStr $\GName[P\!S]$ of Example~\ref{exm:ps} and
			suppose the Scheduler wants to ensure that, whatever process makes a
			request, the resource is eventually granted to it.
			We can represent this specification by means of the formula $\varphi =
			\EExs{\yElm} \AAll{\xElm[1]} \AAll{\xElm[2]} (\SSym, \yElm) (\PSym[1],
			\xElm[1]) (\PSym[2], \xElm[2])(\G((\rSym[1] \rightarrow \F \gSym[1])
			\wedge (\rSym[2] \rightarrow \F \gSym[2])))$.
			It is easy to see that $\GName[P\!S] \models \varphi$.
			Indeed, consider the strategy $\strFun$ for $\SSym$ defined as follows.
			For all tracks of the form $\trkElm \cdot \sttElm'_{i}$, we set a possible
			preemption, \ie, $\strFun(\trkElm \cdot \sttElm'_{i}) = 1$.
			For the tracks of the form $\trkElm \cdot \sttElm[1, 2]$ we set the action
			as prescribed in the sequel: \emph{(i)} if there is no occurrence of
			$\sttElm'_{1}$ and $\sttElm'_{2}$ or the last occurrence is
			$\sttElm'_{1}$, then we release the resource to
			$\PSym[1]$, \ie, $\strFun(\trkElm \cdot \sttElm[1, 2]) = 0$; \emph{(ii)}
			if the last occurrence in $\trkElm$ between $\sttElm'_{1}$ and
			$\sttElm'_{2}$ is $\sttElm'_{1}$, then then we release the resource to
			$\PSym[2]$, \ie, $\strFun(\trkElm \cdot \sttElm[1, 2]) = 1$.

		\end{exa}

	\end{subsection}

\end{section}

% End of file SectionI.tex

%%****************************************************************************%%
%%                                                                            %%
%% Reasoning About Strategies: On the Satisfiability Problem                  %%
%%                                                                            %%
%% SectionII.tex                                                              %%
%%                                                                            %%
%% Revision 0                                                                 %%
%%                                                                            %%
%% Copyright (C) 2014, Fabio Mogavero, Aniello Murano, Giuseppe Perelli, and  %%
%%                     Moshe Y. Vardi.                                        %%
%% All rights reserved.                                                       %%
%%                                                                            %%
%%****************************************************************************%%

% Begin of file SectionII.tex

\begin{section}{Hardness results}
	\label{sec:hard}

  In~\cite{MMPV14} it has been shown that the model-checking for \SL is
  \NonElmH.
  Here, we prove that the satisfiability problem is even harder, \ie,
  undecidable.
  To do this, we first introduce a sentence that is satisfiable only on
  unbounded models.
  Then, by using this result, we prove the undecidability result through a
  reduction of the classic domino problem~\cite{Wan61}.

	\begin{subsection}{Unbounded models}
		\label{sec:hard;sub:unbmod}

		We now show that \SL does not enjoy the bounded-tree model property.
		In general, a classic modal logic satisfies this property if, whenever a
		formula is satisfiable, it is so on a model in which all states have a
		number of successors bounded by an a priori fixed constant.
		However, in the case of \SL, this condition is not sufficient to
		characterize bounded models, since \SL has the power of distinguishing among
		different ways to reach a given state from another one, and the number of
		ways might be infinite, as that of actions can be infinite itself.
		An example of an unbounded model with a finite number of states is given in
		Figure~\ref{fig:ordsat:sl}.
		For this reason, we say that a \ConGamStr is \emph{bounded} if the set of
		actions $\ActSet[\GName]$ is finite.
		Moreover, a model is \emph{finite} if it is bounded and the number of states
		is also finite.
		Clearly, if a logic invariant under unwinding enjoys the finite model
		property, it enjoys the bounded-tree model property as well.
		The other direction may not hold, instead, as exemplified by the \MuCalculus
		with backward modalities~\cite{Var98,Boj03}.
		Unfortunately, \SL does not enjoy either property.

		In order to prove the existence of satisfiable \SL formulas with unbounded
		models only, we introduce, in the following definition, the sentence
		$\varphi^{ord}$ to be used as a counterexample for the bounded-tree model
		property.

		\begin{defi}[Ordering Sentence]
			\label{def:ord}
			Let $x_{1} < x_{2} \defeq \EExs{\ySym} \: \varphi(x_{1},
			x_{2}, \ySym)$ be an agent-closed formula, named \emph{partial order},
			on the sets $\APSet = \{ \pSym \}$ and $\AgnSet = \{ \alpha, \beta \}$,
			where $\varphi(x_{1}, x_{2}, \ySym) \defeq ((\alpha, x_{1})
			(\beta, \ySym) (\X \pSym)) \wedge ((\alpha, x_{2}) (\beta, \ySym) (\X
			\neg \pSym))$.
			Then, the \emph{ordering sentence} $\varphi^{ord} \defeq \varphi^{unb}
			\wedge \varphi^{trn}$ is the conjunction of the following two sentences,
			called \emph{unboundedness} and \emph{transitivity} strategy requirements:
			\begin{enumerate}
				\item
					$\varphi^{unb} \defeq \AAll{x_{1}} \EExs{x_{2}} \: x_{1} <
					x_{2}$;
				\item
					$\varphi^{trn} \defeq \AAll{x_{1}} \AAll{x_{2}} \AAll{x_{3}}
					\: (x_{1} < x_{2} \wedge x_{2} < x_{3}) \rightarrow
					x_{1} < x_{3}$.
			\end{enumerate}
		\end{defi}

		\figrasvii

		Intuitively, $\varphi^{unb}$ asserts that, for each strategy in $x_{1}$,
		there is a different strategy in $x_{2}$ that is in relation of $<$
		\wrt the first one, \ie, $<$ has no upper bound, due to the fact that,
		by the definition of $\varphi(x_{1}, x_{2}, \ySym)$, it is not
		reflexive.
		Moreover, $\varphi^{trn}$ ensures that the relation $<$ is transitive too.
		Consequently, $\varphi^{ord}$ induces a strict partial pre-order on the
		strategies.

		Obviously, in order to be useful, the sentence $\varphi^{ord}$ needs to be
		satisfiable, as reported in the following lemma.

		\begin{lem}[Ordering Satisfiability]
			\label{lmm:ordsat:sl}
			\ \allowbreak The sentence $\varphi^{ord}$ is satisfiable.
		\end{lem}
		\begin{proof}
			To prove that $\varphi^{ord}$ is satisfiable, consider the unbounded \CGS\
			$\GName^{\star}$ in Figure~\ref{fig:ordsat:sl}, where \emph{(i)} $\APSet \defeq \{ p \}$, \emph{(ii)}
			$\AgnSet \defeq \{ \alpha, \beta \}$, \emph{(iii)}
			$\ActSet_{\GName^{\star}} \defeq \SetN$, \emph{(iv)}
			$\SttSet_{\GName^{\star}} \defeq \{ s_{0}, s_{1}, s_{2} \}$, \emph{(v)}
			${s_{0}}_{\GName^{\star}} = s_{0}$, \emph{(vi)}
			$\apFun_{\GName^{\star}}(s_{0}) = \apFun_{\GName^{\star}}(s_{2}) \defeq
			\emptyset$ and $\apFun_{\GName^{\star}}(s_{1}) \defeq  \{ p \}$, \emph{(vii)} $\PSet \defeq \set{ \decElm \in \DecSet_{\GName^{\star}} }{ \decElm(\alpha) \leq \decElm(\beta) }$, and
			\emph{(viii)} $\trnFun_{\GName^{\star}}$ is such that if $\decElm \in \PSet$ then $\trnFun_{\GName^{\star}}(s_{0}, \decElm) = s_{1}$
			else $\trnFun_{\GName^{\star}}(s_{0}, \decElm) = s_{2}$, and
			$\trnFun_{\GName^{\star}}(s, \decElm) = s$, for all $s \in \{ s_{1}, s_{2}
			\}$ and $\decElm \in \DecSet_{\GName^{\star}}$.
% 			\ref{fig:ordsat[sl]}).

			Now, it is easy to see that $\GName^{\star} \models \varphi^{unb}$, since
			for every strategy $\strFun[][x_{1}]$ for $x_{1}$, consisting of picking a
			natural number $n = \strFun[][x_{1}](s_{0})$ as an action at the initial
			state, we can reply with a strategy $\strFun[][x_{2}]$ for $x_{2}$ having
			$\strFun[][x_{2}](s_{0}) > n$ and a strategy $\strFun[][y]$ for $y$ having
			$\strFun[][y](s_{0}) = n$.
			Formally, we have that $\GName^{\star}, \asgFun, s_{0} \models
			\varphi(x_{1}, x_{2}, y)$ iff $\asgFun(x_{1})(s_{0}) \leq
			\asgFun(y)(s_{0}) < \asgFun(x_{2})(s_{0})$, for all assignments $\asgFun
			\in \AsgSet_{\GName^{\star}}(\{ x_{1}, x_{2}, y \}, s_{0})$.

			By a similar reasoning, we can see that $\GName^{\star} \models
			\varphi^{trn}$.
			Indeed, consider three strategies $\strFun[][x_{1}]$, $\strFun[][x_{2}]$,
			and $\strFun[][x_{3}]$ for the variables $x_{1}$, $x_{2}$, and $x_{3}$,
			respectively, which correspond to picking three natural numbers $n_{1} =
			\strFun[][x_{1}](s_{0})$, $n_{2} = \strFun[][x_{2}](s_{0})$, and $n_{3} =
			\strFun[][x_{3}](s_{0})$.
			Now, if $\GName^{\star}, \asgFun, s_{0} \models x_{1} < x_{2}$ and
			$\GName^{\star}, \asgFun, s_{0} \models x_{2} < x_{3}$, for an
			assignments $\asgFun \!\in\! \AsgSet_{\GName^{\star}}(\{ x_{1}, x_{2},
			\allowbreak x_{3} \}, s_{0})$ where $\asgFun(x_{1}) = \strFun[][x_{1}]$,
			$\asgFun(x_{2}) = \strFun[][x_{2}]$, and $\asgFun(x_{3}) =
			\strFun[][x_{3}]$, we have that $n_{1} < n_{2}$ and $n_{2} < n_{3}$.
			Consequently, $n_{1} < n_{3}$.
			Hence, by using a strategy $\strFun[][y]$ for $y$ with
			$\strFun[][y](s_{0}) = \strFun[][x_{1}](s_{0})$, we have $\GName^{\star},
			\asgFun[ {y \mapsto \strFun[][y]} ], s_{0} \models \varphi(x_{1},
			x_{3}, y)$ and thus $\GName^{\star}, \asgFun, s_{0} \models x_{1} <
			x_{3}$.
		\end{proof}

		Next lemmas report two important properties of the sentence $\varphi^{ord}$,
		for the negative statements we want to show.
		Namely, they state that, in order to be satisfied, $\varphi^{ord}$ must
		require the existence of strict partial order relations on strategies and
		actions that do not admit any maximal element.
		From this, as stated in Theorem~\ref{thm:unbprp}, we directly derive that
		$\varphi^{ord}$ needs an infinite chain of actions to be satisfied, \ie, it
		cannot have a bounded model.

		\begin{lem}[Strategy Order]
			\label{lmm:strord}
			Let $\GName$ be a model of $\varphi^{ord}$.
			Moreover, let $\rRel^{<} \subseteq \StrSet_{\GName} \times \StrSet_{\GName}$ be a relation between strategies of $\GName$ such that $\rRel^{<}(\strFun_{1}, \strFun_{2})$ holds iff $\GName, \emptyfun{[x_{1} \mapsto \strFun_{1}][x_{2} \mapsto \strFun_{2}]}, {s_{0}}_{\GName} \models x_{1} < x_{2}$, for all strategies $\strFun_{1}, \strFun_{2} \in \StrSet_{\GName}$.
			Then, $\rRel^{<}$ is a \emph{strict partial order without maximal
			element}.
		\end{lem}
		\begin{proof}
			The proof derives from the fact that $\rRel^{<}$ satisfies the following
			properties:
			\begin{enumerate}
				\item
					\emph{Irreflexivity}: $\forall \strFun \in
					\StrSet_{\GName} .\: \neg \rRel^{<}(\strFun,
					\strFun)$;
				\item
					\emph{Unboundedness}: $\forall \strFun_{1} \in
					\StrSet_{\GName} \: \exists \strFun_{2} \in
					\StrSet_{\GName} .\: \rRel^{<}(\strFun_{1},
					\strFun_{2})$;
				\item
					\emph{Transitivity}: $\forall \strFun_{1}, \strFun_{2}, \strFun[][3]
					\in \StrSet_{\GName} .\: (\rRel^{<}(\strFun_{1},
					\strFun_{2}) \wedge \rRel^{<}(\strFun_{2}, \strFun[][3])) \rightarrow
					\rRel^{<}(\strFun_{1}, \strFun[][3])$.
			\end{enumerate}
			Indeed, Items \emph{(ii)} and \emph{(iii)} are directly derived from the
			strategy unboundedness and transitivity requirements.
			The proof of Item \emph{(i)} derives, instead, from the following
			reasoning.
			By contradiction, suppose that $\rRel^{<}$ is not a strict order, \ie,
			there is a strategy $\strFun \in \StrSet_{\GName}$ for
			which $\rRel^{<}(\strFun, \strFun)$ holds.
			This means that, at the initial state ${s_{0}}_{\GName}$ of $\GName$,
			there exists an assignment $\asgFun \in \AsgSet_{\GName}(\{ x_{1}, x_{2},
			y \}, {s_{0}}_{\GName})$ for which $\GName, \asgFun, {s_{0}}_{\GName}
			\models \varphi(x_{1}, x_{2}, y)$, where $\asgFun(x_{1}) = \asgFun(x_{2})
			= \strFun$.
			The last fact implies the existence of a successor of ${s_{0}}_{\GName}$
			in which both $p$ and $\neg p$ hold, which is clearly impossible.
		\end{proof}

		\begin{lem}[Action Order]
			\label{lmm:acord}
			Let $\GName$ be a model of $\varphi^{ord}$.
			Moreover, let $\sRel^{<} \subseteq \ActSet_{\GName} \times
			\ActSet_{\GName}$ be a relation between actions of $\GName$ such that
			$\sRel^{<}(c_{1}, c_{2})$ holds iff, for all strategies $\strFun_{1},
			\strFun_{2} \in \StrSet_{\GName}$ with $c_{1} =
			\strFun_{1}({s_{0}}_{\GName})$ and $c_{2} =
			\strFun_{2}({s_{0}}_{\GName})$, it holds that $\rRel^{<}(\strFun_{1},
			\strFun_{2})$, where $c_{1}, c_{2} \in \ActSet_{\GName}$.
			Then, $\sRel^{<}$ is a \emph{strict partial order without maximal
			element}.
		\end{lem}
		\begin{proof}
			The irreflexivity and transitivity of $\sRel^{<}$ are directly derived
			from the fact that, by Lemma \ref{lmm:strord}, $\rRel^{<}$ is
			irreflexive and transitive too.
			The proof of the unboundedness property derives, instead, from the
			following reasoning.
			As first thing, observe that, since the formula $x_{1} < x_{2}$ relies on
			$\X p$ and $\X \neg p$ as the only temporal operators, it holds that
			$\rRel^{<}(\strFun_{1}, \strFun_{2})$ implies $\rRel^{<}(\strFun_{1}',
			\strFun_{2}')$, for all strategies $\strFun_{1}, \strFun_{2},
			\strFun_{1}', \strFun_{2}' \in \StrSet_{\GName}$ such
			that $\strFun_{1}({s_{0}}_{\GName}) = \strFun_{1}'({s_{0}}_{\GName})$ and
			$\strFun_{2}({s_{0}}_{\GName}) = \strFun_{2}'({s_{0}}_{\GName})$.
			Now, suppose by contradiction that $\sRel^{<}$ does not satisfy the
			unboundedness property, \ie, there is an action $c \in \ActSet_{\GName}$
			such that, for all actions $c' \in \ActSet_{\GName}$, it does not hold that $\sRel^{<}(c, c')$.
			Then, by the definition of $\sRel^{<}$ and the previous observation, we
			derive the existence of a strategy $\strFun \in
			\StrSet_{\GName}$ with $\strFun({s_{0}}_{\GName}) = c$
			such that $\rRel^{<}(\strFun, \strFun')$ does not hold, for any strategy
			$\strFun' \in \StrSet_{\GName}$, which is clearly
			impossible.
		\end{proof}

		Now, we have all tools to prove that \SL\ lacks of the bounded-tree model
		property, which hold, instead, for several commonly used multi-agent logics,
		such as \ATLS.

% 		\begin{thm}[\SL\ Negative Properties]
% 			\label{thm:negprop[sgsl]}
% 			For \SL, it holds that:
% 			\begin{enumerate}
% 				\item\label{thm:negprop[sgsl](bndtree)}
% 					does not have the bounded-tree model property;
% 				\item\label{thm:negprop[sgsl](finmod)}
% 					does not have the finite-model property.
% 			\end{enumerate}
% 		\end{thm}

		\begin{thm}[\SL\ Unbounded Model Property]
			\label{thm:unbprp}
			\SL does not enjoy the bounded model property.
		\end{thm}

		\begin{proof}
			To prove the statement, we show that the sentence $\varphi^{ord}$ of
			Definition~\ref{def:ord} cannot be satisfied on a bounded \CGS.
			Consider a \CGS\ $\GName$ such that $\GName \models \varphi^{ord}$.
			The existence of such a model is ensured by Lemma
			\ref{lmm:ordsat:sl}.
			Now, consider the strict partial order without maximal element between
			actions $\sRel^{<}$ described in Lemma \ref{lmm:acord}.
			By a classical result on first order logic model theory \cite{EF95}, the
			relation $\sRel^{<}$ cannot be defined on a finite set.
			Hence, $\card{\ActSet} = \infty$.
		\end{proof}

	\end{subsection}

	\begin{subsection}{Undecidable satisfiability}
		\label{sec:hard;sub:undsat}

		We finally show the undecidability of the satisfiability problem
		for \SL through a reduction from the \emph{recurrent domino problem}.
% 		In particular, as we discuss later, the reduction also holds for \CHPSL\
% 		under the concurrent game semantics.

		The \emph{domino problem}, proposed for the first time by Wang~\cite{Wan61},
		consists of placing a given number of tile types on an infinite grid,
		satisfying a predetermined set of constraints on adjacent tiles.
		One of its standard versions asks for a compatible tiling of the whole plane
		$\SetN \times \SetN$.
		The \emph{recurrent domino problem} further requires the existence of a
		distinguished tile type that occurs infinitely often in the first row of the
		grid.
		This problem was proved to be highly undecidable by Harel, and in
		particular, \SCompC{1}{1}~\cite{Har84}.
		The formal definition follows.

		\begin{defi}[Recurrent Domino System]
			\label{def:recdomsys}
			An \emph{$\SetN \times \SetN$ recurrent domino system} $\DName =
			\DSStruct[t_{0}]$ consists of a finite non-empty set $\DSet$ of
			\emph{domino types}, two \emph{horizontal} and \emph{vertical matching
			relations} $\HRel, \VRel \subseteq \DSet \times \DSet$, and a
			\emph{distinguished tile type} $t_{0} \in \DSet$.
			The recurrent domino problem asks for an \emph{admissible tiling} of
			$\SetN \times \SetN$, which is a \emph{solution mapping} $\dsmFun : \SetN
			\times \SetN \to \DSet$ such that, for all $x, y \in \SetN$, it holds that
			\emph{(i)} $(\dsmFun(x, y), \dsmFun(x + 1, y)) \in \HRel$, \emph{(ii)}
			$(\dsmFun(x, y), \dsmFun(x, y + 1)) \in \VRel$, and \emph{(iii)}
			$\card{\set{ x \in \SetN }{ \dsmFun(x, 0) = t_{0} }} = \omega$.
		\end{defi}

		\begin{paragraph}*{\textbf{Grid specification}}

			Consider the sentence $\varphi^{grd} \defeq
			\bigwedge_{a \in \AgnSet} \varphi_{a}^{ord}$, where $\varphi_{a}^{ord} =
			\varphi_{a}^{unb} \wedge \varphi_{a}^{trn}$ are the \emph{order sentences}
			and $\varphi_{a}^{unb}$ and $\varphi_{a}^{trn}$ are the
			\emph{unboundedness} and \emph{transitivity} strategy requirements for
			agents $\alpha$ and $\beta$ defined, similarly to Definition
			\ref{def:ord}, as follows:
			\begin{enumerate}
				\item
					$\varphi_{a}^{unb} \defeq \AAll{z_{1}} \EExs{z_{2}} \: z_{1} <_{a}
					z_{2}$;
				\item
					$\varphi_{a}^{trn} \defeq \AAll{z_{1}} \AAll{z_{2}} \AAll{z_{3}} \:
					(z_{1} <_{a} z_{2} \wedge z_{2} <_{a} z_{3}) \rightarrow z_{1} <_{a}
					z_{3}$;
			\end{enumerate}
			where $x_{1} <_{\alpha} x_{2} \defeq \EExs{y} \varphi_{\alpha}(x_{1},
			x_{2}, y)$ and $y_{1} <_{\beta} y_{2} \defeq \EExs{x}
			\varphi_{\beta}(y_{1}, y_{2}, x)$ are the two \emph{partial order}
			formulas on strategies of $\alpha$ and $\beta$, respectively, with
			$\varphi_{\alpha}(x_{1}, x_{2}, y) \defeq (\beta, y) ((\alpha, \allowbreak
			x_{1}) (\X p) \wedge (\alpha, x_{2}) (\X \neg p))$ and
			$\varphi_{\beta}(y_{1}, y_{2}, x) \defeq (\alpha, x) ((\beta, y_{1}) (\X
			\neg p) \wedge (\beta, y_{2}) (\X p))$.
			Intuitively, $<_{\alpha}$ and $<_{\beta}$ correspond to the horizontal and
			vertical ordering of the positions in the grid, respectively.

			It is easy to show that $\varphi^{grd}$ is satisfiable, by using the same
			candidate model $\GName^{\star}$ (see Figure~\ref{fig:ordsat:sl})and a
			proof argument similar to that proposed in Lemma \ref{lmm:ordsat:sl} for
			the simpler order sentence.
			\begin{lem}[Grid Ordering Satisfiability]
				\label{lmm:grdsat[sl]}
				The sentence $\varphi^{grd}$ is satisfiable.
			\end{lem}
			\begin{proof}
				Let $\GName^{\star}$ be the model described in Lemma
				\ref{lmm:ordsat:sl}.
				On one hand, by using the same lemma, it is evident that $\GName^{\star}
				\models \varphi_{\alpha}^{ord}$.
				On the other hand, in order to prove that $\GName^{\star} \models
				\varphi_{\beta}^{ord}$, first observe that $\GName^{\star}, \asgFun,
				{s_{0}}_{\GName^{\star}} \models \varphi_{\beta}(y_{1}, y_{2}, x)$ iff
				$\asgFun(y_{1})({s_{0}}_{\GName^{\star}}) <
				\asgFun(x)({s_{0}}_{\GName^{\star}}) \leq
				\asgFun(y_{2})({s_{0}}_{\GName^{\star}})$, for all assignments $\asgFun
				\in \AsgSet_{\GName^{\star}}(\{ y_{1}, y_{2}, x \},
				{s_{0}}_{\GName^{\star}})$.
				At this point, the thesis follows by a reasoning similar to the one
				proposed in the lemma.
			\end{proof}

			Consider now a model $\GName$ of $\varphi^{grd}$ and, for all agents $a
			\in \AgnSet$, the relation $\rRel_{a}^{<} \subseteq
			\StrSet_{\GName} \times \StrSet_{\GName}$ between strategies defined as follows: $\rRel_{a}^{<}(\strFun_{1}, \strFun_{2})$
			holds iff $\GName, \emptyfun{[z_{1} \mapsto \strFun_{1}][z_{2} \mapsto
			\strFun_{2}]}, {s_{0}}_{\GName} \models z_{1} <_{a} z_{2}$, for all
			strategies $\strFun_{1}, \strFun_{2} \in
			\StrSet_{\GName}$.
			By using a proof similar to that of Lemma \ref{lmm:strord}, it is
			possible to see that $\rRel_{a}^{<}$ is a \emph{strict partial order
			without maximal element} on $\StrSet_{\GName}$.

      Now, to apply the desired reduction, we need to transform $\rRel_{a}^{<}$
      into a total order over strategies, by using the following two lemmas.
      \begin{lem}[Strategy Equivalence]
        \label{lmm:streqv}
        Let $\rRel_{a}^{\equiv} \subseteq \StrSet_{\GName} \times
        \StrSet_{\GName}$, with $a \in \AgnSet$, be the relation between
        strategies such that $\rRel_{a}^{\equiv}(\strFun_{1}, \strFun_{2})$
        holds iff neither $\rRel_{a}^{<}(\strFun_{1}, \strFun_{2})$ nor
        $\rRel_{a}^{<}(\strFun_{2}, \strFun_{1})$ holds, for all $\strFun_{1},
        \strFun_{2} \in \StrSet_{\GName}$.
        Then $\rRel_{a}^{\equiv}$ is an \emph{equivalence relation}.
      \end{lem}
      \begin{proof}
        It is immediate to see that the relation $\rRel_{a}^{\equiv}$ is
        reflexive, since $\rRel_{a}^{<}$ is not reflexive.
        Moreover, it is symmetric by definition.
        Finally, due to the definition of the partial order formula $<_{a}$, it
        is also transitive and, thus, $\rRel_{a}^{\equiv}$ is an
        \emph{equivalence relation}.
        Indeed, if $\rRel_{\alpha}^{\equiv}(\strFun_{1}, \strFun_{2})$ holds, we
        have that $\GName, \asgFun[12], {s_{0}}_{\GName} \models \AAll{y}
        (\beta, y) ((\alpha, x_{1}) (\X \neg p) \vee (\alpha, x_{2}) (\X p))$
        and $\GName, \asgFun[12], {s_{0}}_{\GName} \models \AAll{y} (\beta, y)
        ((\alpha, x_{2}) (\X \neg p) \vee (\alpha, x_{1}) (\X p))$, for all
        assignments $\asgFun[12] \in \AsgSet(\GName, \{ x_{1}, x_{2} \},
        {s_{0}}_{\GName})$ such that $\asgFun[12](x_{1}) = \strFun_{1}$ and
        $\asgFun[12](x_{2}) = \strFun_{2}$.
        Consequently, $\GName, \asgFun[12], {s_{0}}_{\GName} \models \AAll{y}
        (\beta, y) ((\alpha, x_{1}) (\X \neg p) \vee (\alpha, x_{2}) (\X p))
        \wedge ((\alpha, x_{2}) (\X \neg p) \vee (\alpha, x_{1}) (\X p))$, which
        is equivalent to $\GName, \asgFun[12], {s_{0}}_{\GName} \models \AAll{y}
        (\beta, y) ((\alpha, x_{1}) \allowbreak (\X p) \wedge (\alpha, x_{2})
        (\X p)) \vee ((\alpha, x_{1}) (\X \neg p) \vee (\alpha, x_{2}) (\X \neg
        p))$.
        In other words, for all strategies $\strFun$, either $\GName,
        \asgFun[12]{[y \mapsto \strFun]}, {s_{0}}_{\GName} \models (\beta, y)
        ((\alpha, x_{1}) (\X p) \wedge (\alpha, x_{2}) (\X p))$ or $\GName,
        \asgFun[12]{[y \mapsto \strFun]}, {s_{0}}_{\GName} \models (\beta, y)
        ((\alpha, x_{1}) (\X \allowbreak \neg p) \wedge (\alpha, x_{2}) (\X \neg
        p))$ holds.
        Similarly, from $\rRel_{\alpha}^{\equiv}(\strFun_{2}, \strFun_{3})$, we
        can derive that, for all strategies $\strFun$ and assignments
        $\asgFun[23] \in \AsgSet(\GName, \{ x_{2}, x_{3} \}, {s_{0}}_{\GName})$
        with $\asgFun[23](x_{2}) = \strFun_{2}$ and $\asgFun[23](x_{3}) =
        \strFun_{3}$, either $\GName, \asgFun[23]{[y \mapsto \strFun]},
        {s_{0}}_{\GName} \models (\beta, y) ((\alpha, x_{2}) (\X p) \wedge
        (\alpha, x_{3}) (\X p))$ or $\GName, \asgFun[23]{[y \mapsto \strFun]},
        {s_{0}}_{\GName} \models (\beta, y) ((\alpha, x_{2}) (\X \allowbreak
        \neg p) \wedge ''(\alpha, x_{3}) (\X \neg p))$ holds.
        Therefore, by putting the two deductions together, we have that either
        $\GName, \asgFun[13], {s_{0}}_{\GName} \models (\beta, y) ((\alpha,
        x_{1}) (\X p) \wedge (\alpha, x_{3}) (\X p))$ or $\GName, \asgFun[13],
        {s_{0}}_{\GName} \models (\beta, y) ((\alpha, x_{1}) (\X \neg p) \wedge
        (\alpha, x_{3}) (\X \neg p))$ holds, for all assignments $\asgFun[13]
        \in \AsgSet(\GName, \{ x_{1}, x_{3}, y \}, {s_{0}}_{\GName})$ such that
        $\asgFun[13](x_{1}) = \strFun_{1}$ and $\asgFun[13](x_{3}) =
        \strFun_{3}$.
        Thus, by following the above reasoning at the reverse, we immediately
        derive that $\rRel_{\alpha}^{\equiv}(\strFun_{1}, \strFun_{3})$ holds,
        as well.
        Obviously, the same reasoning applies to $\rRel_{\beta}^{\equiv}$.
      \end{proof}

      Let $\StrSet_{a}^{\equiv} \defeq \class{
      \StrSet_{\GName}}{ \:\rRel_{a}^{\equiv} }$ be the
      quotient set of $\StrSet_{\GName}$ \wrt
      $\rRel_{a}^{\equiv}$, for $a \in \AgnSet$, \ie, the set of the related
      equivalence classes over strategies.
      Then, the following holds.
      \begin{lem}[Strategy Total Order]
        \label{lmm:strtotord}
        Let $\sRel_{a}^{<} \subseteq \StrSet_{a}^{\equiv} \times
        \StrSet_{a}^{\equiv}$, with $a \in \AgnSet$, be the relation between
        classes of strategies such that $\sRel_{a}^{<}(\FSet_{1}, \FSet_{2})$
        holds iff $\rRel_{a}^{<}(\strFun_{1}, \strFun_{2})$ holds, for all
        $\strFun_{1} \in \FSet_{1}$, $\strFun_{2} \in \FSet_{2}$, and
        $\FSet_{1}, \FSet_{2} \in \StrSet_{a}^{\equiv}$.
        Then $\sRel_{a}^{<}$ is a \emph{strict total order with minimal element
        but no maximal element}.
      \end{lem}
      \begin{proof}
        The fact that $\sRel_{a}^{<}$ is a \emph{strict partial order without
        maximal element} derives directly from the same property of
        $\rRel_{a}^{<}$.
        In fact, due to the specific definition of the partial order formula
        $<_{a}$, if $\rRel_{a}^{\equiv}(\strFun', \strFun'')$ and
        $\rRel_{a}^{<}(\strFun', \strFun)$ (resp., $\rRel_{a}^{<}(\strFun,
        \strFun')$) hold, we obtain that $\rRel_{a}^{<}(\strFun'', \strFun)$
        (resp., $\rRel_{a}^{<}(\strFun, \strFun'')$) holds as well.
        Indeed, as shown in the proof of Lemma~\ref{lmm:streqv},
        $\rRel_{\alpha}^{\equiv}(\strFun', \strFun'')$ implies that, for all
        assignments $\asgFun[\equiv] \in \AsgSet(\GName, \{ x', x'', y \},
        {s_{0}}_{\GName})$ with $\asgFun[\equiv](x') = \strFun'$ and
        $\asgFun[\equiv](x'') = \strFun''$, either $\GName, \asgFun[\equiv],
        {s_{0}}_{\GName} \models (\beta, y) ((\alpha, x') (\X p) \wedge (\alpha,
        x'') (\X p))$ or $\GName, \asgFun[\equiv], {s_{0}}_{\GName} \models
        (\beta, y) ((\alpha, x') (\X \neg p) \wedge (\alpha, x'') (\X \neg p))$
        holds.
        Moreover, $\rRel_{\alpha}^{<}(\strFun', \strFun)$ (resp.,
        $\rRel_{\alpha}^{<}(\strFun, \strFun')$) implies that, for all
        assignments $\asgFun[<] \in \AsgSet(\GName, \{ x, x' \},
        {s_{0}}_{\GName})$ with $\asgFun[<](x) = \strFun$ and $\asgFun[<](x') =
        \strFun'$, there exists a strategy $\gFun$ such that $\GName,
        \asgFun[<]{[y \mapsto \gFun]}, {s_{0}}_{\GName} \models (\beta, y)
        ((\alpha, x') (\X \neg p) \wedge (\alpha, x) (\X p))$ (resp., $\GName,
        \asgFun[<]{[y \mapsto \gFun]}, {s_{0}}_{\GName} \models (\beta, y)
        ((\alpha, x) (\X  \neg p) \wedge (\alpha, x') (\X p))$).
        Now, by combining both deductions w.r.t.\ the same strategy
        $\gFun$ assigned to the variable $y$, we obtain that $\GName, \asgFun{[y
        \mapsto \gFun]}, {s_{0}}_{\GName} \models (\beta, y) ((\alpha, x'') (\X
        \neg p) \wedge (\alpha, x) (\X p))$ (resp., $\GName, \asgFun{[y \mapsto
        \gFun]}, {s_{0}}_{\GName} \models (\beta, y) ((\alpha, x) (\X  \neg p)
        \wedge (\alpha, x'') (\X p))$), for all assignments $\asgFun \in
        \AsgSet(\GName, \{ x, x'' \}, {s_{0}}_{\GName})$ with $\asgFun[<](x) =
        \strFun$ and $\asgFun[<](x'') = \strFun''$.
        Hence, $\rRel_{\alpha}^{<}(\strFun'', \strFun)$ (resp.,
        $\rRel_{\alpha}^{<}(\strFun, \strFun'')$).
        Obviously, the same reasoning applies if we assume $a = \beta$ instead
        of $a = \alpha$.
        At this point, if there are $\strFun_{1} \in \FSet_{1}$ and $\strFun_{2}
        \in \FSet_{2}$ such that $\rRel_{a}^{<}(\strFun_{1}, \strFun_{2})$
        holds, we directly obtain that $\sRel_{a}^{<}(\FSet_{1}, \FSet_{2})$
        holds as well, for all $\FSet_{1}, \FSet_{2} \in \StrSet_{a}^{\equiv}$
        and $a \in \AgnSet$.

        Moreover, $\sRel_{a}^{<}$ is total, since $\rRel_{a}^{\equiv}$ is an
        equivalence relation that cluster together all strategies of the agent
        $a$ that are not in relation \wrt either $\rRel_{a}^{<}$ or its
        inverse $(\rRel_{a}^{<})^{-1}$.
        Indeed, suppose by contradiction that there are two different classes
        $\FSet_{1}, \FSet_{2} \in \StrSet_{a}^{\equiv}$ such that neither
        $\sRel_{a}^{<}(\FSet_{1}, \FSet_{2})$ nor $\sRel_{a}^{<}(\FSet_{2},
        \FSet_{1})$ holds.
        This means that, for all $\strFun_{1} \in \FSet_{1}$ and $\strFun_{2}
        \in \FSet_{2}$, neither $\rRel_{a}^{<}(\strFun_{1}, \strFun_{2})$ nor
        $\rRel_{a}^{<}(\strFun_{2}, \strFun_{1})$ holds and, so,
        $\rRel_{a}^{\equiv}(\strFun_{1}, \strFun_{2})$.
        But, this contradicts the fact that $\FSet_{1}$ and $\FSet_{2}$ are
        different equivalences classes.

        Finally, it is important to note that in $\StrSet_{a}^{\equiv}$ there is
        also a minimal element \wrt $\sRel_{a}^{<}$.
        Indeed, for a strategy $\strFun \in \StrSet_{\GName}$
        for $\alpha$ (resp., for $\beta$) that forces the play to reach only
        nodes labeled with $p$ (resp., $\neg p$) as successor of
        ${s_{0}}_{\GName}$, independently from the strategy of $\beta$ (resp.,
        $\alpha$), the relation $\rRel_{\alpha}^{<}(\strFun', \strFun)$ (resp.,
        $\rRel_{\beta}^{<}(\strFun', \strFun)$) cannot hold, for any $\strFun'
        \in \StrSet_{\GName}$.
      \end{proof}

			By a classical result on first-order model theory~\cite{EF95}, the
			relation $\sRel_{a}^{<}$ cannot be defined on a finite set.
			Hence, $\card{\StrSet_{a}^{\equiv}} = \omega$, for all $a \in \AgnSet$.
			Now, let $\sRel_{a}^{\prec} \subseteq \StrSet_{a}^{\equiv} \times
			\StrSet_{a}^{\equiv}$ be the \emph{successor} relation on
			$\StrSet_{a}^{\equiv}$ compatible with the strict total order
			$\sRel_{a}^{<}$, \ie, such that $\sRel_{a}^{\prec}(\FSet_{1}, \FSet_{2})$
			holds iff \emph{(i)} $\sRel_{a}^{<}(\FSet_{1}, \FSet_{2})$ holds and
			\emph{(ii)} there is no $\FSet_{3} \in \StrSet_{a}^{\equiv}$ for which
			both $\sRel_{a}^{<}(\FSet_{1}, \FSet_{3})$ and $\sRel_{a}^{<}(\FSet_{3},
			\FSet_{2})$ hold, for all $\FSet_{1}, \FSet_{2} \in \StrSet_{a}^{\equiv}$.
			Then, we can represent the two sets of classes $\StrSet_{\alpha}^{\equiv}$
			and $\StrSet_{\beta}^{\equiv}$, respectively, as the infinite ordered
			lists $\{ \FSet_{0}^{\alpha}, \FSet_{1}^{\alpha}, \ldots \}$ and $\{
			\FSet_{0}^{\beta}, \FSet_{1}^{\beta}, \ldots \}$ such that
			$\sRel_{a}^{\prec}(\FSet_{i}^{a}, \FSet_{i+1}^{a})$ holds, for all indexes
			$i \in \SetN$.
			Note that $\FSet_{0}^{a}$ is the class of minimal strategies w.r.t the
			relation $\sRel_{a}^{<}$.

			At this point, we have all the machinery to build an embedding of the
			plane $\SetN \times \SetN$ into a model $\GName$ of $\varphi^{grd}$.
			Formally, we consider the \emph{bijective map} $\aleph : \SetN \times
			\SetN \to \StrSet_{\alpha}^{\equiv} \times \StrSet_{\beta}^{\equiv}$ such
			that $\aleph(i, j) = (\FSet_{i}^{\alpha}, \FSet_{j}^{\beta})$, for all $i,
			j \in \SetN$.

		\end{paragraph}

		\begin{paragraph}*{\textbf{Compatible tiling}}

			Given the grid structure built on the model $\GName$ of $\varphi^{grd}$
			through the bijective map $\aleph$, we can express that a tiling of the
			grid is admissible by making use of the formula $z_{1} \prec_{a} z_{2}
			\defeq (z_{1} <_{a} z_{2}) \wedge (\neg \EExs{z_{3}} \: z_{1} <_{a} z_{3})
			\wedge (z_{3} <_{a} z_{2})$ corresponding to the successor relation
			$\sRel_{a}^{\prec}$, for all $a \in \AgnSet$.
			Indeed, it is not hard to see that $\GName, \asgFun, {s_{0}}_{\GName}
			\models z_{1} \prec_{a} z_{2}$ iff $\asgFun(z_{1}) \in \FSet_{i}^{a}$ and
			$\asgFun(z_{2}) \in \FSet_{i+1}^{a}$, for all indexes $i \in \SetN$ and
			assignments $\asgFun \in \AsgSet_{\GName}(\{ z_{1}, z_{2} \},
			{s_{0}}_{\GName})$.
			The idea here is to associate with each domino type $t \in \DSet$ a
			corresponding atomic proposition $t \in \APSet$ and to express the
			horizontal and vertical matching conditions via suitable object labeling.
			In particular, we can express that the tiling is locally compatible, the
			horizontal neighborhoods of a tile satisfy the $\HRel$ or $\VRel$
			requirements, respectively.
			All these constraints can be formulated through the following three
			agent-closed formulas:
			\begin{enumerate}
				\item
					$\varphi^{t, loc}(x, y) \defeq (\alpha, x) (\beta, y) (\X (t \wedge
					\bigwedge_{t' \in \DSet}^{t' \neq t} \neg t'))$;
				\item
					$\varphi^{t, hor}(x, y) \defeq \bigvee_{(t, t') \in \HRel} \AAll{x'}
					(x \prec_{\alpha} x' \rightarrow (\alpha, x') (\beta, y) (\X t'))$;
				\item
					$\varphi^{t, ver}(x, y) \defeq \bigvee_{(t, t') \in \VRel} \AAll{y'}
					(y \prec_{\beta} y' \rightarrow (\alpha, x) (\beta, y') (\X t'))$.
			\end{enumerate}
			Informally, we have the following: $\varphi^{t, loc}(x, y)$ asserts that
			$t$ is the only domino type labeling the successors of the root of the
			model $\GName$ that can be reached using the strategies related to the
			variables $x$ and $y$; $\varphi^{t, hor}(x, y)$ asserts that the tile $t'$
			labeling the successors of the root reachable through the strategies $x'$
			and $y$ is compatible with $t$ \wrt the horizontal requirement $\HRel$,
			for all strategies $x'$ that immediately follow that related to $x$ \wrt
			the order $\rRel_{\alpha}^{<}$; $\varphi^{t, ver}(x, y)$ asserts that the
			tile $t'$ labeling the successors of the root reachable through the
			strategies $x$ and $y'$ is compatible with $t$ \wrt the vertical
			requirement $\VRel$, for all strategies $y'$ that immediately follow that
			related to $y$ \wrt the order $\rRel_{\beta}^{<}$.

			Finally, to express that the whole grid has an admissible tiling, we use
			the sentence $\varphi^{til} \defeq \AAll{x} \AAll{y}
			\: \bigvee_{t \in \DSet} \varphi^{t, loc}(x, y) \wedge \varphi^{t, hor}(x,
			y) \wedge \varphi^{t, ver}(x, y)$ that asserts the existence of a domino
			type $t$ satisfying the three conditions mentioned above, for every point
			identified by the strategies $x$ and $y$.

		\end{paragraph}

		\begin{paragraph}*{\textbf{Recurrent tile}}

			As last task, we impose that the grid embedded into $\GName$ has the
			distinguished domino type $t_{0}$ occurring infinitely often in its first
			row.
			To do this, we describe two formulas that determine if a row or a column
			is the first one \wrt the orders $\sRel_{\alpha}^{<}$ and
			$\sRel_{\beta}^{<}$, respectively.
			Formally, we use $0_{a}(z) \defeq \neg \EExs{z'} \: z' <_{a} z$, for $a
			\in \AgnSet$.
			One can easily prove that $\GName, \asgFun, {s_{0}}_{\GName} \models
			0_{\alpha}(z)$ iff $\asgFun(z) \in \FSet_{0}^{a}$, for all assignments
			$\asgFun \in \AsgSet_{\GName}(\{ z \}, {s_{0}}_{\GName})$.
			Now, the infinite occurrence requirement on $t_{0}$ can be expressed with
			the following sentence: $\varphi^{rec} \defeq
			\AAll{x} \AAll{y} (0_{\beta}(y) \wedge (0_{\alpha}(x) \vee (\alpha, x)
			(\beta, y) (\X t_{0}))) \rightarrow \EExs{x'} \: x <_{\alpha} x' \wedge
			(\alpha, x') (\beta, y) (\X t_{0})$.
			Informally, $\varphi^{rec}$ asserts that, when we are on the first row
			identified by the variable $y$ and at a column pointed by $x$ such
			that it is the first column or the node of the \emph{``intersection''}
			between $x$ and $y$ is labeled by $t_{0}$, we have that there exists a
			greater column identified by $x'$ such that its \emph{``intersection''}
			with $y$ is labeled by $t_{0}$ as well.

		\end{paragraph}

		\begin{paragraph}*{\textbf{Construction correctness}}

			At this point, we have all tools to formally prove the correctness of
			the undecidability reduction, by showing the equivalence between the
			satisfiability of the sentence $\varphi^{dom} \defeq \varphi^{grd} \wedge \varphi^{til} \wedge \varphi^{rec}$ and finding a solution of the recurrent tiling problem.

			\figrasviii\
			\begin{thm}[Satisfiability]
				\label{thm:satsl}
				The satisfiability problem for \SL is highly undecidable.
				In particular, it is \SCompH{1}{1}.
			\end{thm}
			\begin{proof}
				For the direct reduction, assume that there exists a solution mapping
				$\dsmFun : \SetN \times \SetN \to \DSet$ for the given recurrent domino
				system $\DName$.
				Then, we can build a finite \CGS\ $\GName_{\dsmFun}^{\star}$ similar to
				the one used in Lemma \ref{lmm:ordsat:sl}, which satisfies the sentence
				$\varphi^{dom}$:

				\begin{enumerate}[label=(\roman*)]
					\item
						$\ActSet_{{\GName_{\dsmFun}^{\star}}} \defeq \SetN$;

					\item
						there are $2 \cdot \card{\DSet} + 1$ different states
						$\SttSet_{{\GName_{\dsmFun}^{\star}}} \defeq \{ s_{0} \} \cup (\{ p,
						\neg p \} \times \DSet)$ such that $\apFun[
						{\GName_{\dsmFun}^{\star}} ](s_{0}) \defeq \emptyset$,
						$\apFun_{{\GName_{\dsmFun}^{\star}}}((p, t)) \defeq \{ p, t \}$,
						and $\apFun_{{\GName_{\dsmFun}^{\star}}}((\neg p, t)) \defeq \{ t
						\}$, for all $t \in \DSet$;

					\item
						each state $(z, t) \in \{ p, \neg p \} \times \DSet$ has only self
						loops $\trnFun_{{\GName_{\dsmFun}^{\star}}}((z, t), \decElm) \defeq
						(z, t)$ and the initial state ${s_{0}}_{\GName_{\dsmFun}^{\star}}
						\defeq s_{0}$ is connected to $(z, t)$ through the decision
						$\decElm$, \ie, $\trnFun_{{\GName_{\dsmFun}^{\star}}}(s_{0},
						\decElm) \defeq (z, t)$, iff

						\begin{enumerate}
							\item
								$t = \dsmFun(\decElm(\alpha), \decElm(\beta))$ and
							\item
								 $z = p$ iff $\decElm(\alpha) \leq \decElm(\beta)$, for all
								 $\decElm \in \DecSet_{{\GName_{\dsmFun}^{\star}}}$.
						\end{enumerate}

				\end{enumerate}

				\noindent By a simple case analysis on the subformulas of $\varphi^{dom}$, it is
				possible to see that $\GName_{\dsmFun}^{\star} \models \varphi^{dom}$.

				Conversely, let $\GName$ be a model of the sentence $\varphi^{dom}$ and
				$\aleph : \SetN \times \SetN \to \StrSet_{\alpha}^{\equiv} \times
				\StrSet_{\beta}^{\equiv}$ the related bijective map built for the grid
				specification task.
				As first thing, we have to prove the existence of a coloring function
				$\eth : \StrSet_{\alpha}^{\equiv} \times \StrSet_{\beta}^{\equiv} \to
				\DSet$ such that, for all pairs of classes of strategies
				$(\FSet^{\alpha}, \FSet^{\beta}) \in \StrSet_{\alpha}^{\equiv} \times
				\StrSet_{\beta}^{\equiv}$ and assignments $\asgFun \in
				\AsgSet_{\GName}(\{ \alpha, \beta \}, {s_{0}}_{\GName})$ with
				$\asgFun(\alpha) \in \FSet^{\alpha}$ and $\asgFun(\beta) \in
				\FSet^{\beta}$, it holds that $\GName, \asgFun, {s_{0}}_{\GName} \models
				\X \eth(\FSet^{\alpha}, \FSet^{\beta})$.
				Then, it remains to note that the solution mapping $\dsmFun = \eth \circ
				\aleph$ built as a composition of the bijective map $\aleph$ and the
				coloring function $\eth$ is an admissible tiling of the plane $\SetN
				\times \SetN$.

				Due to the $\varphi^{t, loc}$ formula in the sentence $\varphi^{til}$,
				we have that, for all assignments $\asgFun \in \AsgSet_{\GName}(\{
				\alpha, \beta \}, {s_{0}}_{\GName})$, there exists just one domino type
				$t \in \DSet$ satisfying the property $\GName, \asgFun, \allowbreak
				{s_{0}}_{\GName} \models \X t$.
				Let $\der{\eth} \!:\! \StrSet_{\GName} \times \StrSet_{\GName} \to \DSet$ be the function that returns such a type, for all pairs of strategies of $\alpha$ and $\beta$, \ie, such that $\GName, \asgFun, {s_{0}}_{\GName} \models \X \der{\eth}(\asgFun(\alpha), \asgFun(\beta))$, for all assignments $\asgFun \in \AsgSet_{\GName}(\{ \alpha, \beta \}, {s_{0}}_{\GName})$.
				Now, it is not hard to see that, due to the formulas $\varphi^{t, hor}$
				and $\varphi^{t, ver}$ in the sentence $\varphi^{til}$, it holds
				\emph{(i)} $(\der{\eth}(\strFun_{\alpha}, \strFun_{\beta}),
				\der{\eth}(\strFun_{\alpha}', \strFun_{\beta})) \in \HRel$ and
				\emph{(ii)} $(\der{\eth}(\strFun_{\alpha}, \strFun_{\beta}),
				\der{\eth}(\strFun_{\alpha}, \strFun_{\beta}')) \in \VRel$, for all
				$\strFun_{\alpha} \in \FSet_{i}^{\alpha}$, $\strFun_{\alpha}' \in
				\FSet_{i+1}^{\alpha}$, $\strFun_{\beta} \in \FSet_{j}^{\beta}$,
				$\strFun_{\beta}' \in \FSet[j + 1]^{\beta}$, and $i, j \in \SetN$.
				Moreover, the guess of the tile type $t'$ adjacent to $t$ is uniform
				\wrt the choice of the successor strategy.
				Indeed, the disjunctions $\bigvee_{(t, t') \in \HRel}$ and $\bigvee_{(t,
				t') \in \VRel}$ precede the universal quantifications $\AAll{x'}$ and
				$\AAll{y'}$ in the formulas $\varphi^{t, hor}$ and $\varphi^{t, ver}$,
				respectively.
				Thus, we have that, for all $\strFun_{\alpha}', \strFun_{\alpha}'' \in
				\FSet_{i}^{\alpha}$ and $\strFun_{\beta}', \strFun_{\beta}'' \in
				\FSet_{j}^{\beta}$ with $i, j \in \SetN$ and $i + j > 0$, it holds that
				$\der{\eth}(\strFun_{\alpha}', \strFun_{\beta}') =
				\der{\eth}(\strFun_{\alpha}'', \strFun_{\beta}'')$.
				Note that this fact is not necessarily true for strategies belonging to
				the minimal classes $\FSet_{0}^{\alpha}$ and $\FSet_{0}^{\beta}$, since
				the sentence $\varphi^{dom}$ does not contain a relative requirement.
				However, every domino type $\der{\eth}(\strFun_{\alpha},
				\strFun_{\beta})$, with $\strFun_{\alpha} \in \FSet_{0}^{\alpha}$ and
				$\strFun_{\beta} \in \FSet_{0}^{\beta}$, can be used to label the origin
				of the plane $\SetN \times \SetN$ in order to obtain an admissible
				tiling.
				So, we can consider a function $\eth$, defined as follows: \emph{(i)}
				$\eth(\FSet_{0}^{\alpha}, \FSet_{0}^{\beta}) \in \set{
				\der{\eth}(\strFun_{\alpha}, \strFun_{\beta}) }{ \strFun_{\alpha} \in
				\FSet_{0}^{\alpha} \wedge \strFun_{\beta} \in \FSet_{0}^{\beta} }$;
				\emph{(ii)} $\eth(\FSet_{i}^{\alpha}, \FSet_{j}^{\beta}) =
				\der{\eth}(\strFun_{\alpha}, \strFun_{\beta})$, for all
				$\strFun_{\alpha} \in \FSet_{i}^{\alpha}$, $\strFun_{\beta} \in
				\FSet_{j}^{\beta}$, and $i, j \in \SetN$ with $i + j > 0$.

				Clearly, \emph{(i)} $(\eth(\FSet_{i}^{\alpha}, \FSet_{j}^{\beta}),
				\eth(\FSet_{i+1}^{\alpha}, \FSet_{j}^{\beta})) \in \HRel$, \emph{(ii)}
				$(\eth(\FSet_{i}^{\alpha}, \FSet_{j}^{\beta}), \eth(\FSet_{i}^{\alpha},
				\FSet[j + 1]^{\beta})) \in \VRel$, and \emph{(iii)} $\card{\set{ i }{
				\eth(\FSet_{i}^{\alpha}, \FSet_{0}^{\beta}) = t_{0} }} = \omega$, for
				all $i, j \in \SetN$.
				So, $\dsmFun = \eth \circ \aleph$ is an admissible tiling.
			\end{proof}

		\end{paragraph}

	\end{subsection}

\end{section}

% End of file SectionII.tex

%%****************************************************************************%%
%%                                                                            %%
%% Reasoning About Strategies: On the Satisfiability Problem                  %%
%%                                                                            %%
%% SectionIII.tex                                                             %%
%%                                                                            %%
%% Revision 0                                                                 %%
%%                                                                            %%
%% Copyright (C) 2014, Fabio Mogavero, Aniello Murano, Giuseppe Perelli, and  %%
%%                     Moshe Y. Vardi.                                        %%
%% All rights reserved.                                                       %%
%%                                                                            %%
%%****************************************************************************%%

% Begin of file SectionIII.tex

\begin{section}{What makes ATL* decidable?}
	\label{sec:what}

	As just shown, \SL does not have the bounded model property and its satisfiability problem is undecidable.
	On the contrary, it is well-known that the satisfiability problem for \ATLS
	is 2\ExpTimeC~\cite{Sch08}.
	This gap in complexity between \SL and \ATLS gives naturally rise to the
	question of which are the inherent properties of \ATLS that make the problem
	decidable.
	In this section, we answer such question by analyzing two syntactic fragments
	of \SL.
	The first one, called \emph{Boolean-Goal Strategy Logic} (\BGSL, for short),
	still has an undecidable satisfiability problem .
	The second one, called \emph{One-Goal Strategy Logic} (\OGSL, for short),
	retains, instead, all positive properties of \ATLS, such as the decision-tree
	model property (see Definition~\ref{def:decunw} and Theorem~\ref{thm:ogsl:tremodprp}) and the bounded model property (see Theorem~\ref{thm:ogsl:bndmdprp}), which allows us to show that its satisfiability problem is 2\ExpTimeC.
	A fundamental feature used as a tool to prove the announced properties is the
	\emph{behavioral satisfiability}, described for the first time
	in~\cite{MMPV14}.

	The section is organized as follows.
	In Subsection~\ref{sec:what;sub:frm}, we introduce the syntactic fragments of
	\SL mentioned above.
	Then, in Subsection~\ref{sec:what;sub:bvr}, we define the concept of
	behavioral satisfiability and recall the corresponding theorem for \OGSL.
	Finally, in Subsection~\ref{sec:what;sub:treeprp}, we introduce the concept of
	decision-tree model property and prove that it is enjoyed by the latter
	fragment.

	\begin{subsection}{Syntactic fragments}
		\label{sec:what;sub:frm}

		In order to formalize the two syntactic fragments of \SL we want to
		investigate, we first need to define the concepts of \emph{quantification}
		and \emph{binding prefixes}.
		%
% 		\begin{defi}[Prefixes]
% 			\label{def:prf}
		A \emph{quantification prefix} over a set $\VSet \subseteq \VarSet$ of
		variables is a finite word $\qpElm \in \set{ \EExs{\xElm}, \AAll{\xElm} }{
		\xElm \in \VSet }^{\card{\VSet}}$ of length $\card{\VSet}$ such that each
		variable $\xElm \in \VSet$ occurs just once in $\qpElm$, \ie, there is
		exactly one index $i \in \numco{0}{\card{\VSet}}$ such that $(\qpElm)_{i}
		\in \{ \EExs{\xElm}, \AAll{\xElm} \}$.
		A \emph{binding prefix} over a set of variables $\VSet \subseteq \VarSet$
		is a finite word $\bpElm \in \set{ (\aElm, \xElm) }{ \aElm \in \AgnSet
		\land \xElm \in \VSet }^{\card{\AgnSet}}$ of length $\card{\AgnSet}$ such
		that each agent $\aElm \in \AgnSet$ occurs just once in $\bpElm$, \ie,
		there is exactly one index $i \in \numco{0}{\card{\AgnSet}}$ for which
		$(\bpElm)_{i} \in \set{ (\aElm, \xElm) }{ \xElm \in \VSet }$.
		By $\VarSet(\qpElm)$ and $\VarSet(\bpElm)$ we denote, respectively, the set
		of variables on which the quantification and binding prefixes $\qpElm$ and
		$\bpElm$ range.
		Finally, $\QPSet(\VSet) \subseteq \set{ \EExs{\xElm}, \AAll{\xElm} }{
		\xElm \in \VSet }^{\card{\VSet}}$ and $\BPSet(\VSet) \subseteq \set{
		(\aElm, \xElm) }{ \aElm \in \AgnSet \land \xElm \in \VSet
		}^{\card{\AgnSet}}$ denote the sets of all quantification and binding
		prefixes over the variables in $\VSet$.
% 		\end{defi}

		We now have all tools to define the syntactic fragments named
		\emph{Boolean-Goal} and \emph{One-Goal Strategy Logic} (\BGSL and \OGSL, for
		short).
		For a \emph{goal} we mean an \SL\ agent-closed formula of the form $\bpElm
		\varphi$, with $\AgnSet \subseteq \free{\varphi}$, being $\bpElm \in
		\BndSet(\VarSet)$ a binding prefix.
		The idea behind \BGSL is to build sentences having only a Boolean
		combination of goals in the scope of a quantification prefix.
% 		More formally, an \BGSL formula is of the form $\qpElm \varphi$, where
% 		$\qpElm$ is a quantification prefix and $\varphi$ is a Boolean combination
% 		of goals of the form $\bpElm \psi$, with $\bpElm$ being a binding prefix
% 		such that $\VarSet(\bpElm) \subseteq \VarSet(\qpElm)$ and $\psi$ being an
% 		\LTL combination of atomic proposition and \BGSL sentences.
		Moreover, \OGSL forces the use of a different quantification prefix for each
		goal in the formula.
		The formal syntax of \BGSL and \OGSL follows.

		\begin{defi}[\BGSL and \OGSL\ Syntax]
			\label{def:xgsl(syntax)}
			\BGSL\ formulas are built inductively from the sets of atomic propositions
			$\APSet$, quantification prefixes $\QPSet(\VSet)$ for any $\VSet \subseteq
			\VarSet$, and binding prefixes $\BPSet(\VarSet)$, by using the following
			grammar, with $\pElm \in \APSet$, $\qpElm \in \cup_{\VSet \subseteq
			\VarSet} \QPSet(\VSet)$, and $\bpElm \in \BPSet(\VarSet)$:
			\begin{center}
				$\varphi ::= \pElm \mid \neg \varphi \mid \varphi \wedge \varphi \mid
				\varphi \vee \varphi \mid \X \varphi \mid \varphi \:\U \varphi \mid
				\varphi \:\R \varphi \mid \qpElm \psi$,\\
				$\psi ::= \bpElm \varphi \mid \neg \psi \mid \psi \wedge \psi \mid \psi
				\vee \psi$,
			\end{center}
			where in the formation rule of $\qpElm \psi$ it is ensured that $\qpElm \in \QPSet(\free{\psi})$.\\
			Finally, the simpler \OGSL\ formulas are obtained by forcing each goal to
			be coupled with a quantification prefix:
			\begin{center}
				$\varphi ::= \pElm \mid \neg \varphi \mid \varphi \wedge \varphi \mid
				\varphi \vee \varphi \mid \X \varphi \mid \varphi \:\U \varphi \mid
				\varphi \:\R \varphi \mid \qpElm \bpElm \varphi$,
			\end{center}
			where in the formation rule $\qpElm \bpElm \varphi$ it is ensured that
			$\qpElm \in \QPSet(\free{\bpElm \varphi})$.\\
			$\SL \supset \BGSL \supset \OGSL$ denotes the syntactic chain of infinite
			sets of formulas generated by the respective grammars with the associated
			constraints on free variables of goals.
		\end{defi}

		Intuitively, in \BGSL and \OGSL, we force the writing of formulas to use
		atomic blocks of quantifications and bindings, where the related free
		variables are strictly coupled with those that are effectively quantified in
		the prefix just before the binding.
		In a nutshell, we can only write formulas by using sentences of the form
		$\qpElm \psi$ belonging to a kind of \emph{prenex normal form} in which the
		quantifications contained into the \emph{matrix} $\psi$ only belong to the
		prefixes $\qpElm'$ for some inner subsentence $\qpElm' \psi' \in \snt{\qpElm
		\psi}$.

		An \BGSL sentence $\phi$ is \emph{principal} if it is of the form $\phi =
		\qpElm \psi$, where $\psi$ is agent-closed and $\qpElm \in
		\QPSet(\free{\psi})$.
		By $\psnt{\varphi} \subseteq \snt{\varphi}$ we denote the set of all
		principal subsentences of the formula $\varphi$.

		In order to practice with the above fragments, let us consider again the
		sentence $\varphi_{NE}$ of Example~\ref{exm:ne}.
		It is easy to see that it is not an \BGSL formula.
		However, by rearranging quantifications and bindings we can obtain the
		equivalent formula $\varphi'_{NE} = \qpSym \bigwedge_{i = 1}^{n} \bpSym[][i]
		\psi_{i} \rightarrow \bpSym
		\psi_{i}$, where $\qpSym = \EExs{\xSym[1]} \cdots \EExs{\xSym[n]}
		\AAll{\ySym[1]} \cdots \AAll{\ySym[n]}$, $\bpSym = (\alpha_{1}, \xSym[1])
		\cdots (\alpha_{n}, \xSym[n])$, $\bpSym[][i] = (\alpha_{1}, \xSym[1]) \cdots
		(\alpha_{i - 1}, \xSym[i - 1]) (\alpha_{i}, \ySym[i]) (\alpha_{i + 1},
		\xSym[i + 1]) \cdots (\alpha_{n}, \xSym[n])$, and $\free{\psi_{i}} =
		\AgnSet$.
		Now, it is not hard to see that $\varphi'_{NE}$, as well as the equivalent
		formulations of $\varphi_{EG}$ and $\varphi_{AG}$ of Example~\ref{exm:eg}
		and Example~\ref{exm:ag}, respectively, belong to \BGSL but not to \OGSL.

%		\begin{rem}[\BGSL\ Satisfiability]
%			\label{rmk:tbngsl(modchkhrd)}
		In Section~\ref{sec:hard}, we prove the undecidability of the
		satisfiability problem for \SL.
		Now, it is not hard to see that the formula $\varphi^{dom}$ used to reduce
		the domino problem in Theorem~\ref{thm:satsl} actually lies in the \BGSL
		fragment.
		Hence, the satisfiability for this logic is undecidable too.
		On the other hand, later in the paper, we prove that the same problem for
		\OGSL is 2\ExpTimeC, thus not harder than the one for \ATLS.
%		\end{rem}
%
		We have the following theorem.

		\begin{thm}
			\label{thm:satbgsl}
			The \emph{satisfiability} problem for \BGSL  is highly undecidable.
			In particular, it is \SCompH{1}{1}.
		\end{thm}

		In addition to this, we recall that in~\cite{TW12}, the authors prove that the satisfiability problem for a logic called \emph{\ATLS with strategy context}, introduced in~\cite{DLM10} is undecidable also on the class of finite models.
		It is easy to prove that \ATLS with strategy context can be embedded in \SL.
		This implies the following.

		\begin{thm}
			\label{thm:satslfinmod}
			The satisfiability on finite models problem for \SL is undecidable.
		\end{thm}

%		The question whether \ATLS with strategy context can be embedded in \BGSL is
%		still open.
%		For this reason, we cannot deduce the same result on \BGSL and we leave it as a conjecture.
%
%		\begin{conj}
%			\label{thm:satbgslfinmod}
%			The satisfiability on finite models problem for \BGSL is undecidable.
%		\end{conj}
		
		\begin{rem}
			\label{rmk:satslbg}
			It is important to notice that, since \ATLS with strategy context cannot be embedded in \BGSL, the result provided in~\cite{TW12} is not sufficient to prove the undecidability of the satisfiability problem for \BGSL.
			Moreover, the result shown here is stronger, as it proves highly undecidability for \BGSL.
			On the contrary, in~\cite{MMPV14}, we introduce a fragment called \emph{"Nested-Goal Strategy Logic"} (\NGSL, for short), that strictly subsumes \BGSL and in which \ATLS with strategy context can be embedded.
			This fragment, then, is undecidable on finite models and highly undecidable on infinite models.
			In addition to this, notice that in~\cite{LM13} the authors show that the decidability problem for \SL under turn-based \ConGamStr{s} is decidable.
			This means that the undecidability under concurrent \ConGamStr{s} is strict.
		\end{rem}

	\end{subsection}

	\begin{subsection}{Behavioral semantics}
		\label{sec:what;sub:bvr}

		We now recall the fundamental property of \emph{behavioral semantics}
		enjoyed by \OGSL.
		All concepts and results have been already introduced and fully
		investigated in~\cite{MMPV14}.
		We report them here for the sake of completeness.

		We first need to describe the concept of Skolem dependence function
		(\skodepfun, for short) and show how any quantification
		prefix contained into an \SL formula can be represented by an adequate
		choice of a \skodepfun over strategies.
		The main idea here is inspired by the technique proposed by Skolem for the
		first-order logic in order to eliminate all existential quantifications over
		variables, by substituting them with second order existential
		quantifications over functions, whose choice is uniform \wrt the universal
		variables.

		We first introduce some notation regarding the quantification prefixes.
		Let $\qpElm \in \QPSet(\VSet)$ be a quantification prefix over a set
		$\VSet \subseteq \VarSet$ of variables.
		By $\QPEVSet{\qpElm} \defeq \set{ \xElm \in \QPVSet(\qpElm) }{ \exists i \in
		\numco{0}{\card{\qpElm}} .\: (\qpElm)_{i} = \EExs{\xElm} }$ and
		$\QPAVSet{\qpElm} \defeq \QPVSet(\qpElm) \setminus \QPEVSet{\qpElm}$ we
		denote the sets of \emph{existential} and \emph{universal variables}
		quantified in $\qpElm$, respectively.
		For two variables $\xElm, \yElm \in \QPVSet(\qpElm)$, we say that $\xElm$
		\emph{precedes} $\yElm$ in $\qpElm$, in symbols $\xElm \qpordRel_{\qpElm}
		\yElm$, if $\xElm$ occurs before $\yElm$ in $\qpElm$, \ie, there are two
		indexes $i, j \in \numco{0}{\card{\qpElm}}\!$, with $i < j$, such that
		$(\qpElm)_{i} \in \{ \EExs{\xElm}, \AAll{\xElm} \}$ and $(\qpElm)_{j} \in \{
		\EExs{\yElm}, \AAll{\yElm} \}$.
		Moreover, we say that $\yElm$ is \emph{functional dependent} on $\xElm$, in
		symbols $\xElm \qpdepRel_{\qpElm} \yElm$, if $\xElm \in \QPAVSet{\qpElm}$,
		$\yElm \in \QPEVSet{\qpElm}$, and $\xElm \qpordRel_{\qpElm} \yElm$, \ie,
		$\yElm$ is existentially quantified after that $\xElm$ is universally
		quantified, so, there may be a dependence between a value chosen by $\xElm$
		and that chosen by $\yElm$.
		This definition induces the set $\QPDepSet(\qpElm) \defeq \set{ (\xElm,
		\yElm) \in \QPVSet(\qpElm) \times \QPVSet(\qpElm) }{ \xElm
		\qpdepRel_{\qpElm} \yElm }$ of \emph{dependence pairs} and its derived
		version $\QPDepSet(\qpElm, \yElm) \defeq \set{ \xElm \in \QPVSet(\qpElm) }{
		\xElm \qpdepRel_{\qpElm} \yElm }$ containing all variables from which
		$\yElm$ depends.
		Finally, we use $\dual{\qpElm} \in \QPSet(\QPVSet(\qpElm))$ to indicate the
		quantification derived from $\qpElm$ by \emph{dualizing} each quantifier
		contained in it, \ie, for all indexes $i \in \numco{0}{\card{\qpElm}}\!$,
		it holds that $(\dual{\qpElm})_{i} = \EExs{\xElm}$ iff $(\qpElm)_{i} =
		\AAll{\xElm}$, with $\xElm \in \QPVSet(\qpElm)$.
		It is evident that $\QPEVSet{\dual{\qpElm}} = \QPAVSet{\qpElm}$ and
		$\QPAVSet{\dual{\qpElm}} = \QPEVSet{\qpElm}$.
		As an example, let $\qpSym = \AAll{\xSym} \EExs{\ySym} \EExs{\zSym}
		\AAll{\wSym} \EExs{\vSym}$.
		Then, we have $\QPEVSet{\qpSym} = \{ \ySym, \zSym, \vSym \}$,
		$\QPAVSet{\qpSym} = \{ \xSym, \wSym \}$, $\QPDepSet(\qpSym, \xSym) =
		\QPDepSet(\qpSym, \wSym) = \emptyset$, \mbox{$\QPDepSet(\qpSym, \ySym) =
		\QPDepSet(\qpSym, \zSym) = \{ \xSym \}$, $\QPDepSet(\qpSym, \vSym) = \{
		\xSym, \wSym \}$, and $\dual{\qpSym} = \EExs{\xSym} \AAll{\ySym}
		\AAll{\zSym} \EExs{\wSym} \AAll{\vSym}$.}

		Finally, we define the notion of \emph{valuation} of variables over a
		generic set $\DSet$, called \emph{domain}, \ie, a partial function $\valFun
		: \VarSet \pto \DSet$ mapping every variable in its domain to an element in
		$\DSet$.
		By $\ValSet_{\DSet}(\VSet) \defeq \VSet \to \DSet$ we denote the set of all
		valuation functions over $\DSet$ defined on $\VSet \subseteq \VarSet$.

		At this point, we give a general high-level semantics for the quantification
		prefixes by means of the following definition of \emph{Skolem dependence
		function}.
		\begin{defi}[Skolem Dependence Function]
			\label{def:sklmap}
			Let $\qpElm \in \QPSet(\VSet)$ be a quantification prefix over a set
			$\VSet \subseteq \VarSet$ of variables, and $\DSet$ a set.
			Then, a \emph{Skolem dependence function} for $\qpElm$ over $\DSet$ is
			a function $\smFun : \ValSet_{\DSet}(\QPAVSet{\qpElm}) \to
			\ValSet_{\DSet}(\VSet)$ satisfying the following two properties:
			\begin{enumerate}
				\item
					\label{def:sklmap:aqnt}
					$\smFun(\valFun)_{\rst \QPAVSet{\qpElm}} \!=\! \valFun$, for all
					$\valFun \in \ValSet_{\DSet}(\QPAVSet{\qpElm})$;~\footnote{By
					$\gFun_{\rst \ZSet} : (\XSet \cap \ZSet) \to \YSet$ we denote the
					\emph{restriction} of a function $\gFun : \XSet \to \YSet$ to the
					elements in the set $\ZSet$.}
				\item
					\label{def:sklmap:eqnt}
					$\smFun(\valFun[][1])(\xElm) \!=\! \smFun(\valFun[][2])(\xElm)$, for
					all $\valFun[][1], \valFun[][2] \!\in\!
					\ValSet_{\DSet}(\QPAVSet{\qpElm})$ and $\xElm \!\in\!
					\QPEVSet{\qpElm}$ such that $\valFun[][1]_{\rst \QPDepSet(\qpElm,
					\xElm)} \!=\! \valFun[][2]_{\rst \QPDepSet(\qpElm, \xElm)}$.
			\end{enumerate}
			$\SMSet_{\DSet}(\qpElm)$ denotes the set of all \skodepfun{s} for $\qpElm$
			over $\DSet$.
		\end{defi}

		Intuitively, Item~\ref{def:sklmap:aqnt} asserts that $\smFun$ assumes the
		same values of its argument \wrt the universal variables in $\qpElm$, while
		Item~\ref{def:sklmap:eqnt} ensures that the value of $\smFun$ \wrt an
		existential variable $\xElm$ in $\qpElm$ does not depend on variables not in
		$\QPDepSet(\qpElm, \xElm)$.
		To get a better insight, note that a \skodepfun $\smFun$ for $\qpElm$ can be
		considered as a set of classical \emph{Skolem functions} that, given a value
		for each variable in $\QPAVSet{\qpElm}$ returns a possible value for all
		variables in $\QPEVSet{\qpElm}$, in a way that is consistent \wrt the order
		of quantifications.
		Observe that, each $\smFun \in \SMSet_{\DSet}(\qpElm)$ is injective, so,
		$\card{\rng{\smFun}} = \card{\dom{\smFun}} =
		\card{\DSet}^{\card{\QPAVSet{\qpElm}}}$.
		Moreover, $\card{\SMSet_{\DSet}(\qpElm)} = \prod_{\xElm \in
		\QPEVSet{\qpElm}} \card{\DSet}^{\card{\DSet}^{\card{\QPDepSet(\qpElm,
		\xElm)}}}$.
		As an example, let $\DSet = \{ 0, 1 \}$ and $\qpSym = \AAll{\xSym}
		\EExs{\ySym} \AAll{\zSym} \in \QPSet(\VSet)$ be a quantification prefix over
		$\VSet = \{ \xSym, \ySym, \zSym \}$.
		Then, we have that $\card{\SMSet_{\DSet}(\qpSym)} = 4$ and
		$\card{\SMSet_{\DSet}(\dual{\qpSym})} = 8$.
		Moreover, the \skodepfun{s} $\smFun[][i] \in
		\SMSet_{\DSet}(\qpSym)$ with $i \in \numcc{0}{3}$ and $\dual{\smFun[][i]}
		\in \SMSet_{\DSet}(\dual{\qpSym})$ with $i \in \numcc{0}{7}$, for a
		particular fixed order, are such that $\smFun[][0](\valFun)(\ySym) = 0$,
		$\smFun[][1](\valFun)(\ySym) = \valFun(\xSym)$,
		$\smFun[][2](\valFun)(\ySym) = 1 - \valFun(\xSym)$, and
		$\smFun[][3](\valFun)(\ySym) = 1$, for all $\valFun
		\in \ValSet_{\DSet}(\QPAVSet{\qpSym})$, and
		$\dual{\smFun[][i]}(\dual{\valFun})(\xSym) = 0$ with $i \in
		\numcc{0}{3}$, $\dual{\smFun[][i]}(\dual{\valFun})(\xSym) = 1$ with $i \in
		\numcc{4}{7}$, $\dual{\smFun[][0]}(\dual{\valFun})(\zSym) =
		\dual{\smFun[][4]}(\dual{\valFun})(\zSym) = 0$,
		$\dual{\smFun[][1]}(\dual{\valFun})(\zSym) =
		\dual{\smFun[][5]}(\dual{\valFun})(\zSym) = \dual{\valFun}(\ySym)$,
		$\dual{\smFun[][2]}(\dual{\valFun})(\zSym) =
		\dual{\smFun[][6]}(\dual{\valFun})(\zSym) = 1 - \dual{\valFun}(\ySym)$, and
		$\dual{\smFun[][3]}(\dual{\valFun})(\zSym) =
		\dual{\smFun[][7]}(\dual{\valFun})(\zSym) = 1$, for all $\dual{\valFun} \in
		\ValSet_{\DSet}(\QPAVSet{\dual{\qpSym}})$.

		We now report the following fundamental theorem that describes
		how to eliminate the strategy quantifications of an \SL\ formula via a
		choice of a suitable \skodepfun over strategies~\cite{MMPV14}.
		This procedure can be seen as the equivalent of the \emph{Skolemization}
		procedure in first-order logic (see~\cite{Hod93}, for more details).
		%
% 		\begin{thm}[\SL Strategy Quantification~\cite{MMPV14}]
% 			\label{thm:sl:strqnt}
% 			Let $\ConGamStrName$ be a \ConGamStr and $\varphi = \qpElm \psi$ an \SL
% 			formula, being $\qpElm \in \QPSet(\VSet)$ a quantification prefix over a
% 			set $\VSet \subseteq \free{\psi} \cap \VarSet$ of variables.
% 			Then, for all assignments $\asgFun \in \AsgSet(\free{\varphi},
% 			\sttElm[0])$, the following holds: $\ConGamStrName, \asgFun, \sttElm[0]
% 			\models \varphi$ iff there exists a \skodepfun $\smFun
% 			\in \SMSet_{ {\StrSet(\sttElm_{0})}}(\qpElm)$ such that $\ConGamStrName,
% 			\asgFun \umrg \smFun(\asgFun'), \sttElm[0] \models \psi$, for all
% 			$\asgFun' \in \AsgSet(\QPAVSet{\qpElm}, \sttElm[0])$.~\footnote{By
% 			$\gFun[1] \umrg \gFun[2] : (\XSet[1] \cup \XSet[2]) \to (\YSet[1] \cup
% 			\YSet[2])$ we denote the operation of \emph{union} of two functions
% 			$\gFun[1] : \XSet[1] \to \YSet[1]$ and $\gFun[2] : \XSet[2] \to \YSet[2]$
% 			defined on disjoint domains, \ie, $\XSet[1] \cap \XSet[2] = \emptyset$.}
% 		\end{thm}
%
% 		As an immediate consequence of the previous result, we derive the following
% 		corollary that restricts to \SL\ sentences.
		%
		\begin{thm}[\SL Strategy Quantification]
			\label{thm:sl:strqnt}
			Let $\ConGamStrName$ be a \ConGamStr and $\varphi = \qpElm \psi$ an \SL
			sentence, where $\psi$ is agent-closed and $\qpElm \in
			\QPSet(\free{\psi})$.
			Then, $\ConGamStrName \models \varphi$ iff there exists a \skodepfun
			$\smFun \in \SMSet_{{\StrSet}}(\qpElm)$ such that
			$\ConGamStrName, \smFun(\asgFun), \sttElm[0] \models \psi$, for all
			$\asgFun \in \AsgSet(\QPAVSet{\qpElm}, \sttElm[0])$.
		\end{thm}

%		Theorem~\ref{thm:sl:strqnt} allows to define an equivalent semantics for the
%		\SL sentences of the form $\qpElm \psi$.
%		More formally, for a given \ConGamStr $\ConGamStrName$, we may assert that
%		$\ConGamStrName \models \qpElm \psi$ iff there exists an \skodepfun $\smFun
%		\in \SMSet_{\StrSet}(\qpElm)$ such that $\ConGamStrName, \smFun(\asgFun),
%		\sttElm[0] \models \psi$, for all $\asgFun \in
%		\AsgSet_{\StrSet}(\QPAVSet{\qpElm})$.

		We now restrict our attention to a particular subclass of \skodepfun{s}
		defined on strategies called \emph{behavioral Skolem dependence functions}.
		Intuitively, an \skodepfun behavioral on strategies can be
		split into an infinite set of \skodepfun{s} over actions, one per each
		track in the domains of strategies.
		As next definition clarifies, not all the \skodepfun{s} are behavioral.
		This means that the announced simplification applies only under certain
		conditions.

		\begin{defi}[Adjoint Functions]
			\label{def:adjfun}
			Let $\smFun: \ValSet_{\StrSet}(\QPAVSet{\qpElm}) \to \ValSet_{\StrSet}(\VarSet)$ be an \skodepfun on strategies and let $\adj{\smFun}: \TrkSet \to (\ValSet_{\ActSet}(\QPAVSet{\qpElm}) \to \ValSet_{\ActSet}(\VarSet))$ be a function mapping every track into a \skodepfun on actions.
			We say that $\adj{\smFun}$ is the \emph{adjoint} of $\smFun$ if
			$\adj{\smFun}(\trkElm)(\flip{\asgFun}(\trkElm))(\xElm) =
			\smFun(\asgFun)(\xElm)(\trkElm)$, for all $\asgFun \in
			\AsgSet_{\StrSet}(\QPAVSet{\qpElm})$, $\xElm \in \VarSet$, and $\trkElm
			\in \TrkSet$~\footnote{By
			$\flip{\gFun} : \YSet \to \XSet \to \ZSet$ we denote the operation of
			\emph{flipping} of a function $\gFun : \XSet \to \YSet \to \ZSet$.}.
% 			Let $\DSet$, $\TSet$, $\USet$, and $\VSet$ be four sets, and $\mFun :
% 			(\TSet \to \DSet)^{\USet} \to (\TSet \to \DSet)^{\VSet}$ and
% $\adj{\mFun}
% 			: \TSet \to (\DSet^{\USet} \to \DSet^{\VSet})$ two functions.
% 			Then, $\adj{\mFun}$ is the \emph{adjoint} of $\mFun$ if
% 			$\adj{\mFun}(\tElm)(\flip{\gFun}(\tElm))(\xElm) =
% 			\mFun(\gFun)(\xElm)(\tElm)$, for all $\gFun \in (\TSet \to
% 			\DSet)^{\USet}$, $\xElm \in \VSet$, and $\tElm \in \TSet$~\footnote{By
% 			$\flip{\gFun} : \YSet \to \XSet \to \ZSet$ we denote the operation of
% 			\emph{flipping} of a function $\gFun : \XSet \to \YSet \to \ZSet$.}
		\end{defi}

		Intuitively, $\adj{\smFun}$ is the adjoint of $\smFun$ if the dependence
		from tracks in $\TrkSet$ in both domain and codomain of the latter function
		can be extracted and put as a common factor of the former function.
		This implies also that, for every pair of functions  $\asgFun[1], \asgFun[2]
		\in \AsgSet_{\StrSet}(\QPAVSet{\qpElm})$ such that
		$\flip{\asgFun[1]}(\trkElm) = \flip{\asgFun[2]}(\trkElm)$ for some $\trkElm
		\in \TrkSet$, it holds that $\smFun(\asgFun[1])(\xElm)(\trkElm) =
		\smFun(\asgFun[2])(\xElm)(\trkElm)$, for all variables $\xElm \in \VarSet$.
		It is immediate to observe that if a function has an adjoint then that
		adjoint is unique.
		At the same way, from an adjoint function it is possible to determine the
		original function without any ambiguity.
		Thus, it is established a one-to-one correspondence between functions
		admitting an adjoint and the adjoints themselves.
		
		We have the following definition.
		
		\begin{defi}
			\label{def:behavsdf}
			An \skodepfun is called \emph{behavioral} if it admits the adjoint function.
			Moreover, by $\BSMSet_{\StrSet}(\qntElm)$ we denote the set of behavioral \skodepfun{s} for $\qntElm$ over the set of strategies $\StrSet$
		\end{defi}
		
%		Moreover, we call \emph{behavioral} those \skodepfun{s} admitting the adjoint function.
		It is proved in~\cite{MMPV14} that a necessary and sufficient condition for a function $\adj{\smFun}$ to be an adjoint of a certain \skodepfun $\smFun \in \SMSet_{\StrSet}(\qpElm)$ is that $\adj{\smFun}(\trkElm)$ is in $\SMSet_{\ActSet}(\qpElm)$, for all $\trkElm \in \TrkSet$.

		Unfortunately, not every \skodepfun has an adjoint function.
		An easy way to prove this, it is to the number of \skodepfun{s} and
		adjoints.
		Indeed, we have that
		$$\card{\SMSet_{\StrSet}(\qpElm)} =
		\prod_{\xElm \in \QPEVSet{\qpElm}} \card{\ActSet}^{\card{\TrkSet} \cdot
		\card{\ActSet}^{\card{\TrkSet} \cdot \card{\QPDepSet(\qpElm, \xElm)}}},$$
		which is doubly exponential in the set $\TrkSet$ of tracks, while
		$$\card{\TrkSet \to \SMSet_{\ActSet}(\qpElm)} = \prod_{\xElm \in
		\QPEVSet{\qpElm}} \card{\ActSet}^{\card{\TrkSet} \cdot
		\card{\ActSet}^{\card{\QPDepSet(\qpElm, \xElm)}}},$$ which is only singly
		exponential in the same set $\TrkSet$.

% 		This fact gives rise to define an alternative semantics for \BGSL (and so
% 		for \OGSL), which is based on the behavioral \skodepfun{s}.

		\begin{defi}[\BGSL Behavioral Semantics~\cite{MMPV14}]
			\label{def:bgsl:behsem}
			Let $\ConGamStrName$ be a \ConGamStr, $\sttElm \in \SttSet$ one of its
			states, and $\qpElm \psi$ an \BGSL formula, where $\psi$ is agent-closed
			and $\qpElm \in \QPSet(\free{\psi})$.
			Then $\ConGamStrName, \sttElm \bmodels \qpElm \psi$ if there
			exists a behavioral \skodepfun $\smFun \in
			\BSMSet_{\StrSet}(\qpElm)$ for $\qpElm$ over $\StrSet$
			such that $\ConGamStrName, \smFun(\asgFun), \sttElm \bmodels \psi$, for
			all $\asgFun \in \AsgSet(\QPAVSet{\qpElm}, \sttElm)$.
		\end{defi}

		Clearly, from the previous definition and Theorem~\ref{thm:sl:strqnt}, we
		have that $\ConGamStrName \bmodels \varphi$ implies $\ConGamStrName \models
		\varphi$.
		In~\cite{MMPV14} it has been shown that the converse may not hold in
		general, \ie, there exists a \ConGamStr $\ConGamStrName$ and a \BGSL formula
		$\varphi$ such that $\ConGamStrName \models \varphi$ but $\ConGamStrName
		\not\bmodels \varphi$.
		However, as a fundamental result for the \OGSL fragment, in~\cite{MMPV14} it
		has also been proved that the behavioral semantics is equivalent to the
		classic one.
    This fact is derived by means of a reduction from the verification problem
    of a \OGSL sentence against a \ConGamStr to the wining problem of a Borelian
    two-player game.

		\begin{thm}[\OGSL\ Behavioral~\cite{MMPV14}]
			\label{thm:ogsl:behsem}
			Let $\ConGamStrName$ be a \ConGamStr and $\varphi$ an \OGSL\ sentence.
			Then, $\ConGamStrName \models \varphi$ iff $\ConGamStrName \bmodels
			\varphi$.
		\end{thm}

		It is important to note that the behavioral property of \OGSL is fundamental
		in proving many positive properties of the logic, as the bounded model
		property, which lead to a decidable procedure for the satisfiability
		problem, as we show later in the paper.

	\end{subsection}

	\begin{subsection}{Tree-model property}
		\label{sec:what;sub:treeprp}

		The satisfiability procedure we propose later in the paper is based on the
		use of alternating tree automata.
		Consequently, we need to establish a kind of \emph{tree model property},
		which is based on a special sub-class of \ConGamStr{s}, namely, the
		\emph{concurrent game-trees} (\ConGamTree, for short), whose structure of
		the underlying graph is a tree.

		\begin{defi}[Concurrent Game Trees]
			\label{def:cgt}
			A \emph{concurrent game tree} (\CGT, for short) is a \CGS\ $\TName \defeq
			\ConGamStrStr$, where \emph{(i)} $\SttSet \subseteq \DirSet^{*}$ is
			a $\DirSet$-tree for a given set $\DirSet$ of directions and \emph{(ii)}
			if $\tElm \cdot \eElm \in \SttSet$ then there is a decision $\decElm \in
			\DecSet$ such that $\trnFun(\tElm, \decElm) = \tElm \cdot \eElm$, for all
			$\tElm \in \SttSet$ and $\eElm \in \DirSet$.
			Furthermore, $\TName$ is a \emph{decision tree} (DT, for short) if
			\emph{(i)} $\SttSet = \DecSet^{*}$ and \emph{(ii)} if $\tElm \cdot \decElm
			\in \SttSet$ then $\trnFun(\tElm, \decElm) = \tElm \cdot \decElm$, for all
			$\tElm \in \SttSet$ and $\decElm \in \DecSet$.
		\end{defi}

		Intuitively, \CGT{s} are \CGS{s} having a transition relation with a tree
		shape and \DT{s} have, in addition, the states that uniquely determine the
		history of the computation leading to them.
		Observe that, for each non trivial track $\trkElm$ (resp., path $\pthElm$) of a \CGT, there exists a unique finite (resp., infinite) sequence of decisions $\decElm[0] \cdot \ldots \cdot \decElm[\card{\trkElm} - 2] \in \DecSet[][*]$ (resp., $\decElm[0] \cdot \decElm[1] \cdot \ldots \in \DecSet[][\omega]$) such that $\trkElm[(i + 1)] = \trnFun(\trkElm[(i)], \decElm[i])$ (resp., $\pthElm[(i + 1)] = \trnFun(\pthElm[(i)], \decElm[i])$), for all $i \in \numco{0}{\card{\trkElm} - 1}$ (resp., $i \in \SetN$).

		We now define a generalization for \CGS{s} of the classic concept of
		\emph{unwinding} of labeled transition systems, namely the
		\emph{decision-unwinding} (see Figure~\ref{fig:def:decunw}, for an example),
		that allows to show that \OGSL enjoys the decision-tree model property.

		\figdefdecunw

		\begin{defi}[Decision-Unwinding]
			\label{def:decunw}
			Let $\GName = \CGSStr$ be a \CGS.
			Then, the \emph{decision-un\-winding} of $\GName$ is the \DT\ $\GName[DU]
			\defeq \tupleg {\APSet} {\AgnSet} {\ActSet} {\DecSet[][*]} {\apFun'}
			{\trnFun'} {\root}$ for which there is a surjective function
			$\unwFun : \DecSet^{*} \to \SttSet$ such that
			\emph{(i)} $\unwFun(\root) = \sElm[0]$,
			\emph{(ii)} $\unwFun(\trnFun'(\tElm, \decElm)) = \trnFun(\unwFun(\tElm),
			\decElm)$, and
			\emph{(iii)} $\apFun'(\tElm) = \apFun(\unwFun(\tElm))$, for all
			$\tElm \in \DecSet^{*}$ and $\decElm \in \DecSet$.
		\end{defi}

		Observe that, due to its construction, each \CGS\ $\GName$ has a unique
		associated decision-unwinding $\GName[DU]$.

		We are now able to prove that \OGSL satisfies the decision-tree model
		property.

		\begin{thm}[\OGSL Decision-Tree Model Property]
			\label{thm:ogsl:tremodprp}
			Let $\varphi$ be a satisfiable \OGSL sentence.
			Then, there exists a \DT $\TName$ such that $\TName \models \varphi$.
		\end{thm}

		\begin{proof}

			The proof proceeds by structural induction on the sentence \OGSL.
			For the Boolean combination of principal sentences, the induction is
			trivial.
			For the case of a principal sentence $\varphi$ of the form $\qpElm
			\bndElm \psi$, by Theorem~\ref{thm:ogsl:behsem}, we derive that there
			exists a behavioral \skodepfun $\smFun \in \BSMSet[
			{\StrSet[\GName]} ](\qpElm)$ such that $\GName
			\models_{\smFun} \qpElm \bpElm \psi$.
			Furthermore, there exists the adjoint function tracking $\adj{\smFun}$ into $\smFun$.
			Now, consider the decision unwinding $\TName = \GName[DU]$ of $\GName$ and
			the lifting $\Gamma: \TrkSet[\TName] \to \TrkSet[\GName]$ of the unwinding
			function $\unwFun$ such that $\Gamma(\trkElm) = \unwFun(\trkElm[0]) \cdot
			\ldots \unwFun(\trkElm[\card{\trkElm} - 1])$, for all $\trkElm \in
			\TrkSet[\TName]$.
			At this point, consider the function $\adj{\smFun'}$ such that
			$\adj{\smFun'}(\trkElm') \defeq \adj{\smFun}(\Gamma(\trkElm'))$, for all
			$\trkElm' \in \TrkSet[\TName]$.
			Clearly, since $\adj{\smFun}$ is a \skodepfun over actions, so
			$\adj{\smFun'}$ is as well.
			Then, consider the \skodepfun $\smFun'$ for which the function
			$\adj{\smFun'}$ is its adjoint.
			By induction on the nesting of principal subsentences in
			$\varphi = \qpElm \bpElm \psi$, we now prove that $\TName \models
			_{\smFun'} \qpElm \bpElm \psi$.
			As base case, \ie, when $\psi$ is an \LTL formula, consider an assignment
			$\asgFun' \in \AsgSet[\TName](\QPAVSet{\qpElm}, \root)$ and the induced
			play $\playElm' = \playFun(\smFun'(\asgFun') \cmp \bpElm, \root)$ over
			$\TName$.
			Moreover, consider an assignment $\asgFun \in
			\AsgSet[\GName](\QPAVSet{\qpElm}, \sElm[0])$ such that, for all
			placeholders $\lElm \in \dom{\asgFun}$ and tracks $\trkElm' \in
			\TrkSet[\TName]$, it holds that $\asgFun(\lElm)(\Gamma(\trkElm')) =
			\asgFun'(\lElm)(\trkElm')$.
			From the satisfiability of $\varphi$ on $\GName$, we derive that
			$\playElm \models \psi$, where $\playElm = \playFun(\smFun(\asgFun) \cmp
			\bpElm, \sElm[0])$.
			Indeed, assume for a while that $(\playElm)_{\leq k} =
			\Gamma((\playElm)_{\leq k}^{'})$.
			Then, from the definition of the labeling in the decision-tree
			unwinding, it easily follows that $\apFun((\playElm)_{k}) =
			\apFun'((\playElm')_{k})$, and, so we derive $\playElm' \models \psi$,
			from $\playElm \models \psi$.
			Consequently, this holds for all $\asgFun' \in \AsgSet[
			{\StrSet[\TName]} ](\QPAVSet{\bpElm})$ and, so, we have that
			$\TName \models_{\smFun'} \qpElm \bpElm \psi$.
			The inductive case, \ie, when $\psi$ contains some subsentence, easily
			follows by considering the principal subsentences as fresh atomic
			propositions.

			It remains to prove, by induction on $k$, that $(\playElm)_{\leq k} =
			\Gamma((\playElm')_{\leq k})$, for all $k \in \SetN$.
			As base case, we have that $(\playElm)_{0} = \sElm[0] = \Gamma(\root) =
			(\playElm')_{0}$.
			As inductive case, assume that $(\playElm)_{\leq k} =
			\Gamma((\playElm^{'})_{\leq k})$.
			Then, in particular, we have that $(\playElm)_{k} =
			\Gamma((\playElm^{'})_{k}) = \unwFun((\playElm^{'})_{k})$.
			By definition of play, it holds that $(\playElm)_{k + 1} =
			\trnFun((\playElm)_{k}, (\adj{\smFun}((\playElm)_{\leq
			k}))(\flip{\asgFun})) \cmp \bpElm)$, which is, by inductive hypothesis,
			equal to $\trnFun(\unwFun((\playElm')_{k}),
			(\adj{\smFun}(\Gamma((\playElm)_{\leq k})))(\flip{\asgFun})) \cmp
			\bpElm)$.
			Now, by the definition of $\adj{\smFun'}$ and
			$\asgFun$, we obtain $\trnFun(\unwFun((\playElm')_{k}),
			(\adj{\smFun}(\Gamma((\playElm)_{\leq k})))(\flip{\asgFun})) \cmp
			\bpElm) = \trnFun(\Gamma((\playElm')_{k}),
			(\adj{\smFun'}((\playElm')_{\leq k}))(\flip{\asgFun'})) \cmp \bpElm)$.
			Finally, by the definition of $\unwFun$ and $\Gamma$, we have that
			$\trnFun(\Gamma((\playElm')_{k}), (\adj{\smFun'}((\playElm')_{\leq
			k}))(\flip{\asgFun'})) \cmp \bpElm) = \Gamma(\playElm')_{k + 1}$.

		\end{proof}

	\end{subsection}

\end{section}

% End of file SectionIII.tex

%%****************************************************************************%%
%%                                                                            %%
%% Reasoning About Strategies: On the Satisfiability Problem                  %%
%%                                                                            %%
%% SectionIV.tex                                                              %%
%%                                                                            %%
%% Revision 0                                                                 %%
%%                                                                            %%
%% Copyright (C) 2014, Fabio Mogavero, Aniello Murano, Giuseppe Perelli, and  %%
%%                     Moshe Y. Vardi.                                        %%
%% All rights reserved.                                                       %%
%%                                                                            %%
%%****************************************************************************%%

% Begin of file SectionIV.tex

\begin{section}{Decidability of SL[1G]}
	\label{sec:dec}

	In this section, we finally provide a 2\ExpTimeC procedure for the \OGSL
	satisfiability problem.
	Before doing this, we have to prove the \emph{bounded-tree model property},
	which results to be crucial for the automata-theoretic approach later
	described.

	\begin{subsection}{Bounded model property}
		\label{sec:dec;sub:bndprp}

		In order to prove the bounded-tree model property for \OGSL, we first need
		to introduce the new concept of \emph{disjoint satisfiability}, which
		concerns the verification of different instances of the same subsentence of
		the original specification.
		Intuitively, it asserts that either these instances can be checked on
		disjoint subtrees of the tree model or, if two instances use part of the
		same subtree, they are forced to use the same dependence map as well.
		This notion is a reformulation of the notion of explicit model introduced
		for \ATLS in~\cite{Sch08}.
		This intrinsic characteristic of \OGSL is fundamental for the building of a
		unique automaton that checks the truth of all subsentences, by simply
		merging their respective automata, without using a projection operation to
		eliminate their own alphabets, which otherwise may be in conflict.
		In this way, we are also able to avoid an exponential blow-up.
		A deeper discussion on this point is reported later in the paper.

		\begin{defi}[Disjoint Satisfiability]
			\label{def:dsjsat}
			Let $\TName$ be a \DT and $\varphi = \qpElm \bpElm \psi$ be a \OGSL
			principal sentence.
			Moreover, let $\SSet \defeq \set{\sttElm \in \SttSet[\TName]}{\TName,
			\sttElm \models \varphi}$.
			Then, $\TName$ satisfies $\varphi$ \emph{disjointly} over $\SSet$ if
			there exist two functions $\headFun: \SSet \to \SMSet_{\ActSet}(\qpElm)$
			and $\bodyFun: \TrkSet(\root) \to \SMSet_{\ActSet}(\qpElm)$ such that,
			for all $\sttElm \in \SSet$ and $\asgFun \in
			\AsgSet[\StrSet](\QPAVSet{\qpElm})$ it holds that $\TName,
			\smFun(\asgFun), \sttElm \models \bpElm \psi$, where the
			behavioral \skodepfun $\smFun \in \BSMSet_{\StrSet}$ is defined,
			by means of its adjoint, as follows:

			\begin{enumerate}[label=(\roman*)]
				\item
					$\adj{\smFun}(\sttElm) \defeq \headFun(\sttElm)$;

				\item
					$\adj{\smFun}(\trkElm) \defeq \bodyFun(\trkElm' \cdot \trkElm)$, for
					all $\trkElm \in \TrkSet(\sttElm)$ with $\card{\trkElm} > 1$, where
					$\trkElm' \in \TrkSet(\root)$ is the unique track such that
					$\trkElm' \cdot \trkElm \in \TrkSet(\root)$~\footnote{Existence and
					uniqueness of $\trkElm'$ is guaranteed by the fact that $\TName$ is a
					\DT.}.
			\end{enumerate}

		\end{defi}

		\noindent The disjoint satisfiability holds for all \OGSL formulas.
		To prove this fact, we first introduce the preliminary definition of
		\emph{twin decision-tree}.
		Intuitively, in such a kind of tree, each action is flanked by a twin one
		having the same purpose of the original.
		This allows to satisfy two sentences requiring the same actions in a given
		state in two different branches of the tree itself, which is what the
		disjoint satisfiability precisely requires.

		\begin{defi}[Twin Decision Tree]
			\label{doubactDT}
			Let $\TName = \ConGamStrStr[][\root]$ be a \DT.
			Then, the \emph{twin decision tree} of $\TName$ is the \DT $\TName'
			\defeq \tupleg{\APSet} {\AgnSet} {\ActSet'} {\SttSet'} {\trnFun'}
			{\apFun'} {\root'}$ with $\ActSet' = \ActSet \times \{new, cont \}$ and
			$\root' = (\root, new)$.
			The labeling and the transition function are defined by means of a set
			of projection functions introduced below:

			\begin{itemize}

				\item
					the function $\prj[\ActSet]: \ActSet' \to \ActSet$ returns the first
					component of the action in $\ActSet'$, \ie, $\prj((\actElm, \iota)) =
					\actElm$, for all $(\actElm, \iota) \in \ActSet'$;

				\item
					the function $\prj[\DecSet]: \DecSet' \to \DecSet$ projects out the
					flags on all the actions in the decision, returning a corresponding
					decision in $\TName$, \ie,
					$\prj[\DecSet](\decElm')(\agnElm) = \prj[\ActSet](\decElm'(\agnElm))$,
					for all $\decElm' \in \DecSet'$ and $\agnElm \in \AgnSet$;

				\item
					the function $\prj[\SttSet]: \SttSet' \to \SttSet$ returns the
					corresponding state in $\TName$, according to the projection made on
					the decisions, \ie, $\prj[\SttSet](\root') = \root$ and
					$\prj[\SttSet](\sttElm' \cdot \decElm') = \prj[\SttSet](\sttElm')
					\cdot \prj[\DecSet](\decElm')$, for all $\sttElm' \in \SttSet'$ and
					$\decElm' \in \DecSet'$;

				\item
					analogously, the function $\prj[\TrkSet]: \TrkSet' \to \TrkSet$,
					returns the concatenation of the projected states, \ie,
					$\prj[\TrkSet](\sttElm') = \prj[\SttSet](\sttElm')$, for all $\sttElm'
					\in \SttSet'$, and $\prj[\TrkSet](\trkElm' \cdot \sttElm') =
					\prj[\TrkSet](\trkElm') \cdot \prj[\SttSet](\sttElm')$, for all
					$\trkElm' \in \TrkSet'$ and $\sttElm' \in \SttSet'$.

			\end{itemize}

			\noindent Then, $\apFun'(\sttElm') \defeq \apFun(\prj[\SttSet](\sttElm'))$.

		\end{defi}

		Observe that $\prj[\SttSet](\trnFun'(\sttElm', \decElm')) =
		\trnFun(\prj[\SttSet](\sttElm'), \prj[\DecSet](\decElm'))$, for all
		$\sttElm' \in \SttSet'$ and $\decElm' \in \DecSet'$.
		We can now prove the disjoint satisfiability property for \OGSL.

		\begin{thm}[Disjoint Satisfiability]
			\label{thm:ogsl:disjsat}
			Let $\varphi = \qpElm \bpElm \psi$ be an \OGSL principal sentence and
			$\TName = \ConGamStrStr[][\root]$ a \DT.
			Moreover, let $\SSet \defeq \set{\sttElm \in \SttSet}{\TName, \sttElm
			\models \varphi}$.
			Then the twin decision tree $\TName'$ of $\TName$ disjointly satisfies
			$\varphi$ over $\SSet' \defeq \set{\sttElm' \in \SttSet'}{\prj(\sttElm')
			\in \SSet}$.
		\end{thm}

		\begin{proof}[Proof idea]
			Starting from the fact that $\TName, \sttElm \models \varphi$,
			for all $\sttElm \in \SSet$, by means of Theorem~\ref{thm:ogsl:behsem},
			we derive the existence of a behavioral \skodepfun $\smFun[\sttElm]$.
			Such a $\smFun[\sttElm]$ is used to define a behavioral \skodepfun
			$\smFun[\sttElm]'$ in $\TName'$ in which the existential agents suitably
			select either $new$ or $cont$ as second component, in order to guarantee
			the satisfaction of different instances over different branches of the
			twin decision tree.
			Indeed, it allows to properly define the two functions $\headFun$ and
			$\bodyFun$ and, consequently, the behavioral \skodepfun $\smFun'$ for
			which we finally prove that $\TName', \sttElm' \models \varphi$, for all
			$\sttElm' \in \SSet'$.
			Since $\smFun'$ has been built from the head and body functions, the
			disjoint satisfiability is immediately derived.
		\end{proof}

		\begin{proof}
			Let $\sttElm \in \SSet$ be one of the states on which $\varphi$ is
			satisfied.
			Since $\TName, \sttElm \models \varphi$, by Theorem~\ref{thm:ogsl:behsem},
			we have that there exists $\smFun[\sttElm] \in
			\BSMSet[\StrSet](\qpElm)$ such that $\TName,
			\smFun[\sttElm](\asgFun), \sttElm \models \bpElm \psi$, for all states
			assignments $\asgFun \in \AsgSet[ {\StrSet[\TName]} ](\QPAVSet{\qpElm})$.
			Then, consider the adjoint function $\adj{\smFun[\sttElm]}': \TrkSet' \pto
			\SMSet_{\ActSet'}(\qpElm)$ defined, for all states $\sttElm' \in
			\SttSet'$, decisions $\decElm' \in \DecSet'$, and tracks $\trkElm' \in
			\TrkSet'$ as follows:
			\begin{itemize}

				\item
					$\adj{\smFun[\sttElm]}'(\sttElm')(\decElm')(\xElm) =
					(\adj{\smFun[\sttElm]}
					(\prj[\SttSet](\sttElm'))(\prj[\DecSet](\decElm'))(\xElm), new)$, if
					$\prj[\SttSet](\sttElm') = \sttElm$;

				\item
					$\adj{\smFun[\sttElm]}'(\trkElm')(\decElm')(\xElm) =
					(\adj{\smFun[\sttElm]}(\prj[\TrkSet]
					(\trkElm'))(\prj[\DecSet](\decElm'))(\xElm), cont)$, otherwise.
			\end{itemize}

			At this point, we assume the function $\headFun: \SttSet' \to
			\SMSet_{\ActSet'}(\qpElm)$ to be defined as follows: $\headFun(\sttElm')
			\defeq \adj{\smFun[\sttElm]}'(\sttElm')$.
			Moreover, we set the function $\bodyFun: \TrkSet'(\root) \to
			\SMSet_{\ActSet'}(\qpElm)$ in such a way that it agrees with
			$\adj{\smFun[\sttElm]}'$ on all tracks $\trkElm' = \sttElm[0]' \cdot
			\ldots \cdot \sttElm[n]'$ for which there is an index $i \in \{0, \ldots,
			n\}$ such that, for all agents $\agnElm \in \AgnSet$ with $\bpElm(\agnElm)
			\in \QPEVSet{\qpElm}$, it holds that:
			\begin{itemize}
				\item
					$\lst{(\trkElm')_{i}}(\agnElm) = (\cElm[\agnElm], new)$, for some
					$\cElm[\agnElm] \in \ActSet$, and
				\item
					$\lst{(\trkElm')_{j}}(\agnElm) = (\cElm[\agnElm], cont)$, for all $j
					\in \{ i + 1, \ldots, n \}$ and for some $\cElm[\agnElm] \in \ActSet$.
					
			\end{itemize}

			Note that the tracks of this form are such that the players bound to an
			existentially quantified variable have selected an action flagged by
			$new$ on the $i$-th step of the game and then keep playing with the $cont$
			flag.
			Intuitively, they are starting the verification of a subsentence right in
			the $i$-th state of the track, by keeping it separated from the
			verification of the other subsentences, which are addressed with the
			$cont$ flag.
			
			For all the other tracks $\trkElm'$, instead, the value of
			$\bodyFun(\trkElm')$ may be arbitrary.
			
			Now, consider the behavioral \skodepfun $\smFun' \in
			\BSMSet_{\StrSet}(\qpElm)$ defined by means of the functions $\headFun$
			and $\bodyFun$ as prescribed by Definition~\ref{def:dsjsat}.
			It remains to prove that $\TName', \sttElm' \models_{\smFun'} \qpElm
			\bpElm \psi$, for all $\sttElm' \in \SSet'$.
			We proceed by induction on the nesting of the principal subsentences of
			$\varphi$.
			As base case, assume that such nesting is $0$. This means that $\psi$ is
			an \LTL formula.
			Now, let $\asgFun' \in \AsgSet[\StrSet'](\QPAVSet{\qpElm})$.
			By construction, it is not hard to see that there exists an assignment
			$\asgFun \in \AsgSet[\StrSet](\QPAVSet{\qpElm})$ for which
			the play $\playElm' \defeq \playFun'(\smFun'(\asgFun')\cmp \bpElm,
			\sttElm')$ satisfies the equality $\prj[\PthSet](\playElm') =
			\playFun(\smFun(\asgFun) \cmp \bpElm, \prj(\sttElm')) =
			\playElm$~\footnote{By $\prj[\PthSet]$ we are denoting the natural lifting
			of the function $\prj[\TrkSet]$ to paths.}.
			Thus, since $\TName, \sttElm \models_{\smFun} \qpElm \bpElm \psi$, we have
			that $\playElm \models \psi$.
			Moreover, it holds that $\apFun'((\playElm')_{i}) =
			\apFun((\playElm)_{i})$, for all $i \in \SetN$, which implies that
			$\playElm' \models \psi$.
			Consequently, we can conclude that $\TName', \sttElm' \models_{\smFun'}
			\qpElm \bpElm \psi$.
			The inductive case, easily follows by considering the inner principal
			subsentences as fresh atomic propositions.
		\end{proof}

		We now have all tools to prove the bounded model property of \OGSL.

		\begin{thm}[Bounded Model Property of \OGSL]
		\label{thm:ogsl:bndmdprp}
			Let $\varphi$ be a \OGSL sentence and $\TName$ be a \DT such that $\TName
			\models \varphi$.
			Then, there exists a bounded \DT $\TName'$ such that $\TName' \models
			\varphi$.
		\end{thm}

		The proof makes use of some instruments and formalisms for \emph{First-Order
		Logic} (\FOL, for short) that are introduced in~\cite{MP15b}.
		For the sake of completeness, here we give an informal discussion of such
		object.
		A \emph{language signature} is a tuple $\LanSigName = \LanSigStr$ in which
		$\ArgSet$ and $\RelSet$ are two finite non-empty sets of \emph{arguments}
		and \emph{relations}, respectively, and $\argFun: \RelSet \to \pow{\ArgSet}
		\setminus \{ \emptyset \}$ is a function mapping each relation in $\RelSet$
		to its non-empty set of arguments.
		Language signatures are used to reformulate \FOL syntax in terms of
		\emph{binding forms}, which are a way to associate variables to relations by
		means of bindings.
% 		This particular formalism allows to introduce a new class of fragments,
% 		based on the binding-form, which have been studied in~\cite{MP15b}.
		The interpretation of \FOL formulas is given on \emph{relational
		structures}, which are tuples $\RelStrName = \RelStrStr$ with $\DomSet$
		being a non-empty domain and where $\relFun(\relElm) \subseteq
		\argFun(\relElm) \to \DomSet$ is a set of functions, representing the tuples
		on which the relation $\relElm \in \RelSet$ is interpreted as true.

		\begin{proof}[Proof Idea]
			The key idea used to prove the theorem is based on the \emph{finite model
			property} of the \emph{One-Binding} fragment of \FOL (\OBFOL, for short),
			proved in~\cite{MP15b}, which allows to define a bounded-tree model
			$\TName'$, which preserves the satisfiability of $\varphi$.
			In particular, for each state $\sttElm[][*]$ of a tree $\TName$ satisfying
			$\varphi$, we build a first-order structure and a \OBFOL formula
			$\eta_{\sttElm[][*]}$ that characterizes the topology of the successors of
			$\sttElm[*]$ in $\TName$.
			Then, since \OBFOL enjoys the finite model property, we are able to build
			a finite first-order structure for $\eta_{\sttElm[][*]}$ from which we can
			build the bounded model $\TName'$ of $\varphi$.
			Such a construction is based on both the disjoint and behavioral
			satisfiability of \OGSL.
			For each state $\sttElm[][*]$ in $\TName$, we consider a set given
			by pairs of subsentences $\eta$ of $\varphi$ and states $\sttElm$, on
			which it holds that $\TName, \sttElm \models_{\smFun} \eta$, where such
			satisfaction is forced to pass through $\sttElm[][*]$ for at least one
			universal assignment fed to $\smFun$.
			This means that at least one play used to satisfy $\eta$ passes through
			$\sttElm[][*]$. 
			By the disjoint satisfiability, we have to cope with at most two
			\skodepfun{s} for each subsentence $\eta$, those given by
			$\headFun[][\eta]$ and $\bodyFun[][\eta]$, implying that the total number
			of \skodepfun{s} to take into account, for all $\sttElm[][*]$, is finite.
			From that, we define a related \OBFOL sentence $\eta_{\sttElm[][*]}$,
			having a model derived from the topology of the successors of
			$\sttElm[][*]$ whose elements are constituted by the actions of $\TName$.
			Now, by applying the finite model property to $\eta_{\sttElm[][*]}$, we
			derive the existence of a model for the formula $\eta_{\sttElm[][*]}$ with
			a finite domain $\ActSet'_{\eta, \sttElm[][*]}$.
			Exploiting the finite model built for all states $\sttElm[][*]$, we are
			able to define the labeling of
			$\TName'$ and a behavioral \skodepfun $\smFun'$ in such a way that
			$\TName', \root \models_{\smFun'} \varphi$.
			
		\end{proof}

		\begin{proof}[Proof of Theorem~\ref{thm:ogsl:bndmdprp}]

			We give the proof for the case of $\varphi = \qpElm \bpElm \psi$, since
			the Boolean combination of principal sentences easily follows from this
			one.
			Given a tree-model $\TName$ for $\varphi$, derived by the tree-model
			property of \OGSL of Theorem~\ref{thm:ogsl:tremodprp}, for each state
			$\sttElm[][*] \in \SttSet$, consider the set $\Phi_{\sttElm[][*]}
			\subseteq \SttSet \times \psnt{\varphi}$ of states of $\TName$ and
			principal subsentences of $\varphi$ such that $(\sttElm, \eta) \in
			\Phi_{\sttElm[][*]}$ iff \emph{(i)} $\TName, \sttElm \models \eta$ and
			\emph{(ii)} there exists an assignment $\asgFun \in
			\AsgSet[\StrSet](\QPAVSet{\qpElm[][\eta]})$ such that
			$\sttElm[][*] = (\playFun(\smFun[(\sttElm, \eta)][ {\sttElm[][*]}
			](\asgFun)) \cmp \bpElm[][\eta], \sttElm)_{n}$, for some $n \in \SetN$,
			where the behavioral \skodepfun $\smFun[(\sttElm, \eta)][ {\sttElm[][*]}
			]$ is defined by means of its adjoint, which is in its turn built from the
			functions $\headFun[][\eta]$ and $\bodyFun[][\eta]$, given by
			Theorem~\ref{thm:ogsl:disjsat}, applied on $\eta = \qpElm[\eta]
			\bpElm[\eta] \psi_{\eta}$.
			Observe that, for a fixed $\eta$, if $\sttElm[1], \sttElm[2] \in \SttSet$,
			with $\sttElm[1] \neq \sttElm[][*]$ and $\sttElm[2] \neq \sttElm[][*]$,
			$\adj{\smFun[(\sttElm, \eta)][ {\sttElm[][*]} ]}(\trkElm[ {\sttElm[1]} ])
			= \headFun[][\eta](\trkElm[ {\sttElm[1]} ]['] \cdot \trkElm[ {\sttElm[1]}
			]) = \headFun[][\eta](\trkElm[ {\sttElm[2]} ]['] \cdot \trkElm[
			{\sttElm[2]} ]) = \adj{\smFun[(\sttElm, \eta)][ {\sttElm[][*]} ]}(\trkElm[
			{\sttElm[2]} ])$, where $\trkElm[ {\sttElm[1]} ]$ and $\trkElm[
			{\sttElm[2]} ]$ are the unique tracks ending in $\sttElm[][*]$ and
			starting in $\sttElm[1]$ and $\sttElm[2]$, respectively, while $\trkElm[
			{\sttElm[1]} ][']$ and $\trkElm[ {\sttElm[2]} ][']$ are the unique tracks
			such that $\trkElm[ {\sttElm[1]} ]['] \cdot \trkElm[ {\sttElm[1]} ]$ and
			$\trkElm[ {\sttElm[2]} ]['] \cdot \trkElm[ {\sttElm[2]} ]$ start from
			$\root$.
			Now, for a given \skodepfun over Actions $\adj{\smFun} \in
			\SMSet_{\ActSet}(\qpElm)$ and a given state $\sttElm \in \SttSet$, define
			the set $Succ_{\vartheta, \bpElm}(\sttElm) \defeq \set{\sttElm'
			\in \SttSet}{\exists \valFun \in \ActSet[][\QPAVSet{\qpElm}].
			\trnFun(\sttElm, \vartheta(\valFun) \cmp \bpElm) = \sttElm'}$, where
			$\vartheta \in \ActSet[][\QPAVSet{\qpElm}] \to \ActSet[][\VarSet(\qpElm)]$
			is a \skodepfun for $\qpElm$ over actions.
			Intuitively, the set $Succ_{\vartheta, \bpElm}(\sttElm)$ defines the set
			of states that can be reached in one step from $\sttElm$ by prescribing
			the agents that are bound by $\bpElm$ to an existential variable to move
			according to the \skodepfun $\vartheta$.
			
			At this point, consider the language signature $\LName =
			\tuplec{\ArgSet}{\RelSet}{\argFun} = \tuplec{\AgnSet}{\APSet}{\argFun}$
			with $\argFun(\pElm) = \AgnSet$, for all $\pElm \in \APSet$, where each
			atomic proposition is viewed as a relation having the agents as arguments
			and, so, the decisions as elements of its interpretation.
			Moreover, for all sets $\PSet \subseteq \APSet$, let
			$\mask^{\PSet} = \bigwedge_{\pElm \in \PSet} \pElm \wedge \bigwedge_{\qElm
			\in \APSet \setminus \PSet} \neg \qElm$ be the \OBFOL formula asserting
			that only the relations in $\PSet$ hold.
			Finally, for all $(\sttElm, \eta) \in \Phi_{\sttElm[][*]}$, consider the
			\OBFOL sentence
			$\eta_{\sttElm}^{*} = \qpElm_{\eta} \bpElm_{\eta} \bigvee_{\sttElm \in
			Succ_{\smFun_{(\sttElm, \eta))}^{\sttElm[][*]}(trk(\sttElm[][*]),
			\bpElm[][\eta]}} \mask^{\apFun(\sttElm)}$~\footnote{By $trk(\sttElm)$ we
			are denoting the unique track starting from $\root$ and ending in
			$\sttElm$.}.
			Clearly, by definition, each $\eta_{\sttElm}^{*}$ is satisfied by the
			relational structure $\RelStrName_{\sttElm[][*]} =
			\tupleb{\ActSet}{\relFun_{\sttElm[][*]}}$ with
			$\relFun_{\sttElm[][*]}(\pElm) \defeq \set{\decElm \in \DecSet}{\pElm \in
			\apFun(\trnFun(\sttElm[][*], \decElm)}$, where a relation $\pElm$ is
			interpreted as true on all decisions that allow $\sttElm[][*]$ to reach a
			state in which $\pElm$ holds.
			Indeed, for each partial valuation $\valFun \in
			\ActSet[][\QPAVSet{\qpElm[][\eta]}]$, it holds that either $\sttElm' =
			\trnFun(\sttElm, \headFun[][\eta](\sttElm)(\valFun)) \in
			Succ_{\headFun[][\eta](\sttElm), \bpElm[][\eta]}(\sttElm)$ or $\sttElm'' =
			\trnFun(\sttElm, \bodyFun[][\eta](trk(\sttElm))(\valFun)) \in
			Succ_{\bodyFun[][\eta](trk(\sttElm)), \bpElm[][\eta]}(\sttElm)$, which
			implies that either $\RelStrName[ {\sttElm[][*]} ],
			\headFun[][\eta](\sttElm)(\valFun) \models mask^{\apFun(\sttElm')}$ or
			$\RelStrName[ {\sttElm[][*]} ], \bodyFun[][\eta](trk(\sttElm))(\valFun)
			\models mask^{\apFun(\sttElm'')}$.
			Hence, we have that $\RelStrName[ {\sttElm[][*]} ] \models
			\eta^{*}_{\sttElm}$ and, since this is true for all $(\sttElm, \eta) \in
			\Phi_{\sttElm[][*]}$, we derive that $\RelStrName[ {\sttElm[][*]} ]
			\models \bigwedge_{(\sttElm, \eta) \in \Phi_{\sttElm[][*]}}
			\eta^{*}_{\sttElm}$.

			At this point, from the finite model property of \OBFOL, we derive that
			there exists a finite relational structure $\RelStrName[ {\sttElm[][*]}
			]['] = \RelStrStr[ {\sttElm[][*]} ][']$ such that $\RelStrName[
			{\sttElm[][*]} ]['] \models \bigwedge_{(\sttElm, \eta) \in
			\Phi_{\sttElm[][*]}} \eta^{*}_{\sttElm}$.
			Moreover, we define $\adj{\smFun[(\sttElm, \eta)][' {\sttElm[][*]} ]}$
			to be such that $\RelStrName[ {\sttElm[][*]} ][']
			\models_{\adj{\smFun[(\sttElm, \eta)][' {\sttElm[][*]} ]}}
			\eta^{*}_{\sttElm}$.
			Observe that, in the proof of finite model property for
			\OBFOL~\cite{MP15b}, the bound on $\DomSet[ {\sttElm[][*]} ][']$ only
			depends on the quantification and binding prefixes given in the formulas
			$\eta^{*}_{\sttElm}$, which all occur in $\varphi$.
			Thus, the size of $\DomSet[ {\sttElm[][*]} ][']$ does not depend on
			$\eta$ and, \wlogx, we can assume that $\DomSet[ {\sttElm[][*]} ]['] =
			\ActSet'$ for all $\sttElm[][*] \in \SttSet$.
			Moreover, again from the finite model property proof of \OBFOL, there
			exists a function $\MFun_{\sttElm[][*]}: \DecSet' \to \DecSet$, with
			$\DecSet' = \ActSet'^{\AgnSet}$, such that, for all $(\sttElm, \eta) \in
			\Phi_{\sttElm[][*]}$ and $\decElm' \in \rng{\adj{\smFun[(\sttElm, \eta)]['
			{\sttElm[][*]} ]}}$, we have that $\MFun_{\sttElm[][*]}(\decElm') \in
			\rng{\adj{\smFun[(\sttElm, \eta)][ {\sttElm[][*]} ]}}$, where
			$\adj{\smFun[(\sttElm, \eta)][ {\sttElm[][*]} ]}$ is the \skodepfun used
			to satisfy $\eta_{\sttElm}^{*}$ on $\RelStrName$.
			At this point, we define a \DT $\TName'$ having $\ActSet'$ as set of
			actions.
			In order to define the labeling function $\apFun'$, first consider the
			mapping $\Gamma: \SttSet' \to \SttSet$ recursively defined as follows:

			\begin{itemize}
				\item
					$\Gamma(\root) = \root$;
				\item
					$\Gamma(\sttElm' \cdot \decElm') = \Gamma(\sttElm') \cdot
					\MFun_{\Gamma(\sttElm')}(\decElm')$.
			\end{itemize}

			By means of $\Gamma$, define $\apFun'(\sttElm') \defeq
			\apFun(\Gamma(\sttElm'))$ for all $\sttElm' \in \SttSet'$.
			It remains to prove that $\TName' \models \varphi$.
			We do this by using a \skodepfun $\smFun' \in
			\SMSet_{\StrSet'}(\qpElm)$ defined from the adjoint
			$\adj{\smFun'}$ introduced in the following.

			Let $\trkElm'$ be a track in $\TName'$ and consider $\sttElm' =
			\lst{\trkElm'}$.
			If $(\varepsilon, \varphi) \in \Phi_{\Gamma(\sttElm')}$, define
			$\adj{\smFun'}(\trkElm') = \smFun_{(\sttElm, \eta)}'^{\sttElm[][*]}$.
			For all other tracks, the value of $\adj{\smFun'}$ may be arbitrary.
			We now show that $\TName, \emptyfun, \varepsilon \models_{\smFun'}
			\varphi$, by induction on the nesting of principal subsentences.
			As base case, suppose that $\varphi$ has nesting $0$.
			This implies that it is of the form $\qpElm \bpElm \psi$ with $\psi$
			being an \LTL formula.
			Then, consider a universal assignment $\asgFun' \in
			\AsgSet_{\StrSet'}(\QPAVSet{\qpElm})$ and then the assignment $\smFun'(\asgFun')$.
			This determines a play $\playElm' = \playFun(\smFun'(\asgFun'),
			\varepsilon)$.
			Now, consider a universal assignment $\asgFun \in
			\AsgSet_{\StrSet}(\QPAVSet{\qpElm})$ such that, for all $\xElm \in
			\QPAVSet{\qpElm}$ and $\trkElm' \in \TrkSet'(\root)$, it holds that
			$\asgFun'(\xElm)(\trkElm') = \asgFun(\xElm)(\Gamma(\trkElm'))$, where
			$\Gamma$ is the lifting over tracks of the mapping over states defined
			above, \ie, by $\Gamma(\trkElm')$ is the track in $\TName$ obtained from
			$\trkElm'$ by mapping each state $\sttElm'$ in $\trkElm'$ into $\sttElm =
			\Gamma(\sttElm')$.
			It holds that $\playElm = \playFun(\smFun(\asgFun) \cmp \bpElm, \root)$ is
			such that, for all $i \in \SetN$, $(\playElm)_{\leq i} =
			\Gamma((\playElm')_{\leq i})$.
			Indeed, by induction on $i$, as base case, we have that $(\playElm)_{\leq
			0} = \root = \Gamma(\root) = (\playElm')_{\leq 0}$.
			As inductive case, suppose that $(\playElm)_{\leq i} =
			\Gamma((\playElm')_{\leq i})$.
			Then, $(\playElm)_{i + 1} = \trnFun((\playElm)_{i},
			\adj{\smFun}((\playElm)_{\leq i}) \cmp \bpElm) =
			\trnFun(\Gamma((\playElm')_{i}), \adj{\smFun}(\Gamma((\playElm')_{\leq
			i})) \cmp \bpElm) = \Gamma(\trnFun'((\playElm')_{i},
			\adj{\smFun'}((\playElm')_{\leq i}) \cmp \bpElm)) =
			\Gamma((\playElm')_{i + 1})$.
			Thus, according to the definition of $\apFun'$, we have that
			$\apFun'((\playElm')_{i}) = \apFun((\playElm)_{i})$, for all $i \in
			\SetN$.
			Since $\playElm \models \psi$, we derive that $\playElm' \models \psi$.
			This holds for all possible universal assignments $\asgFun' \in
			\AsgSet_{\StrSet'}(\QPAVSet{\qpElm})$.
			Hence, it holds that $\TName' \models_{\smFun'} \qpElm \bpElm \psi$.
			The inductive case follows by considering the principal subsentences of
			$\varphi$ as fresh atomic propositions.
		\end{proof}

	\end{subsection}

	\begin{subsection}{Alternating tree automata}
		\label{subsec:altaut}

		\emph{Nondeterministic tree automata} are a generalization to infinite trees
		of the classical \emph{nondeterministic word automata} on infinite words
		(see~\cite{Tho90}, for an introduction).
		\emph{Alternating tree automata} are a further generalization of
		nondeterministic tree automata~\cite{MS87}.
		Intuitively, on visiting a node of the input tree, while the latter sends
		exactly one copy of itself to each of the successors of the node, the former
		can send several own copies to the same successor.
		Here we use, in particular, \emph{alternating parity tree automata}, which
		are alternating tree automata along with a \emph{parity acceptance
		condition} (see~\cite{GTW02}, for a survey).

		We now give the formal definition of alternating tree automata.
		\begin{defi}[Alternating Tree Automata]
			\label{def:ata}
			An \emph{alternating tree automaton} (\emph{\ATA}, for short) is a tuple
			$\AName \defeq \ATAStruct$, where $\LabSet$, $\DirSet$, and $\QSet$ are,
			respectively, non-empty finite sets of \emph{input symbols},
			\emph{directions}, and \emph{states}, $\qElm[0] \in \QSet$ is an
			\emph{initial state}, $\aleph$ is an \emph{acceptance condition} to be
			defined later, and $\delta : \QSet \times \LabSet \to \PBoolSet(\DirSet
			\times \QSet)$ is an \emph{alternating transition function} that maps each
			pair of states and input symbols to a positive Boolean combination on the
			set of propositions of the form $(\dElm, \qElm) \in \DirSet \times \QSet$,
			a.k.a. \emph{moves}.
		\end{defi}

		On one hand, a \emph{nondeterministic tree automaton} (\emph{\NTA}, for
		short) is a special case of \ATA\ in which each conjunction in the
		transition function $\delta$ has exactly one move $(\dElm, \qElm)$
		associated with each direction $\dElm$.
		This means that, for all states $\qElm \in \QSet$ and symbols $\sigma \in
		\LabSet$, we have that $\atFun(\qElm, \sigma)$ is equivalent to a Boolean
		formula of the form $\bigvee_{i} \bigwedge_{\dElm \in \DirSet} (\dElm,
		\qElm[i, \dElm])$.
		On the other hand, a \emph{universal tree automaton} (\emph{\UTA}, for
		short) is a special case of \ATA\ in which all the Boolean combinations that
		appear in $\delta$ are conjunctions of moves.
		Thus, we have that $\atFun(\qElm, \sigma) = \bigwedge_{i} (\dElm[i],
		\qElm[i])$, for all states $\qElm \in \QSet$ and symbols $\sigma \in
		\LabSet$.

		The semantics of the \ATA s is given through the following concept of run.
		\begin{defi}[\ATA\ Run]
			\label{def:ata(run)}
			A \emph{run} of an \ATA\ $\AName = \ATAStruct$ on a $\LabSet$-labeled
			$\DirSet$-tree $\TName = \LTStruct$ is a $(\DirSet \times \QSet)$-tree
			$\RSet$ such that, for all nodes $\xElm \in \RSet$, where $\xElm =
			\prod_{i = 1}^{n} (\dElm[i], \qElm[i])$ and $\yElm \defeq \prod_{i =
			1}^{n} \dElm[i]$ with $n \in \numco{0}{\omega}$, it holds that \emph{(i)}
			$\yElm \in \TSet$ and \emph{(ii)}, there is a set of moves $\SSet
			\subseteq \DirSet \times \QSet$ with $\SSet \models \delta(\qElm[n],
			\vFun(\yElm))$ such that $\xElm \cdot (\dElm, \qElm) \in \RSet$, for all
			$(\dElm, \qElm) \in \SSet$.
		\end{defi}

		In the following, we consider \ATA s along with the \emph{parity acceptance
		condition} (\emph{\APT}, for short) $\aleph \defeq (\FSet_{1}, \ldots,
		\FSet_{k}) \in (\pow{\QSet})^{+}$ with $\FSet_{1} \subseteq \ldots \subseteq
		\FSet_{k} = \QSet$ (see~\cite{KVW00}, for more).
		The number $k$ of sets in the tuple $\aleph$ is called the \emph{index} of
		the automaton.
		We also consider \ATA s with the \emph{co-B\"uchi acceptance condition}
		(\emph{\ACT}, for short) that is the special parity condition with index
		$2$.

		Let $\RSet$ be a run of an \ATA\ $\AName$ on a tree $\TName$ and $\wElm$ one
		of its branches.
		Then, by $\infFun(\wElm) \defeq \set{ \qElm \in \QSet }{ \card{\set{ i \in
		\SetN }{ \exists \dElm \in \DirSet . (w)_{i} = (\dElm, \qElm) }} = \omega }$
		we denote the set of states that occur infinitely often as the second
		component of the letters along the branch $w$.
		Moreover, we say that $w$ satisfies the parity acceptance condition $\aleph
		\!=\! (\FSet_{1}, \ldots, \FSet_{k})$ if the least index $i \!\in\!
		\numcc{1}{k}$ for which $\infFun(w) \cap \FSet_{i} \neq \emptyset$ is even.

		At this point, we can define the concept of language accepted by an \ATA.
		\begin{defi}[\ATA\ Acceptance]
			\label{def:ata(acp)}
			An \ATA\ $\AName = \ATAStruct$ \emph{accepts} a $\LabSet$-labeled
			$\DirSet$-tree $\TName$ iff is there exists a run $\RSet$ of $\AName$ on
			$\TName$ such that all its infinite branches satisfy the acceptance
			condition $\aleph$.
		\end{defi}
		By $\LangSet(\AName)$ we denote the language accepted by the \ATA\ $\AName$,
		i.e., the set of trees $\TName$ accepted by $\AName$.
		Moreover, $\AName$ is said to be \emph{empty} if $\LangSet(\AName) =
		\emptyset$.
		The \emph{emptiness problem} for $\AName$ is to decide whether
		$\LangSet(\AName) = \emptyset$.

	\end{subsection}

	\begin{subsection}{Satisfiability procedure}
		\label{sec:dec;sub:sat}

		We finally solve the satisfiability problem for \OGSL\ and show that it is
		2\ExpTimeC, as for \ATLS.
		The algorithmic procedures is based on an automata-theoretic approach, which
		reduces the decision problem for the logic to the emptiness problem of a
		suitable universal Co-B\"uchi tree automaton (\UCT, for short)~\cite{GTW02}.
		From an high-level point of view, the automaton construction seems similar
		to what was proposed the in literature for \CTLS~\cite{KVW00} and
		\ATLS~\cite{Sch08}.
		However, our technique is completely new, since it is based on the novel
		notions of behavioral semantics and disjoint satisfiability.

		\noindent \textbf{Principal sentences.}
			To proceed, we first have to introduce the concept of encoding for an
			assignment and the labeling of a \DT.
			\begin{defi}[Assignment-Labeling Encoding]
				\label{def:asglabenc}
				Let $\TName$ be a \DT, $\tElm \in \SttSet[\TName]$ one of its states,
				and $\asgFun \in \AsgSet[\TName](\VSet, \tElm)$ an assignment defined on the set $\VSet \subseteq \VarSet$.
				A $(\ValSet_{\ActSet[\TName]}(\VSet) \times \pow{\APSet})$-labeled
				$\DecSet[\TName]$-tree $\TName' \defeq
				\LTTuple{}{}{\SttSet[\TName]}{\uFun}$ is an \emph{assignment-labeling
				encoding} for $\asgFun$ on $\TName$ if $\uFun(\lst{(\trkElm)_{\geq 1}})
				\!=\! (\flip{\asgFun}(\trkElm), \apFun[\TName](\lst{\trkElm}))$, for
				all $\trkElm \in \TrkSet[\TName](\tElm)$~\footnote{Note that
				$\lst{\root} = \root$.}.
			\end{defi}
			\noindent
			Observe that there is a unique assignment-labeling encoding for each
			assignment over a given \DT.

			Now, we prove the existence of a \UCT\ $\UName[\bpElm \psi][\ActSet]$
			for each \OGSL\ goal $\bpElm \psi$ having no principal subsentences.
			The $\UName[\bpElm \psi][\ActSet]$ recognizes all the assignment-labeling
			encodings $\TName'$ of an a priori given assignment $\asgFun$ over a
			generic \DT\ $\TName$, whenever the goal is satisfied on $\TName$ under
			$\asgFun$.
			Intuitively, we start with a \UCW, recognizing all infinite words on the
			alphabet $\pow{\APSet}$ that satisfy the \LTL\ formula $\psi$, obtained by
			a simple variation of the Vardi-Wolper construction~\cite{VW86b}.
			Then, we run it on the encoding tree $\TName'$ by following the directions
			identified by the assignment in its labeling.
			\begin{lem}[\OGSL\ Goal Automaton]
				\label{lmm:ogsl:golaut}
				Let $\bpElm \psi$ an \OGSL\ goal without principal subsentences and
				$\ActSet$ a finite set of actions.
				Then, there exists an \UCT\ $\UName[\bpElm \psi][\ActSet] \defeq
				\TATuple {\ValSet_{\ActSet}(\free{\bpElm \psi}) \times \pow{\APSet}}
				{\DecSet} {\QSet_{\bpElm \psi}} {\atFun_{\bpElm \psi}}
				{\qElm_{0\bpElm\psi}} {\aleph_{\bpElm \psi}}$ such that, for all \DT s
				$\TName$ with $\ActSet_{\TName} = \ActSet$, states $\tElm \in
				\SttSet_{\TName}$, and $\tElm$-total assignments $\asgFun \in
				\AsgSet[\TName](\free{\bpElm \psi}, \tElm)$, it holds that $\TName,
				\asgFun, \tElm \models \bpElm \psi$ iff $\TName' \in
				\LangSet(\UName[\bpElm \psi][\ActSet])$, where $\TName'$ is the
				assignment-labeling encoding for $\asgFun$ on $\TName$.
			\end{lem}

			\begin{proof}
				A first step in the construction of the \UCT\ $\UName[\bpElm
				\psi][\ActSet]$, is to consider the \UCW\ $\UName[\psi] \defeq \WATuple
				{\pow{\APSet}} {\QSet[\psi]} {\atFun[][\psi]} {\QSet[0\psi]}
				{\aleph_{\psi}}$ obtained by dualizing the \NBW\ resulting from the
				application of the classic Vardi-Wolper construction to the \LTL\
				formula $\neg \psi$~\cite{VW86b}.
				Observe that $\LangSet(\UName[\psi]) = \LangSet(\psi)$, \ie, this
				automaton recognizes all infinite words on the alphabet $\pow{\APSet}$
				that satisfy the \LTL\ formula $\psi$.
				Then, define the components of $\UName[\bpElm \psi][\ActSet] \defeq
				\TATuple {\ValSet_{\ActSet}(\free{\bpElm \psi}) \times \pow{\APSet}}
				{\DecSet} {\QSet_{\bpElm \psi}} {\atFun_{\bpElm \psi}}
				{\qElm[0\bpElm\psi]} {\aleph_{\bpElm \psi}}$, as follows:
				\begin{itemize}
					\item
						$\QSet[\bpElm \psi] \defeq \{ \qElm[0\bpElm\psi] \} \cup
						\QSet[\psi]$, with $\qElm[0\bpElm\psi] \not\in \QSet[\psi]$;
					\item
						$\atFun_{\bpElm \psi}(\qElm[0\bpElm\psi], (\valFun, \sigma)) \defeq
						\bigwedge_{\qElm \in \QSet[0\psi]} \atFun_{\bpElm \psi}(\qElm,
						(\valFun, \sigma))$, for all $(\valFun, \sigma) \in
						\ValSet_{\ActSet}(\free{\bpElm \psi}) \times \pow{\APSet}$;
					\item
						$\atFun_{\bpElm \psi}(\qElm, (\valFun, \sigma)) \!\defeq\!
						\bigwedge_{\qElm' \!\in\! \atFun[][\psi](\qElm, \sigma)}
						(\valFun \cmp \bpElm, \qElm')$, for all $\qElm \!\in\!
						\QSet[\psi]$ and $(\valFun, \sigma) \in
						\ValSet_{\ActSet}(\free{\bpElm \psi}) \times \pow{\APSet}$;
					\item
						$\aleph_{\bpElm \psi} \defeq \aleph_{\psi}$.
				\end{itemize}
				Intuitively, the \UCT\ $\UName[\bpElm \psi][\ActSet]$ simply runs
				the \UCW\ $\UName[\psi]$ on the branch of the encoding individuated by
				the assignment in input.
				Thus, it is easy to see that, for all states $\tElm \in \SttSet[\TName]$
				and $\tElm$-total assignments $\asgFun \in \AsgSet[\TName](\free{\bpElm
				\psi}, \tElm)$, it holds that $\TName, \asgFun, \tElm \models \bpElm
				\psi$ iff $\TName' \in \LangSet(\UName[\bpElm \psi][\ActSet])$, where
				$\TName'$ is the assignment-labeling encoding for $\asgFun$ on $\TName$.
			\end{proof}

			We now introduce a new concept of encoding regarding the behavioral
			dependence maps over strategies.
			\begin{defi}[Behavioral Dependence-Labeling Encoding]
				\label{def:elmspclabenc}
				Let $\TName$ be a \DT, $\tElm \in \SttSet[\TName]$ one of its states,
				and $\smFun \in \BSMSet_{\StrSet_{\TName}}(\qpElm)$ a
				behavioral dependence map over strategies for a quantification prefix
				$\qpElm \in \QPSet(\VSet)$ over the set $\VSet \subseteq \VarSet$.
				A $(\SMSet_{\ActSet[\TName]}(\qpElm) \times \pow{\APSet})$-labeled
				$\DirSet$-tree $\TName' \defeq \LTTuple{}{}{\SttSet[\TName]}{\uFun}$ is
				a \emph{behavioral dependence-labeling encoding} for $\smFun$ on
				$\TName$ if $\uFun(\lst{(\trkElm)_{\geq 1}}) \!=\!
				(\adj{\smFun}(\trkElm), \apFun[\TName](\lst{\trkElm}))$, for all
				$\trkElm \!\in\! \TrkSet[\TName](\tElm)$.
			\end{defi}
			\noindent
			Observe that also in this case there exists a unique behavioral
			dependence-labeling encoding for each behavioral dependence map over
			strategies.

			Finally, in the next lemma, we show how to locally handle the strategy
			quantifications on each state of the model, by simply using a
			quantification over actions modeled by the choice of an action dependence
			map.
			Intuitively, we guess in the labeling what is the right part of the
			dependence map over strategies for each node of the tree and then verify
			that, for all assignments of universal variables, the corresponding
			complete assignment satisfies the inner formula.
			\begin{lem}[\OGSL\ Sentence Automaton]
				\label{lmm:ogsl(sntaut)}
				Let $\qpElm \bpElm \psi$ be an \OGSL\ principal sentence without
				principal
				subsentences and $\ActSet$ a finite set of actions.
				Then, there exists an \UCT\ $\UName[\qpElm \bpElm \psi][{\ActSet}]
				\defeq \TATuple {\SMSet_{\ActSet}(\qpElm) \times \pow{\APSet}}
				{\DecSet} {\QSet[\qpElm \bpElm \psi]} {\atFun_{\qpElm \bpElm \psi}}
				{\qElm[0\qpElm\bpElm\psi]} {\aleph_{\qpElm \bpElm \psi}}$ such that, for
				all \DT s $\TName$ with $\ActSet[\TName] = \ActSet$, states $\tElm \in
				\SttSet[\TName]$, and behavioral dependence maps over strategies
				$\smFun \in \BSMSet_{\StrSet[\TName]}(\qpElm)$, it holds that
				$\TName, \smFun(\asgFun), \tElm \bmodels \bpElm \psi$, for all $\asgFun
				\in \AsgSet[\TName](\QPAVSet{\qpElm}, \tElm)$, iff $\TName' \in
				\LangSet(\UName[\bpElm \psi][\ActSet])$, where $\TName'$ is
				the behavioral dependence-labeling encoding for $\smFun$ on $\TName$.
			\end{lem}

			\begin{proof}
				By Lemma~\ref{lmm:ogsl:golaut} of \OGSL\ goal automaton, there is an
				\UCT\  $\UName[\bpElm \psi][\ActSet] \defeq
				\TATuple {\ValSet_{\ActSet}(\free{\bpElm \psi}) \times \pow{\APSet}}
				{\DecSet} {\QSet[\bpElm \psi]} {\atFun_{\bpElm \psi}}
				{\qElm_{0\bpElm\psi}} {\aleph_{\bpElm \psi}}$ such that, for all \DT{s}
				$\TName$ with $\ActSet[\TName] = \ActSet$, states $\tElm \in
				\SttSet[\TName]$, and
				assignments $\asgFun \in \AsgSet[\TName](\free{\bpElm \psi}, \tElm)$, it
				holds that $\TName, \asgFun, \tElm \models \bpElm \psi$ iff $\TName' \in
				\LangSet(\UName[\bpElm \psi][\ActSet])$, where $\TName'$ is the
				assignment-labeling encoding for $\asgFun$ on $\TName$.

				Now, transform $\UName[\bpElm \psi][\ActSet]$ into the new \UCT\
				$\UName[\bpElm \psi][\ActSet] \defeq \TATuple
				{\SMSet_{\ActSet}(\qpElm) \times \pow{\APSet}} {\DecSet} {\QSet_{\qpElm
				\bpElm \psi}} {\atFun_{\qpElm \bpElm \psi}} {\qElm_{0\qpElm\bpElm\psi}}
				{\aleph_{\qpElm \bpElm \psi}}$, with $\QSet_{\qpElm \bpElm \psi} \defeq
				\QSet[\bpElm \psi]$, $\qElm[0\qpElm\bpElm\psi] \defeq
				\qElm[0\bpElm\psi]$, and $\aleph_{\qpElm \bpElm \psi} \defeq
				\aleph_{\bpElm \psi}$, which is used to handle the quantification prefix
				$\qpElm$ atomically, where the transition function is defined as
				follows: $\atFun_{\qpElm \bpElm \psi}(\qElm, (\smFun, \sigma)) \defeq
				\bigwedge_{\valFun \in \ValSet_{\ActSet}(\QPAVSet{\qpElm})}
				\atFun_{\bpElm \psi}(\qElm, (\smFun(\valFun), \sigma))$, for all
				$\qElm \in \QSet[\qpElm \bpElm \psi]$ and $(\smFun, \sigma) \in
				\SMSet_{\ActSet}(\qpElm) \times \pow{\APSet}$.
				Intuitively, $\UName[\qpElm \bpElm \psi][\ActSet]$ reads an action
				dependence map $\smFun$ on each node of the input tree $\TName'$
				labeled with a set of atomic propositions $\sigma$ and simulates the
				execution of the transition function $\atFun_{\bpElm \psi}(\qElm,
				(\valFun, \sigma))$ of $\UName[\bpElm \psi][\ActSet]$, for each
				possible valuation $\valFun = \smFun(\valFun')$ on $\free{\bpElm \psi}$
				obtained from $\smFun$ via a universal valuation $\valFun' \in
				\ValSet_{\ActSet}(\QPAVSet{\qpElm})$.
				It is worth observing that we cannot move the component set
				$\SMSet_{\ActSet}(\qpElm)$ from the input alphabet to the states of
				$\UName[\qpElm \bpElm \psi][\ActSet]$ by making a related guessing
				of the dependence map $\smFun$ in the transition function, since the
				automaton is universal and we have to ensure that all states in a given
				node of the tree $\TName'$, \ie, in each track of the original model
				$\TName$, \mbox{make the same choice for $\smFun$.}

				Finally, it remains to prove that, for all states $\tElm \in
				\SttSet[\TName]$ and behavioral dependence maps over strategies $\smFun
				\in \BSMSet_{\StrSet[\TName]t}(\qpElm)$, it holds that
				$\TName, \smFun(\asgFun), \tElm \bmodels \bpElm \psi$, for all $\asgFun
				\in \AsgSet[\TName](\QPAVSet{\qpElm}, \tElm)$, iff $\TName' \in
				\LangSet(\UName[\qpElm \bpElm \psi][\ActSet])$, where $\TName'$ is
				the behavioral dependence-labeling encoding for $\smFun$ on $\TName$.

				\emph{[Only if].}
				Suppose that $\TName, \smFun(\asgFun), \tElm \bmodels \bpElm \psi$, for
				all $\asgFun \in \AsgSet[\TName](\QPAVSet{\qpElm}, \tElm)$.
				Since $\psi$ does not contain principal subsentences, we have that
				$\TName, \smFun(\asgFun), \tElm \models \bpElm \psi$.
				So, due to the property of $\UName[\bpElm \psi][\ActSet]$,
				%Nello: dire quale proprietà
				it follows that there exists an assignment-labeling encoding
				$\TName[\asgFun]['] \in \LangSet(\UName[\bpElm \psi][\ActSet])$,
				which implies the existence of a $(\DecSet \times \QSet[\bpElm
				\psi])$-tree $\RSet[\asgFun]$ that is an accepting run for
				$\UName[\bpElm \psi][\ActSet]$ on $\TName[\asgFun][']$.
				At this point, let $\RSet \defeq \bigcup_{\asgFun \in
				\AsgSet[\TName](\QPAVSet{\qpElm}, \tElm)} \RSet[\asgFun]$ be the union
				of all runs.
				Then, due to the particular definition of the transition function of
				$\UName[\qpElm \bpElm \psi][\ActSet]$, it is not hard to see that
				$\RSet$ is an accepting run for $\UName[\qpElm \bpElm \psi][\ActSet]$
				on $\TName'$ defined as above.
				Hence, $\TName' \in \LangSet(\UName[\qpElm \bpElm \psi][\ActSet])$.

				\emph{[If].}
				Suppose that $\TName' \in \LangSet(\UName[\qpElm \bpElm
				\psi][\ActSet])$.
				Then, there exists a $(\DecSet \times \QSet[\qpElm \bpElm \psi])$-tree
				$\RSet$ that is an accepting run for $\UName[\qpElm \bpElm
				\psi][\ActSet]$ on $\TName'$.
				Now, for each $\asgFun \in \AsgSet[\TName](\QPAVSet{\qpElm}, \tElm)$,
				let $\RSet[\asgFun]$ be the run for $\UName[\bpElm \psi][\ActSet]$
				on the assignment-state encoding $\TName[\asgFun][']$ for
				$\smFun(\asgFun)$ on $\TName$.
				Due to the particular definition of the transition function of
				$\UName[\qpElm \bpElm \psi][\ActSet]$, it is not hard to see that
				$\RSet[\asgFun] \subseteq \RSet$.
				Thus, since $\RSet$ is accepting, we have that $\RSet[\asgFun]$ is
				accepting as well.
				So, $\TName[\asgFun]['] \in \LangSet(\UName[\bpElm \psi][\ActSet])$.
				At this point, due to the property of $\UName[\bpElm \psi][\ActSet]$,
				it follows that $\TName, \smFun(\asgFun), \tElm \models \bpElm \psi$.
				Since $\psi$ does not contain principal subsentences, we have that
				$\TName, \smFun(\asgFun), \tElm \!\bmodels\! \bpElm \psi$, for all
				$\asgFun \!\in\! \AsgSet[\TName](\QPAVSet{\qpElm}, \tElm)$.
			\end{proof}

			\noindent \textbf{Full sentences.}
			By summing up all previous results, we are now able to solve the
			satisfiability problem for the full \OGSL\ fragment.

			To construct the automaton for a given \OGSL\ sentence $\varphi$, we first
			consider all \UCT\ $\UName[\phi][\ActSet]$, for an assigned bounded
			set $\ActSet$, previously described for the principal sentences $\phi \in
			\psnt{\varphi}$, in which the inner subsentences are considered as atomic
			propositions.
			Then, thanks to the disjoint satisfiability property of
			Definition~\ref{def:dsjsat}, we can merge them into a unique \UCT\
			$\UName[\varphi]$ that supplies the dependence map labeling of internal
			components $\UName[\phi][\ActSet]$, by using the two functions
			$\headFun$ and $\bodyFun$ contained into its labeling.
			Moreover, observe that the final automaton runs on a $b$-bounded
			decision tree, where $b$ is obtained from
			Theorem~\ref{thm:ogsl:bndmdprp} on the bounded-tree model property.
			\begin{thm}[\OGSL\ Automaton]
				\label{thm:ogsl(aut)}
				Let $\varphi$ be an \OGSL\ sentence.
				Then, there exists an \UCT\ $\UName[\varphi]$ such that $\varphi$ is
				satisfiable iff $\LangSet(\UName[\varphi]) \neq \emptyset$.
			\end{thm}

			Finally, by a simple calculation of the size of $\UName[\varphi]$ and the
			complexity of the related emptiness problem, we state in the next theorem
			the precise computational complexity of the satisfiability problem for
			\OGSL.
			\begin{thm}[\OGSL\ Satisfiability]
				\label{thm:ogsl(sat)}
				The satisfiability problem for \OGSL\ is 2\ExpTimeC.
			\end{thm}

			\begin{proof}
				By Theorem~\ref{thm:ogsl(aut)} of \OGSL\ automaton, to verify whether an
				\OGSL\ sentence $\varphi$ is satisfiable we can calculate the emptiness
				of the \UPT\ $\UName[\varphi]$.
				This automaton is obtained by merging all \UCT{s}
				$\UName[\phi][\ActSet]$, with $\phi = \qpElm \bpElm \psi \in
				\psnt{\varphi}$, which in turn are based on the \UCT{s} $\UName[\bpElm
				\psi][\ActSet]$ that embed the \UCW{s} $\UName[\psi]$.
				By a simple calculation, it is easy to see that $\UName[\varphi]$ has
				$2^{\AOmicron{\card{\varphi}}}$ states.
				Indeed, by the Vardi-Wolper construction, all the \UCW{s} $\UName[\psi]$ are of size bouded by $2^{\card{\psi}}$.
				Consequently, due to Lemma~\ref{lmm:ogsl:golaut}, also the \UCW{s} $\UName[\bndElm \psi]$ have the same bound on the state space.
				Therefore, due to the construction of Lemma~\ref{lmm:ogsl(sntaut)}, the cardinality of the state space of the \UCW{s} $\UName[\qntElm \bndElm \psi]$ is $\AOmicron{2^{\card{\qntElm \bndElm \psi}}}$.
				Finally, since the all the $\qntElm \bndElm \psi$ occur into $\varphi$, we obtain that the size of the \UCW\ $\UName[\varphi]$ is bounded by $2^{\card{\varphi}}$.
				
				Now, by using a well-known nondeterminization procedure for \APT
				s~\cite{MS95}, we obtain an equivalent \NPT\ $\NName[\varphi]$ with
				$2^{2^{\AOmicron{\card{\varphi}}}}$ states and index
				$2^{\AOmicron{\card{\varphi}}}$.

				The emptiness problem for such a kind of automaton with $n$ states and
				index $h$ is solvable in time $\AOmicron{n^{h}}$.%~\cite{KV98}.
				Thus, we get that the time complexity of checking whether $\varphi$ is
				satisfiable is $2^{2^{\AOmicron{\card{\varphi}}}}$.
				Hence, the membership of the satisfiability problem for \OGSL in
				2\ExpTime\ directly follows.
				Finally the thesis is proved, by getting the relative lower bound from
				the same problem for \CTLS~\cite{VS85}.
			\end{proof}

	\end{subsection}

\end{section}

% End of file SectionIV.tex

% 
% 	\input{SectionVI}

% 	\input{Conclusion}

%	 \input{Acknowledgments}

% 	\appendix
% 
% 	\input{AppendixA}

	% \input{AppendixB}

	% \input{AppendixC}

	% \input{AppendixD}

	\footnotesize
	\bibliographystyle{alpha}
	\bibliography{References}

\end{document}